\documentclass[12pt,a4paper,twoside]{report}
\usepackage{mathpazo}
\usepackage{amsmath}
\usepackage{amssymb}
\usepackage{mathrsfs}
\usepackage{latexsym}
\usepackage{graphics}
\usepackage{epsfig}
\usepackage{lscape}
\usepackage{fancyhdr}
\usepackage[usenames]{color}
\usepackage{multirow}
\usepackage{textcomp}
\usepackage{subfigure}
\usepackage[small,bf]{caption2}

\renewcommand{\chaptermark}[1]{ \markboth{\chaptername\ \thechapter.\ #1}{}}
\renewcommand{\sectionmark}[1]{ \markright{\thesection.\ #1}}

\newcommand{\degrees}{^{\circ}}
\newcommand{\bfr}{\mathbf{r}}
\newcommand{\bfrp}{\mathbf{r'}}
\newcommand{\geant}{G\textsc{eant}4 }
\newcommand{\ROOT}{\textsf{ROOT}}
\newcommand{\matrixR}{\overrightarrow{\mathbf{R}}}
\newcommand{\matrixT}{\overrightarrow{\mathbf{T}}}
\newcommand{\matrixP}{\overrightarrow{\mathbf{P}}}
\newcommand{\matrixQ}{\overrightarrow{\mathbf{Q}}}
\newcommand{\red}  {\textcolor{red}}
\newcommand{\green}{\textcolor{PineGreen}}
\newcommand{\blue} {\textcolor{blue}}

\begin{document}
  
  \pagestyle{empty}

\newpage
\thispagestyle{empty}


\vspace*{1cm}
\begin{center}
  {\Large\bf The Acoustic Detection of Ultra High Energy Neutrinos\\}
    \vspace{2cm}
	 {\large 
	    Jonathan David Perkin\\
	      \vspace{0.5cm}
	      Department of Physics and Astronomy\\
	      The University of Sheffield}
\end{center}

\vspace{3cm}
\begin{figure}[h]
  \begin{center}
    \includegraphics[width=0.2\textwidth]{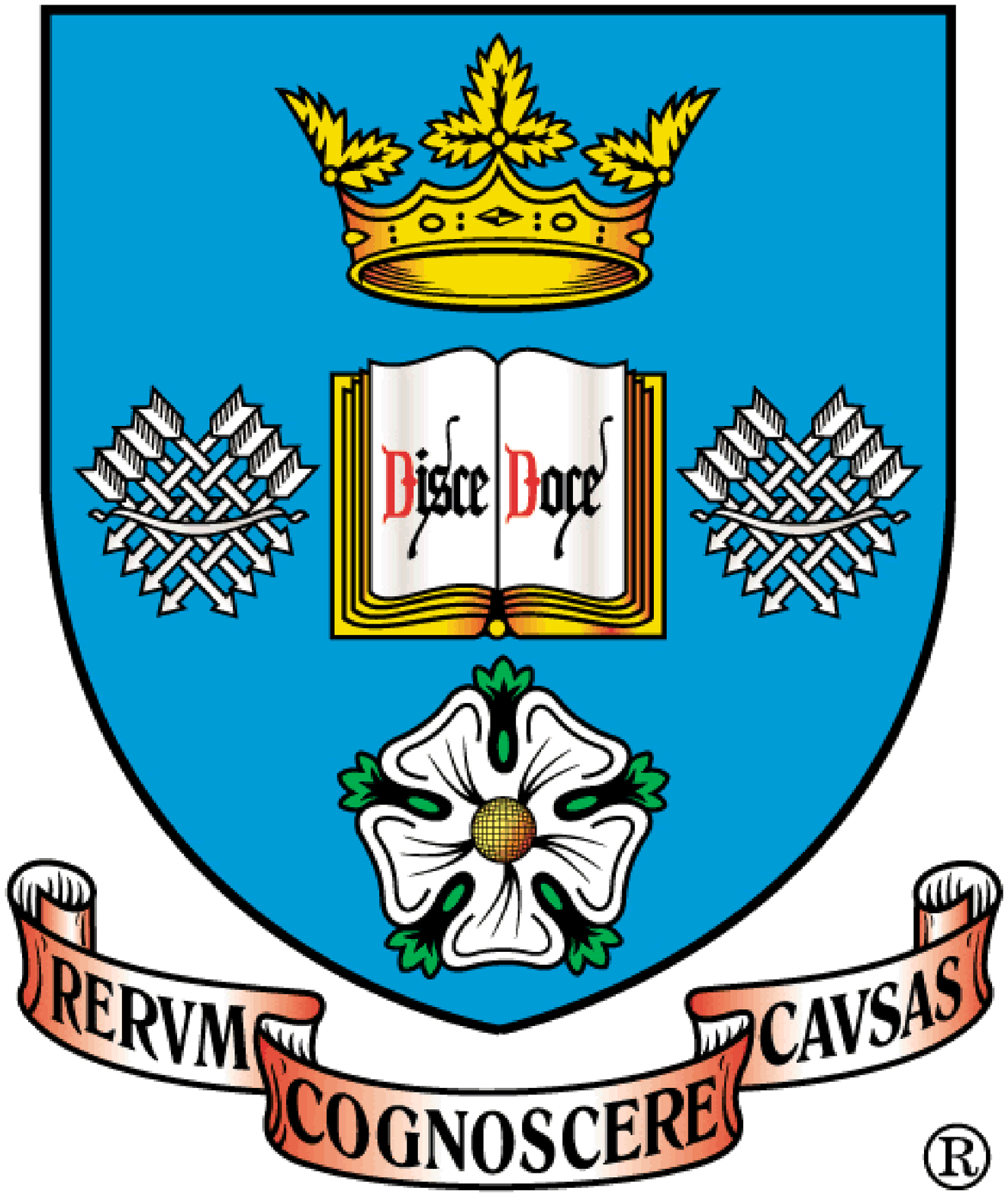}
   \end{center}
\end{figure}

\vspace{3cm}
\begin{center}
  {\large  
    Thesis submitted for the Degree of Doctor of Philosophy
    in the\\
    University of Sheffield\\
    \vspace{0.3cm}
    $\cdot$ April 2007 $\cdot$}
\end{center}

  \cleardoublepage
  
  \pagestyle{fancy}
  \pagenumbering{roman}
  
  \chapter*{Abstract}
Attempts have been made to parameterise the thermoacoustic emission of
particle cascades induced by EeV neutrinos interacting in the
sea. Understanding the characteristic radiation from such an event
allows us to predict the pressure pulse observed by underwater
acoustic sensors distributed in kilometre scale arrays. We find that
detectors encompassing thousands of cubic kilometres are required,
with a minimum of 100 hydrophones per kilometre cubed, in order to
observe the flux of neutrinos predicted by the attenuation of ultra
high energy cosmic rays on cosmic microwave background photons. The
pressure threshold of such an array must be in the range $5 \to
10$\,mPa and the said detector will have to operate for five years or
more. Additionally a qualitative analysis of the first acoustic data
recorded by the Rona hydrophone array off the north-west coast of
Scotland is reported.

  \include{acknowledgements}
  \include{author}
  
  \cleardoublepage
  
  \chapter*{}
  \begin{center}
    \it{For}\\ \it{Mum and Dad}
  \end{center}
  
  \cleardoublepage
  
  \tableofcontents
  \listoffigures
  \listoftables
  
  \cleardoublepage
  
  \pagenumbering{arabic}

  \chapter{Introduction}
\label{chp:introduction}
Chapters \ref{chp:introduction} and \ref{chp:detection} of this thesis
summarise the history of our understanding of neutrinos and the
methods by which they can be detected. Chapters \ref{chp:neutrino}, 
\ref{chp:propagation}  and \ref{chp:largescale} discuss the potential
performance of hypothetical large scale neutrino detectors. Chapter
\ref{chp:experiment} reports  an analysis of data
from an underwater acoustic sensor array situated in the north west of
Scotland in the United Kingdom. Finally, in Chapter \ref{chp:discussion}
a discussion summarising this work and making predictions for
future work is given.
\section{A Brief History of the Neutrino} 
\label{sec:briefhistory}
Neutrinos are the second most abundant known particles in the Universe
after cosmological photons. They are produced copiously, for example, by
stellar bodies, like the Sun, throughout their considerable lifetimes
and during their spectacular deaths. There are, on average, three
hundred million neutrinos per cubic metre of space, billions penetrate
our bodies unnoticed every second. However, neutrinos remain poorly
understood.

It is because neutrino interactions are mediated only by the weak
nuclear and gravitational forces that they acquire their
mystery. However, it is this very same reason that makes them
attractive to the astronomer. Whereas charged particles are deflected
by magnetic fields, photons are absorbed
by inter-stellar matter and softened by radiation fields, 
and more massive particles fall deeper into gravitational wells; the
neutrino traverses the Cosmos retaining its
energy and directionality until it eventually undergoes a
collision far away from its place of origin.
Should such a collision occur on Earth it could yield
information from parts of the Universe no other form of light or
matter can reach.

Whilst detection of neutrinos has been achieved, there has been no direct
measurement of their rest mass. In fact, evidence of neutrino mass has
only recently come to light. The development of our understanding of
neutrinos, pertaining to neutrino astronomy, is illustrated in the
following time-line (adapted from \cite{timeline}):
\begin{itemize}
\item[{\bf 1920-1927}] Charles Drummond Ellis and James Chadwick {\it et al}
  firmly establish that the beta decay spectrum is continuous
  \cite{beta:chadwick},\cite{beta:ellis},
  contrary to the predictions of a two body decay: $n \to p + e^-$.
\item[{\bf 1930}] Wolfgang Pauli postulates the existence of a third, neutral
  particle present in beta decay to explain the observed  $\beta$ energy
  spectrum \cite{postulation:pauli}: $n \to p + e^- + \nu_{neutral}$.
\item[{\bf 1933}] Enrico Fermi incorporates Pauli's neutral particle into
  beta decay theory and bestows it the name ``neutrino'' ({\it little
  neutral one}) \cite{beta:fermi}. The continuous beta decay energy
  spectrum is explained.
\item[{\bf 1953}] Fred Reines and Clyde Cowan detect neutrinos from the
  Hanford Nuclear Reactor using a delayed-coincidence technique on the
  reaction: $\overline{\nu} + p \to n + \beta^+$ \cite{nu:cowan}.
  Neutrinos were registered through observation of the photons emitted
  simultaneously by capture of the neutron $n$ and annihilation of the
  positron $\beta^+$. 
\item[{\bf 1957}] Bruno Pontecorvo makes the first hypothesis of neutrino
  oscillation, in this instance between neutrino and anti-neutrino
  states \cite{oscillation:pontecorvo}.
\item[{\bf 1968}] Ray Davis Jr measures the Solar Neutrino Flux in the
  Homestake Mine. He observes a deficit in the number of interactions
  compared to predictions by John Bahcall's Standard Solar 
  Model. This deficit becomes known as 
  the ``Solar Neutrino Problem'' \cite{homestake:davis}.
\item[{\bf 1976}] The tau lepton is discovered at SLAC and through
  analysis of its decay it is concluded that the tau is accompanied by
  its own unique flavour of neutrino, the tau neutrino \cite{tau:slac}.
\item[{\bf 1986-1987}] The Kamiokande experiment confirms the solar neutrino
  deficit. Then, along with IMB, the first measurements of neutrinos
  from a supernova (SN) are made \cite{nu:kamiokande}, \cite{nu:imb}.
  Neutrino Astronomy is born.
\item[{\bf 1989}] LEP constrains the number of light neutrino species to
  three: electron, muon and tau \cite{nuflavours:lep}.
\item[{\bf 1998}] Super-Kamiokande reports on the flavour oscillation of
  atmospheric neutrinos, direct evidence for three different neutrino
  mass states \cite{mass:superk}.
\item[{\bf 2000}] Twenty four years after the tau lepton is discovered,
  DONUT reports on the first direct observation of the tau neutrino
  \cite{nutau:donut}.
\item[{\bf 2002}] The Sudbury Neutrino Observatory detects both neutral and
  charged current interactions from solar neutrinos, evidence is
  mounting in support of oscillations as a solution to the Solar
  Neutrino Problem \cite{nu:sno}.
\item[{\bf 2002}] KAMLAND confirms neutrino oscillations consistent with the
  observed solar neutrino deficit, this time with reactor neutrinos
  \cite{nu:kamland}.
\item[{\bf 2002}] AMANDA produces the most detailed neutrino sky-map
  ever, with no evidence (yet) of point sources \cite{skymap:amanda}.
\item[{\bf 2006-7}] The cubic kilometre extension of AMANDA, codenamed
  IceCube is under construction at the South pole
  \cite{amandatoicecube:halzen}. ANTARES nears completion on the
  Mediterranean sea bed \cite{antares:becherini}. Neutrino Astronomy
  is in full flight.
\end{itemize}
\section{Neutrinos in the Standard Model and \mbox{Beyond}}
\label{sec:neutrinosin}
Neutrinos belong to a family of particles called ``Leptons'' (from
the Greek word {\it leptos} meaning small or fine). Leptons exist in three
flavours and each forms a couplet of one charged and one neutral particle as
illustrated in Table \ref{tab:leptons}:
\begin{table}[h]
  \begin{center}
    \begin{tabular}{| c | c | c | c }
      $e^-$   &  $\mu^-$  &  $\tau^-$  & $q=e$  \\
      $\nu_e$ & $\nu_\mu$ & $\nu_\tau$ & $q=0$  \\
    \end{tabular}
    \caption[The leptons]{ \small The leptons. \label{tab:leptons}}
  \end{center}
\end{table}

The Standard Model (SM) of particle physics predicts that neutrinos
are massless, spin half, fermions interacting only via the weak
nuclear force. This imposes the condition that all neutrinos are
left-handed (and all anti-neutrinos are right-handed), since, in the
massless limit it is impossible to boost to a frame of reference where
a neutrino with left handed helicity (projection of spin in the
direction of momentum) has a right handed chirality (the sign of the
helicity).

The SM has been constructed from Quantum Field Theories (QFTs) that
explain the unification of the electromagnetic and weak nuclear forces into
the electro-weak force via Quantum Electro Dynamics (QED) and the
interpretation of
strong nuclear interactions via Quantum Chromo Dynamics (QCD). Within
these QFTs there is no gravitational component, so, although the SM
describes to great precision a wealth of experimental data, it remains
for this reason, amongst many others, an incomplete
theory. Furthermore, as has already been intimated, it does not
contain a neutrino mass term.

Although it was not realised at the time, the Solar Neutrino Problem
was evidence for neutrino flavour oscillations. A free neutrino
propagating through space exists in a superposition of fundamental mass
eigenstates $\nu_1$, $\nu_2$ and $\nu_3$ the mixing of which gives
rise to the three neutrino flavours $\nu_e$, $\nu_{\mu}$ and
$\nu_{\tau}$. The neutrino flavours are in fact manifestations of weak
nuclear eigenstates. This can be illustrated in the form of a matrix
equation:
\begin{equation}
  \left( \begin{array}{l}
    \nu_{e}    \\
    \nu_{\mu}  \\
    \nu_{\tau} \\
  \end{array} \right)
  =
  \left( \begin{array}{lll}
    U_{e1}    & U_{e2}    & U_{e3}    \\
    U_{\mu1}  & U_{\mu2}  & U_{\mu3}  \\
    U_{\tau1} & U_{\tau2} & U_{\tau3} \\
  \end{array} \right)
  \left( \begin{array}{c}
    \nu_1 \\
    \nu_2 \\
    \nu_3 \\  
  \end{array} \right)
  \label{eq:numixing}
\end{equation}
where $U$ is the $3 \times 3$ unitary Maki-Nakagawa-Sakata (MNS)
matrix \cite{matrix:mns}. This is analogous to mixing in the quark
sector as described by the Cabibbo Kobayashi Maskawa (CKM) matrix
\cite{databook:pdg}. The unitary matrix $U$ can be decomposed into
three rotations:
\begin{equation}
  U=
  \left( \begin{array}{ccc}
               1           &      0      &          0              \\
               0           &    c_{23}   &        s_{23}           \\
               0           &   -s_{23}   &        c_{23}           \\ 
  \end{array} \right)  
  \left( \begin{array}{ccc}
              c_{13}       &      0      & e^{-i\delta_{CP}}s_{13} \\
               0           &      1      &          0              \\
   -e^{i\delta_{CP}}s_{13} &      0      &        c_{13}           \\ 
  \end{array} \right)
  \left( \begin{array}{ccc}
             c_{12}        &    s_{12}   &          0              \\
            -s_{12}        &    c_{12}   &          0              \\
               0           &      0      &          1              \\ 
  \end{array} \right)  
  \label{eq:mnsmatrix}
\end{equation}
where the abbreviations $s_{ij} \equiv sin \theta_{ij} $ and
$c_{ij} \equiv cos \theta_{ij} $ are employed. This decomposition allows us
to describe three-flavour neutrino mixing in terms of six parameters:
three mixing angles $\theta_{12}$, $\theta_{23}$ and $\theta_{13}$; a
complex, {\it CP} violating phase $\delta_{CP}$; and, two mass-squared
differences $\Delta m^2_{12} = m^2_2 - m^2_1$ and 
$\Delta m^2_{23} = m^2_3 - m^2_2$. The sign of
$\Delta m^2_{23}$ determines the neutrino mass hierarchy; it is said
to be ``normal'' if $\Delta m^2_{23}>0$, or, ``inverted'' if
$\Delta m^2_{23}<0$. Experiments in neutrino oscillation look for an
appearance or disappearance in one of the neutrino flavours as a
result of mixing. $\theta_{23}$ and $\Delta m_{23}$ are traditionally
known as the ``atmospheric'' mixing parameters describing
$\nu_{\mu} \to \nu_{\tau}$ oscillations; and, $\theta_{12}$ and
$\Delta m_{12}$ are the ``solar'' mixing parameters, concerned with 
$\nu_{e} \to \nu_{\mu}$ flavour oscillations. In each case a
deficit (or disappearance) of the expected flux of the first flavour
is seen as a result of oscillation into the second kind.

Neutrino oscillations have implications for the neutrino
astronomer. The flux of neutrinos from cosmic accelerators is
predominantly the result of pion decays into muons and neutrinos,
essentially producing no $\nu_{\tau}$, i.e.
{$\phi_{\nu_e} : \phi_{\nu_{\mu}} : \phi_{\nu_{\tau}} = 1 : 2 :0$}.
However, since the length of propagation from source to observer is
very much greater than the baseline for oscillation, it is expected
that by the time extra-terrestrial neutrinos arrive at the Earth they
will be evenly mixed:
{$\phi_{\nu_e} : \phi_{\nu_{\mu}} : \phi_{\nu_{\tau}} = 1 : 1 : 1$}.
\section{Sources of Ultra High Energy Neutrinos}
\label{sec:sourcesof}
Many theoretical models have been proposed that predict fluxes of
Ultra High Energy Neutrinos (UHE$\nu$s). A recent, short review is
presented in \cite{sources:berezinsky}; for the interested reader an
exhaustive review is given by \cite{sources:torres}.
Mechanisms in which Cosmic
Rays (CRs) are accelerated to UHE {($E > 1$\,EeV, where
  $1$\,EeV$=10^{18}$\,eV)} are described as ``Bottom-Up''
scenarios. The subsequent weak decay of such CRs through $pp$
collisions and interaction with ambient, or intergalactic, radiation
fields produces an associated flux of UHE neutrinos. So-called
``Top-Down'' neutrino production can occur through various exotic
mechanisms such as: the weak decay of GUT 
scale\footnote{ above $10^{16}$\,GeV Grand Unified Theories (GUT)
  propose that the couplings of the strong, weak and electromagnetic
  forces converge.}
particles produced during early cosmological epochs; annihilation of
superheavy Dark Matter (DM) particles; and effects due to the existence of
Quantum Gravity. Furthermore they can be a signature of ``beyond the
standard model'' processes such as violation of Lorentz Invariance and
enhancements to the neutrino-nucleon cross section. A brief discussion
of Bottom-Up and Top-Down processes follows.
\subsection{Bottom-Up neutrino production}
\label{subsec:bottomup}
%
%
Bottom-Up neutrino production relies exclusively on the acceleration
of charged particles of cosmological origin as an injector to the
neutrino flux. For instance, once a CR proton has been accelerated to
UHE, it can disintegrate into pions following a collision with
another proton; this subsequently produces neutrinos when the pions
decay.
\subsubsection{The cosmic ray spectrum}
\label{subsubsec:crspectrum}
\begin{figure}[htbp]
  \centering
    \includegraphics[width=0.85\textwidth]{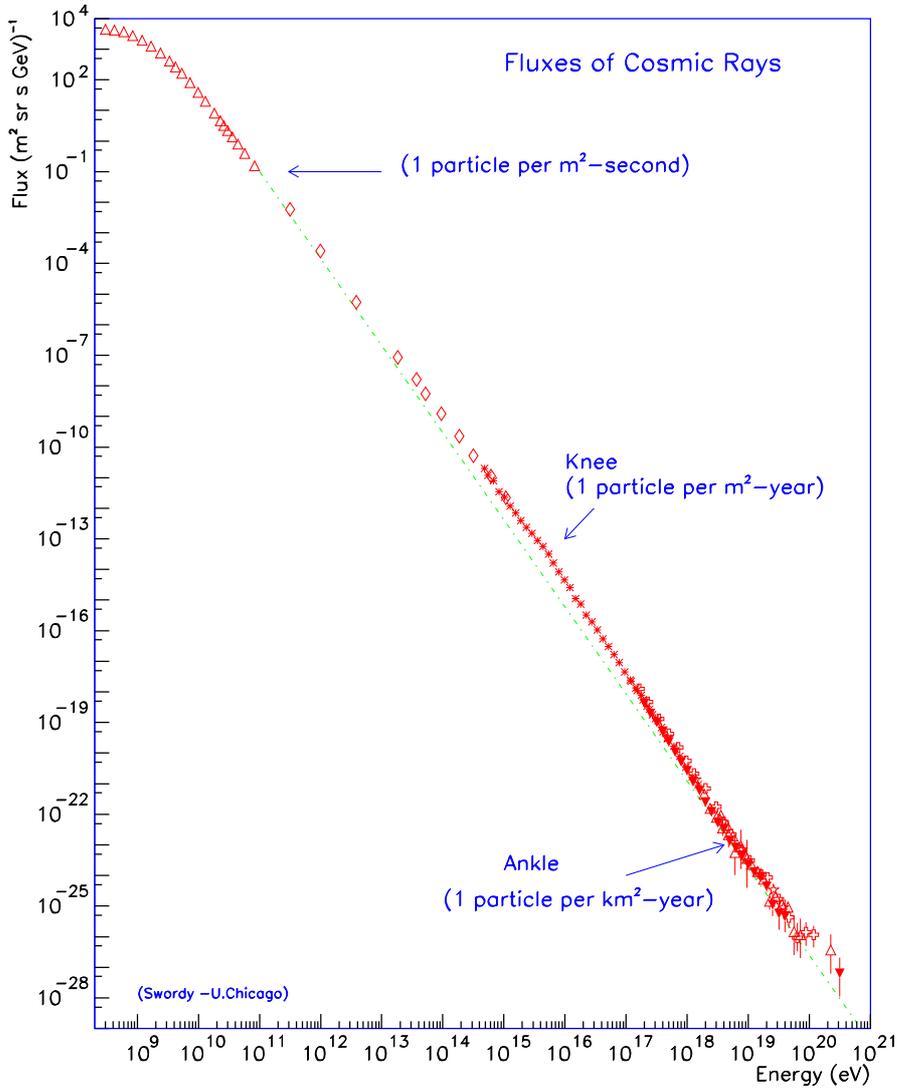}
    \caption[The cosmic ray spectrum]
	    {
	      \small
	      Above $1$\,GeV the cosmic ray spectrum (from
	      \cite{crspectrum:swordy}) is well described by a broken
	      $E^{-\gamma}$ power law, with $\gamma = 2.7$ and two
 	      distinguishing features, caused by a change in spectral
	      index $\gamma$, at $10^{14}$\,eV, labelled `knee' and at
	      $10^{18}$\,eV, labelled `ankle'.
	      \label{fig:crspectrum}
	    }  
\end{figure}

The high energy ($E>1$\,GeV) CR spectrum, as plotted in Figure
\ref{fig:crspectrum}, extends a further twelve decades of energy
throughout which the intensity reduces by $33$ orders of
magnitude. Some $94$ years since Victor Hess first discovered CRs
their origin is still widely debated. The CR flux below $E>1$\,GeV is
dominated by the Solar Wind: coronal ejecta that are readily captured
in the Earth's magnetic field, giving rise to the aurora
phenomenon. The interstellar low energy CR flux is thus very difficult
to determine. Despite being of great interest, low energy CRs will not
be discussed further.

Traditionally the CR spectrum has been well described in terms of two
main features, namely the ``knee'' and the ``ankle''. Both are due to
a deviation from the underlying power law distribution, due to a
change in spectral index (see caption of Figure \ref{fig:crspectrum}).
The causes of each of them remain a controversial topic.
The origin of the ``knee'', seen as a bump in the spectrum at a CR
energy around $10^{14}$\,eV, remains an unsolved problem, to which some 
of the proposed answers include: a change in CR composition, interactions
with a Galactic DM halo, collisions with massive neutrinos,
and different acceleration mechanisms to produce the CRs.
The change in flux at the ankle has traditionally been interpreted as
a transition from Galactic to extragalactic CRs. More recently it has
been suggested that the contribution of extragalactic CRs is of
importance at lower energies, not far above $10^{17}$\,eV
\cite{cosmicrays:hillas}. Furthermore, extra detail in the spectrum
may yet be resolved as more experimental data is acquired, not least a
potential second knee and perhaps a toe.

\begin{figure}[h!]
  \centering
  \includegraphics[width=0.75\textwidth]{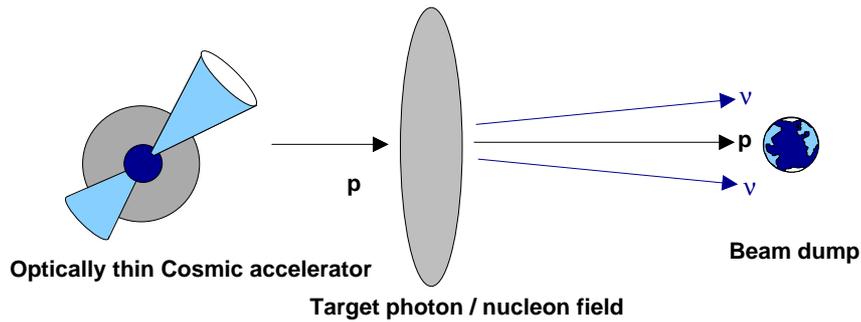}
  \caption[Cosmic acceleration]
	  {
	    \small
	    Protons are accelerated to UHEs at acceleration sites
	    across the Universe. Matter and radiation
	    fields provide target material for $pp$ and $p\gamma$
	    collisions, producing an associated flux of UHE
	    neutrinos.
	    \label{fig:cosmicaccelerator}
	  }
\end{figure}

The concept behind Bottom-Up UHECR and UHE$\nu$ generation is illustrated in
{Figure \ref{fig:cosmicaccelerator}}. The Waxman-Bahcall (WB)
\cite{waxmann:wblimit} bound on the neutrino flux assumes that all
neutrinos result from CRs accelerated at optically thin (i.e
transparent to CRs) sites. Subsequently these CRs undergo $pp$ and
$p\gamma$ collisions after they leave the acceleration site. This
bound is however speculative, as it assumes that all the energy of
the proton is transferred to the resultant pion following a collision;
in reality the pions, and consequently the neutrinos, will have lower
energy. The WB bound can be exceeded by non-accelerator neutrino
sources, or if they exist, accelerator sites that are opaque to CRs
such that they do not appear in the observed CR flux.
\subsubsection{Shock acceleration mechanisms}
\label{subsubsec:shockacceleration}
``Fermi'' acceleration is one of the driving forces that underpins the
generation of UHECRs. The process of Fermi acceleration involves a
transference of the bulk kinetic energy of a plasma to the individual
ions it contains. Ions are accelerated, over long periods of time, by
shock fronts - regions of compression in the plasma at the interface
of two areas at different pressures.
Successive head on collisions with a shock front tend to
increase the kinetic energy of an ion.
This mechanism is sufficient to produce the observed power law
spectrum of CRs. However, one must add some non-linearity, to
describe the shock reaction to ion acceleration, and thus ensure the
energy spectrum does not diverge. 
Acceleration occurs in the relativistic outflows that surround a
powerful central engine, such as a super-massive Black Hole (BH) in an
Active Galactic Nucleus (AGN).
Local inhomogeneities in the turbulent magnetic field
structures confine ions initiating a random walk. Whilst the ion
velocities are changed by magnetic confinement, the kinetic energy is
not. It is possible for an ion to receive multiple kicks
from a single shock if its random walk overtakes the compression front
many times, eventually leading to relativistic ion velocities.

A second method of acceleration, resulting from bulk magnetised
plasma inhomogeneities has been motivated by observational
evidence. 
It is known as ``collisionless shock acceleration'' because ions are
accelerated by the electromagnetic field of the plasma rather than by
particle-particle collisions. 
The details of magnetic fields throughout the Cosmos remain elusive
and continue to be actively studied; there is
mounting evidence that collisionless shock acceleration can describe
the observed CR spectrum from multiple sources.
\cite{collisionlessshocks:waxman}.
\subsubsection{Acceleration sites throughout the Cosmos}
\label{subsubsec:accelerationsites}
Traditionally Supernova Remnants (SNR) have been regarded as the primary
Galactic acceleration sites for high energy CRs below the
knee. The observation of radio, optical and X-ray emission
accompanying the emission of charged particles from such regions has
long provided evidence for electron acceleration up to multi-GeV
energies. Under the assumption that protons and other ions will be
accelerated in places where electron acceleration occurs, the
difficult task of determining the leptonic and hadronic components of
such regions is underway. In particular, X-ray images from the CHANDRA
satellite support CR acceleration in the forward blast wave of a SNR,
a thin shell in the outer shock \cite{chandraobs:warren}. This has
been determined from various details such as the shape and form of the
emission, and the observed spectra.

If SNRs are typical accelerators for Galactic CRs then Active
Galactic Nuclei are the traditional sites of CR acceleration
beyond the Milky Way. The standard model of an AGN is comprised of a
central super-massive BH surrounded by a dusty torus of
matter which forms an accretion disk in the equatorial plane. Axial
relativistic jets of charged particles are emitted as accreted matter
falls on to the central BH. It is in the AGN jets that shock
acceleration takes place. In the case where the relativistic outflow
is directed towards the Earth an AGN is detected distinctly as a gamma
ray source and is classified as a Blazar.

An alternative candidate accelerator is the Gamma Ray Burst
(GRB). GRBs are the most energetic events observed in the
Universe. There are two breeds of GRB: short, typically lasting
less than two seconds; and long,  typically lasting tens of seconds.
The luminosity of a GRB is of the order $10^{45}$\,J and can exceed
that of the combined luminosity of all the stars contained within its
host galaxy. The sudden gamma ray emission is now thought to be 
the result of a rapid mass accretion onto a compact body, resulting
from events such as core collapse supernovae in stars of a few solar
masses, or binary mergers \footnote{any combination of black hole,
  neutrino star, white dwarf   etc. collisions may be
  viable}. The result of such an event is a relativistic outflow of
charged particles that penetrates the surrounding medium, sweeping up
and accelerating ions, however the mechanism by which these
relativistic outflows are produced remains a hot topic.

\begin{figure}[htbp]
  \centering
    \includegraphics[width=0.45\textwidth]{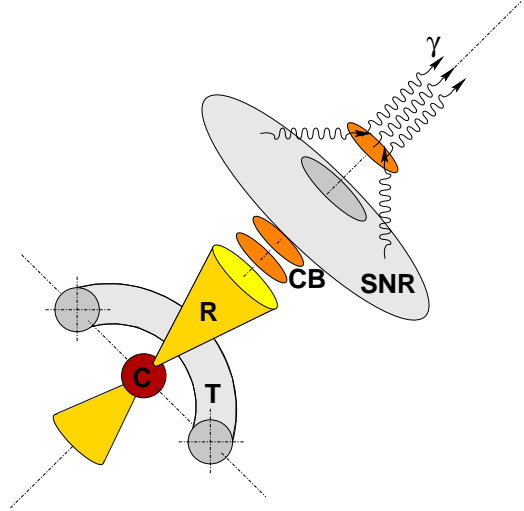}
    \caption[GRB Cannonball]
	    {
	      \small
	      Schematic of the GRB cannonball model: A ``plasmoid'' is
	      seeded by an abrupt collapse of matter from the torus
	      (T) onto the compact body (C). A cannonball (CB) is
	      ejected along the relativistic outflow (R), and, as it
	      passes through the remnants of the parent supernova (SNR)
	      photons ($\gamma$) Compton Scatter up to GRB energies.
	      \label{fig:cannonball}	    
	    }
\end{figure}

A novel, cannonball (CB) model, of GRBs has been
formulated from the hypothesis that following a core collapse SN into
a neutron star or BH, an accretion disk is formed around the compact
body \cite{originofcrs:dar}. 
A CB is emitted when a large chunk of the accretion matter falls
abruptly onto the compact object. The GRB photons are produced as the
CB falls through the ambient radiation of the host SN and photons are
Compton scattered up to GRB frequencies, see Figure
\ref{fig:cannonball}. Again CR acceleration occurs in the shock fronts
of the relativistic outflow.
\subsubsection{The GZK mechanism}
\label{subsubsec:gzkmechanism}
\begin{figure}[htb]
  \centering
  \includegraphics[width=0.7\textwidth]{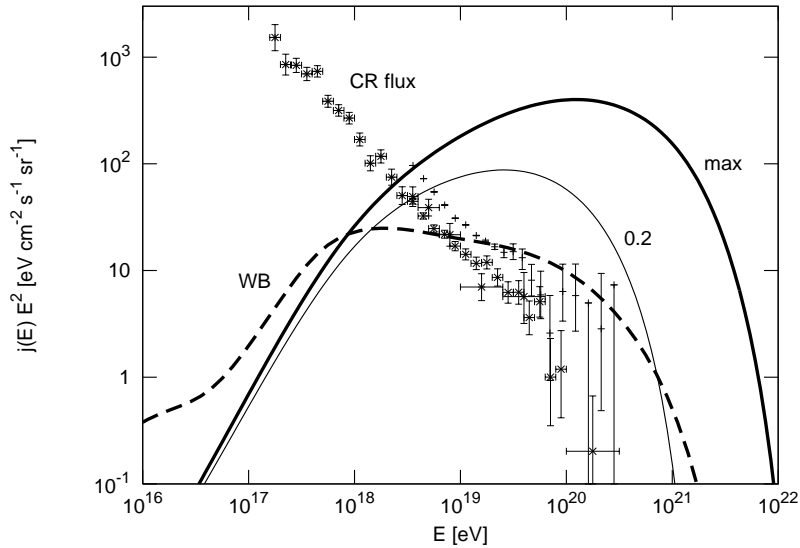}
  \caption[GZK neutrino fluxes]
	  {
	    \small
	    Cosmogenic neutrino fluxes (from \cite{auger:incgsknus}) 
	    as a function of energy. The
	    thick solid line (max) is for an initial proton flux
	    $\propto 1/E$, assuming that the EGRET flux is entirely
	    due to pion-photoproduction. The thin solid line ($0.2$)
	    shows the neutrino flux when the associated photons
	    contribute only up to $20\%$ of the EGRET flux. The dashed
	    line indicates an initial proton flux $\propto 1/E^2$.
	    The experimental points represent the UHECR flux measured
	    by AGASA (crosses) and HiRes (asterisks) - see Section
	    \ref{sec:transgzk} for further discussion.
	    \label{fig:gzknus}
	  }
\end{figure}

A golden channel for neutrino production exists through pion
photo-production from UHECR protons. This takes place via excitation
of the $\Delta^{+}$ resonance resulting from interaction with $2.7$\,K
Cosmic Microwave Background (CMB) photons, as formulated in
Equation \ref{eq:cosmogenics}:
\begin{equation}\label{eq:cosmogenics}
  p + \gamma_{\textrm{CMB}} \to \Delta^{+} \to n + \pi^+
\end{equation}
where
\begin{eqnarray*}
  n\phantom{x} &\to&  p + e^{-} + \overline{\nu_{e}}                  \\
  \pi^+        &\to&  \mu^+     + \nu_{\mu}                           \\
  \mu^+        &\to&  e^{+}     + \nu_{e}      + \overline{\nu_{\mu}} \\
\end{eqnarray*}
This reaction was postulated first by Kenneth Greisen
\cite{gzk:greisen} and later, independently by Zatsepin and Kuzmin
\cite{gzk:zatsepinkuzmin}, and has been dubbed the GZK mechanism. The
threshold energy for pion production by CR protons on photons of
energy $7 \times 10^{-4}$\,eV (the mean energy of black-body radiation
at $2.7$\,K) is $\approx 5 \times 10^{19}$\,eV. Some pion
photo-production occurs below this threshold because of the
high-frequency tail of the black-body photon spectrum. The consequence
of this process is to limit the source distance of UHECR protons to within
$\approx 50$\,MPc, (within the extent of our local group of galaxies)
the typical attenuation length in the CMB photon field at this energy. 
The flux of GZK neutrinos appears in Figure \ref{fig:saundresults}.
It is highly desirable that a neutrino
telescope be sensitive to the flux of cosmogenic neutrinos produced
via the GZK mechanism, since this represents what is essentially a
guaranteed signal, and a smoking gun for EeV scale CR proton
acceleration. Predictions of the flux of GZK neutrinos must satisfy
constraints on the diffuse gamma ray background due to the process:
$\Delta^{+} \to p + \pi^0$, where $\pi^0 \to \gamma \gamma$, which
competes with the process in Equation \ref{eq:cosmogenics}.
Such limits are presently set by the Energetic
Gamma Ray Experiment Telescope (EGRET), a satellite borne detector
sensitive to gamma rays in the energy range $20$\,MeV to
$30$\,GeV. Furthermore, they are dependent on an assumed initial proton
flux. An example of different GZK neutrino fluxes is presented in
Figure \ref{fig:gzknus}. Also noteworthy is the photonic analogue of
this mechanism, whereby CR photons are attenuated by the cosmic
infrared, microwave and radio backgrounds.
\subsection{Top-Down neutrino production}
\label{subsec:topdown}
The apparent lack of suitable candidate sites for particle
acceleration up to UHE, within our neighbourhood, has motivated the
investigation of possible non-accelerator sources.
Here, UHE neutrinos constitute some of the decay products of arbitrary,
super-massive, $X$ particles. $X$ particle production can occur
through a number of mechanisms, including the decay of topological
defects (a spontaneous break in symmetry\footnote{an intrinsic
  property that renders an object invariant under certain
  transformations} during a transition in
phase); metastable super heavy (i.e. $m > 10^{21}$\,eV) relic
particles left over from the Big Bang; or very massive Dark Matter
(DM) particles. Furthermore, extensions beyond SM physics, such as
extra-dimensional regimes, can provide scenarios for the production of
UHE neutrinos.
%
\subsubsection{$X$ particles from topological defects}
\label{subsubsec:topologicaldefects}
\begin{figure}[htbp]
  \begin{center}
    \includegraphics[width=0.85\textwidth]{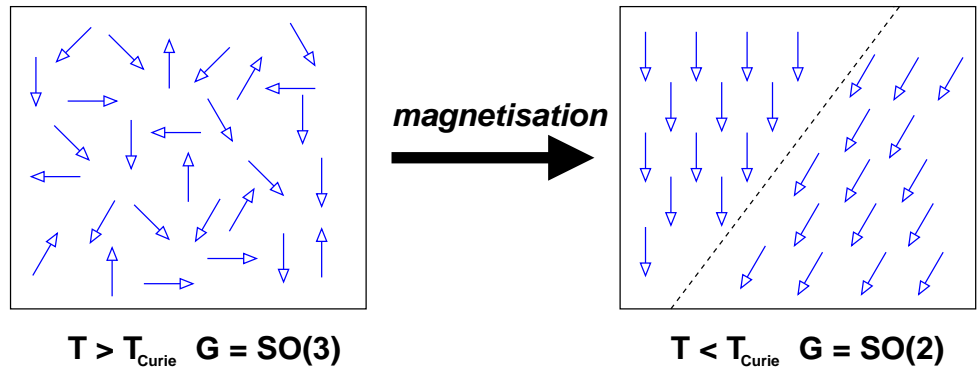}
    \caption[Domain structures in a ferromagnet]
	    {
	      \small
	      As a ferromagnet is cooled through its Curie temperature
	      it forms domain regions of aligned magnetic field
	      separated by Bloch walls (dotted line). Spontaneous
	      symmetry breaking occurs as one requires the
	      introduction of a new degree of freedom:
	      {\it net magnetisation} to describe the state of the
	      system. Hence there is a transition from the SO(3) to
	      SO(2) group symmetry.
	      \label{fig:ferromagnet}
	    }
  \end{center}
\end{figure}

Topological Defects (TD) in states of matter occur as a result of
symmetry breaking phase transitions. Terrestrial examples include
vortex lines in superfluid helium, magnetic flux tubes in type II
semiconductors and disinclination lines in liquid crystals. One
usually envisages the formation of TDs as a result of thermal phase
transitions. For instance, a ferromagnet acquires domain structures as
it is cooled through its Curie point and a symmetry is spontaneously
broken (see Figure \ref{fig:ferromagnet}). An extensive review of TD
models, including their generation and topologies, is presented in
Bhattacharjee and Sigl's 2000 Paper {``Origin and Propagation of
  Extremely High Energy Cosmic Rays''}
\cite{originofehecr:bhattacharjee}.
 After the Big Bang it is natural to assume that
there may have been some temperature through which the Universe cooled
and underwent a similar process, acquiring domain structures
\cite{topology:kibble}.
This method of formation, however, appears to contradict
the inflationary paradigm, which is to dilute the concentration of
unwanted TDs by introducing an exponential expansion of the Universe through
an inflationary phase. It has, though, been realised that TD formation can
take place \emph{after} inflation by way of non-thermal phase transitions.
Such models have been created that allow for abundances of quasi-stable TDs
that can exist at present in the Universe \cite{phasetransitions:khlebnikov}.
In summary, even if the Universe undergoes a period of inflation
during an early cosmological epoch, TD formation can form at a later time, due
to non-thermal phase transitions, thus remaining in concordance with
the inflationary paradigm.

Topological Defects can be described in terms of their Higgs
field, the underlying quantum field  predicted by theory, with which
ordinary matter interacts in order to acquire mass (an analogous
process is that by which massive objects moving through fluids
experience drag). Generally, if $\eta$ is the vacuum expectation
value (VEV) of a Higgs field in a broken symmetry phase, then an
associated  TD has core size $\sim \eta^{-1}$. At the centre of a
core the Higgs field vanishes, the topologies of core centres are
given in Table \ref{tab:coretopologies}.

\begin{table}[htb]
  \begin{center}
    \begin{tabular}{c|c}
      Core Topology & Topological Defect \\
      \hline
      point         & monopole           \\
      line          & cosmic string      \\
      surface       & domain wall        \\
    \end{tabular}
    \caption[Core topologies for different defects]
	    {\small Core topologies for different defects.
	     \label{tab:coretopologies}}
  \end{center}
\end{table}

Far outside of the core, symmetry is broken and the Higgs fields are in their
proper ground states. Hence the `defect' is a core region of unbroken symmetry 
(``false vacuum'') surrounded by broken symmetry regions (``true vacuum'').
The energy densities of the gauge and Higgs fields within the defect are higher
than outside and they are stable due to a `winding' of the Higgs fields
around the cores. Energy is therefore trapped inside the cores and it is in
this way that the TDs acquire mass. In general the mass of the TD, 
$m_{TD}$, is proportional to $T_c$ where $T_c$ is the critical
temperature of the defect
forming phase transition. For a monopole $m_{TD} \sim T_c$, for a
cosmic string the mass per unit length $m_{TD} \sim T_c^2$ and for
a domain wall the mass per unit area $m_{TD} \sim T_c^3$.
For generic symmetry breaking potentials of the Higgs field, $T_c \sim
\eta$ \cite{originofehecr:bhattacharjee}.
Topological Defects can be envisaged as trapped quanta of
massive gauge and Higgs fields of the underlying spontaneously broken gauge
theory.
Sometimes there are quanta of fermion fields trapped in the defect
cores because of their coupling to the massive gauge and Higgs fields, this
combination of fields effectively constitutes a massive object
contained within the defect - this is the $X$ particle.
%
\subsubsection{Metastable superheavy relic particles as $X$ particles}
\label{subsubsec:MSRPs}
Metastable Superheavy Relic Particles (MSRPs) are
an expected bi-product of inflation. 
In certain models MSRPs can exist in the present
with sufficient abundance as to act as a source of non-thermal,
superheavy dark matter \cite{originofehecr:bhattacharjee}. 
A well cited description of CR production from relics
of inflation is given in \cite{inflationrelics:kuzmin}.
One has to overcome the problem of producing a particle that has a
lifetime that is both finite and long enough for it to survive to the
present cosmological epoch. This paves the way for the introduction of
exotic physics such as quantum gravity and instantons. Any model put
forward must adhere to present day limits set by experiment. It
suffices to say that until there are higher statistics for the most
energetic events, there will exist an abundance of theories proposing
MSRP progenitors to the UHECR and UHE neutrino fluxes.
\section{Observations of Trans-GZK CRs and Limits on the UHECR Flux}
\label{sec:transgzk}
\begin{figure}[htb]
  \begin{center}
    \includegraphics[width=0.8\textwidth]{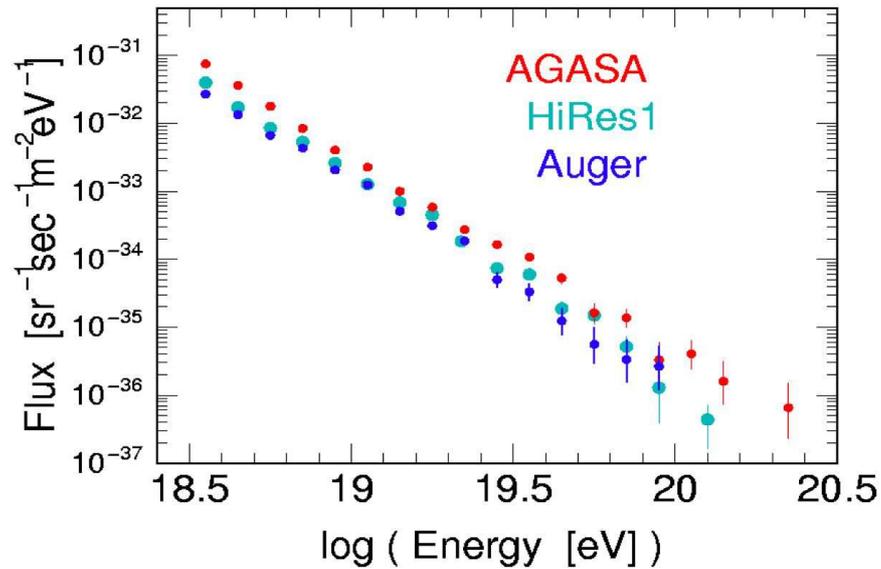}
    \caption[The measured UHECR flux]
	    {
	      \small
	      The flux of UHE Cosmic Rays (from \cite{auger:firstdata})
	      as measured by AGASA, AUGER
	      and HiRes. The highest energy AUGER event ($>140$\,EeV)
	      does not appear in this plot because the core of the
	      shower fell outside the surface detector and as such it
	      did not pass all quality cuts for flux calculations.
	      \label{fig:augerflux}
	    }
  \end{center}
\end{figure}

As stated in Section \ref{subsec:bottomup} the threshold energy for the
GZK mechanism is $\approx 5 \times 10^{19}$\,eV.
To date, AGASA, AUGER and HiRes have all
reported the existence of particles above this threshold.  The UHECR
flux they have measured is shown in Figure \ref{fig:augerflux}. The
appearance of CRs above the GZK threshold suggests that the origin of
such CRs is within our local group of galaxies despite there being no
known sources, or more specifically acceleration regions contained
therein, that can produce CRs at such energies. As the number of
events recorded at these energies slowly increases it should become
apparent if they are a component of the diffuse CR background or they
originate from point sources.
\section{Summary}
\label{sec:chp1summary}
It has been shown in this chapter how neutrinos come to reach the Earth
from various cosmological origins, most of which are not fully
understood.
The discovery of neutrino mass and subsequently flavour oscillations
means that although one expects muon type neutrinos to outnumber
electron type neutrinos by two to one, with no tau neutrino component;
in fact, by the time of their arrival at Earth they are evenly mixed
into equal fractions.

The flux of UHE neutrinos in which the Earth bathes is intimately
linked to the acceleration and subsequent decay of UHECRs. Wherever
proton acceleration has occurred, one can expect to encounter the
production of neutrinos with comparable vigour. The electromagnetic
and hadronic components of cosmological accelerator outflow are 
retarded by inter galactic matter and radiation. This presents a limit
on how far we can observe them through the cosmic molasses. Neutrinos
however can reach us from the furthest depths of the Universe.

The UHE neutrino flux may not rely exclusively on particle
acceleration. The lack of suitable acceleration sites within our
vicinity has motivated the idea that the highest energy CR events that
have been observed resulted from the decay of very massive entities
known only as \mbox{``$X$-particles''}

  \cleardoublepage

  \chapter{Neutrino Detection Methods}
\label{chp:detection}
\section{Introduction}
The detection medium for Ray Davis Jr's pioneering Homestake experiment was
a large tank of tetrachloroethene \cite{homestake:davis}.
This chemical, essentially a cleaning fluid, is sensitive to neutrino
capture through the reaction {$\nu_{solar} + {_{17}\mathrm{Cl}^{37}}
  \to e^- + {_{18}\mathrm{Ar}^{37}}$}. Neutrinos are emitted by the
Sun through the conversion of hydrogen to helium and in the decay of
Be$^7$, B$^9$, N$^{13}$ and O$^{15}$. Amazingly it was by measuring
the number of individual Argon atoms present after flushing out the
detector that it was possible to quantify the neutrino flux of solar
neutrinos through the apparatus.

Today, there are numerous experiments across
the world either under construction or in operation, dedicated to the
detection of astrophysical neutrinos. The principal method by which they are
registered is through {\v C}erenkov emission. This can be at optical
wavelengths via their muon daughters, or at radio frequencies via the
Askaryan Effect \cite{askaryaneffect:askaryan}.
Large, natural bodies of transparent\footnote{i.e. transparent to the
secondary radiation used to observe the neutrino interaction} 
dielectric media such as Mediterranean Sea
water, Antarctic ice, subterranean salt domes and even the lunar
regolith can serve as natural neutrino calorimeters.
\section{Deep Inelastic Scattering}
\label{sec:dis}
\begin{figure}[h]
  \begin{center}
    \includegraphics[width=0.4\textwidth]{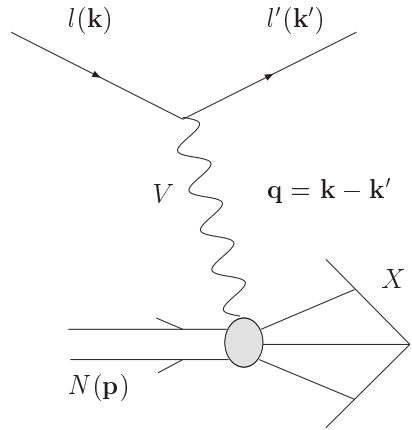}
    \caption[Deep inelastic scattering]
	    {
	      \small
	      Deep inelastic scattering. $l$ is the incoming neutrino,
	      $l'$ the outgoing lepton; $V$ is the vector boson ($Z$
	      or $W^{\pm}$) and $N$ is the target nucleon. $X$ is the
	      arbitrary excited hadronic final state. $k$, $k'$ and
	      $p$ are the 4-momenta of the incoming neutrino $l$, the
	      outgoing lepton $l'$ and the nucleon $N$ respectively.
	    }
	    \label{fig:dis}
  \end{center}
\end{figure}

Figure \ref{fig:dis} illustrates how high energy neutrinos $l$,
interact with a target nucleon $N$, by deep inelastic scattering (DIS)
on its constituent quarks. The basic interaction is given by Equation
\ref{eq:nuinteraction}:
\begin{equation}
  l + N \to l' + X
  \label{eq:nuinteraction}
\end{equation}
where $N$ is a nucleon, $l'$ is the outgoing lepton and $X$ is one or
more excited hadrons. Interactions can be neutral current (NC), as
mediated by the weak-neutral $Z$ boson, or charged current (CC) as
mediated by the weak-charged $W^{\pm}$ bosons. The contributions 
of these two components to the total neutrino-nucleon cross section
are illustrated in Figure \ref{nunucleonxsection} in Appendix
\ref{appendix:nucrosssection}.

The square momentum transfer of the interaction, also known as
hardness, is defined as $Q^2 = -q^2 = -(\mathbf{k} - \mathbf{k'})^2$.
Two dimensionless scaling variables can be used to describe the
kinematics of DIS, each of which represents a physical characteristic
of the event. Firstly the Bjorken-x variable, defined as:
\begin{equation}
  x = \frac{Q^2}{2(\mathbf{p} \cdot \mathbf{q})}
  \label{eq:bjorkenx}
\end{equation}
which is the fraction of the nucleon momentum carried by the
participating (valence or sea, depending on $E_{\nu}$)
quark. Second is the Bjorken-y variable, defined as:
\begin{equation}
  y = \frac{\mathbf{p} \cdot \mathbf{q}}{\mathbf{p} \cdot \mathbf{k}}
  \label{eq:bjorkeny}
\end{equation}
which is the fraction of the lepton energy transferred to the hadronic
system. In the case of a neutrino DIS then this is the fraction of the
neutrino energy that goes into the hadronic cascade, the behaviour of
which will be discussed further in Chapter \ref{chp:neutrino}.

Calorimetric registration of a neutrino DIS can result through
the development of the hadronic cascade and, following CC
interactions, through propagation of the charged lepton. Electrons
induce an electromagnetic shower that develops collinearly to the
hadronic shower, whereas muons, having a longer interaction length
travel through the detector for several metres before decaying.
In the case of CC tau neutrino interactions a characteristic ``double
bang'' signal can be observed: first a hadronic cascade is initiated
by the DIS which is closely followed by a displaced secondary cascade
induced by the decay of the tau lepton. 

Particle cascades tend to excite atoms in a calorimeter via excitation
and ionisation, furthermore there is {\v C}erenkov emission which can
be observed in the visible or radio frequency band. This is explained
in detail for each case below.
\section{Optical {\v C}erenkov Neutrino Telescopes}
\label{sec:opticalcerenkov}
Optical neutrino telescopes are optimally sensitive to
the weak interaction of $\nu_{\mu}$ neutrinos into muons via the process
{$\nu_{\mu} + N \to \mu^- + X$}. The energy of the neutrino is shared
between the resultant muon, $\mu$, and the hadronic cascade, $X$, with
the muon taking between a half and three quarters of the total
neutrino energy; it then continues along a path that is
effectively collinear ($< 0.5$\,$\degrees$ separation for $E_{\nu} >
1$\,TeV) to the bearing of the incident neutrino.
The charge of the traversing muon causes the surrounding medium to become
polarised. Subsequent depolarisation of the medium results in the
emission of {\v C}erenkov photons along the relativistic charged
particle track. Interference between photons can occur if the
wavefronts overlap, which is only possible if the muon travels faster
than light does in the detection medium. Hence the net polarisation of
the medium is assymetric. Wavelets interfere constructively
to produce the {\v C}erenkov wavefront if $\beta(=v/c) >1/n$, where
$v$ is the muon speed and $n$ is the refractive index of the
medium. The range of the muon in water extends from $\sim 5$\,m at
$1$\,GeV to $\sim 7$\,km at $10$\,TeV\cite{databook:pdg}.
%
%
\begin{figure}[ht]
  \begin{center}
    \includegraphics[width=0.5\textwidth]{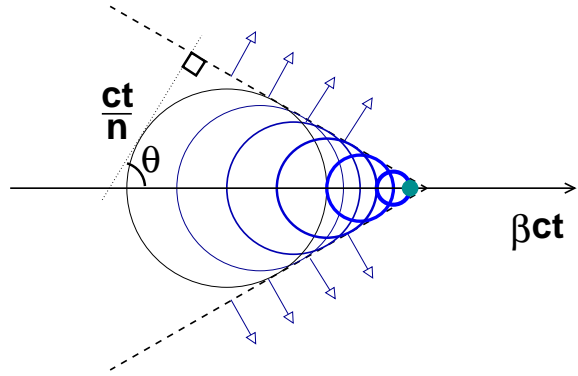}
    \caption[{\v C}erenkov light cone]
	    {
	      \small
	      Schematic representation of {\v C}erenkov light emission.
	    }
	    \label{fig:cerenkov}
  \end{center}
\end{figure}

The number of {\v C}erenkov photons produced, $N$, within a wavelength
interval $d\lambda$ and unit distance $dx$ is given by \cite{databook:pdg}:
\begin{equation}
  \frac{d^2N}{dxd\lambda} = \frac{2\pi \alpha}{\lambda^2}\times
  (1 - \frac{1}{\beta^2 n^2})
  \label{eq:cerenkovphotons}
\end{equation}

The wavefront formed by coherent emission of {\v C}erenkov photons
forms a cone of radiation with opening angle dependent on the
velocity of the emitting particle, given by
Equation \ref{eq:cerenkovangle}.
Thus if one can reconstruct the angular aperture of a {\v C}erenkov
light cone, it can be used to infer the velocity of the transient
particle. One must then assume a rest mass in order to resolve the
particle momentum and hence it's energy.
%
%
%
\begin{equation}
  \cos \theta = \frac{1}{\beta n}
  \label{eq:cerenkovangle}
\end{equation}

Detection of those photons emitted is facilitated by photomultiplier
tubes (PMTs), devices that convert optical quanta into charge quanta
(conceptually the inverse of a light bulb). PMTs tend to be
optimally sensitive at wavelengths around $450$\,nm and one has to be
mindful to choose a detector medium in which the scattering and
absorption of such light is kept to a minimum.

Traditionally, a neutrino ``telescope'' views the Cosmos using the
Earth as a filter to those muons produced in CR induced air showers
high in the atmosphere. Hence a neutrino telescope built in the
northern hemisphere, looking downward through the centre of the
Earth, will see a view of the southern sky, and vice
versa. Muon neutrino interactions can occur in the transparent medium
surrounding the instrumentation or in the Earth's crust below, so long
as the muon has sufficient collision length to allow it to propagate
through the detector, preserving the directionality of its parent
neutrino. 
The performance of an optical neutrino telescope is limited at low
energies by the short length of the muon tracks and at high energies
by the opacity of the Earth. Since the neutrino-nucleon cross section
increases with energy, as one looks for neutrinos with greater energy
the Earth ceases to act merely as a filter and instead becomes
impenetrable.

A summary of existing projects and proposed extensions follows,
including, at the end of this section, the current limits on the
neutrino flux as set by optical neutrino telescopes.
\subsection{AMANDA}
\label{subsec:amanda}
\begin{figure}[htbp]
  \begin{center}
    \includegraphics[width=0.85\textwidth]{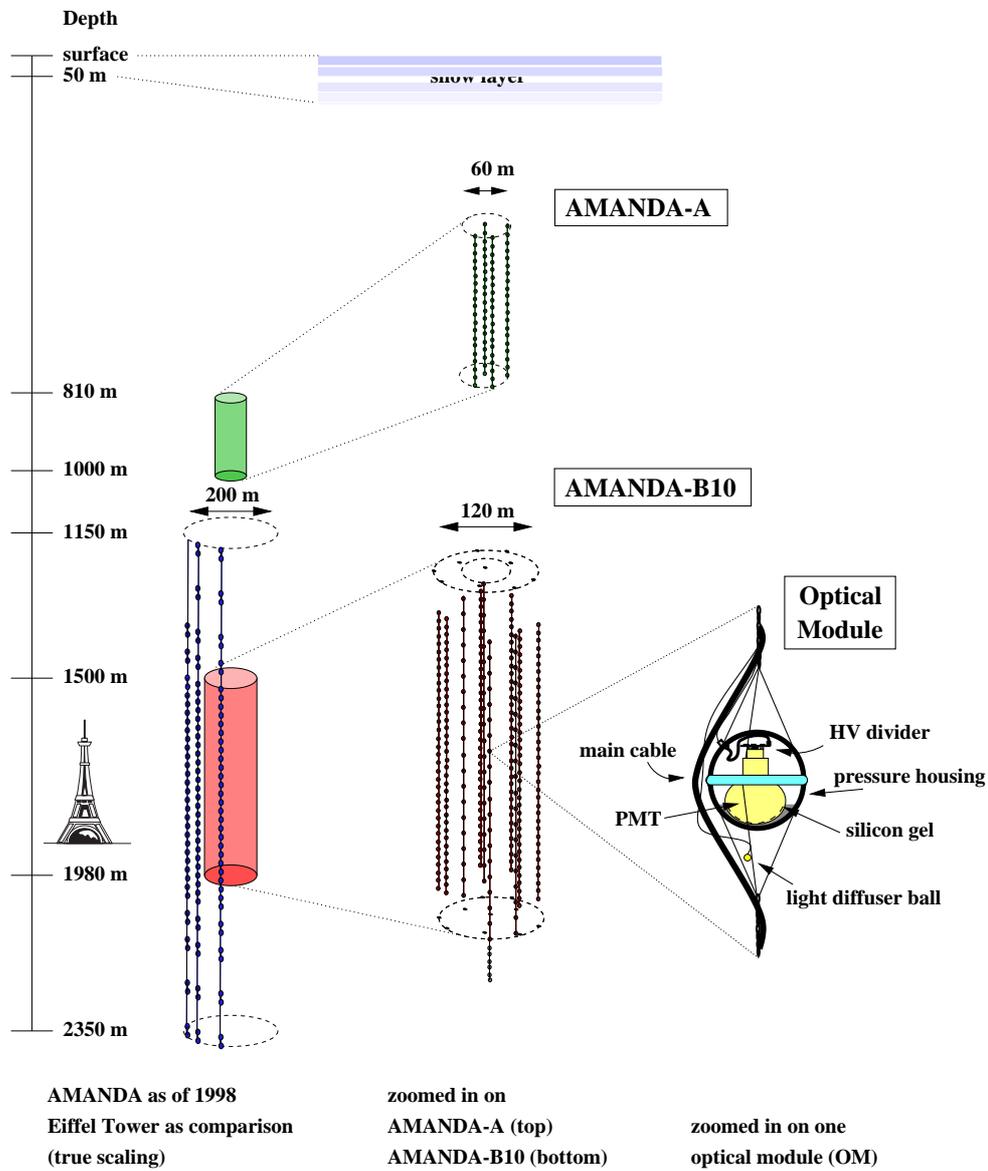}
    \caption[Antarctic Muon And Neutrino Detector Array]
    	    {
	      \small
	      Schematic of AMANDA phases A $(1993/94)$ and B-10
    	      $(1997/98)$ including graphic of an AMANDA optical module
    	      (OM) containing PMT. From \cite{amanda:andres}.
	    }
	    \label{fig:amanda}
  \end{center}
\end{figure}

Undoubtedly the most advanced optical neutrino telescope to date
is the Antarctic Muon And Neutrino Detector Array
(AMANDA)\cite{amanda:andres}. The most recent second-phase AMANDA-II
\cite{amandaii:andres}
is an array of $667$ downward pointing PMTs, occupying a cylinder of
one kilometre height and $200$\,m diameter, located $1.5$\,km deep under
the $3$\,km thick, southern polar ice cap. Construction of AMANDA-II was
completed during the austral summer of 1999-2000 following the
initial phases of the shallow ice array AMANDA-A in 1993/94,
AMANDA-B4 ($80$ PMTs on $4$ strings) in 1995/96 and AMANDA-B10 (an
additional $216$ PMTs on $6$ strings) in 1997/98 as illustrated in
Figure \ref{fig:amanda}. Deployment of
detector strings is made possible by drilling the ice with heated jets
of $75$\,$\degrees$C water. Operating at a speed of $1$\,cm\,s$^{-1}$, it
takes approximately three and a half days to bore a $50 \to 60$\,cm
diameter hole to a depth of $2$\,km. After a bore hole is formed, the
strings are dropped into place and the ice re-freezes over a period
of about $35$\,hours.

\begin{figure}[htbp]
  \begin{center}
    \includegraphics[width=0.6\textwidth]{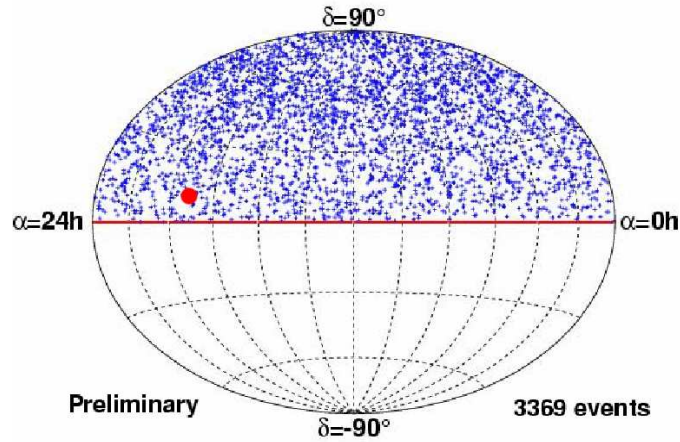}
    \caption[AMANDA sky plot]
    	    {
	      \small
	      The sky as viewed through the AMANDA neutrino telescope
	      (from \cite{amandatoicecube:halzen}) with the
    	      coordinates of the crab $\alpha = 19.2h$, $\delta =
    	      9.2$\,$\degrees$ highlighted by the
    	      \red{$\bullet$} marker.
	    }
	    \label{fig:amandaskyplot}
  \end{center}
\end{figure}

AMANDA has been in operation for over $5$\,years in its latest,
phase-II, configuration collecting between $7$ and $10$ neutrinos
{\it  per diem}
\cite{amandatoicecube:halzen}. A background rate of $\sim 80$\,Hz
is present due to downward going muons from atmospheric CR
interactions and must be rejected through angular cuts and event
reconstruction. Point source searches have been performed using the
first four years of AMANDA neutrino data and limits on the neutrino
flux have been estimated using $800$ days of data
\cite{amanda:andres}.
Further analysis is currently underway. The most significant point
source excess, located in the direction of the Crab Nebula, has a
confidence level of only $3.4\sigma$ \cite{amandatoicecube:halzen}. A
view of the sky as viewed through AMANDA is plotted in Figure
\ref{fig:amandaskyplot}. Clearly there is a case for a kilometre cubed
extension to AMANDA if decisive identification of point sources is to be
made. IceCube \cite{amandatoicecube:halzen}
will incorporate the existing AMANDA-II instrumentation
and extend to a total of $4800$ active PMTs and furthermore include
IceTop - an array of surface based scintillator detectors for the
purpose of calibration, background rejection and CR studies.
\subsection{ANTARES}
\label{subsec:antares}
The ANTARES (Astronomy with a Neutrino Telescope and Abyssal
Environmental RESearch) neutrino telescope is
currently under construction in the Mediterranean Sea, off the coast of
southern France \cite{antares:becherini} .
It consists of twelve $480$\,m long strings each bearing
$75$ ten inch Hammamatsu PMTs arranged three per storey, each enclosed
in a pressure resistant glass sphere (PMT$+$sphere$=$optical module
(OM)), with a declination angle of $45$\,$\degrees$. This declination
helps limit the loss of transparency due to the build up of
sedimentation on the glass housing to less than $1\%$ per
year. Each storey has a box of controlling electronics, the Local
Control Module (LCM) along with the $3$ OMs. There are $5$ storeys per
sector with a vertical spacing of $14.5$\,m and $5$ sectors per line,
with a horizontal line spacing of between $60$\,m and $75$\,m.
The pointing accuracy is $\sim 1.5$\,$\degrees$ at $E_{\nu} = 100$\,GeV
but improves to less than $0.5$\,$\degrees$ for energies greater than
$1$\,TeV due to the increased length of the muon track and the
subsequent increase in {\v C}erenkov photons \cite{antares:firstresults}.

Since March 2005 a Mini test Instrumentation Line with Optical Module
(MILOM) has been recording data such as environmental conditions and
background rates due to bio-luminescence and the decay of solute
$^{40}$K. The MILOM has detected a prevailing current of
$10$\,cm\,s$^{-1}$ in the west-east direction, a sea temperature of
between $13.2$\,$\degrees$C and $13.3$\,$\degrees$C and a sound
velocity of about $1545$\,m\,s$^{-1}$. In June 2005 the baseline rates
for light background were measured to be between $90$\,kHz and
$140$\,kHz at a trigger threshold of $0.5$ photoelectrons
\cite{antares:firstresults}.
%
%
The angular performance of the
ANTARES telescope relies  heavily on knowledge of the positioning of
its OMs. This is provided by a sophisticated acoustic positioning
system that has been tested on board the MILOM and has so far met the
required $10$\,cm spatial resolution. PMT timing resolutions of $\sim
0.5$\,ns have been measured by illuminating OMs with pulsed light from LED
beacons mounted at the base of the MILOM \cite{antares:firstresults}.
The first fully instrumented line started recording downward going,
atmospheric muons as of March 2nd, 2006. By April 2007 seven fully
instrumented lines were in operation. It is expected that full
deployment of the completed $12$ string detector will be complete
before the end of 2007. 
\subsection{Lake Baikal} 
\label{subsec:baikal}
Baikal is the  oldest study group still active; in the late 1970s
they trail blazed their way in tandem with the ill fated DUMAND (Deep
Underwater Muon And Neutrino Detector) project \cite{dumand}
providing today's experimentalists with a wealth of information.
Located at a depth of just over a kilometre, in Lake Baikal, Siberia,
is the NT200 neutrino telescope \cite{baikal:aynutdinov}.
In operation since 1998, NT200 comprises $192$ OMs
arranged in pairs on $8$ vertical strings attached to a fixed metallic
frame. Each OM consists of a $37$\,cm {\it QUASAR} PMT encased in a
transparent, pressure resistant housing. Readout and control
electronics sit in a similar housing and are located midway between
each storey. All OMs face downward with the exception of the eleventh
and second OM on each string, which are oriented in the upward
direction. Successive OM pairs are separated by a $6.25$\,m vertical
gap, whilst the $8$ strings sit on a regular heptagon, of side
$18.6$\,m, with one string at the centre. Since 1998 no neutrino
events have passed the NT200 cut selection criteria
\cite{baikal:aynutdinov}. An upgrade to
NT200+, which will incorporate three new strings and increase
sensitivity to higher energy events, is underway.
\subsection{NESTOR}
\label{subsec:nestor}
The Neutrino Extended Submarine Telescope with Oceanographic Research
(NESTOR) Project works at a mean depth of $4$\,km in the Ionian Sea, off
the southwestern tip of the Peloponnesus
\cite{nestor:aggouras}. Briefly, the apparatus is composed of rigid
titanium girders arranged in a six point star formation, with one
upward and one downward facing OM at each tip. Here, an OM
corresponds to one $15$\,inch diameter PMT along with on board DC-DC
converter, surrounded by a spherical glass
housing that is pressure resistant up to $630$\,atm. The diameter of
each hexagonal floor (star) is $32$\,m. A $1$\,m sphere at the centre of
each star houses the local electronics. A completed NESTOR tower
consists of twelve such elements in a vertical arrangement with a
separation of $30$\,m. To date one star element has been deployed for
testing purposes accumulating over two million triggers in the longest
period of continuous operation. Proof of principle has been achieved:
calibration via LED flashers; detection of backgrounds from the decay
of K$^{40}$ and bio-luminescence; and reconstruction of atmospheric muon
tracks all support the continued development of the full scale NESTOR
detector.
\subsection{Kilometre-Cubed optical {\v C}erenkov neutrino telescopes}
\label{subsec:km3telescopes}
Even before the first generation of large scale neutrino telescopes
are complete a second wave of activity emerges, this time on an almost
unimaginable scale. The experiments described previously
have become test benches for the world's largest particle detector
experiments - cubic kilometre neutrino telescopes:
\subsubsection{IceCube}
\label{subsubsec:icecube}
IceCube is the natural successor to AMANDA and, as mentioned in
Section \ref{subsec:amanda}, will incorporate the existing AMANDA-II
apparatus as a means to upgrading to an enormous $4800$ OMs and
encompassing a cubic kilometre of Antarctic ice.

By the end of summer 2006 IceCube had $9$ new strings and $16$ of
the IceTop surface {\v Cerenkov} detector stations; a further $858$
DOMs had been approved for the next stages of fitting.
Each of the deployed strings, already acquiring data, consists of $60$
Digital Optical Modules (DOM) on a $2.5$\,km cable
\cite{amandatoicecube:halzen}.
\subsubsection{KM3NeT}
\label{subsubsec:km3net}
Funding from the European Union is now in place to undertake the
KM3NeT design study phase. Building on the experiences of ANTARES,
NEMO and NESTOR, the KM3NeT collaboration seeks to develop a
water based partner to IceCube, at a site in the
Mediterranean, yet to be determined.
 Such a telescope, in partnership with its austral
cousin, would complete a full sky coverage, allowing for some
overlap in the plane of the Galactic centre. The Galactic centre
itself is viewable exclusively from the northern hemisphere.
\subsubsection{NEMO}
\label{subsubsec:nemo}
The NEutrino Mediterranean Observatory (NEMO) project represents an
alternative ongoing effort to pilot a cubic kilometre optical
{\v C}erenkov neutrino telescope in the Mediterranean Sea. Whilst full
optimisation of detector geometry has yet to be performed, a
demonstration model incorporating some of the 
features of a kilometre cubed detector was deployed at a site off
the coast of Sicily in December 2006 \cite{nemo:piattelli}
\subsection{Limits on the neutrino flux from optical {\v C}erenkov 
neutrino telescopes}
\label{subsec:limitsfromoptical}
\begin{figure}[h!]
  \centering
  \includegraphics[angle=90, width=0.75\textwidth]{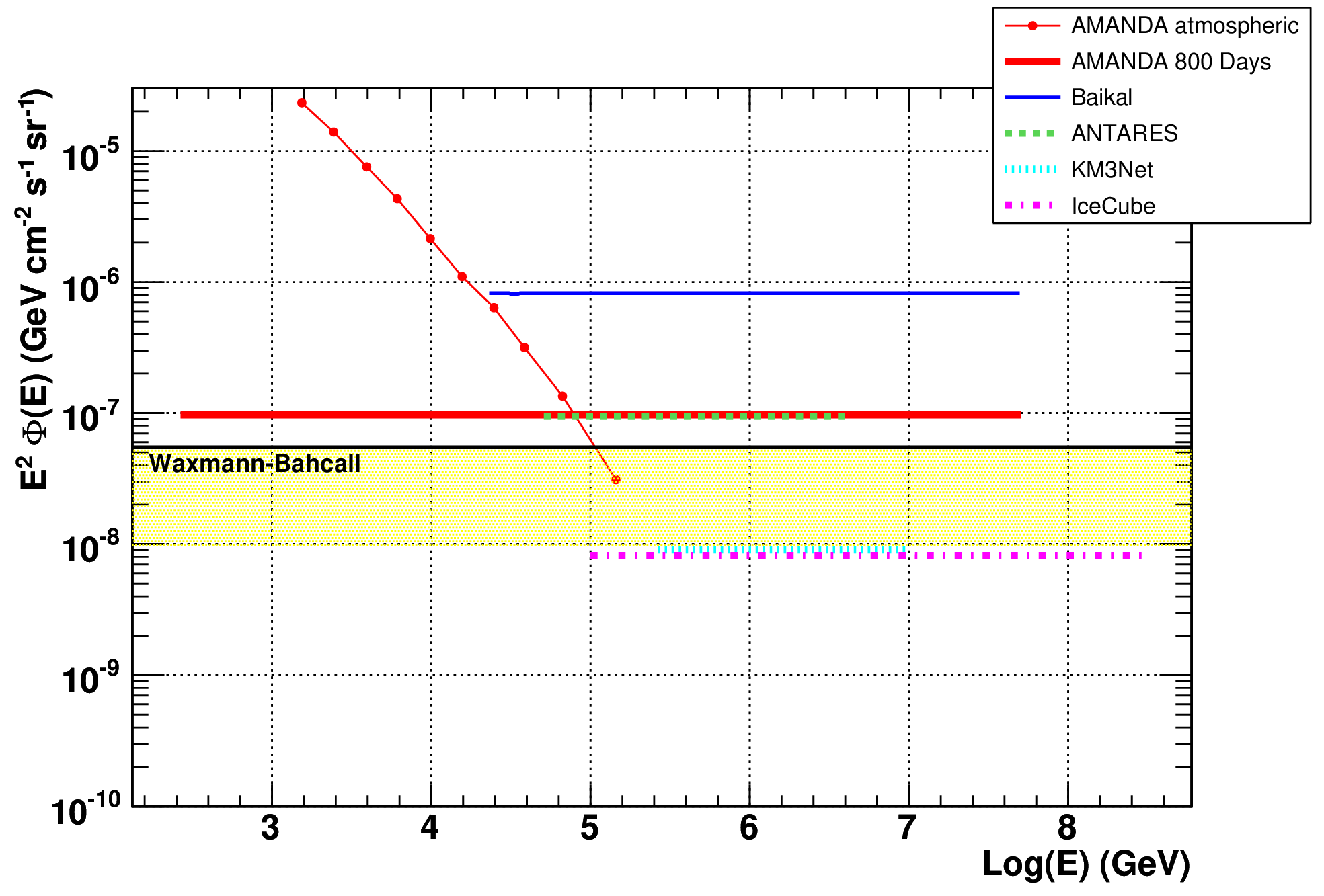}
  \caption[Limits on the neutrino flux from optical telescopes]
      {
	\small
	Constraints on the neutrino flux set by optical neutrino
	telescopes. The \red{$\bullet$} markers represent
	the AMANDAII atmospheric neutrino data set, the
	\red{thick solid line} is the
	$800$\,days AMANDAII sensitivity and from the same paper is an
	estimate of the Waxman-Bahcall neutrino flux predicted from
	acceleration of the highest energy cosmic rays
	\cite{amandatoicecube:halzen}. The latest limits from the
	Lake Baikal experiment are plotted as the
	\blue{thin solid line}
	\cite{baikal:aynutdinov}. The \green{dashed},
	\textcolor{cyan}{dotted} and \textcolor{magenta}{dash-dotted}
	curves represent the projected sensitivities from one year of
	ANTARES \cite{sokalski:antareslimits}, KM3NeT
	\cite{katz:km3limits}, and IceCube
	\cite{amandatoicecube:halzen} respectively.
      }
  \label{fig:opticallimits}
\end{figure}

Current limits on the neutrino flux from optical {\v C}erenkov based
detectors appear in Figure \ref{fig:opticallimits}. Some predicted
limits are also included based on proposed extensions to existing
experiments and from new projects. Plotted on the same figure are some
model neutrino fluxes predicted by theory.
\section{Radio {\v C}erenkov Detectors}
\label{sec:radiocerenkov}
Radio {\v C}erenkov detectors are sensitive, not to the passage of
muons, but to the electromagnetic component of neutrino induced particle
cascades. The optimum process for detection is a charged current $\nu_e$
interaction. This simultaneously initiates an electromagnetic and a
hadronic particle shower from the emergent electron and hadron
respectively. The two showers are in principle
superimposed. In 1962 G. A. Askaryan realised that the net charge of
the particle cascade, in a dense medium, will not be neutral - since
positrons emitted by pair-production will annihilate in flight, leaving a
negatively charged excess. Additional electro-negativity is provided
by the accumulation of delta rays and Compton scattered electrons. In
his paper \cite{askaryaneffect:askaryan} Askaryan predicts that at the
maximum of an $E = 10^{18}$\,eV shower there is a $10\%$ electron charge
excess; one can expect to observe a net $20\% - 30\%$ negative charge
excess.

Almost forty years after this effect was postulated came
experimental confirmation.
Formerly, radio emission from cosmic-ray induced particle showers was
shown to be measured as a result of the geomagnetic separation of
electric charges. Through the detection of photon induced cascades in
$3200$\,kg of silica sand, evidence for the observation of the Askaryan
effect was given by correlation of the radio pulse size with the
energy of a shower. Furthermore it was shown that the experimental
observations were inconsistent with geomagnetic charge separation and
consistent with {\v C}erenkov emission because of the alignment of the
polarisation along the shower axis and not the local geomagnetic dip 
\cite{askaryanobservation:saltzberg}.

A brief report of existing radio {\v C}erenkov neutrino telescopes
follows, ended by a summary of experimental limits they impose on
the neutrino flux.
\subsection{ANITA}
\label{subsec:anita}
The balloon borne ANtarctic Impulsive Transient Antenna (ANITA)
experiment is a recent and innovative apparatus designed for the radio
detection of CR neutrinos. An $18.4$ day flight has been
achieved by a prototype, named ``ANITA-lite'' early in 2004.
\cite{anita:barwick}
The principal ANITA design features a $2\pi$ array of antennas
sensitive over the frequency range
$230 - 1200$\,MHz, observing $\sim 2$Mkm$^3$ of
Antarctic ice, where the depth is equivalent to the radio attenuation
length, from an altitude of $37$\,km. The ANITA-lite prototype, whilst
only consisting of two ANITA antennas with $12\%$ of the viewing area
of the full scale detector, still observed a considerable
$0.17$\,Mkm$^3$ of ice, producing competitive constraints on the UHE
neutrino flux that ruled out the remaining ``Z-Burst'' ($\nu \overline{\nu}
\to Z^0$) flux that was not already constrained by previous
experiments, as a mechanism for the production of the highest energy
CRs. Simulations of the full scale detector project a sensitivity down
to sub-GZK fluxes, thus opening a window to cosmogenic neutrinos and
possibly never before seen, extra galactic cosmic accelerators. A two
month mission successfully launched in December 2006.
\subsection{FORTE}
\label{subsec:forte}
Developed by the Los Alamos and Sandia National Laboratories the Fast
On-orbit Recording of Transient Events (FORTE) satellite sits in a
circular orbit at an altitude of $800$\,km and inclined
$70$\,$\degrees$ from the Earth's equator. Launched in 1997, its
on board optical and radio frequency receivers can be exploited for the
study of hadronic particle cascades in the Greenland ice sheets as
well as lightning storms and for its intended use as a nuclear
detonation detector. The most recent analysis of the FORTE data set
contains over $4$ million events, recorded from September 1997 to
December 1999\cite{forte:lehtinen}. Only a single event passed through
each level of cuts, so the calculated flux limits for the experiment
assume this to be a representation of the background
level. Backgrounds are present in the form of radio emissions from
lightning strikes. Certain artifacts of a lightning event can be used
to reject backgrounds of this nature such as signal grouping, random
polarisations and pulse pairs resulting from reflections off the
ground. Lightning activity over the Greenland is however extremely
rare and thus good knowledge of the geographic location of a signal
can further reduce the background activity. Tentative limits have been
placed on the flux of astronomical neutrinos in the energy range
$10^{13} \to 10^{16}$\,GeV. One treats these with caution noting that
they extend beyond the scale for GUT particle masses and are
calculated in the absence of neutrino-nucleon cross section data above
$30$\,TeV.
\subsection{GLUE}
\label{subsec:glue}
\begin{figure}[htbp]
  \begin{center}
    \includegraphics[width=0.5\textwidth]{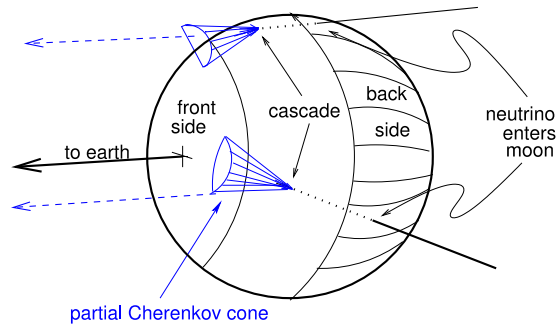}
    \caption[UHE cascades in the lunar regolith]
    	    {
	      \small
	      UHE cascades in the lunar regolith (from \cite{glue:gorham}).
	    }
	    \label{fig:glueschem}
  \end{center}
\end{figure}

As stated at the beginning of this chapter, it is possible to observe
neutrino induced cascades that occur not in sea, ice or salt on Earth
but in the lunar regolith; this is illustrated schematically in Figure 
\ref{fig:glueschem}. The Goldstone Lunar Ultra-high energy neutrino
Experiment (GLUE) looks for radio signals from this source. This
technique is facilitated by two large antennas, one
$70$\,m and one $34$\,m in size, separated by a baseline of $22$\,km. By
utilising the two receivers in coincidence, with a large separation,
one can in principle reduce the anthropogenic radio background. On
average the background due to random coincidences and thermal noise is
reported to be $3 \times 10^{-3}$\,Hz, resulting in a $\geq 95\%$
livetime during operation \cite{glue:gorham}.

The reported analysis incorporates a livetime of approximately $120$ hours
which, although quite small, employs up to
$0.1$\,Mkm$^3$ of detector material, which is enough to constrain some
of the aforementioned Z-Burst and TD models. No neutrino candidate
events are present in the $120$ hours data set \cite{glue:gorham}.
\subsection{LOFAR/LOPES}
\label{subsec:lofar}
LOFAR, the LOw Frequency ARray, including the LOFAR PrototypE Station
(LOPES), is a large scale radio telescope, being built for a broad range of
astrophysical studies. One goal is the detection of
CRs and neutrinos through geo-synchrotron emission and radio
fluorescence in the Earth's atmosphere and the Askaryan effect in
the terrestrial and lunar regoliths \cite{lofarlopes}.
The operable range of frequencies is from $15 \to 240$\,MHz.
The cleanest channel for direct neutrino detection is through
{\v C}erenkov emission in the lunar regolith \cite{lofar:falcke} in
the same way as illustrated by Figure \ref{fig:glueschem}. It is
predicted that the efficiency of the technique should allow for detection
of neutrinos an order of magnitude below predicted GZK fluxes, after only
$30$\,days observation at a signal detection threshold of 
$500 \times 10^{-26}\,\mathrm{Wm}^{-2}\mathrm{Hz}^{-1} $
\cite{lofar:scholten}.
%
%
%
%
\subsection{RICE}
\label{subsec:rice}
RICE is the Radio Ice Cherenkov Experiment, at the South Pole. It was
deployed in the Antarctic ice in tandem with AMANDA
(section \ref{subsec:amanda}) and consists of an $18$ channel array of radio
receivers distributed about a cube of length $200$\,m. The depth of each
receiver varies between $100-300$\,m. Each of the dipole antennas has a
dedicated, in-ice pre-amplifier before the signal transmission along a
cable to the surface where background noise from local
electronics and more importantly the AMANDA PMTs is filtered
away. Event triggering initially requires that at least four channels
register a peak above a threshold set beyond the level of thermal and
background fluctuations, or that any channel registers a peak above
threshold in coincidence with a $30-$fold AMANDA-B trigger. A horn
antenna at the surface provides an active veto against surface
generated background transient events. Raw trigger rates of $10$\,Hz are
reported \cite{rice:kravchenko} after veto. There is a reported
livetime of $\sim 90\%$ after discrimination of surface transients of
anthropogenic origin. Event vertices are reconstructed via the
difference in arrival times of signals at successive pairs as well as
a lattice interpolation algorithm; a {\v C}erenkov cone of opening
angle $57$\,$\degrees$ is then superimposed on the vertex. A location
accuracy of $\sim 5$\,m is reported. Data analysis spanning records from
1999-2005 has been performed \cite{ricelimits:kravchenko} yielding
no events consistent with a neutrino induced cascade.
\subsection{SALSA}
\label{subsec:salsa}
The SALt dome Shower Array (SALSA) collaboration represent the
greatest interest in the instrumentation of subterranean salt domes
with radio antennae for the detection of ultra-high energy, neutrino
induced particle cascades. To date preliminary investigations of
United States sites have been undertaken as a means to surveying the
suitability of naturally occurring halite deposits for use as
calorimeters. Indications are that in terms of
enclosed mass within one attenuation length radius, salt domes provide
a competitive material to ice. Furthermore, the
simple cubic-lattice structure of halite is potentially less causative
of distortion through depolarisation and rotation of the signal about
the plane of polarisation. Additionally halite is less birefringent
than ice too. Results from a Monte Carlo (MC) simulation of an
instrumented salt dome have been reported  \cite{salsa:gorham}. A
sensitivity of the order $10$ events per year from a minimal GZK flux
is expected from a $10 \times 10 \times 10$ antenna array on a
$200$\,m grid spacing. Perhaps more important is the recent
experimental work that has confirmed the generation of coherent
Askaryan radio pulses, in salt, as a result of a net charge excess,
synthesised during the development of particle
cascades \cite{askaryanobservation:saltzberg}. As discussed at the
beginning of this section, this work clearly upholds the production of
radio pulses via the Askaryan effect and refutes the hypothesis of
geometric charge separation as their source.
\subsection{Limits on the neutrino flux from radio {\v C}erenkov\\
  neutrino telescopes}
\label{subsec:radiolimits}
\begin{figure}[h!]
  \centering
  \includegraphics[angle=90, width=0.75\textwidth]{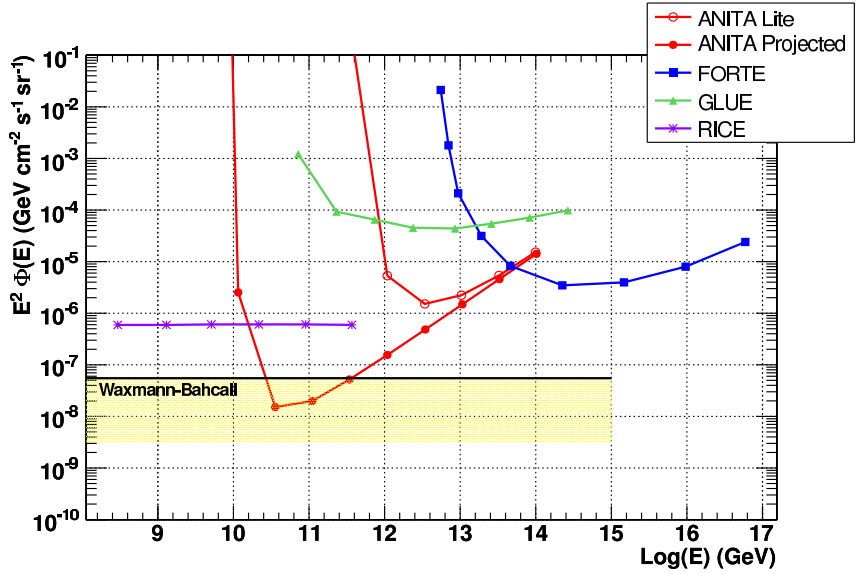}
  \caption[Limits on the neutrino flux from radio telescopes]
      {
	\small
	Constraints on the neutrino flux limits from radio {\v
	  C}erenkov detectors. The sensitivity of ANITA Lite is shown
	by the \red{$\circ$} markers and the projected
	sensitivity of the full ANITA detector by the
	\red{$\bullet$} markers
	\cite{anita:barwick}. The FORTE sensitivity is shown by the
	\blue{\scriptsize$\blacksquare$} markers
	\cite{forte:lehtinen}. GLUE \cite{glue:gorham} and RICE
	\cite{ricelimits:kravchenko} are indicated by the
	\green{$\blacktriangle$} and \textcolor{Purple}{$\times$}
	markers respectively. Again the WB flux of neutrinos from
	\cite{amandatoicecube:halzen} is shown.
      }
      \label{fig:radiolimits}
\end{figure}

Current limits on the neutrino flux from radio {\v C}erenkov based
detectors appear in Figure \ref{fig:radiolimits}. Some predicted
limits are also included based on proposed extensions to existing
experiments and from new projects. Plotted on the same figure are some
model neutrino fluxes predicted by theory.
\section{Acoustic Detection of UHE Neutrinos}
\label{sec:acousticdetectors}
Acoustic detection of high energy particles was first postulated in 1957
\cite{acoustic:askaryan} and subsequently full theoretical analyses of
the signal production mechanism were performed
(e.g. \cite{acoustic:learned}). The hadronic 
particle cascade induced at the interaction vertex by an UHE neutrino
has enough thermal energy
to locally heat the surrounding medium, causing it to rapidly expand. This
produces a pressure pulse, measurable on a suitable acoustic receiver.
The minimum neutrino energy required for this 
technique lies around $10$\,EeV ($10^{19}$\,eV). Whilst the
structure of the hadronic showers is the same for all three flavours
of neutrino, the behaviour of the lepton differs.
For $\nu_e$ type neutrinos the
lepton energy is deposited as an electromagnetic cascade that is
effectively detected along with the  hadron shower. In the case of
$\nu_{\mu}$s the resulting muon is virtually undetectable
acoustically because the mean free path for
catastrophic\footnote{meaning nearly all the particle energy is lost
  in one collision} Bremsstrahlung and pair-production is of the order
of several km.  $\nu_{\tau}$ interactions produce a tau lepton which may
or may not decay within the fiducial volume of a detector, initiating
a second cascade. 
\subsection{The LPM effect}
\label{subsec:lpm}
Before discussing the acoustic detection of neutrinos it is necessary
to consider the dynamics of such energetic events.
At EeV energies quantum interference affects the
passage of leptons through a medium. The Landau Pomeranchuck Migdal
(LPM) effect \cite{lpm:landau},\cite{lpm:migdal} describes a
suppression of the cross sections for pair production and
Bremsstrahlung. This effect dominates the development of the leptonic
component of a neutrino DIS event as its interaction lengths become 
comparable to the interatomic distances of the medium through which it
propagates. An EeV electromagnetic cascade that is typically a few
metres in length at TeV energies can extend to hundreds of metres
because of the LPM effect. The discussion of thermoacoustic emission
that follows is limited to the hadronic component of the
neutrino interaction. Any acoustic emission from the electromagnetic
cascade is assumed to be inaudible due to the extended nature of the
source and is thus neglected.
\subsection{Formation of the acoustic signal}
\label{subsec:acousticsignal}
\begin{figure}[ht]
  \begin{center}
    \includegraphics[angle=15, width=0.5\textwidth]{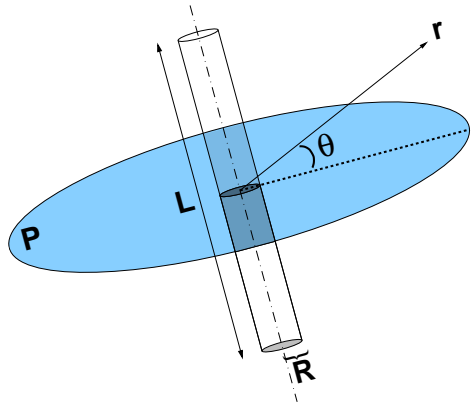}
    \caption[Schematic of the energy deposition]
	    {
	      \small
	      Schematic of the energy deposition. At $E_{\nu} =
	      10^{20}$\,eV, in water, a hadronic shower deposits $99 \%$
	      of its energy in a cylinder of length $L = 20$\,m and
	      radius $R = 20$\,cm. The energy deposition of the shower
	      can be considered as a series of coherent sources lying
	      along the shower axis.
	    }
	\label{fig:pancake}
  \end{center}
\end{figure}

The thermal energy deposition along the cascade axis occurs via
ionisation and excitation of the surrounding medium and is
instantaneous on the acoustic and thermal diffusion timescale.
The cascade can be thought of as a series of discrete regions in which a
Gaussian heat deposition occurs, resulting in an instantaneous step
in temperature.
The corresponding pressure wave is simply the second derivative with
respect to time of the change in temperature, yielding a distinctive
bipolar acoustic pulse.
Because of the instantaneous development of the cascade, the acoustic
radiation is emitted in phase such that individual pulses
interfere. This is analogous to a line of sources emitting Huygens'
wavelets.
In the far field these sources have undergone Fraunhofer diffraction
such that the total acoustic radiation is confined to a narrow,
so-called ``pancake''
with an opening angle $\theta$ of about $5$\,$\degrees$.
The geometry of the cascade and the acoustic emission is illustrated
in Figure \ref{fig:pancake}. 

Following the workings of \cite{acoustic:learned} and
\cite{sensitivity:lehtinen} the acoustic signal production is
analysed. If the energy per unit volume per unit time is given as a
function $E(\bfr,t)$, then the total neutrino energy is $E_0 = \int_V
E(\bfr) d^3 \bfr$. The wave equation for the pressure pulse produced
$p$ is:
\begin{equation}
  \nabla^2 ( p + \frac{1}{\omega_0} \dot{p}) - \frac{1}{c^2} \ddot{p} =
  - \frac{\beta}{C_p} \frac{\partial E}{\partial t}
  \label{eq:waveequation}
\end{equation}
Where for seawater the parameters are:
\begin{description}
\item[$c$,] the speed of sound in water $\sim 1500$\,ms$^{-1}$,
\item[$\beta$,] the bulk coefficient of thermal expansion $\sim
  2.0 \times 10^{-4}$\,K$^{-1}$,
\item[$C_p$,] the specific heat capacity at constant pressure
  $\sim 3.8 \times 10^3$\,Jkg$^{-1}$K$^{-1}$,
\item[$\omega_0$,] the characteristic attenuation frequency
  $\sim 25$\,GHz.
\end{description}
It should be noted that $\omega_0$ is actually a function of the frequency of 
the radiation (it is, in fact, dependent on the sound attenuation
coefficient, which is itself frequency dependent) but to keep these
calculations simple it is taken as a constant for the frequency range
of interest $f = 10 - 100$\,kHz.
One can express the instantaneous nature of the heat deposition by
$E(\bfrp,t) = E(\bfrp) \delta (t)$, where $\bfrp$ is the location
where the instantaneous heat deposition takes place. The resultant
pressure wave some other location $\bfr$ as a function of time $t$ is
therefore:
\begin{equation}
  p(\bfr,t) = \int_V E(\bfrp)G(\bfr - \bfrp,t)d^3\bfrp
  \label{eq:pressurewave}
\end{equation}
$G(\bfr,t)$ is the pressure pulse generated by a point source
$E = \delta (\bfr) \delta (t)$, taking attenuation into account this is:
\begin{equation}
  G(\bfr,t) = - \frac{\beta}{4\pi C_p} \frac{(t-r/c)}{r \sqrt{2\pi}\tau^3}
  e^{-(t-r/c)^2 / (2\tau^2)}
  \label{eq:pressurepulse}
\end{equation}
where $\tau=\sqrt{r/(\omega_0c)}$.
\subsection{Laboratory based measurements of thermoacoustic\\emission}
\label{subsec:laboratorymeasurements}
Since the thermoacoustic mechanism for detection of particle cascades
was proposed \cite{acoustic:learned}
a number of experiments have attempted to measure the
effect in the laboratory, either through the use of particle
accelerator beams or high energy light sources. A discussion of two
such experiments follows.
\subsubsection{The 1979 Brookhaven-Harvard experiments}
\label{subsubsec:brookhaven}
Some twenty-two years after Askaryan's proposal of the thermoacoustic
mechanism for high energy particle detection, the first experiments
were undertaken at the Brookhaven National Laboratory (BNL) and
Harvard University cyclotron, to record the acoustic signal emitted due
to proton beams traversing fluid media. The following summary is based
on the material reported in \cite{brookhaven:sulak}. Three
experimental apparatus were used: 
\begin{enumerate}
\item
  The BNL $200$\,MeV proton linear accelerator (LINAC) produced a
  beam of protons with total bunch energies between $10^{19} \to
  10^{21}$\,eV. The range of the beam in water was $\sim 30$\,cm, with
  a fixed   $4.5$\,cm diameter. The spill time was variable between $3
  \to 200$\,\textmu s. The energy of the beam was not tunable.
\item
  The Harvard University cyclotron accelerated protons up to $158$\,MeV
  with a beam energy that could be tuned down to $10^{15}$\,eV (close
  to threshold for the acoustic signal). The beam had a range of
  $16$\,cm in water with a minimum $50$\,\textmu s deposition time, which
  was long in comparison to the sound transit time across the diameter of
  the beam and thus dominated the temporal structure of the emitted acoustic
  signal. The beam diameter could be varied between $1 \to 10$\,cm.
\item
  A second fast extracted beam (FEB) from a $28$\,GeV proton
  accelerator was utilised at BNL. This had a fixed beam energy of
  $10^{19}$\,eV, with a range of $20$\,cm. The diameter of the beam was
  variable between $5 \to 20$\,mm and the deposition time was short at
  $2$\,\textmu s. In contrast to the cyclotron experiment, where the
  beam spill time was the dominant effect, the sound transit time
  across the beam diameter dominated the temporal structure of the signal.
\end{enumerate}
Each of the above apparatus had sufficient dimension to allow for
resolution of the initial signal and subsequent reflections (much
greater than the beam diameter or cascade length). The primary source
of error was due to the uncertainty in the measurement of signal
amplitudes and beam intensities. It is reported that audible clicks
could be heard by the unaided ear, but this in itself did not prove
the thermoacoustic mechanism. It still remained to disprove acoustic
emission due to molecular dissociation and microbubble formation.

\begin{figure}[htbp]
  \centering
  \begin{minipage}[h!]{\textwidth}
    \centering
    \includegraphics[width=0.5\textwidth]{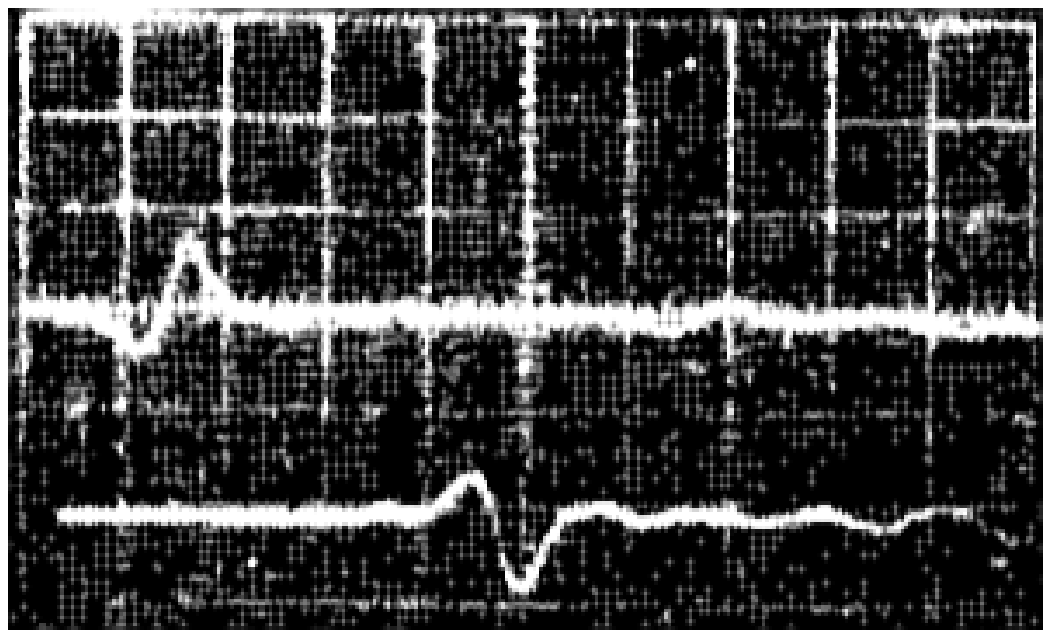}
    \caption[Measured acoustic pulse from a fast proton beam]
	    {
	      \small
	      The pressure pulse observed from a $10$\,\textmu s proton beam
	      spill (from \cite{brookhaven:sulak}).
	      The hydrophone producing the lower trace is
	      $180$\,$\degrees$ out of phase with the upper one.	    
	    }
	    \label{fig:protonbeams1}
  \end{minipage}
  \hfill
  \begin{minipage}[h!]{\textwidth}
    \centering
    \includegraphics[width=0.5\textwidth]{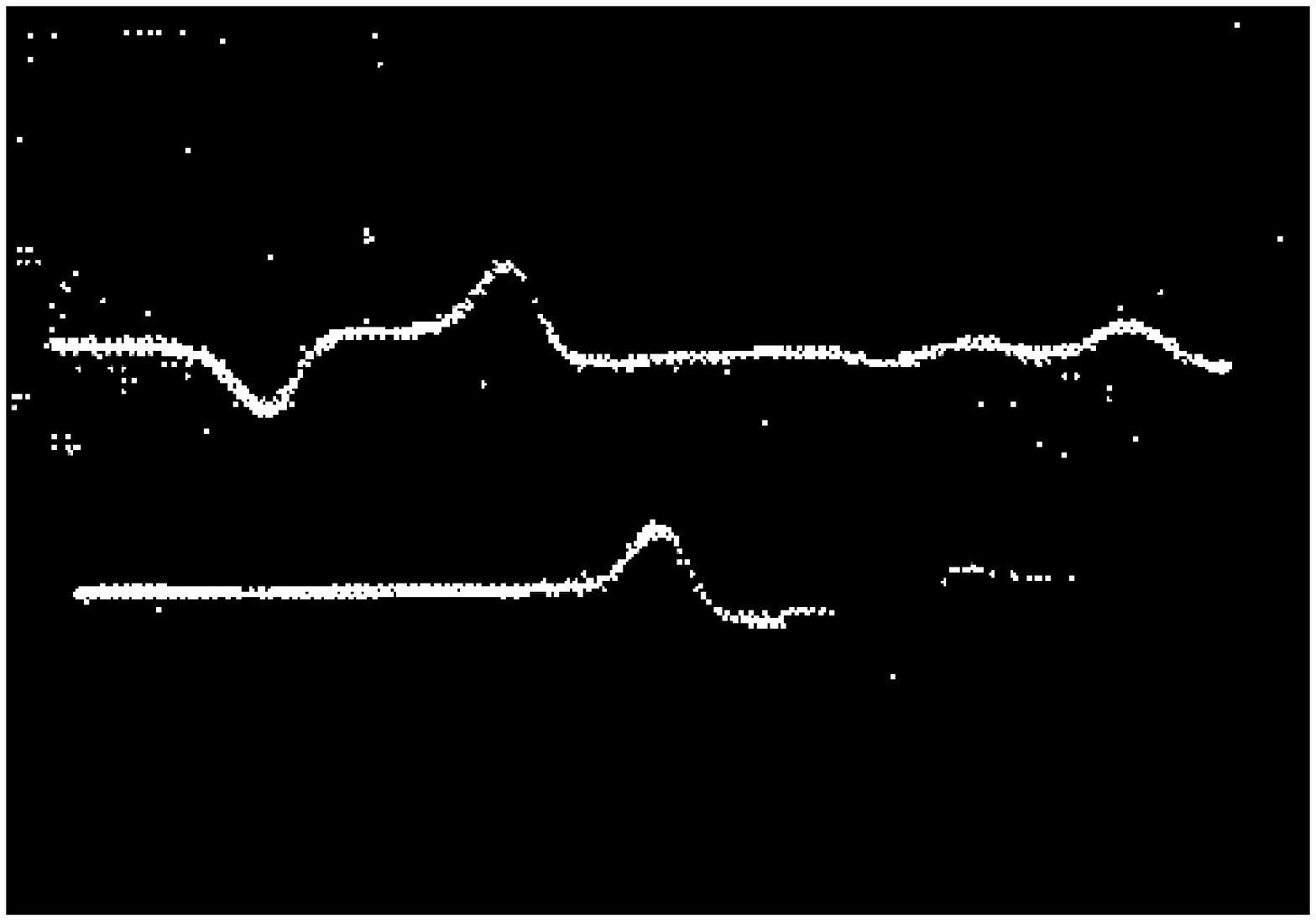}
    \caption[Measured acoustic pulse from a slow proton beam]
	    {
	      \small
	      The pressure pulse observed from a $100$\,\textmu s proton beam
	      spill (from \cite{brookhaven:sulak}).
	      The hydrophone producing the lower trace is
	      $180$\,$\degrees$ out of phase with the upper one.	    
	    }
	    \label{fig:protonbeams2}
  \end{minipage}
  \hfill
  \begin{minipage}[h!]{\textwidth}
    \centering
    \includegraphics[width=0.5\textwidth]{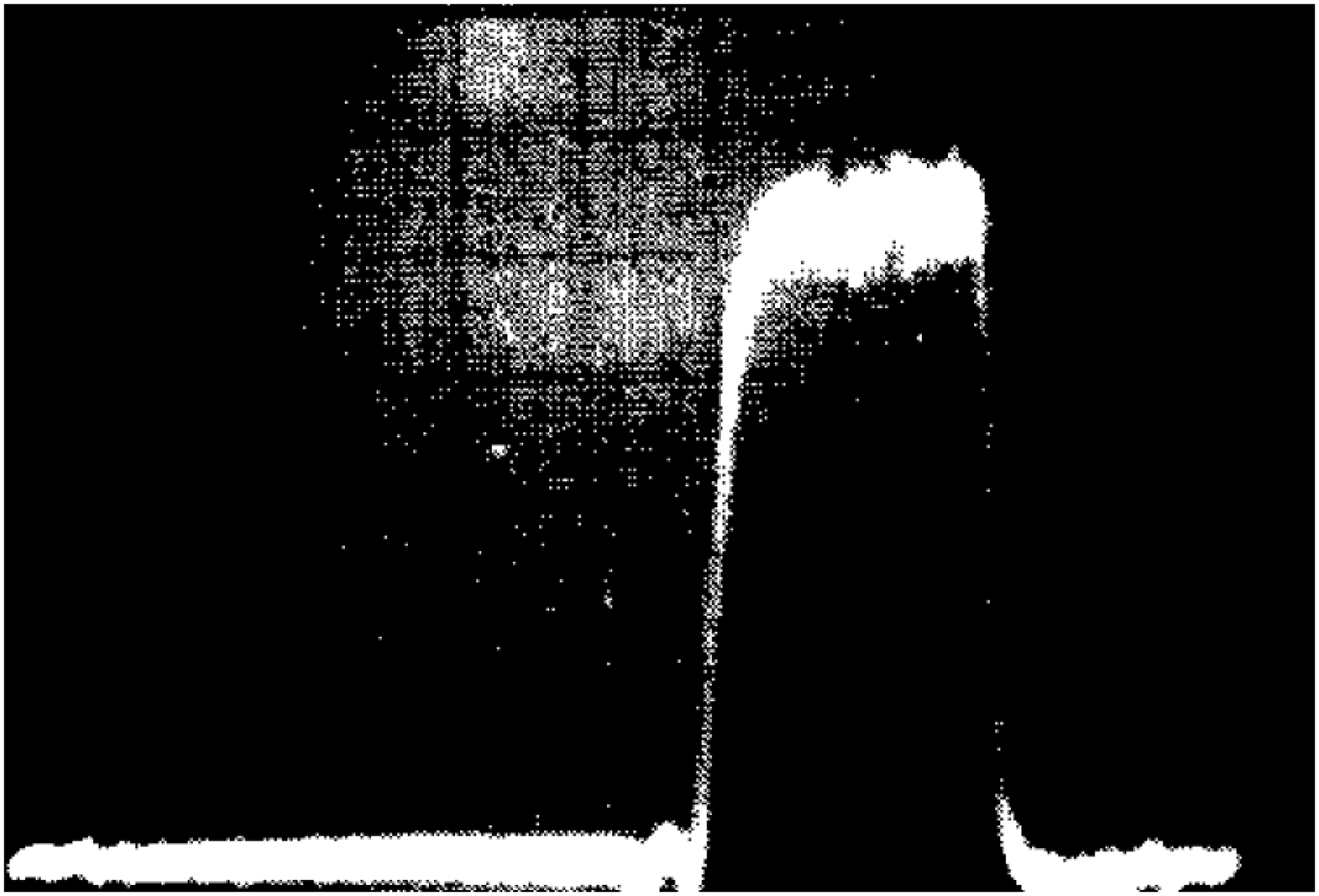}
    \caption[Time profile of $100$\,\textmu s beam intensity]
	    {
	      \small
	      Time profile of $100$\,\textmu s beam intensity
	      (from \cite{brookhaven:sulak}).
	    }
	    \label{fig:protonbeams3}
  \end{minipage}
  \hfill
\end{figure}

In the case of a short beam spill time ($10$\,\textmu s), where the signal
transit time was $d/c = 4.5$\,cm/$1.5$\,mm \textmu s$^{-1}=30$\,\textmu s
in water at $20$\,$\degrees$C, a clear bipolar signal was registered, this
is plotted in Figure {\ref{fig:protonbeams1}}.
The separation between this and the subsequent first reflection
displayed the appropriate delay for the given geometry. The half width
of the pulse was directly proportional to the transit time across the
diameter of the beam and the time delay between hydrophones at two
different locations was consistent with their respective path
lengths. A leading compression was confirmed by placing two
hydrophones $180$\,$\degrees$ out of phase. 
When the beam spill time was significantly longer ($100$\,\textmu s) than
the signal transit time ($30$\,\textmu s) then there was a clear separation
between the leading compression and the subsequent, final
rarefaction. The observed signal (Figure {\ref{fig:protonbeams2}}) is
calculated as the convolution of a quasi-instantaneous pulse
(Figure {\ref{fig:protonbeams1}}), and the beam spill intensity
(Figure {\ref{fig:protonbeams3}}) as a function of time.

The signal period as a function of spill time served to act as a proof
of principle for the thermoacoustic mechanism and its underlying
linear theory. The level of agreement was limited to $2\sigma$
($85\%$) due to the large errors imposed by the uncertainties in the
measurements of the beam intensities and pulse heights.

Repeated tests were performed with different media such as olive oil
and CCl$_4$ to look for deviations from the linear models as a result
of excitation of ions and/or molecular dissociation. 
Microbubble formation was also examined by using the different media
under varying combinations of temperature and pressure (so as to alter
the microbubble diameters). No deviation was observed as a result of
either ion excitation, molecular dissociation or microbubble formation
within the $25\%$ experimental uncertainty. Furthermore the spatial
dependence of signal amplitudes was shown to vary as the reciprocal of
the distance away from the cascade as anticipated by the model of
linear superposition in the far-field condition.

The aforementioned experiment serves as a proof of principle for the
thermoacoustic model. There were large inherent uncertainties to the
technique used for measuring the thermoacoustic emission from each of
the accelerator beams used, constraining the agreement between the
observed data and the predictions of the underlying theory only to
within $2\sigma$.
\subsubsection{Uppsala proton beam and Erlangen laser beam studies}
\label{subsubsec:uppsala-erlangen}
The most recent attempt to test the predictions of the thermoacoustic
mechanism \cite{uppsala-erlangen:graf} have utilised both a high
energy ($10 \to 400$\,PeV total bunch energy) proton beam and an EeV
pulsed infra-red ($1064$\,nm) Nd:YAG laser. The proton beam, being
charged, both ionises and excites the target medium but the laser beam,
being neutral, will only excite target atoms. The density of pure
water is maximal at a temperature of $4$\,$\degrees$C; below this the
coefficient of thermal expansion ($\beta$ in Equation
\ref{eq:waveequation}) is negative. Hence an increase in temperature
in pure water below $4$\,$\degrees$C produces a compression and not an
expansion. The coefficient for sea water is greater than that for pure
water and increases with increasing pressure, temperature and salinity
\cite{oceanphysics}.
In this experiment the temperature of the target water tank
was varied between $1$\,$\degrees$C and $20.0$\,$\degrees$C $\pm
0.1$\,$\degrees$C. The proton beam, delivered from the $177$\,MeV
cyclotron at the Theodor Svedberg Laboratory in Uppsala, Sweden,
had a diameter of $1$\,cm and a spill time of $30$\,\textmu s. It
produced a shower with approximately uniform energy deposition in
water for $20$\,cm, ending in a Bragg Peak at approximately
$22$\,cm. The Erlangen Physics Institute laser energy was adjusted
between $0.1 \to 10$\,EeV with a fixed pulse length of $9$\,ns and
beam diameter of a few mm. The energy deposition of the beam, along
the beam axis, decayed exponentially with an absorption length of $6.0
\pm 0.1$\,cm. 

Acoustic pulse data from the laser produced the expected inversion,
predicted by the thermoacoustic model, at $4.0$\,$\degrees$C to within
$0.1$\,$\degrees$C. The same measurement using the proton beam however
revealed an apparently temperature independent contribution to the
overall amplitude, present at a level of $5\%$ of the signal at
$15$\,$\degrees$C. The results predicted by the thermoacoustic mechanism
were obtained following subtraction of this artifact from the proton
beam data. The source of this contribution to the acoustic signal was
not verified and was identified as a topic for further
investigation. A hypothesis was proposed that the extra signal was a
result of the charge of the proton beam, hence it was not apparent in
the laser induced pressure waves.

Whilst the results of the Erlangen-Uppsala tests still leave some
unanswered questions, they appear to strengthen the support of the
thermoacoustic model. It still however remains to identify the extra
source of sound apparently resulting from the charged nature of the
proton beam.
\subsection{Experimental results from SAUND}
\label{subsec:currentlimits}
The SAUND collaboration \cite{website:saund} have published
\cite{saund:vandenbroucke} the first diffuse
neutrino flux limit based on their work on a United States Navy hydrophone
array located in the Tongue of the Ocean, a deep basin situated near the 
Bahamas. They use a seven hydrophone subset of the total $52$
hydrophone AUTEC array,
arranged in a hexagonal pattern at depths between $1570$\,m and
$1600$\,m. The spacing between hydrophones is between $1.50$\,km and $1.57$\,km
and they are mounted upon vertically standing booms that are anchored to
the seabed.
\begin{figure}[h]
  \begin{center}
    \includegraphics[width=0.5\textwidth]{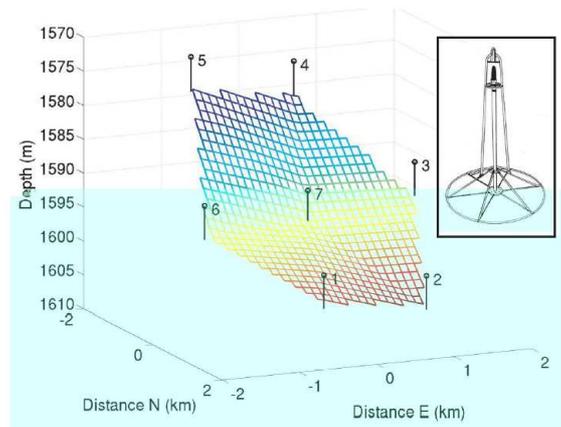}
    \caption[SAUND schematic]
	    {
	      \small
	      Schematic view of the hydrophone array used by the SAUND 
	      collaboration (from \cite{saund:vandenbroucke}).
	    }
	    \label{fig:saundschematic}
  \end{center}
\end{figure}
\begin{figure}[h]
  \begin{center}
    \includegraphics[width=0.85\textwidth]{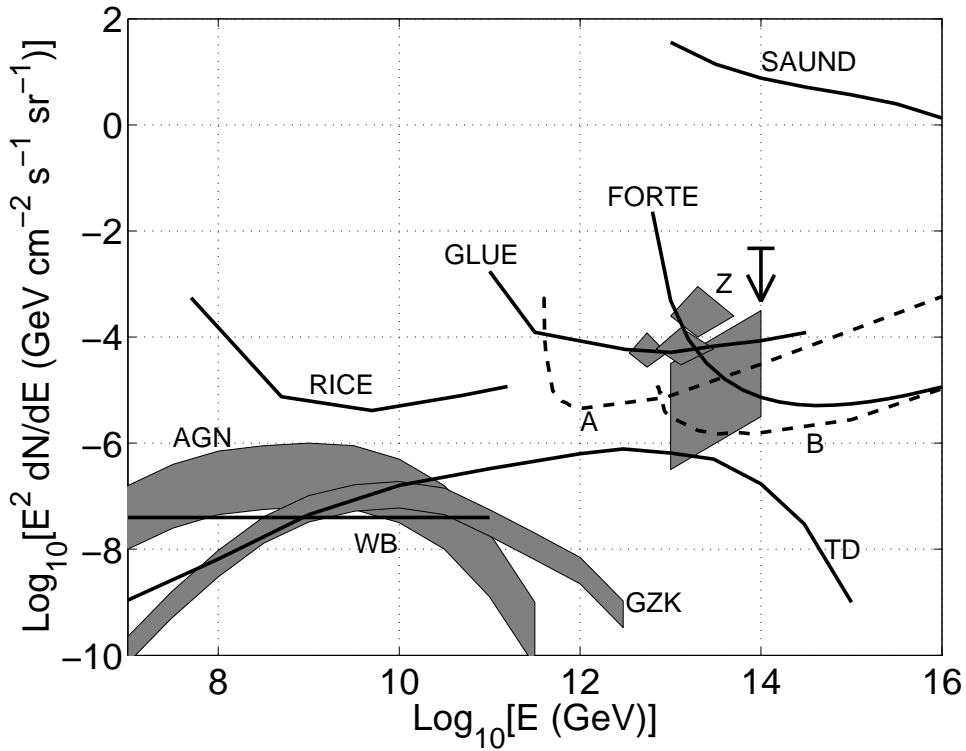}
    \caption[Limits on the neutrino flux from SAUND]
	    {
	      \small
	      Flux limit curves for the AUTEC array (solid line labelled SAUND)
	      and the hypothetical arrays (A and B) including the predicted
	      flux of several theoretical models: ``Z'' for Z-Bursts, ``TD''for
	      Topological Defects, ``WB'' for the Waxman and Bahcall flux and
	      ``GZK'' for the GZK theoretical flux. Some experimental limits 
	      (FORTE, GLUE 
	      and RICE) are also indicated (from \cite{saund:vandenbroucke}).
	    }
	    \label{fig:saundresults}
  \end{center}
\end{figure}

A novel technique was employed by which calibration of the hydrophones
was achieved by dropping weighted light bulbs into the sea from a
stationary boat at the surface. Measuring the signal as the bulb
imploded under pressure gave a rough estimation of the energy
sensitivity of the array to a known source and allowed for timing
calibration.

In addition to the AUTEC array performance evaluation a computer
simulation was developed to test the sensitivity of two hypothetical arrays.
Each was comprised of hexagonal lattices of 1.5km long strings, the strings
being modelled to have continuous pressure sensitivity along their
entire length. Array ``A'' is bounded by circle of radius 5km with
500m nearest neighbour spacing. Array ``B'' is bounded by a circle of
radius 50km, with 5km spacing between strings. The results of this
work are illustrated in Fig \ref{fig:saundresults}, the next phase of
this experiment ``SAUND II'' is currently underway.
\section{Practical Motivation for the Acoustic \mbox{Technique}}
\label{sec:practical}
The introduction hints why an astronomer would look for neutrino
light: neutrinos interact only via the weak nuclear and gravitational
forces. This means they can reach us from the furthest depths of the
Cosmos without being deflected, illuminating parts of the universe
no other particles can. The very fact that CRs exist with energies
greater than $10^{20}$\,eV is the greatest motivation for UHE neutrino
astronomy, since there must be a neutrino counterpart to such
emission. One must however convince oneself that acoustic neutrino
detection can compete with existing methods and support the detection
of neutrinos at the highest energies.
Table \ref{tab:attenuationlengths} shows
a comparison of signal attenuation lengths for different methods of detection.
\begin{table}[htb]
\centering
\small{\begin{tabular}{ l l | c | c | c }
&               & {\bf Water}  & {\bf Ice}    & {\bf Salt}  \\
\hline
{\bf EM Optical}&({\v C}erenkov) &$\sim 50$\,m &$\sim 100$\,m &$\sim 0$\,m  \\
\hline
{\bf EM Radio}  &(0.1$\to$1.0GHz)&$\sim 0$\,m  &$\sim$ few km&$\sim 1$\,km  \\
\hline
{\bf Acoustic}  &(10kHz)&$\gtrsim 10$\,km&$\gtrsim 10$\,km&$\gtrsim 10$\,km \\
\end{tabular}}
\caption[Attenuation lengths for different techniques]
	  {\small Attenuation lengths for different techniques
	  \label{tab:attenuationlengths}}
\end{table}
Acoustic detection methods can offer the largest instrumented and
effective volumes since the signal propagation lengths are orders of
magnitude greater than for {\v C}erenkov detectors.
\subsection{``Back of the envelope'' comparison of effective volumes for an optical neutrino telescope and an acoustic telescope with 1000 sensors}
\label{subsec:comparison}
\begin{figure}[htb]
  \centering
  \includegraphics[width=0.6\textwidth]{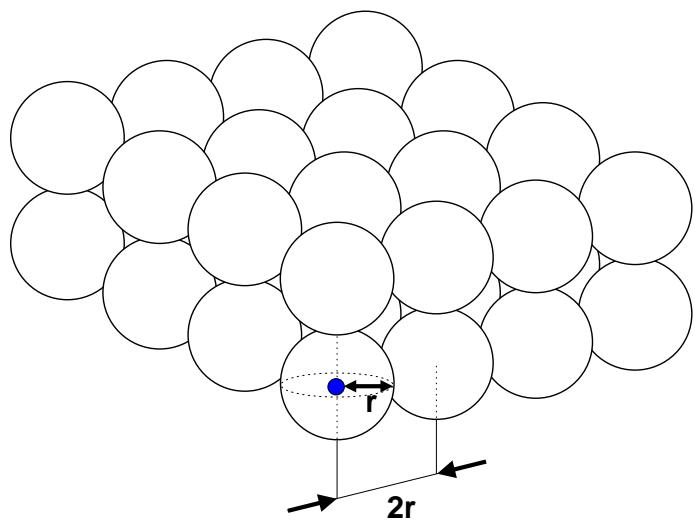}
  \caption[Simple cubic lattice]
	  {
	    \small
	    The instrumented volume is divided equally amongst each
	    sensor. In a simple cubic lattice each sensor sits at the
	    centre of a sphere of radius $\mathbf{r}$.
	  }
	  \label{fig:simplecubic}
\end{figure}

The maximal effective volume $V_{\mathrm{eff}}$ of ANTARES to $\nu_{\mu} +
\overline{\nu_{\mu}}$ is $\sim 1$\,km$^3$ \cite{antares:effective}.
ANTARES comprises approximately $1000$ sensors, instrumented in a volume of
$0.03$\,km$^3$.
If one assumes a regular spacing between sensors then the volume
occupied per sensor is approximately
$0.03/1000 = 3\times10^{-5}$\,km$^3$/sensor.
The simplest arrangement of these volumes is a simple cubic lattice as
illustrated in Figure \ref{fig:simplecubic}.

This gives a separation
between sensors of:
\begin{eqnarray*}
  \frac{4}{3}\pi r^3 &=&  3\times10^{-5} \\
                   r &=& 0.0192 \,\mathrm{km} \\
                  2r &=& \sim 40\,\mathrm{m}
\end{eqnarray*}
which corresponds to $80\%$ of the $\sim 50$\,m attenuation length of
$450$\,nm {\v C}erenkov light.

The most conservative estimate for $V_{\mathrm{eff}}$ of an acoustic array is
to assume that $V_{\mathrm{eff}} = V_{\mathrm{inst}}$ where
$V_{\mathrm{inst}}$ is the volume instrumented by the detector. 
If one populates a volume
equivalent to $1000$ acoustic sensors separated by $80\%$ of an again
conservative $10$\,km attenuation length for sound:
\begin{eqnarray*}
  \frac{4}{3}\pi r^3 &=& \frac{V_{\mathrm{eff}}}{1000} \\
  V_{\mathrm{eff}} &=& 1000 \times \frac{4}{3} \pi (8/2)^3 \\
  V_{\mathrm{eff}} &=& 85333\mathrm{km}^3
\end{eqnarray*}

Remember, we have already been conservative in our approach, and have
shown that if $V_{\mathrm{eff}} = V_{\mathrm{inst}}$ for an acoustic
sensor array, it can have an effective volume $\sim 85000$ times greater
than an optical array, with the same number of sensors. It has, however,
been suggested that an acoustic array can experience
$V_{\mathrm{eff}}$ up to one hundred times greater than
{$V_{\mathrm{inst}}$} \cite{timo:thesis}, which would then give a $85
\times 10^5$ times larger effective volume for an acoustic array over
an optical.

\begin{description}
\item[caveat]
      The Optical Modules in ANTARES are not spaced on a regular cubic
      lattice, the exact geometry is complex and is dependent on
      reconstruction requirements, but since the reconstruction
      requirements of our hypothetical acoustic array are unknown, we give
      both detectors an equivalent geometry and an equivalent sensor
      separation as a function of signal attenuation. Similarly
      positioning acoustic sensors with a separation of $10$\,km, just
      because of the long attenuation lengths of sound, neglects any
      requirement for reconstruction.
\end{description}

In conclusion, an acoustic array can potentially have an effective
volume of the order $85 \times 10^3 \to 85 \times 10^5$
times greater than an optical array with an equivalent number of
sensor elements, neglecting requirements for reconstruction. In the
following chapters, for historical reasons, typically one thousand
acoustic sensors are considered in a volume of one cubic kilometre, a
considerably denser population than argued here. However, as the
discussion progresses a sparser optimum sensor density, taking into
consideration reconstruction requirements will be considered.

\section{Summary}
\label{sec:chp2summary}
\begin{figure}[htb]
  \includegraphics[width=0.85\textwidth]{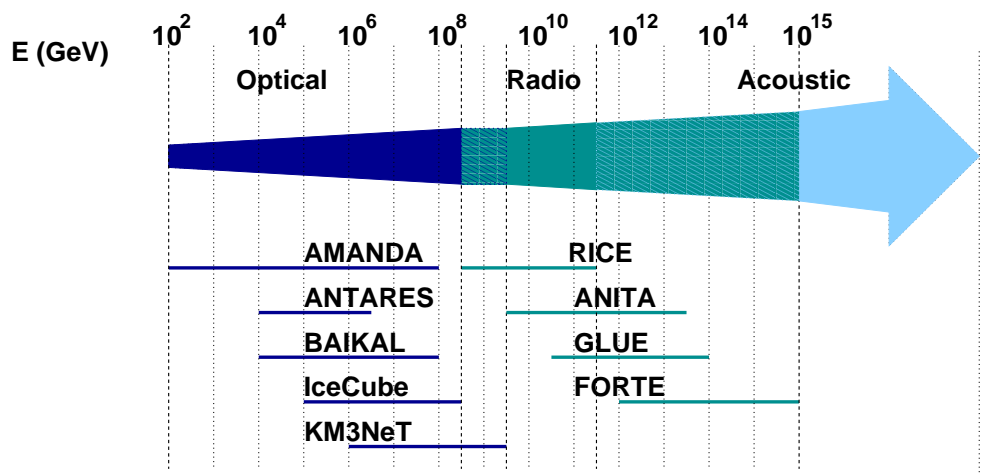}
  \caption[Energy windows for neutrino detection techniques]
  {
    \small
    The range of neutrino energies detectable by each technique
    illustrated in terms of the existing and projected sensitivities
    of several neutrino detection experiments.
  }
  \label{fig:energyranges}
\end{figure}

Three methods of detecting cosmological neutrinos have been
discussed. Each one is at a different level of maturity yet all are
proving fruitful. Figure \ref{fig:energyranges} incorporates the
present and predicted sensitivities from Figures
\ref{fig:opticallimits} and \ref{fig:radiolimits} into an
illustration of the energy ranges for each method of
detection. The expected energy range of acoustic detection is also
included. The case for the development of acoustic detection
lies firstly in its sensitivity to the highest energy neutrinos, thus
completing the study of the entire neutrino energy spectrum, and secondly, the
ability of the acoustic technique to deliver detectors with vast
effective volumes.

  \cleardoublepage

  \chapter{Simulating Neutrino Interactions}
\label{chp:neutrino}
\section{Introduction}
\label{sec:nuintro}
This chapter deals with the simulation of neutrinos colliding into
the Earth at energies of several Joules. It describes the first links
in the simulation chain that ultimately predicts the sensitivity of a
hypothetical array of underwater acoustic sensors to a flux of ultra
high energy neutrinos. The first section discusses the underlying
event - the neutrino deep inelastic scatter, the second section
describes simulation of the induced particle shower and the final
section describes how the acoustic signal is generated as a result of
the thermal energy deposition of the hadronic cascade.

\section{Neutrino Event Generation} 
\label{sec:nuevent}
%
The kinematics of neutrino deep inelastic scattering were
introduced in the opening paragraphs of Chapter
\ref{chp:detection}. In order to determine the contribution of the
neutrino energy to the induced hadronic cascade it is necessary to
compute the Bjorken-$y$ dimensionless scaling variable. Physically
$y$ describes the fraction of the incident neutrino energy that is
carried away by the hadronic system.
%

\begin{figure}[htb]
  \centering
  \hfill
  \begin{minipage}[h!]{0.47\textwidth}
    {\includegraphics[width=\textwidth]{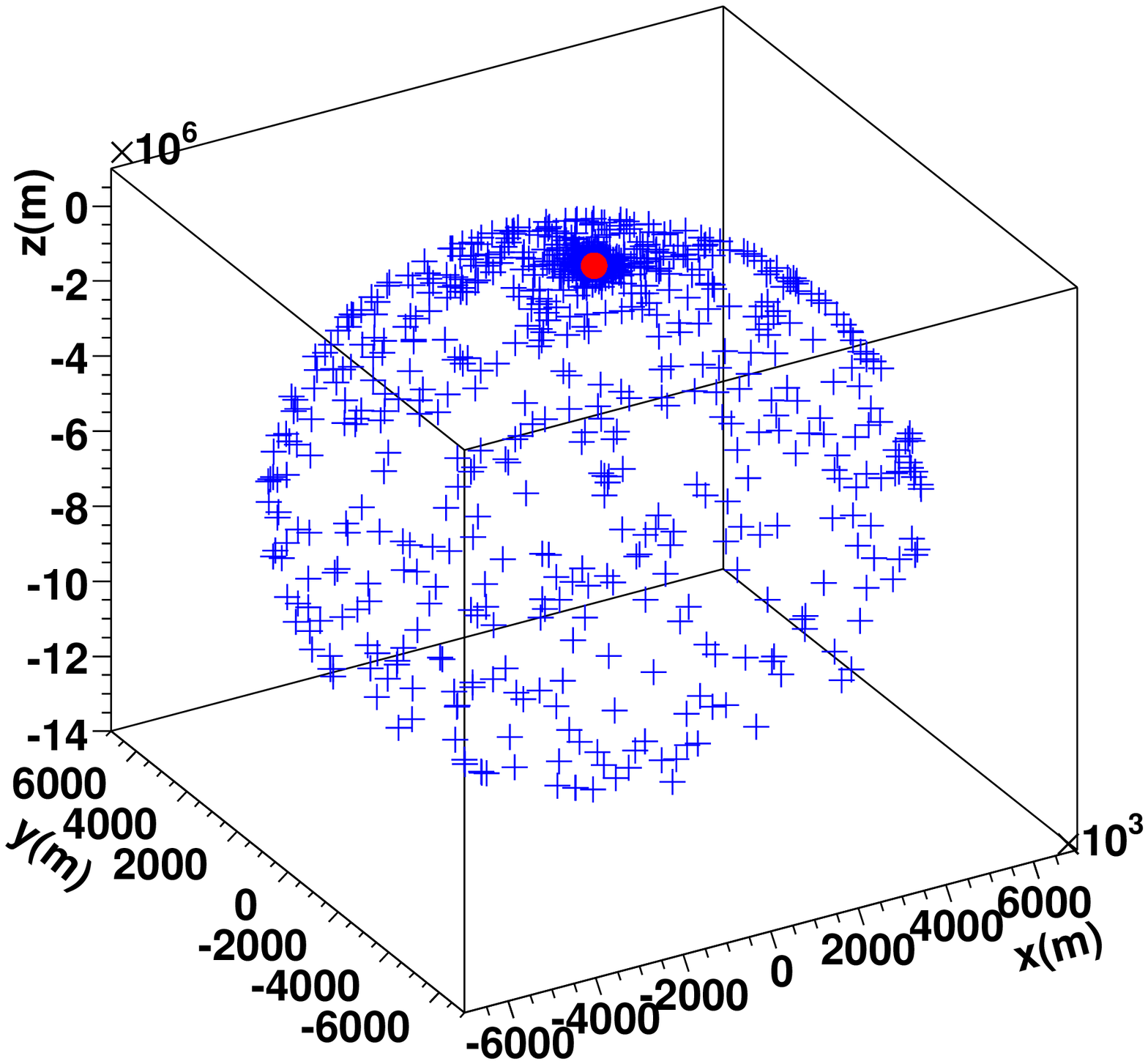}}
  \end{minipage}
  \hfill
  \begin{minipage}[h!]{0.47\textwidth}
    {\includegraphics[width=\textwidth]{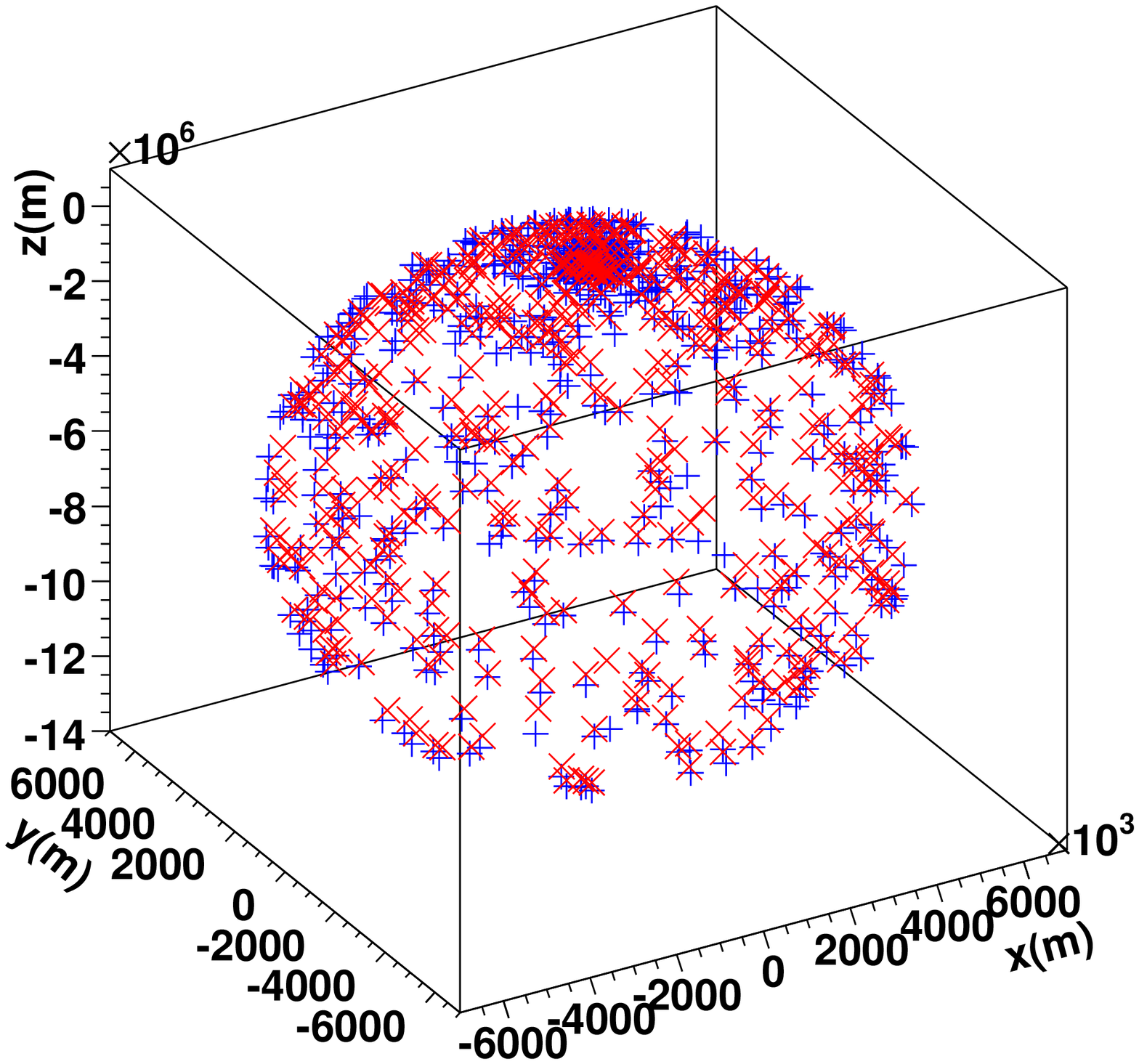}}
  \end{minipage}
  \hfill
  \caption[Propagation of neutrinos through the Earth's crust]
	  {
	\small
	Low energy neutrinos ({\it Left}) propagate
	freely from the surface of the Earth (\blue{$+$}
	markers), through the crust to the fiducial volume
	(\red{$\bullet$} marker)
	where they are forced to interact. High energy neutrinos
	({\it Right}) do not travel far from their point
	of origin (\blue{$+$} markers) before they are attenuated by
	the crust (\red{$\times$} markers). This
	demonstrates the increase in the neutrino nucleon cross
	section with neutrino energy.
      }
      \label{fig:anisattenuationcrust}
\end{figure}

\begin{figure}[htbp]
  \centering
  \includegraphics[width=0.8\textwidth]{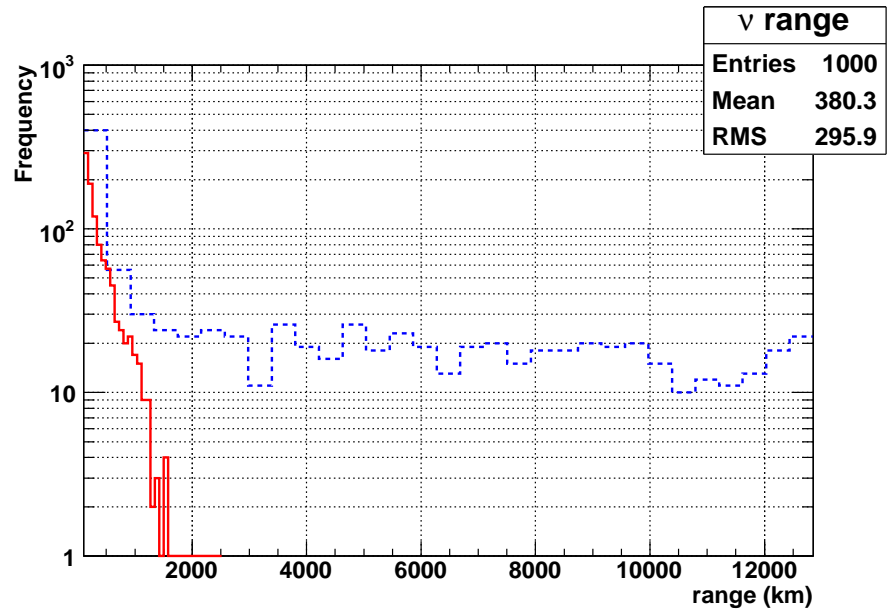}
  \caption[Neutrino range in Earth's crust]
	  {
	    \small
	    Range of neutrinos in ANIS at $1 < E_{\nu} < 10^3$\,GeV
	    (\blue{dotted}) 
	    and $10^9 < E_{\nu} < 10^{12}$\,GeV
	    (\red{solid}). The low energy
	    neutrinos can penetrate up to the full diameter of the Earth
	    $\sim 12000$\,km before they are forced to interact, whereas
	    high energy neutrinos are attenuated by about $2000$\,km of rock.
	  }
	  \label{fig:rangecombined}
\end{figure}

The All Neutrino Interaction Simulation (ANIS) \cite{anis:gazizov} has
been developed for analysis of data from the AMANDA \cite{amanda:andres}
neutrino telescope. It is a fully object oriented C++ toolkit,
utilising the CLHEP vector class and HepMC event records \cite{clhep}.
The program
generates neutrinos of any flavour according to a specified flux and
propagates them through the Earth eventually forcing them to interact
in a specified volume should they not be attenuated en route. The
propagation of low energy {($< 10^3$\,GeV)} neutrinos through the
Earth's crust and the attenuation of neutrinos at high {($> 10^9$\,GeV)}
energies is plotted in Figure \ref{fig:anisattenuationcrust}. The
range of neutrinos, through the Earth's crust, at low and high energy
is plotted in Figure \ref{fig:rangecombined}.
%

It is assumed that all neutrinos interact in a fiducial volume, a
cylindrical ``can'' surrounding the volume instrumented by
hydrophones. The energy spectrum is assumed to be flat in $\log E$,
hence  $dN/dE \propto E^{-1}$. Only neutrinos originating from a
positive hemispherical shell (with its
origin at the centre of the instrumented volume) are considered. Hence
the ANIS program need only propagate neutrinos through the
detection medium (water) and not through the Earth's crust.

%
%

\begin{figure}[htbp]
  \centering
  \hfill
  \subfigure[Energy dependence]
	    {
	      \includegraphics[width=0.45\textwidth]{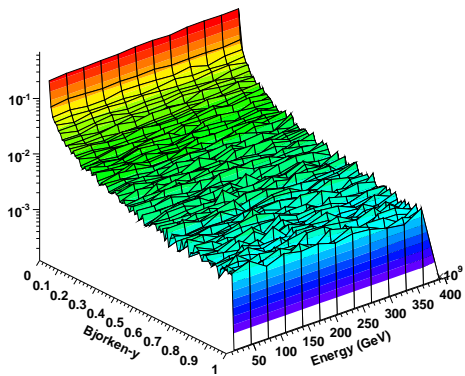}
	      \label{fig:anis:a}
	    }
  \hfill
  \subfigure[Bjorken-$y$ probability density for all energies]
	    {
	      \includegraphics[width=0.45\textwidth]{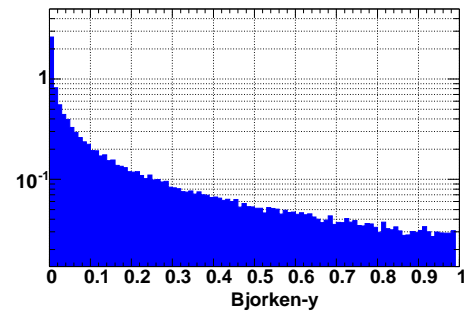}
	      \label{fig:anis:b}
	    }
  \hfill
  
  \caption[Three flavour Bjorken-$y$ distribution]
	  {
	    \small
	    Bjorken-$y$ probability density for three flavours of
	    neutrino in the energy range $10^8 \to 10^{12}$\,GeV,
	    interacting in water. As indicated by plot
	    \ref{fig:anis:a} the shape of the probability density is
	    independent of the neutrino energy. Plot \ref{fig:anis:b}
	    is the three flavour $y$-distribution averaged over all
	    energies between $10^8 \to 10^{12}$\,GeV (i.e. a
	    projection of plot \ref{fig:anis:a} along the energy axis).
	  }
	  \label{fig:anis}
	    
\end{figure}

The mean fraction of the neutrino energy imparted to the hadronic
cascade is approximately $25\%$, however, the actual value can vary
from $0-100\%$. This is shown in Figure \ref{fig:anis}, and is assumed
to be independent of neutrino energy. 
Initially in this simulation all hadronic cascades are
assumed to contain $25\%$ of the energy of the incident neutrino,
since vertex reconstruction relies on the timing and not the magnitude
of a pressure pulse. The effect of a non constant $y$-value will be
discussed in Section \ref{sec:energyrecon} later when describing
attempts to reconstruct the energy of a neutrino from a given event.
There is one remaining issue to be addressed at this
stage of the simulation, that of event multiplicity. In the
section that follows it is assumed that the hadronic cascade is
initiated by a single excited hadron that, following the neutrino DIS,
carries $25\%$ of the energy of the incident neutrino. More accurately
there will be some small number of excited partons\footnote{i.e. some
admixture of valence quarks, sea quarks and gluons} between which this
percentage of the neutrino energy is shared; however, ANIS restricts
itself to calculation of the energy of the excited hadronic final
state and not the parton content.

\begin{figure}[htb]
  \centering
  \includegraphics[width=0.5\textwidth]{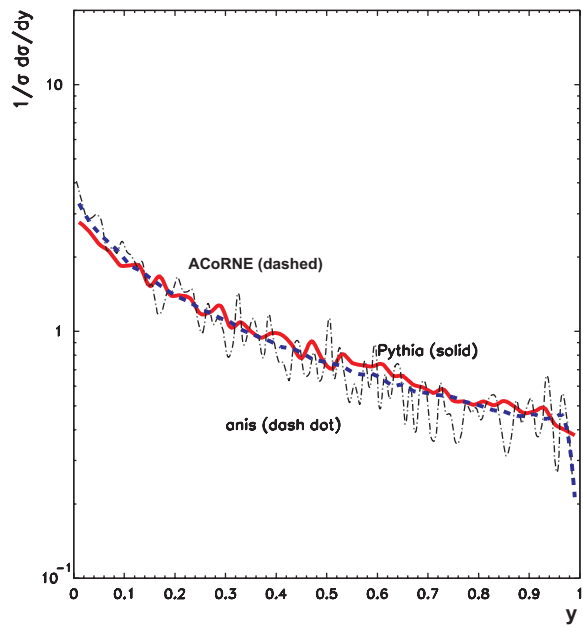}
  \caption[Comparison of neutrino event generators]
	  {
	    \small
	    Bjorken-$y$ distributions at $10^5$\,GeV from different
	    event generators.
	  }
	  \label{fig:aniscompare}
\end{figure}

A comparison of Bjorken-$y$ distributions from ANIS and two other
event generators, namely PYTHIA and a bespoke program developed for
the ACoRNE collaboration, (see \cite{terry:corsika}) has been made. The
results in Figure \ref{fig:aniscompare} indicate that the produced
distributions are consistent.
\section{Simulating Neutrino Induced Particle Cascades}
\label{sec:cascades}
%
\subsection{Physics processes}
\label{subsec:processes}
Given an expression for the energy of the hadronic system one can
proceed to simulate the development of the subsequent particle cascade
within a given material. A suitable particle physics toolkit for
simulating the passage of particles through matter is \geant
\footnote{results from version $8.0$ are presented throughout}
\cite{GEANT}. There are three categories of physics model at the heart
of this program: those driven by theory; parameterised models which
combine theory and data; and empirical physics models driven purely by
data. The user is required to choose which physics models and which
cross-section data are used in a given energy range. Two models can
overlap in energy so long as one model does not fully occupy the same
total energy range as another. A set of ready made high energy
calorimetry ``physics lists'' are distributed with the \geant source
code. Conceptually the formula for construction of a physics list is
as follows:
\begin{equation}
  \textsf{physics list}  = \textsf{particles} + \textsf{processes} + 
  \textsf{cross-sections}
  \label{eq:physicslist}
\end{equation}
Unless a particle is assigned a process it will do nothing in the
simulation. If a particle is assigned more than one process then the
processes compete. One process may invoke many models
and each model has a default cross-section.
It is first determined when and where a process should occur, which
depends on interaction length and cross-section; secondly the final
state is generated, which is dependent on the model invoked.
\subsection{Particle production thresholds}
\label{subsec:thresholds}
A particle propagates through the simulated
detector losing energy via the production of secondary particles. There
must be some energy below which a particle no longer produces
secondaries otherwise the program will suffer from infrared
divergence. Hence the user must impose an energy threshold
cut. However, such a cut may result in a poor estimation of  the
stopping location and energy deposition of the particle, so, the cut
is made on the particle's range instead. The range cut is the same for
all materials but the corresponding energy threshold is material
dependent. Once a particle reaches the energy below which no further
secondaries are produced it is tracked to zero energy through a
continuous energy loss mechanism.

In order to generate the thermal energy density $E(\mathbf{r}')$
required by the pressure field integral in Equation
\ref{eq:pressurewave} one initiates a shower simulation with a single
proton at an energy corresponding to  $25\%$ of the neutrino
energy. The energy deposited by each successive interaction in the
simulation is recorded in a \ROOT \cite{ROOT} ntuple that can be
analysed offline. As the shower ages its  composition tends toward
electrons and gamma rays through pion decay;  the range cuts chosen
for water correspond to a $3$\,MeV energy  threshold for tracking of
e$^\pm$ and $\gamma$. Below this energy no further secondaries are
produced (a gamma for instance will no longer pair-produce) and the
particle is forced to deposit the remainder of its energy
continuously. This dramatically reduces CPU time and prevents the
ntuple file size from diverging.
%
%
%
\begin{figure}[htbp]
  \subfigure[Radial Energy Deposition]
	    { \centering
	      \hfill
	      \includegraphics[width=0.7\textwidth]{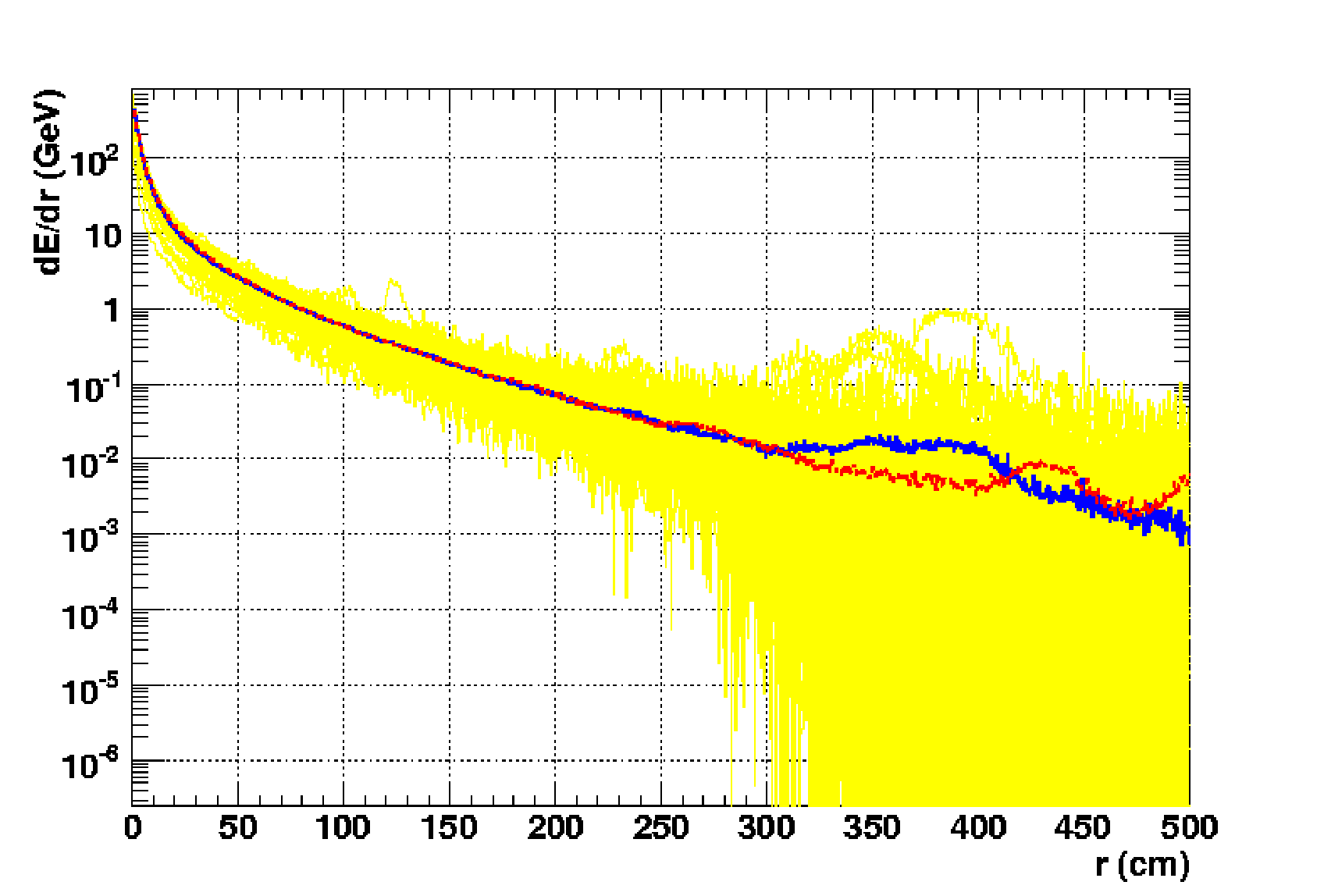}
	      \label{fig:geantdedrcuts}
	      \hfill
	    }
  \subfigure[Longitudinal Energy Deposition]
	    { \centering
	      \hfill
	      \includegraphics[width=0.7\textwidth]{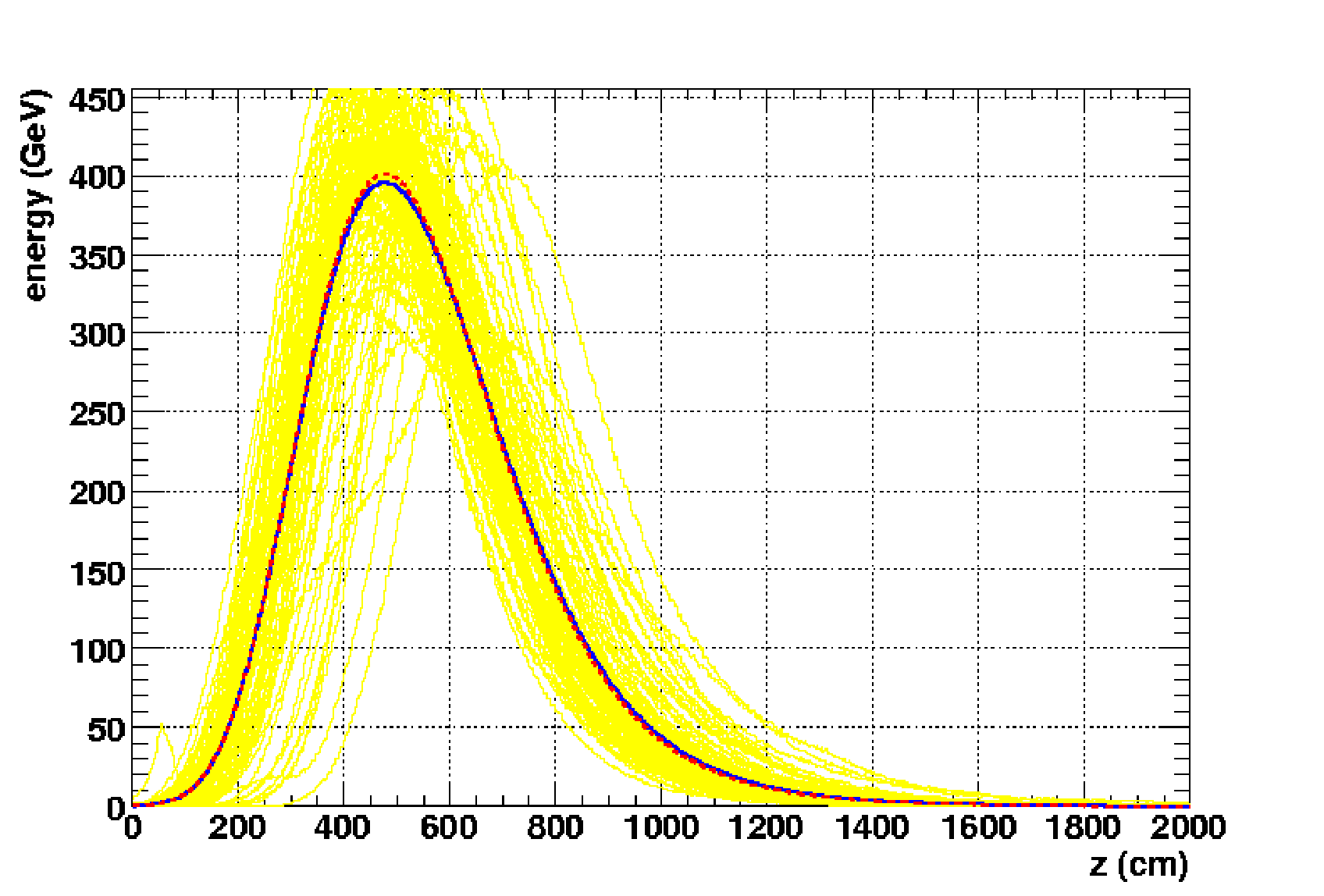}
	      \label{fig:geantdedzcuts}
	      \hfill
	    }
  \caption[Average energy deposited by one hundred $100$\,TeV proton showers]
	  {
	    \small
	    Average radial (Figure \ref{fig:geantdedrcuts}) and
	    longitudinal (Figure \ref{fig:geantdedrcuts}) energy
	    profiles for one hundred $100$\,TeV \geant proton showers
	    with (\blue{solid}) and without
	    (\red{dotted}) range cuts.
	  }
	  \label{fig:geantcuts}
\end{figure}
The plots in Figure \ref{fig:geantcuts} illustrate the shape of the
energy deposition with or without cuts applied. Whilst the effect on
the shower shapes is negligible when compared to individual shower
fluctuations (on top of which the average distributions are
superimposed) the CPU time per event decreases from
$\mathcal{O}\,20$\,hours to  $\mathcal{O}\,20$\,minutes at
$100$\,TeV. Furthermore the low energy extremes of the shower (where
the agreement is less good) have little effect on the
resulting acoustic pulse, the shower core being the dominant
contributor.
%

\begin{figure}[htbp]
  \centering
  \includegraphics[width=0.7\textwidth]{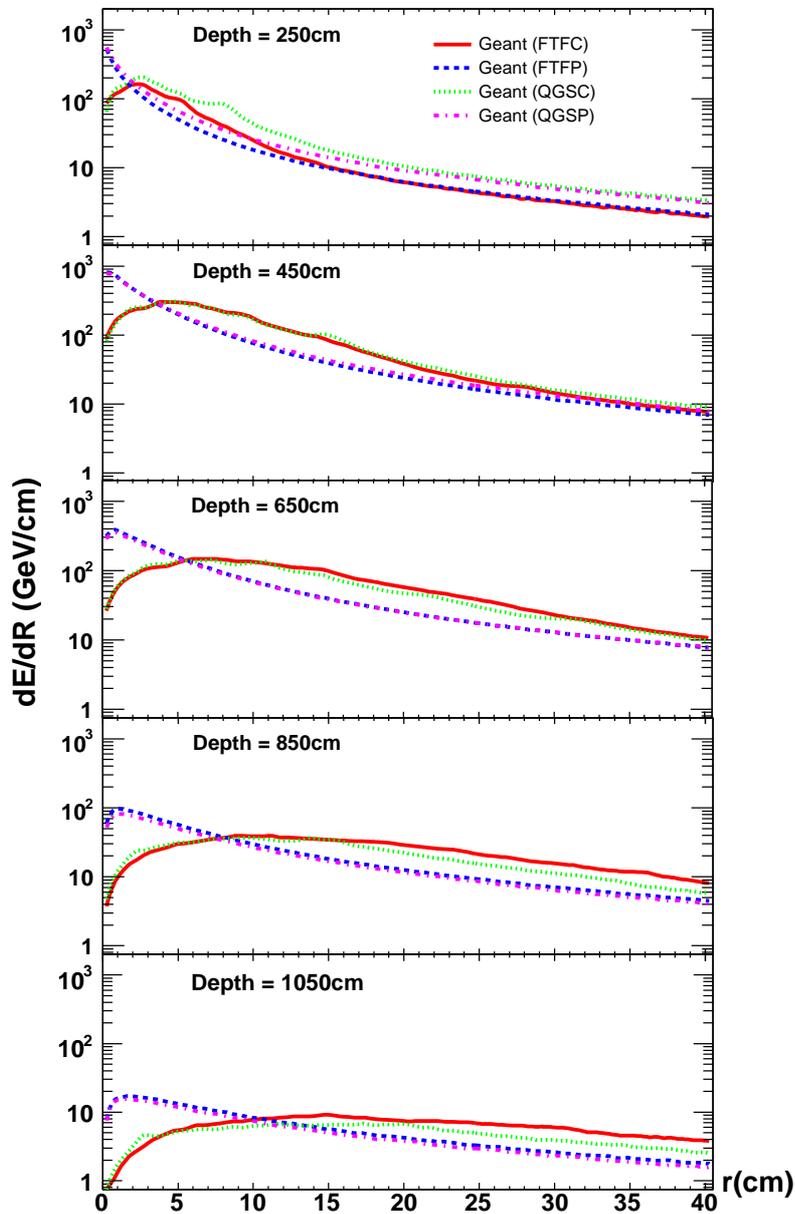}
  \caption[Effect of physics lists on radial shower shape]
	  {
	    \small
	    Average radial energy deposited per $20$\,g\,cm$^{-2}$
	    vertical slice per unit radial distance for one hundred
	    proton showers at  $100$\,TeV, using each of
	    the high energy hadronic calorimetry physics lists. Depth
	    refers to the distance along the shower axis.
	  }
	  \label{fig:physicslists1}
\end{figure}

\begin{figure}[htbp]
  \centering
  \includegraphics[width=0.7\textwidth]{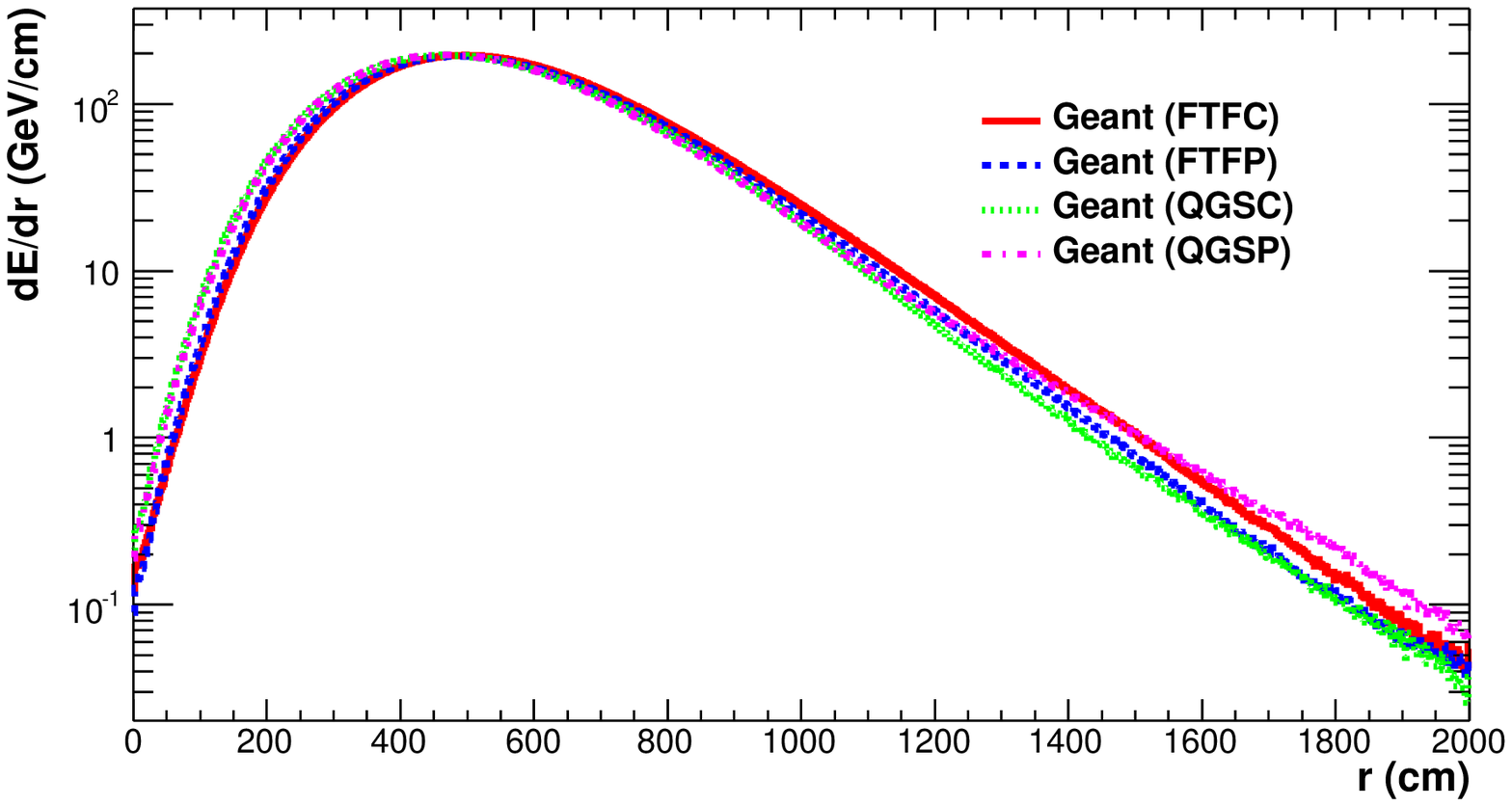}
  \caption[Effect of physics lists on longitudinal shower shape]
	  {
	    \small
	    Average longitudinal energy deposition of one hundred
	    $100$\,TeV proton showers, at various depths along the
	    longitudinal axis of the shower, using each of the high
	    energy hadronic calorimetry physics lists.
	  }
	  \label{fig:physicslists2}
\end{figure}

Four theory driven high energy hadronic calorimetry ``use-by-case''
physics lists are provided with the \geant distribution. They are
composed of the Quark Gluon String (QGS) model or the FRITIOF (Lund
string dynamics) model \cite{FRITIOF}
combined with either Pre-Equilibrium or Chiral Invariant
Phase-space (CHIPS) decay modes. The range of validity for each
model extends from $10$\,GeV to $100$\,TeV. The effect of each physics
list on the radial and longitudinal shower shapes is shown in Figures
\ref{fig:physicslists1} and \ref{fig:physicslists2} respectively.
%
%
The CHIPS decay mode tends to produce a broader, older shower, with a
central energy ``hole'' in comparison to the  pre-equilibrium
decays, whilst there is little to distinguish  between the QGS and
FRITIOF hadron interaction schemes. References for these models can be
found, for example in \cite{CHIPSI}, \cite{CHIPSII}, \cite{QGS} and
\cite{FRAGMENTATION}, as cited in the \geant documentation. The
Pre-Equilibrium model is favoured over CHIPS since it is more
consistent with the distributions seen in UHECR air showers and it is
the model preferred by other groups undertaking acoustic
studies. Comparisons of the longitudinal and radial shower shapes
produced by \geant and a modified version of the CORSIKA air shower
program are plotted in Appendix \ref{appendix:geantvscorsika}.
%
%
%
%
%
\section{Formation of the Acoustic Signal}
\label{sec:formation}
\begin{figure}[h!]
  \centering
  \includegraphics[width=0.7\textwidth]{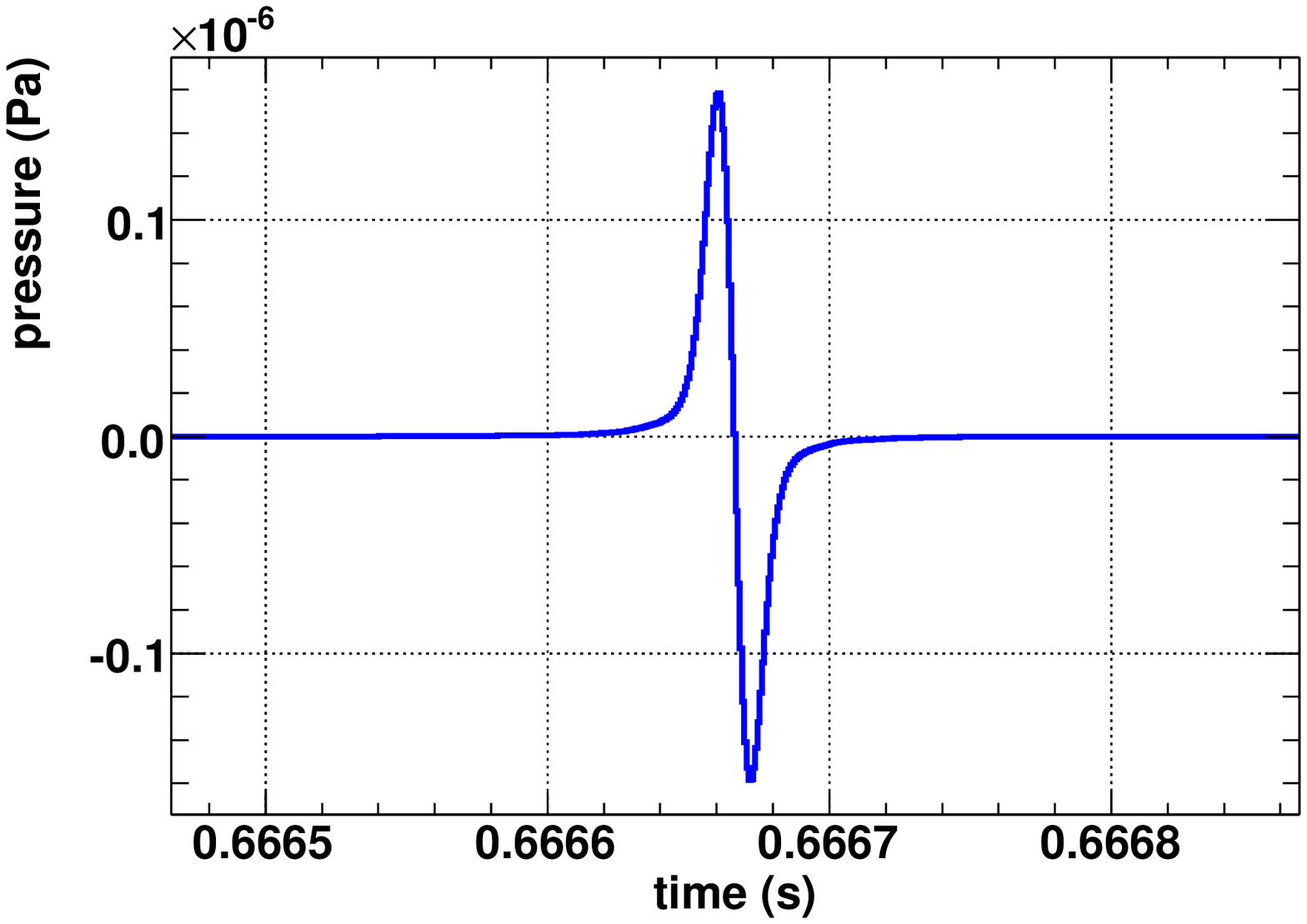}
  \caption[Acoustic pressure signal from a $100$\,TeV proton induced shower]
	  {
	    \small
	    The pressure pulse for a $100$\,TeV proton induced \geant
	    shower in seawater, at a distance of 1km from the
	    cascade. The angle with the plane transverse to the 
	    shower direction is $0$\,$\degrees$.
	  }
	  \label{fig:singlepulse}
\end{figure}

%
The acoustic signal resulting from a hadronic cascade is computed by
integrating the energy contained in the \ROOT ntuple as produced by the
\geant simulation according to Equation \ref{eq:pressurewave}. A
typical pulse resulting from a $100$\,TeV proton induced shower is
plotted in Figure \ref{fig:singlepulse}.

\begin{figure}[htb]
  \centering
  \includegraphics[width=0.7\textwidth]{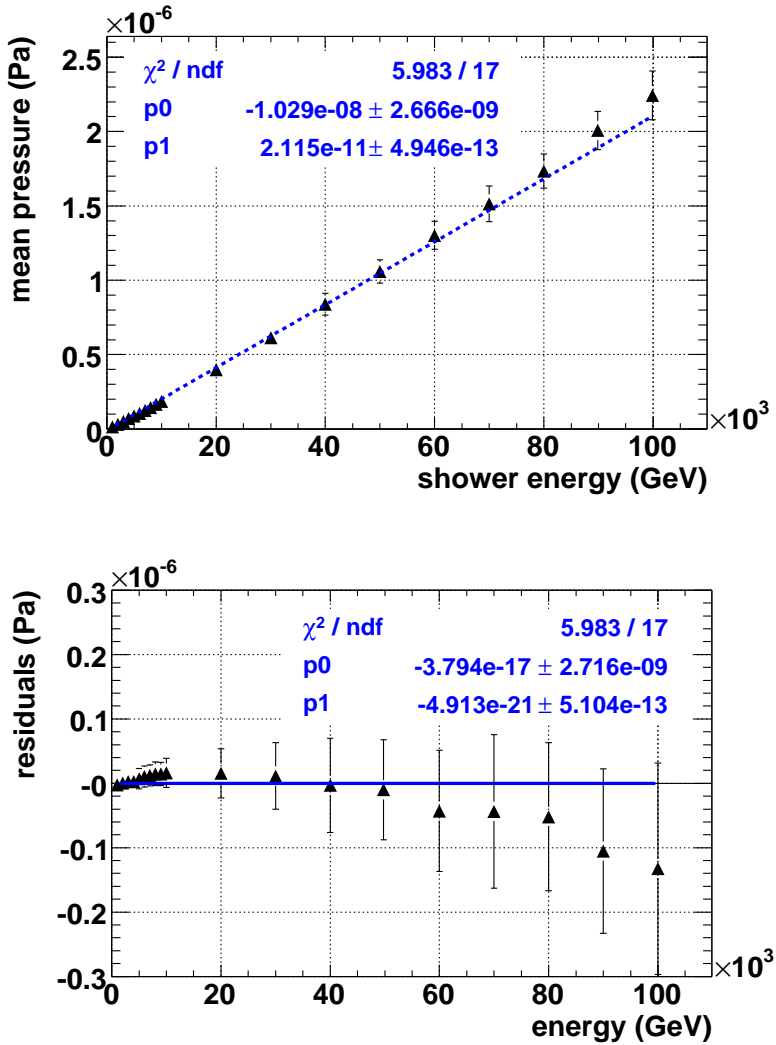}
  \caption[Peak pulse pressure vs. shower energy]
	  {
	    \small
	    Plot of mean peak pulse pressure against energy of the
	    hadronic shower demonstrating the linear relationship
	    between the two. Error bars are of magnitude $1
	    \sigma_{mean}$. The residuals plot shows the deviation of
	    the data points from the fitted function, there is an
	    inherent instability at higher energy due to the limited
	    accuracies of the models used (see Appendix
	    \ref{appendix:geantvscorsika}).
	  }
	  \label{fig:evsp}
\end{figure}

For the purpose of simulating the performance of large scale
hydrophone arrays it is necessary to determine a relationship between
the energy of a neutrino and the resulting pressure amplitude at a
given location. The shape of the cascade energy density
becomes more uniform with increasing energy. The integrated energy of
the cascade scales linearly with the energy of the neutrino; hence,
the magnitude of the resultant pressure signal is assumed to scale
with the energy of the neutrino. Consequently an analytical
parameterisation is sought to relate peak pressure amplitude to
neutrino energy. The details of how this parameterised pressure is
attenuated as it propagates from the cascade to a given hydrophone is
discussed in the next chapter.
The peak pressures at $1$\,km from various proton induced \geant
cascades at energies in the range $10^3 \to 10^5$\,GeV are plotted as
a function of the proton energy in Figure
\ref{fig:evsp}. A linear function is fitted to the values
allowing for a direct determination of the peak pressure at $1$\,km
from an arbitrary shower for a neutrino of a given energy. The
gradient indicates a mean pulse peak pressure of
$2.12\times10^{-11}$\,Pa per GeV of thermoacoustic energy. The
subsequent pulse at each hydrophone location is suitably scaled and then
the effects of attenuation are applied.

\section{Summary}
\label{sec:nusummary}
From the simulation of UHE neutrino interactions we have learnt that
approximately $25\%$ of the energy of a neutrino is carried away by
the hadronic component of neutrino-nucleon deep inelastic
scatter. It is therefore assumed that the particle cascade, and
subsequent thermoacoustic energy deposition, induced by a neutrino
with energy $E_{\nu}$, is equivalent to that produced by a proton at
$E_{\nu}/4$. Furthermore, from the simulation of the evolution of
TeV proton showers in the sea, we have learnt that the peak pressure
of an acoustic pulse resulting from such a shower at a given
distance, scales linearly with the energy contained in the
shower. These findings equip us with a parameterisation allowing us to
estimate the peak pressure $p$ of a pulse at $1$\,km from a particle
cascade initiated by a neutrino of energy $E_{\nu}$ given by:
\begin{equation}
  p = 0.25 \times k_{\nu} \times E_{\nu}
  \label{eq:enuparam}
\end{equation}
where $k_{\nu}$ is the conversion constant and is equal to $21.15
\times 10^{-21}$\,Pa\,eV$^{-1}$. The parameterisation in Equation
\ref{eq:enuparam} will be used in the remainder of the simulation chain
discussed in the chapters that follow.

  \cleardoublepage

  \chapter{Simulating the Propagation of the Acoustic Signal}
\label{chp:propagation}
%
%
%
\section{Introduction}
\label{sec:propagationintro}
Thus far it has been reported how the mechanisms which give rise to
the distinctive bipolar pressure pulse, expected following an UHE
neutrino interaction in the sea, are computed. This Chapter
subsequently introduces parameterisations of signal attenuation
resulting from its propagation through seawater. Incorporated into
this attenuation are the effects of refraction, resulting from
variation in sound velocity in the sea, and the effects of ambient
noise and filtering.
\section{Attenuation of the Acoustic Signal}
\label{sec:attenuation}
A hydroacoustic pressure wave is attenuated as it propagates away from
its point of origin. There are three contributions to this
attenuation: a geometric factor whereby the pulse diminishes the
farther it travels, a diffractive attenuation that is a consequence of
the coherent nature of the acoustic emission, and the attenuation of
the pulse due to the bulk mechanical properties of the medium through
which it propagates. Each of these effects is discussed below.
\subsection{Geometric attenuation}
\label{subsec:geometricattenuation}
As described in Section \ref{subsec:acousticsignal} the development of
the particle cascade occurs instantaneously on the thermal diffusion
timescale. The result of this is coherent emission along the shower
axis. In the far-field ($r \gg l$ where $r$ is the distance from the
interaction and $l$ is the cascade length) the geometric attenuation
of such emission goes as the reciprocal of the distance travelled
\cite{acoustic:learned}.
\subsection{Attenuation by the medium}
\label{subsec:attenuationbymedium}
\begin{figure}[htb]
  \centering
  \includegraphics[width=0.75\textwidth]{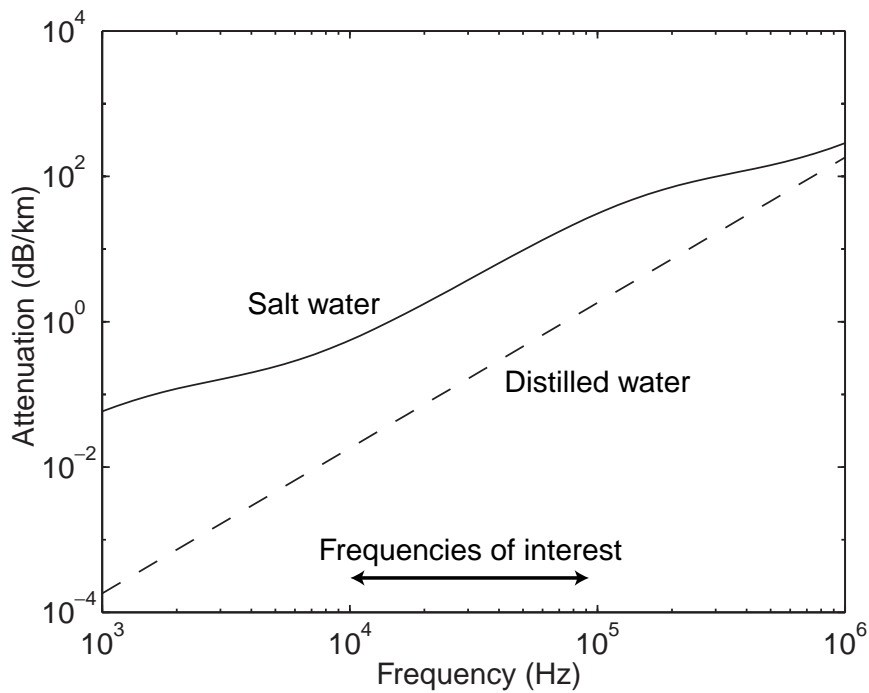}
  \caption[Sound attenuation coefficient for seawater]
	  {
	    \small
	    The sound attenuation coefficient for seawater 
	    (from \cite{sensitivity:lehtinen}). Also shown
	    is the absorption coefficient for distilled water.
	  }
	  \label{fig:absorptioncoefficient}
\end{figure}

\begin{figure}[htb]
  \centering
  \includegraphics[width=0.75\textwidth]{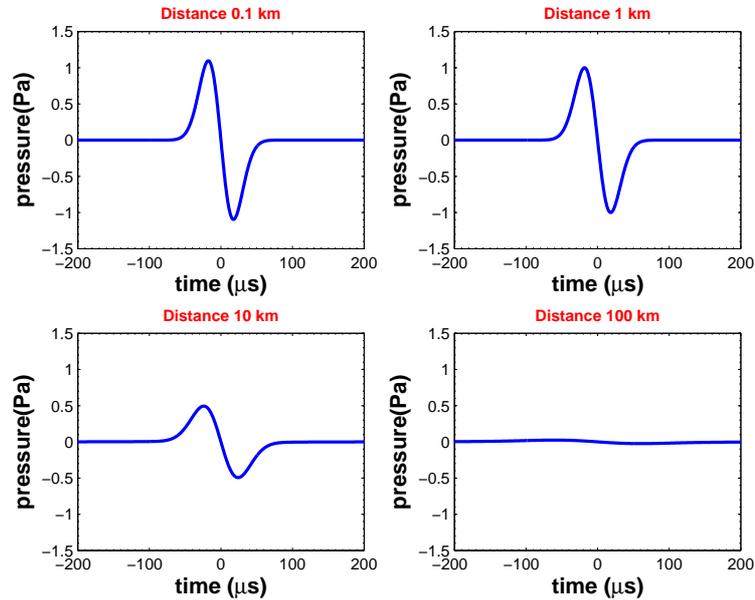}
  \caption[Attenuation by seawater]
	  {
	    \small
	    Variation of pulse shape with respect to distance, due to
	    seawater absorption, for a signal normalised to $1$\,Pa at
	    $100$\,m from the particle cascade.
	  }
	  \label{fig:attenuationbymedium}
\end{figure}

The second contribution to the attenuation of the acoustic radiation,
produced by a neutrino induced particle shower, is governed by the
bulk mechanical properties of the medium through which it
traverses. Water is a low-pass filter, high frequency components being
readily attenuated, this can be seen in the sound attenuation co-efficient
plot from Lehtinen {\it et al} \cite{sensitivity:lehtinen} shown in Figure
\ref{fig:absorptioncoefficient}. For the purposes of the large scale
detector simulation the attenuation  was parameterised by a fourth order
polynomial and recomputed for a Mediterranean\footnote{The Tyrrenian
  Sea located north of Sicily was used in the ACoRNE funding proposal
  \cite{acorne:proposal}} water temperature of $25$\,$\degrees$C. The
effect of this attenuation on a bipolar signal is shown in Figure
\ref{fig:attenuationbymedium}.

The primary reason for the difference in absorption coefficient
between seawater and distilled water is the excitation of
magnesium sulfate and boric acid molecules that are present in
seawater. Such molecules can have multiple stable states and
transitions between states can be induced by a change of pressure,
effectively converting hydroacoustic pressure into heat.
%
%
%
\subsection{Angular spread}
\label{subsec:angularspread}
\begin{figure}[htbp]
  \centering
  \hfill
  \begin{minipage}[h!]{0.45\textwidth}
    {\includegraphics[width=\textwidth]{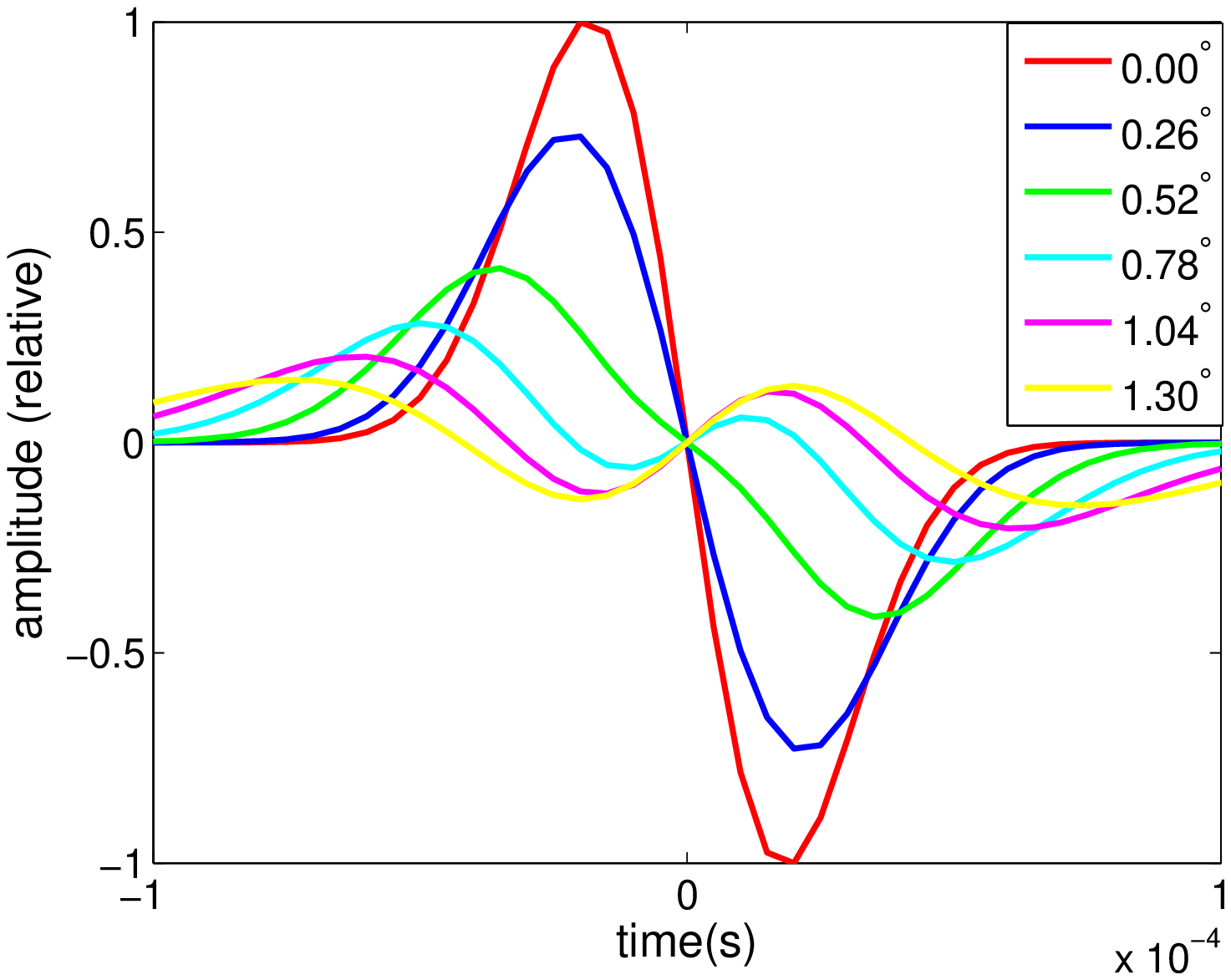}}
  \end{minipage}
  \hfill
  \begin{minipage}[h!]{0.45\textwidth}
    {\includegraphics[width=\textwidth]{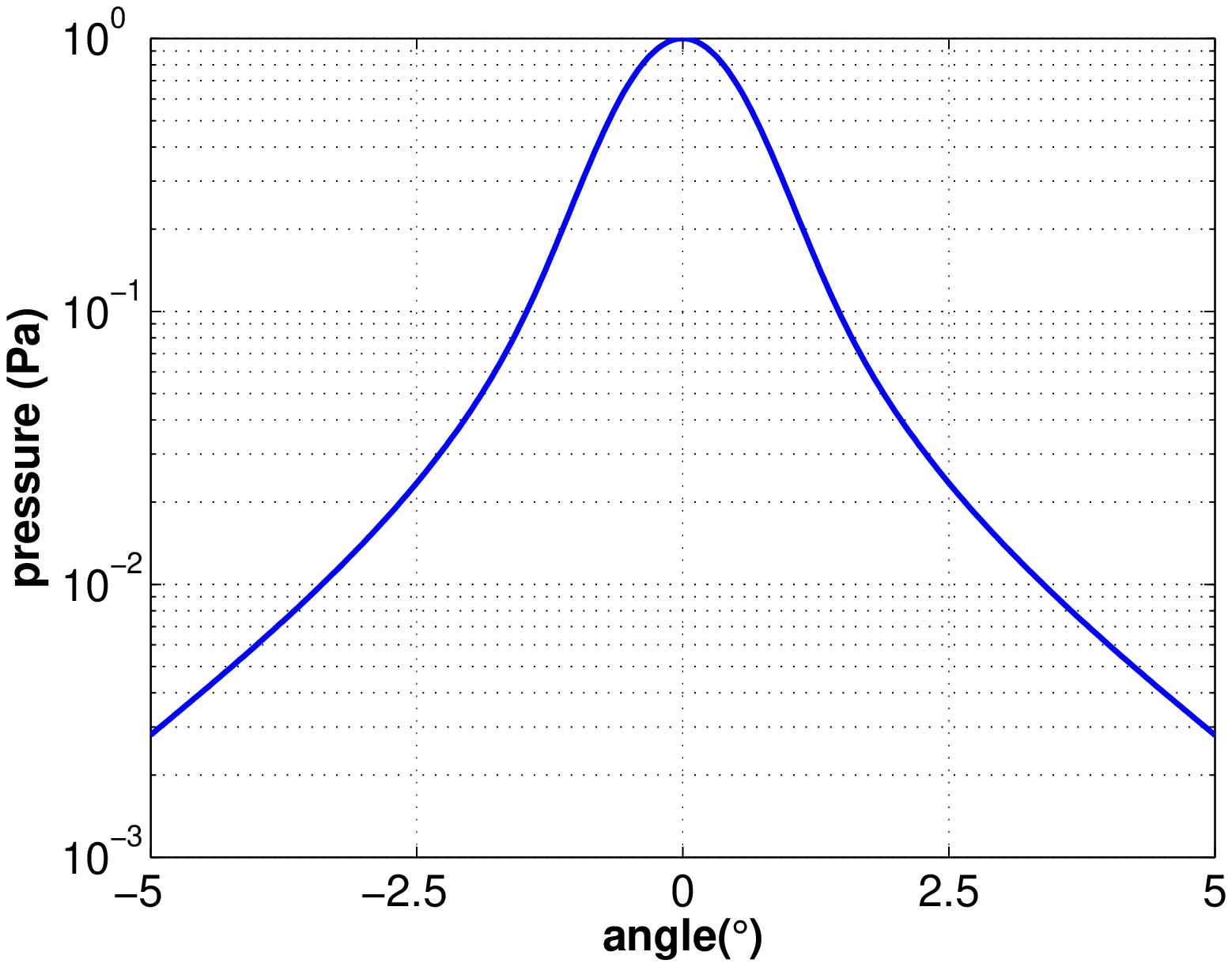}}
  \end{minipage}
  \caption[Angular spreading of acoustic pulses]
      {
	\small
	Variation of far-field pulse shapes ({\it Left})
	and signal amplitudes ({\it Right}) as a function of
	angle out of the plane of the pancake (from
	\cite{privcomm:danaher}). The step in angle of
	$0.26$\,$\degrees$ derives from the digitisation of the sample
	into $1024$ bins, which was chosen because it facilitates a
	faster IFT algorithm in MATLAB \footnotemark.
      }
      \label{fig:angularspread}
\end{figure}

As one moves out of the plane perpendicular to the shower axis, that
is, the plane of the pancake, the acoustic radiation, and hence pulse
pressure, is attenuated due to diffractive effects. This frequency
dependent process both diminishes the pulse height and broadens the
pulse width. 
For each frequency component of the attenuated pulse,
there is a corresponding $\sin (x)/x$ diffraction pattern, in analogy
to the diffraction of light through a narrow slit. The length of the
shower corresponds to the size of the aperture.
The pulse at zero degrees out of the pancake plane is computed
by taking the Inverse Fourier Transform (IFT) at zero angle. The pulse
at a given angle out of the pancake plane is thus computed by taking
the IFT of the respective $\sin (x)/x$ function for that angle
\cite{privcomm:danaher}. This effect is plotted in Figure
\ref{fig:angularspread}. The pulse amplitude is attenuated to nearly
one thousandth of its original magnitude at $5$\,$\degrees$ out of the
pancake plane for a $10$\,m long shower.
\footnotetext{http://www.mathworks.com/ Accessed 02/05/2006}
%
%
%
The angular distribution plotted in Figure \ref{fig:angularspread} is
parameterised and then used to scale the peak pressure recorded on a
given receiver as a function of its angle out of the plane
perpendicular to the shower.
%
%

The three components of attenuation described above are applied in
succession to the unattenuated peak pressure amplitude as calculated
via the linear energy to pressure relation formulated in Section
\ref{sec:formation}. Given a formalism for calculating the attenuation
of an acoustic pulse, as a function of the distance it propagates
through the sea, a method for determining the distance travelled is
desired.
\section{Refraction}
\label{sec:refraction}
\begin{figure}[htb]
  \centering
  \includegraphics[width=0.4\textwidth]{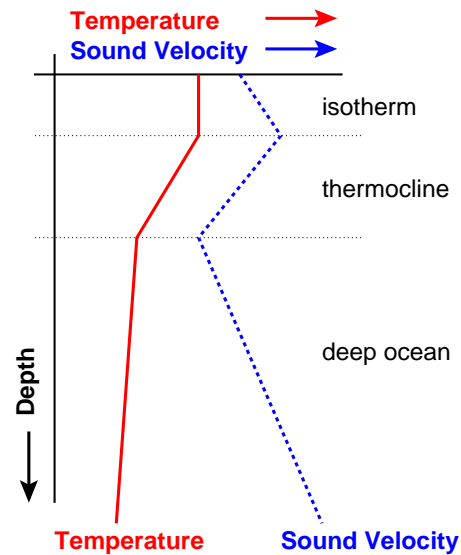}
  \caption[Oceanic depths of constant sound gradient]
	  {
	    \small
	    Three layers representing regions of constant sound
	    gradient in the sea: the isotherm, the thermocline and the
	    deep.
	  }
	  \label{fig:thermocline}
\end{figure}

The velocity of sound in seawater is dependent on temperature,
pressure, and salinity. A simple means to describe the stratification
of the sea into horizontal layers of constant sound gradient is
illustrated in Figure \ref{fig:thermocline}. There are three zones of
sound propagation to consider: near the surface there exists an
isothermal layer which is well mixed by surface waves and wind, within
the isotherm the sound velocity increases with depth due to a slight
increase in pressure; below the isotherm is a region known as the
thermocline through which the sound velocity decreases with depth due to
the corresponding decrease in temperature; below the base of the
thermocline, in the deep ocean zone, the sound velocity gradient
inverts and sound velocity increases with depth due to the corresponding
increase in pressure.
Sound rays tend toward the region of lowest sound velocity. 
Where there is an inversion in the sound
velocity gradient sound rays are refracted away from each other,
leading to the formation of ``shadow zones'', regions from where
acoustic signals may be inaudible. When building an acoustic neutrino
telescope it is desirable to be far away from such environments since
this will contribute to a loss in effective volume. 
%
\begin{figure}[htb]
  \centering
  \includegraphics[width=0.4\textwidth]{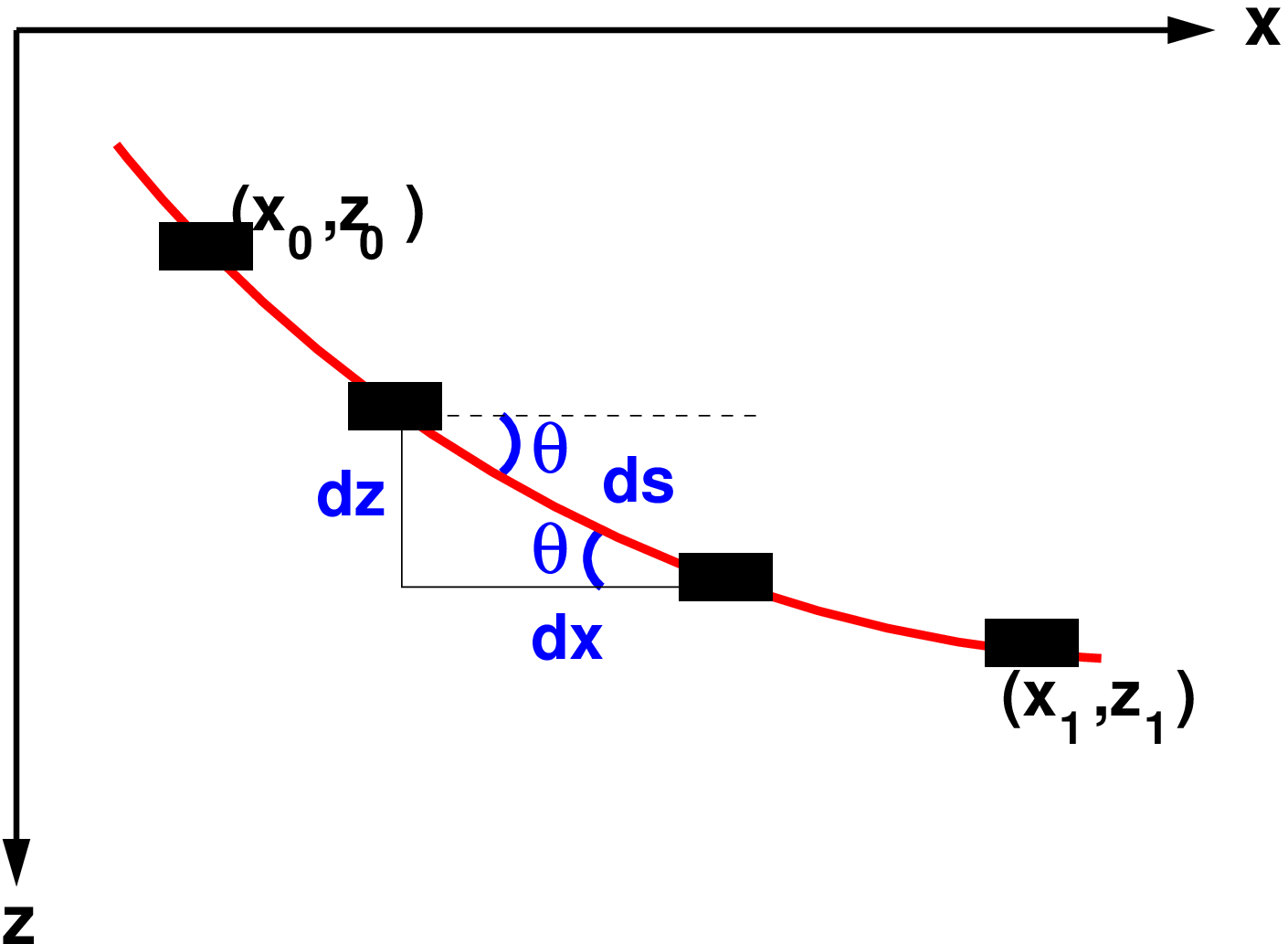}
  \caption[Refraction geometry]
	  {
	    \small
	    Geometry of a single curved ray in a medium with varying
	    sound velocity.
	  }
	  \label{fig:refractiongeom}
\end{figure}
A full derivation of the ray tracing equations in the general case,
where the sound velocity gradient is not linear, and for the case of a
linear sound velocity gradient can be found in Appendix
\ref{appendix:raytracing}. For the geometry illustrated in Figure
\ref{fig:refractiongeom} with a constant sound velocity gradient
$g = dc(z)/dz$, it can be shown that:
\begin{eqnarray}
x_1-x_0&=&\frac{c_0}{g}\Big(\frac{\sin\theta_0-\sin\theta_1}{\cos\theta_0}\Big)\\\nonumber\\
\label{eq:refraction:range}
s_1-s_0&=&\frac{c_0}{g\cos\theta_0}(\theta_0-\theta_1)\\\nonumber\\
\label{eq:refraction:length}
z_1-z_0&=&\frac{c_0}{g\cos\theta_0}(\cos\theta_1-\cos\theta_0)
\label{eq:refraction:depth}\\\nonumber\\
t_1-t_0&=&-\frac{1}{g}\ln\Big\{\frac{\cos\theta_0(1+\sin\theta_1)}{\cos\theta_1(1+\sin\theta_0)}\Big\}
\label{eq:refraction:time}
\end{eqnarray}
where $c_0 = c(z=0)$ is the sound velocity at the origin depth,
$t_0$ is the time at the point of origin and $t_1$ is the time
at point $(x_1,z_1)$. Angles $\theta_0$ and $\theta_1$ are the angles
the ray makes with the horizontal at points $(x_0,z_0)$ and
$(x_1,z_1)$ respectively; $s_0$ and $s_1$ define the arc length along
the ray. For simplicity the problem has been reduced into two dimensions
defining the $xz$-plane (since refraction only occurs in the vertical ($z$)
direction one can arbitrarily align rays along one of the
horizontal axes). The ray tracing algorithm iterates in steps of $x$
for a given path between source (cascade) and receiver
(hydrophone). Equation \ref{eq:refraction:range} can be solved for the
deflection angle $\theta_1$ which then forms the input to Equation
\ref{eq:refraction:depth} along with $\theta_0$ which is known.

In order to compute the trajectory of the real refracted ray $R$ that
intersects source and receiver at $(x_0,z_0)$ and $(x_1,z_1)$
respectively one must first compute the angle of deflection $\theta_d$
of the deflected ray $d$ that results from sending a ray along the
imaginary linear path $i$. This is because only the sound velocity as
a function of depth, the coordinates $(x_0,z_0)$ and $(x_1,z_1)$;
and the angle $\theta_i$ are known {\it a priori}.
\begin{figure}[htb]
  \centering
  \includegraphics[width=0.75\textwidth]{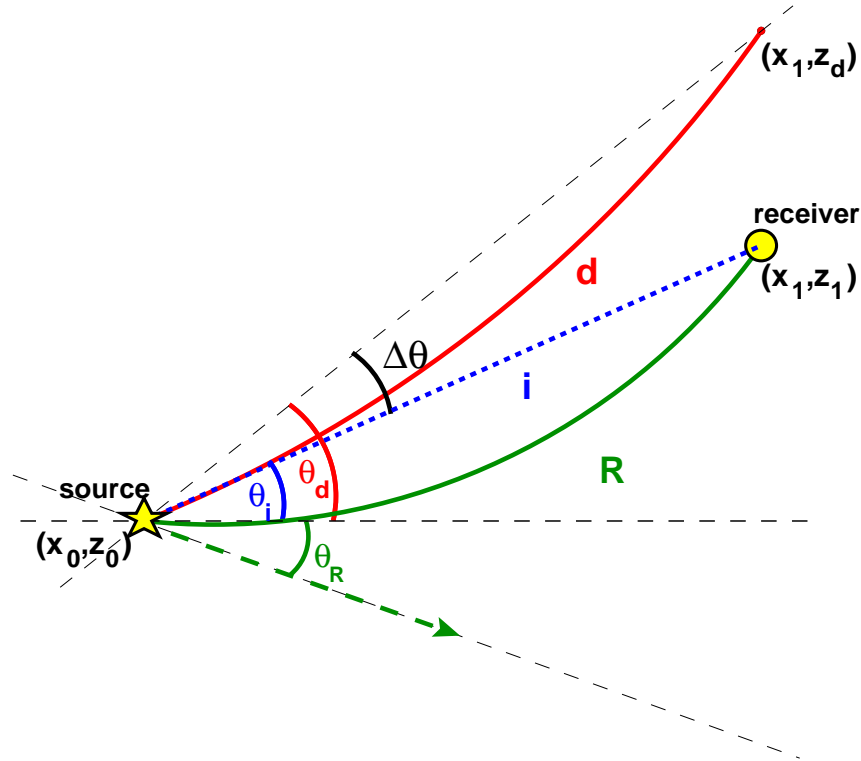}
  \caption[Computing the ray start angle]
	  {
	    \small
	    The difference between the angle of deflection
	    \red{$\theta_d$} of the deflected ray \red{$d$} and the
	    start angle \blue{$\theta_i$} of the undeflected imaginary
	    ray \blue{$i$} is $\Delta\theta =$
	    \red{$\theta_d$}$-$\blue{$\theta_i$}.
	    The start angle for the real refracted ray $R$ is
	    therefore given by
	    \green{$\theta_R$}$=$\blue{$\theta_i$}$-\Delta\theta$.
	  }
	  \label{fig:startangle}
\end{figure}

\begin{figure}[htb]
  \centering
  \includegraphics[width=0.75\textwidth]{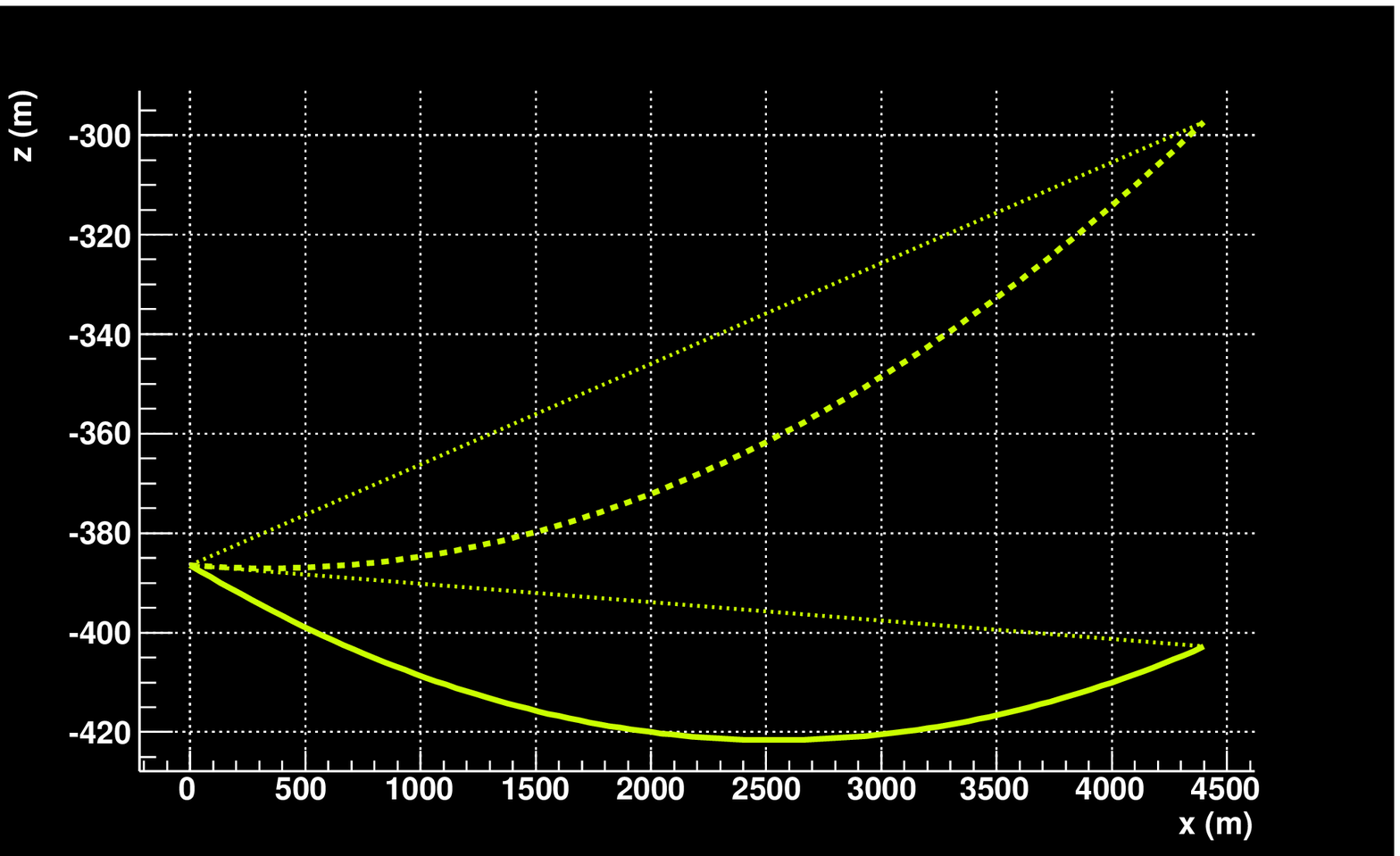} 
  \caption[Ray trace from detector simulation]
	  {
	    \small
	    A ray trace from the large scale detector simulation.
	    The solid curved ray and the dashed curved ray correspond
	    to the rays \green{$R$} and \red{$d$} respectively in
	    Figure \ref{fig:startangle}.
	  }
          \label{fig:npraytrace}
\end{figure}

\begin{figure}[htb]
  \centering
  \includegraphics[width=0.75\textwidth]{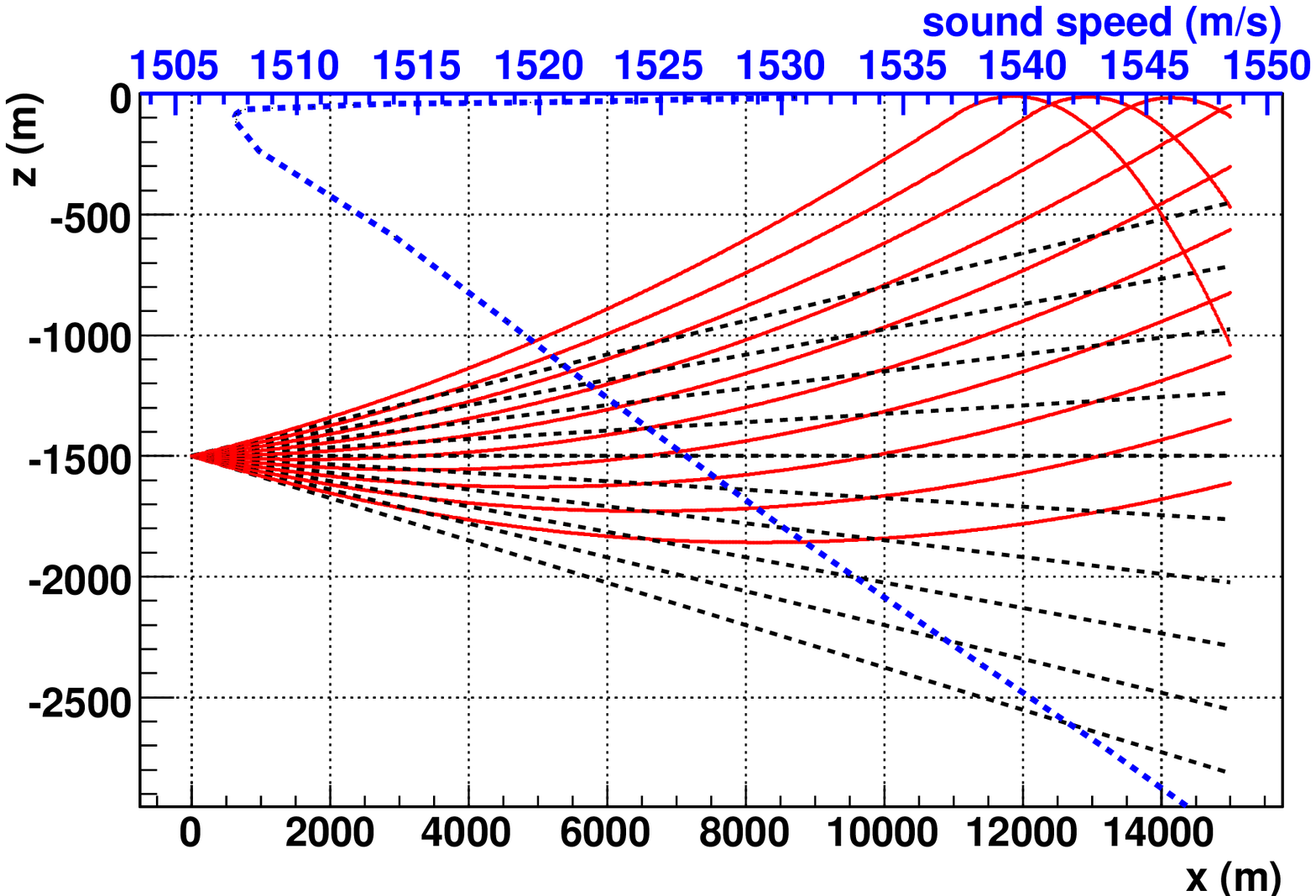}
  \caption[Ray trace for ANTARES SVP]
	  {
	    \small
	    Ray trace for a source at $1500$\,m depth. Refracted rays
	    (solid) and unrefracted rays (dashed) are shown. The SVP
	    for the ANTARES site is also shown, with sound speed across
	    the upper $x$ axis. The ANTARES SVP is essentially
	    linear. Neglecting the inversion in the sound speed
	    gradient equates to extrapolating the linear section
	    directly to the upper $x$-axis (This yields 
	    $c_0 = 1506$\,ms$^{-1}$ and $g = (-) 0.0156$\,s$^{-1}$).
	    }
	    \label{fig:antaresraytrace}
\end{figure}

\begin{figure}[htb]
  \centering
  \includegraphics[width=\textwidth]{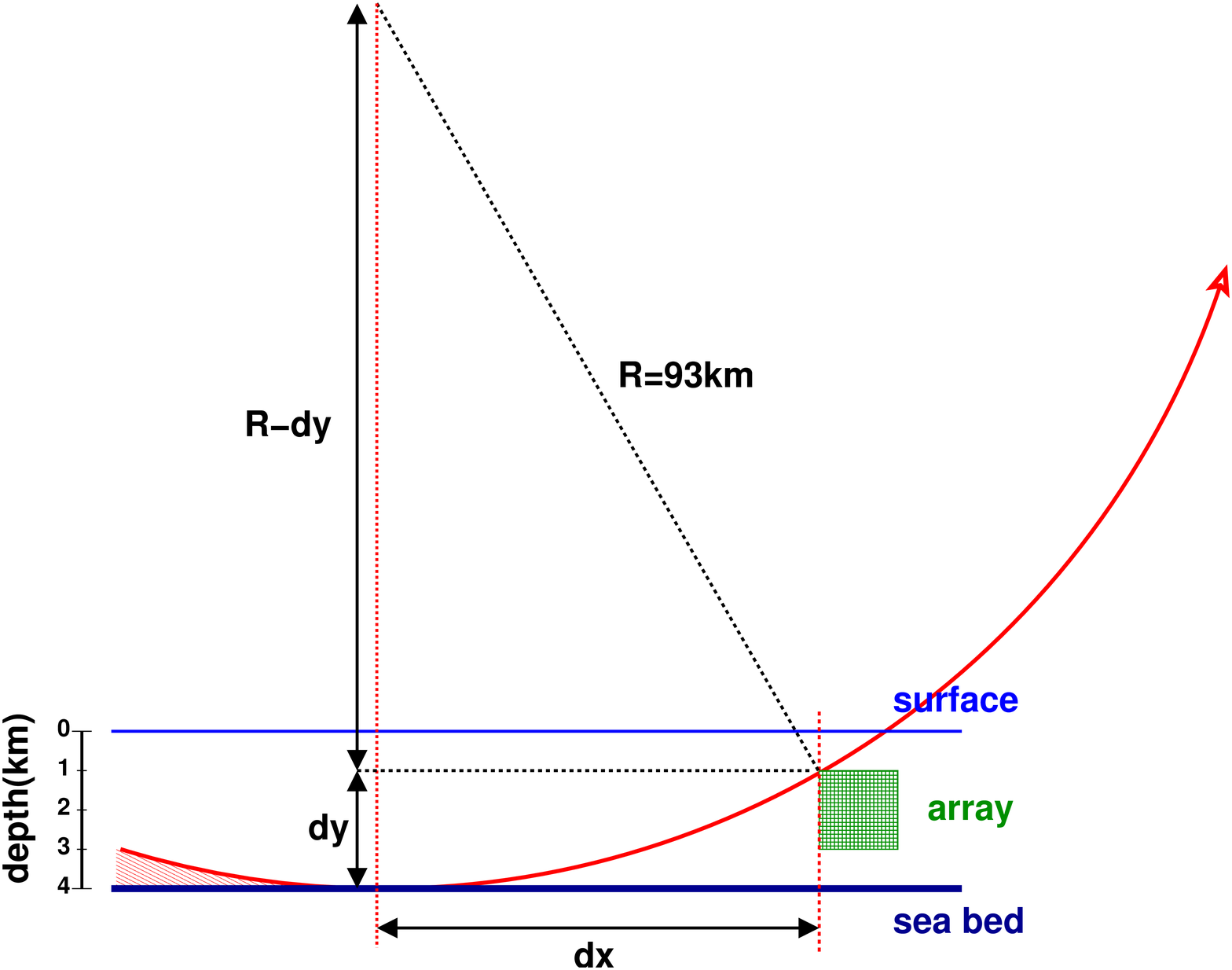}
  \caption[A natural limit on source distance imposed by refraction]
	  {
	    \small
	    Rays following circular trajectories of radius $R=93$\,km
	    and originating from the \red{shaded} zone and beyond cannot
	    reach the \green{array}. Hence there is a natural
	    distance $D = 2\,dx$ beyond which sources cannot be
	    heard (so long as rays are not reflected and any inversion
	    in the SVP is neglected, as is the case here).
	    The Mediterranean has a
	    maximal depth of about $4$\,km, ideally an array would
	    sit equidistant between the base of the thermocline and the
	    sea bed. In this simulation the isotherm and thermocline
	    are neglected along with surface and sea-bed reflections,
	    and the array sits at mid depth. 
	    $dx = \sqrt{R^2-(R-dy)^2}$, yielding a value for
	    $D = 2 \times dx = 2 \times 24 = 48$\,km.
	  }
	  \label{fig:sourcelimit}
\end{figure}

A plot of the ray trace as computed in the large scale detector
simulation as described in Chapter \ref{chp:largescale} is shown
in Figure \ref{fig:npraytrace}.
In the case of a constant sound velocity gradient:
\begin{eqnarray*}
 g &=& \textrm{constant} \\
 c &=& c_0 + gz 
\end{eqnarray*}
where $c$ is the sound speed at depth $z$ and $c_0$ is the sound
speed at $z=0$. The values for $c_0$ and $g$ are obtained from a
sound velocity profile (SVP) for a  particular location in the
sea. The SVP at the site of the ANTARES neutrino telescope is shown
in Figure \ref{fig:antaresraytrace} superimposed on a ray trace for a
source emitting from a depth of $1500$\,m at various angles to the
horizontal. The effect of the  inversion in the sound velocity
gradient at the base of the thermocline is included, hence rays are
refracted back toward the sea bed as they approach the surface. In
the hydrophone array simulation only the deep sea zone, with
approximated linear SVP, is considered, such that there is no
inversion in the sound velocity gradient and rays are terminated at
the surface without reflection. Similarly rays that hit the sea floor
are terminated without reflection. This represents a more
conservative approach than one where boundary reflections are
considered, since the effective volume will increase if rays can reach
a given array via surface or sea floor reflections. Detailed
information of the boundary conditions is required to effectively
model ray reflection, hence it is presently neglected.
In the simulation the hydrophone array sits at $-2$\,km in the 
deep ocean zone, far away from the thermocline in a region where the
sound velocity gradient is linear. The effect of a linear sound
velocity profile is to send acoustic rays along circular trajectories
\cite{acoustic:boyles} of radius $R = c_0/g = 1506/0.0156 \sim
93$\,km. There exists a natural upper limit on source distance, due
to refraction, in a sea of finite depth and constant sound velocity
gradient. The concept is illustrated in Figure \ref{fig:sourcelimit};
sources greater than $48$\,km beyond an array located at mid depth in
a $4000$\,m deep sea cannot be detected. This is the absolute maximum
distance from which a source can be heard due to the geometrical
constraints of refraction. With respect to the performance of an
acoustic array, by demanding several signals in coincidence, which are
themselves emitted subject to angular constraints, it is unlikely a
neutrino induced particle cascade could be successfully reconstructed
at a distance of this magnitude. The exact limit on the distance from
which cascades can be reconstructed is of importance when calculating
the effective volume of a given hydrophone array and will be discussed
further in Section \ref{sec:effectivevolume}.

Successful computation of curved ray trajectories thus allows the
calculation of hit times and ray path lengths from Equations
\ref{eq:refraction:range} and \ref{eq:refraction:time}
respectively. The effect of attenuation as discussed in Sections
\ref{subsec:geometricattenuation} to \ref{subsec:attenuationbymedium}
is thus applied to the refracted times and ray paths.

\section{Ambient Noise, Filtering and Thresholds}
\label{sec:noisefilterandthresholds}
\begin{figure}[htb]
  \centering
  \includegraphics[width=\textwidth]{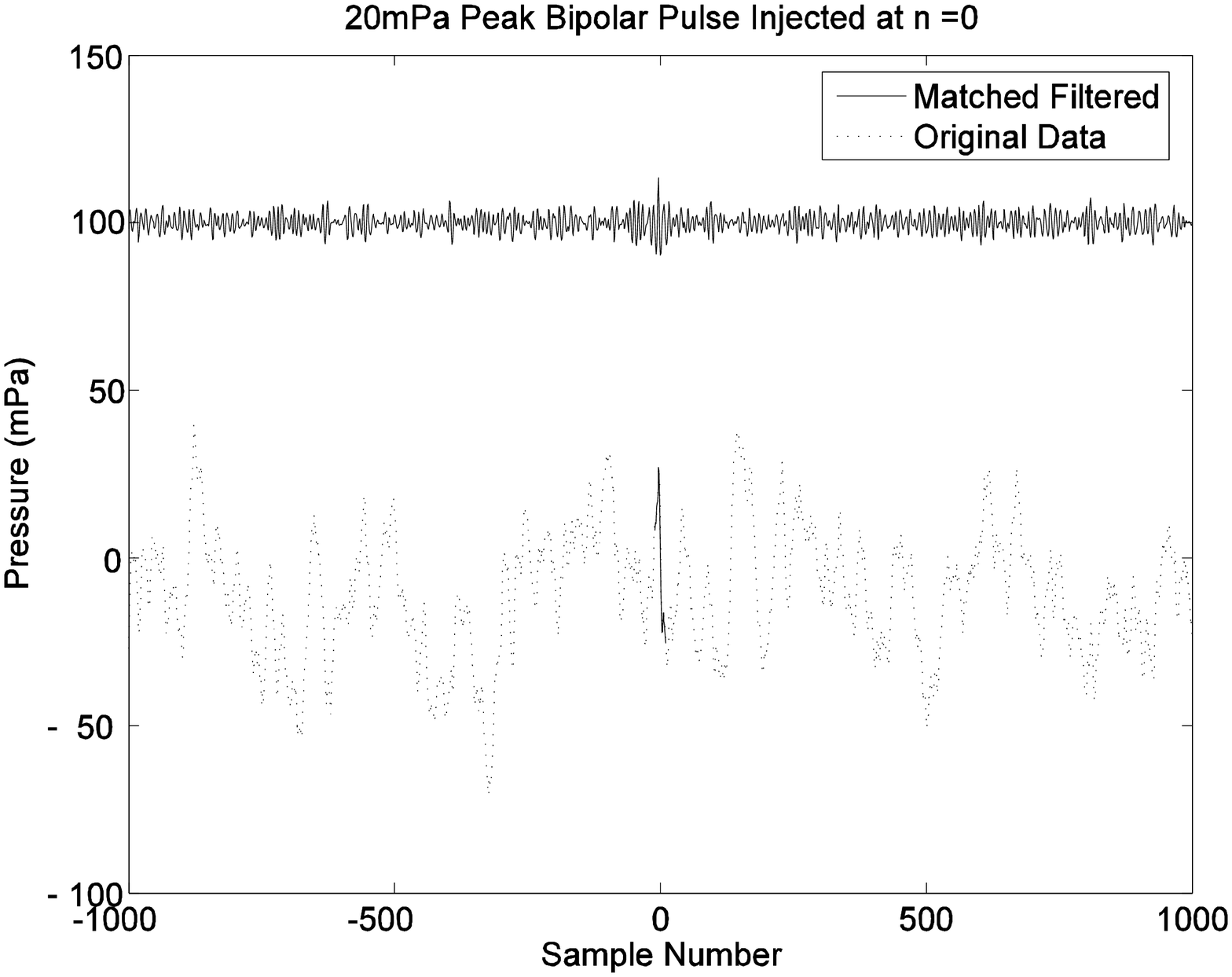}
  \caption[Matched filter performance]
	  {
	    \small
	    The effect of a matched filter on a bipolar signal
	    embedded in some Rona data (from
	    \cite{danaher:matchedfilter} which contains a full description
	    of the matched filtering technique).
	    The effect of such a filter on a signal embedded in a
	    flat, Gaussian, white noise background is to enhance signal
	    to noise by approximately a factor of three; this is the
	    case in the large scale detector simulation.
	    The lower plot is the raw input with the bipolar signal
	    emboldened because it is difficult to distinguish by eye
	    from the noise. Offset above is the filtered output,
	    where the bipolar signal is clear relative to the
	    background noise.
	  }
	  \label{fig:matchedfilter}
\end{figure}

Estimation of the sensitivity of a hydrophone array relies on
reconstruction of the neutrino interaction vertex from the acoustic
signals received. Since it is the timing of the signals that is of
importance in the detector simulation, a parameterisation of neutrino
energy to peak pressure can be used. In reality event triggers will
depend on knowledge of the bipolar pulse shape. The performance of a
matched filter is coupled to the attenuation calculation introduced in
Section \ref{subsec:angularspread} as a means of simulating the
response of a hydrophone to bipolar signals. The matched filter is applied
to a flat, Gaussian white noise background. The noise level is
determined from Knudsen curves (see Appendix \ref{appendix:knudsen})
which relate the surface wind speeds to
the level of ambient noise. The matched filter enhances the signal by
approximately a factor of three. The simulated hydrophones are
modelled to have omnidirectional sensitivity; an example of matched
filter performance is plotted in Figure
\ref{fig:matchedfilter}. Naturally the precise shape of a bipolar
signal in the sea varies as a result of the attenuation described in
Section \ref{sec:attenuation}. As such a particular matched filter
will be optimal for some function of distance and angle from a given
source. In reality several matched filters will have to be used to
fully optimise neutrino signal identification. This ensemble
of filters can be represented in the large scale simulation by a
single entity, optimised for all frequencies, with a probability for
correct identification of a neutrino induced signal equal to unity.

The signal detection threshold, in the detector simulation,
is set at $35$\,mPa, which corresponds to a probability of false alarm
(PFA) of one false trigger in ten years, due to noise exclusively
resulting from surface wind speed, with a
five-fold coincidence. A threshold of $35$\,mPa indicates a mean noise
level of $35.9$\,dB. The mean noise level from June to August in the
Tyrrenian Sea for example is $35$\,dB with an average wind speed of
$5.6$\,knots\cite{acorne:proposal}.
\section{Summary}
\label{sec:propagationsummary}
We have seen that acoustic radiation is modified by its transmission
through the sea. This leads to a set of constraints on the simulation of a
large scale acoustic array. Hydrophones that are greater than
$5$\,$\degrees$ out of the plane defined by the acoustic pancake are
neglected because of the signal attenuation resulting from
diffraction. There is a natural, geometry dependent, limit on the
distance from which a source can be heard by a hydrophone located in a
sea of finite depth because of the refraction of sound rays towards
the surface. Finally, a threshold for detection of $35$\,mPa has been
established based on the ambient noise in the sea and the probability
that this noise could emulate a neutrino like acoustic signal.

  \cleardoublepage

  \chapter{Large Scale Detector Simulation}
\label{chp:largescale}
%
%
%
%
%
%
\section{Introduction}
\label{largeintro}
All the ingredients are now available for the recipe that 
computes the sensitivity of a hypothetical array of hydrophones
to a flux of UHE neutrinos. In summary the following steps have been
described: 
\begin{itemize}
\item Generation of the mean fraction of the incident neutrino energy
  taken away by the hadronic particle cascade.
\item Simulation of the evolution of a hadronic cascade through
  seawater, yielding the thermal energy density deposited therein.
\item Integration of the cascade energy and subsequent production of
  the bipolar acoustic signal.
\item Parameterisation of the pulse formation process for fast
  calculation of peak acoustic pressure using the far-field
  approximation\footnote{A near-field pulse will obviously contain
  much higher frequency components, but the assumption that the
  matched filter is optimal across all frequencies of interest takes
  this into consideration}.
\item Modelling of the attenuation of acoustic radiation as it permeates
  the surrounding seawater thus diminishing the peak pressure.
\item Calculation of the trajectories and signal times for sound rays
  undergoing refraction due to a gradient in the sound velocity as a
  function of ray depth.
\end{itemize}
The following discussion explains the procedures undertaken in a
bespoke simulation for the purpose of estimating hypothetical
hydrophone array performances.
\section{The Sensitivity Estimation Procedure}
\label{sec:procedure}
The steps taken by the large scale detector simulation are as follows:
\begin{description}
\item[Step 1.]
  An ensemble of downward going neutrinos is generated with an energy
  spectrum linear in $\log_{10} E_{\nu}$.
\item[Step 2.]
  A hydrophone array is ``constructed'' in a deep sea environment from a
  file containing the coordinates of typically one thousand
  hydrophones distributed at random in a volume of one cubic kilometre.
\item[Step 3.]
  The program then samples the neutrino spectrum and calculates the
  detector response on an event by event basis:
  \begin{enumerate}
    \renewcommand{\labelenumii}{\roman{enumii}.}
  \item 
    Each neutrino is forced to interact in a fiducial volume, called
    the ``can'', that surrounds the volume defined by the hydrophone
    locations.
  \item
    The peak pressure as a function of neutrino energy and attenuation
    at each hydrophone is calculated for both
    refracted and unrefracted sound rays assuming an omnidirectional
    sensitivity.
  \item 
    The pressure amplitude and arrival time for all those hydrophones that
    register a signal above threshold is recorded, taking into
    consideration the performance of a matched filter.
  \end{enumerate}
\item[Step 4.]
  For those events that register hits above threshold reconstruction
  of the interaction vertex and the neutrino trajectory is
  attempted.
\item[Step 5.]
  If the vertex reconstruction algorithm returns an unphysical vertex
  location the event is discarded, otherwise the event is classified
  as detected.
\item[Step 6.]
  The sensitivity of the hydrophone array under examination is then
  calculated from the number of successfully reconstructed
  events. This is translated into a limit on the neutrino flux
  assuming that the hydrophone array detected no events for a given
  period of observation.
\end{description}
An example of the source distributions from which the event data are
sampled can be found in Appendix \ref{appendix:nusourcedata}. Since we
are operating close to threshold, a hard cut can be placed on the
angle of a given receiver with respect to the plane of the
pancake such that above $5$\,$\degrees$ or so the receivers are
neglected since the angular attenuation has diminished the signal by
several orders of magnitude (see Section \ref{subsec:angularspread}).
\section{Neutrino Spectra and Geometries Used}
\label{sec:nuspecandgeom}
\begin{figure}[htb]
  \centering
  \includegraphics[width=0.75\textwidth]{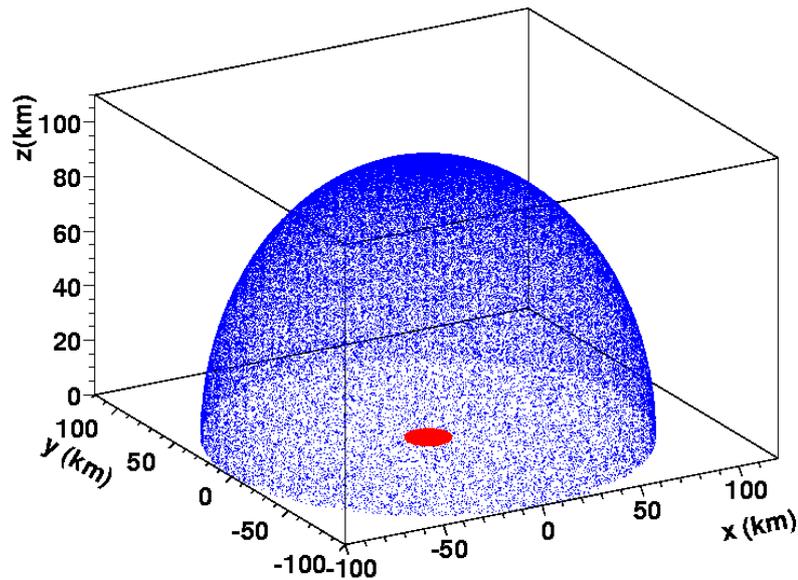}
  \caption[Neutrino source locations]
	  {
	    \small
	    The spatial distribution of neutrino source locations
	    isotropic over a positive hemispherical shell centred on the
	    instrumented volume. Surrounding the instrumented volume
	    is the cylindrical can. The radius of the shell is
	    somewhat arbitrary as long as the shell itself encompasses
	    the can. In this plot it is $100$\,km.
	  }
	  \label{fig:nushell}
\end{figure}

\begin{figure}[htbp]
  \centering
  \includegraphics[width=0.5\textwidth]{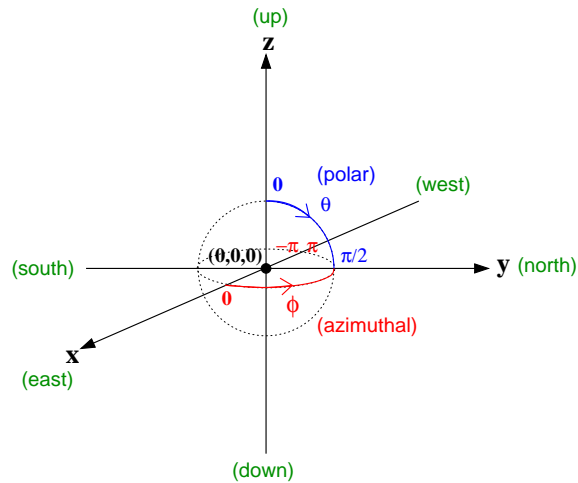}
  \caption[Coordinate systems]
      {
	\small
	Coordinate system employed in the hydrophone array sensitivity
	estimation program.
      }
      \label{fig:coordinates}
\end{figure}

As with the neutrino event generation program discussed in Section
\ref{sec:nuevent} the neutrino energy spectrum is arbitrarily modelled
to be flat in $\log E$ since the ensuing sensitivity calculation will
incorporate a model independent neutrino flux. Therefore:
\begin{equation}
  \frac{dN}{d(\log E_{\nu})} = \textrm{constant} \Leftrightarrow 
  \frac{dN}{dE} \propto E^{-1}
\end{equation}
The neutrinos are isotropic over a hemispherical shell that is
centred at the origin of the instrumented volume. All neutrinos
propagate freely from the shell to a point in the cylindrical,
fiducial volume, known as the ``can'' that envelops the volume defined
by the hydrophone locations. The distribution of generated neutrino
events is plotted in Figure \ref{fig:nushell}.
The origin of the coordinate system that is employed in this
simulation is defined to be at the centre of the instrumented volume at
a depth of $2000$\,m in a sea of depth $4000$\,m. A conventional
right-handed set of axes is fixed thus, with the $z$-axis running
vertically and the $xy$-plane defining the horizontal. This coordinate
system is illustrated in Figure \ref{fig:coordinates}. Typically
$1000$ hydrophones are placed, at random, in a volume of one
kilometre cubed, centred at the origin.
\section{Event Reconstruction}
\label{sec:eventreconstruction}
Signal arrival times are used to resolve interaction vertices. The
formula for vertex reconstruction of sound rays that have undergone
refraction is different from the formula that is used for rays that
have linear trajectories.
\subsection{Vertex location with linear rays}
\label{subsec:linearvertexreconstruction}
In a homogeneous medium the velocity of sound $c$ is constant. Point
source location from differences in signal arrival times is well
documented and has been implemented by the RICE
\cite{rice:kravchenko}, SAUND \cite{saund:vandenbroucke}, Erlangen
\cite{timo:thesis} and ACoRNE groups. The standard formalism is
described below. 

\begin{figure}[htbp]
  \centering
  \hfill
  \begin{minipage}[h!]{0.45\textwidth}
    \includegraphics[width=\textwidth]{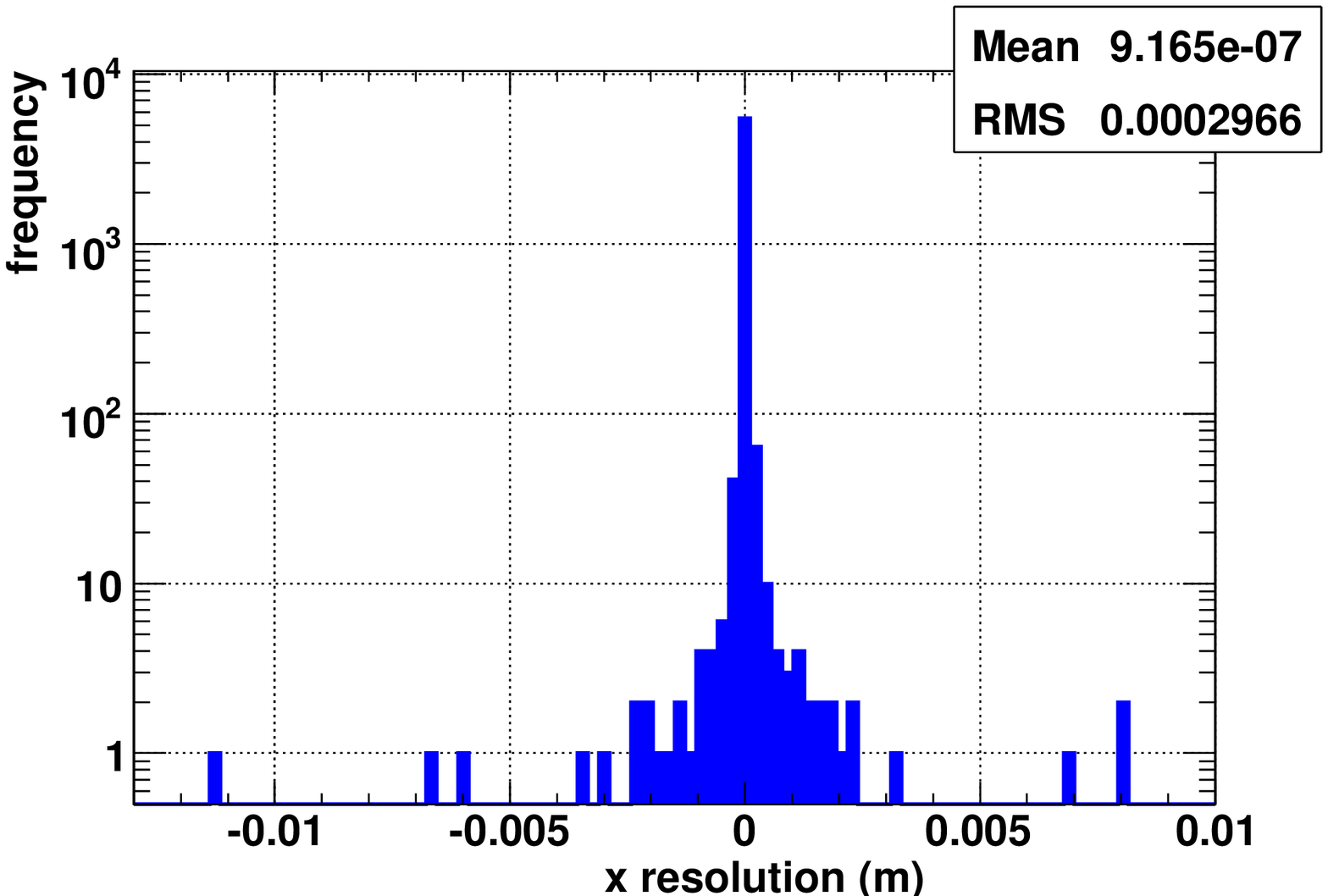}
  \end{minipage}
  \hfill
  \begin{minipage}[h!]{0.45\textwidth}
    \includegraphics[width=\textwidth]{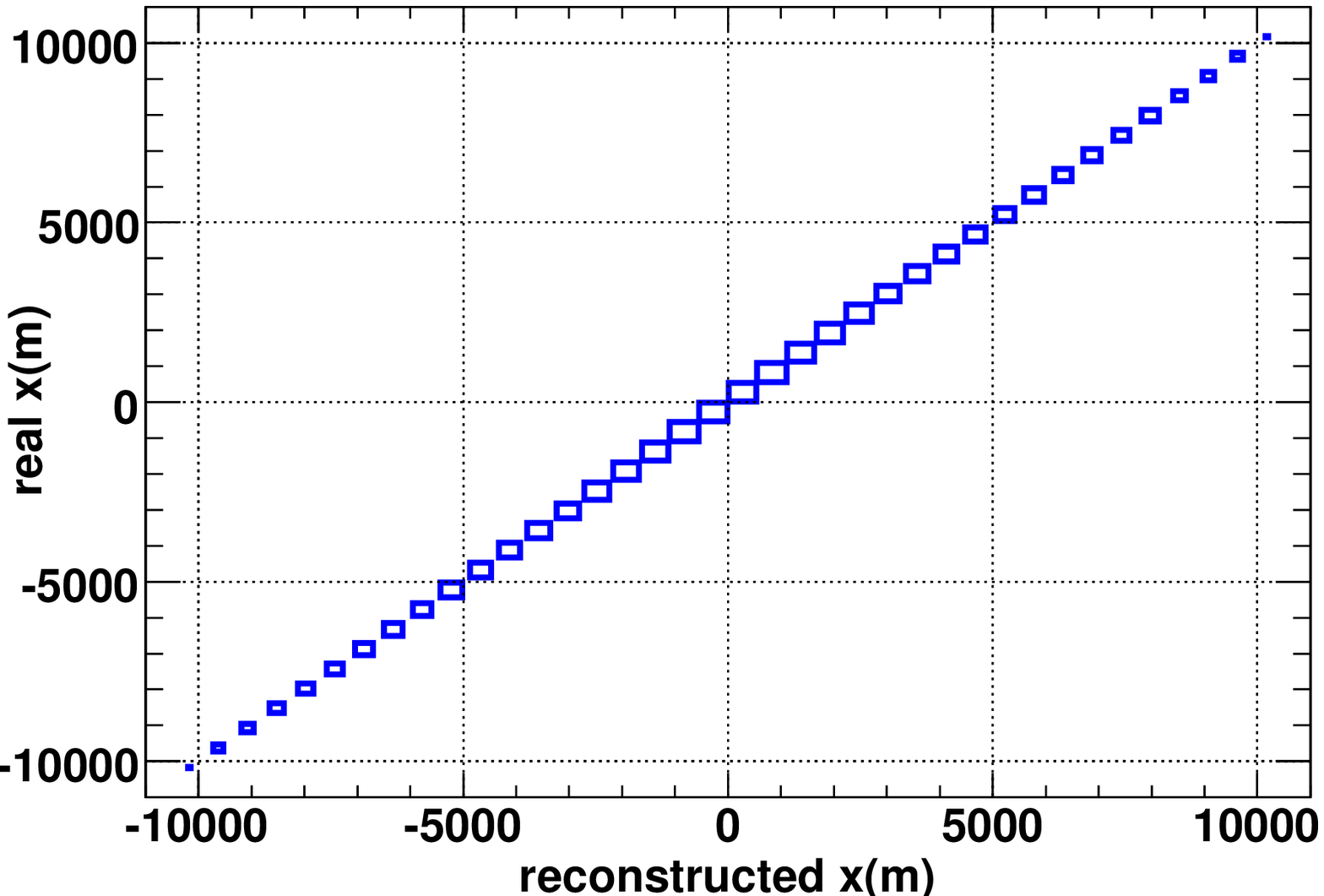}
  \end{minipage}
  \hfill
  \caption[Unsmeared $x$-coordinate resolution]
	  {
	    \small
	    The resolution of the vertex $x$-coordinate, in the case
	    of a constant sound velocity and no smearing on the
	    hydrophone location, has a mean value consistent with zero
	    and a standard deviation\footnotemark of less than $1$\,mm.
	   
	  }
	  \label{fig:unsmearedxres}
\end{figure}

\begin{figure}[htb]
\centering
\includegraphics[width=0.45\textwidth]{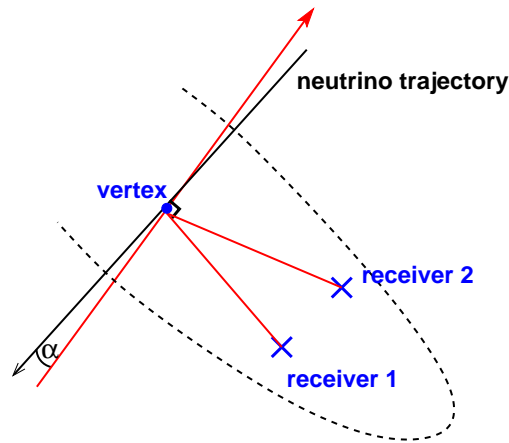}
\caption[Finding the pointing vector]
	{
	  \small
	  The vector normal to the plane defined by a pair of
	  hydrophones lying in the acoustic pancake and the
	  reconstructed vertex points back along the axis of
	  the hadronic cascade. Hence this vector is collinear with
	  the trajectory of the incident
	  neutrino. The pointing resolution is defined by the angle
	  $\alpha$ between the pointing vector and the incoming
	  neutrino trajectory.
	}
	\label{fig:pointing}
\end{figure}

A receiver at location $\vec r_i$, hears a signal at time
$t_i$, from a source at location $\vec v$, originating at time $t_0$,
according to the following equation
\begin{equation}
|\vec v - \vec r_i|^2 = c^2(t_i - t_0)^2
\label{eq:linearrecon}
\end{equation}
Expanding the bracketed terms and solving for $\vec v$ gives:
\begin{eqnarray*}
  v^2 + r_i^2 - 2\vec v \cdot \vec r_i & = & c^2(t_i - t_0)^2 \\
  (r_i^2 - r_j^2) - c^2[(t_i - t_0)^2 - (t_j - t_0)^2]  & = & 
  2\vec v \cdot (\vec r_i - \vec r_j) \\
  (r_i^2 - r_j^2) - c^2[t_i^2 - t_j^2 - 2t_0(t_i - t_j)] & = & 
  2\vec v \cdot (\vec r_i - \vec r_j) \\
  (r_i^2 - r_j^2) - c^2[t_i^2 - t_j^2] + 2c^2t_0(t_i - t_j) & = &
  2\vec v \cdot (\vec r_i - \vec r_j) \\
\end{eqnarray*}
which can be expressed in the form of a matrix equation:
\begin{equation}
  \matrixR + t_0 \matrixT = \mathbf{M} \vec v
  \label{eq:linearrecon:matrixequation}
\end{equation}
where, in the limiting case of four receivers:
\begin{eqnarray*}
  \matrixR   &=& {r_i^2 - r_j^2 - c^2(t_i^2 - t_j^2)} 
                 \quad (i,j = 1,2;1,3;1,4)\\
  \matrixT   &=& {2c^2(t_i - t_j)} \quad \qquad \qquad (i,j = 1,2;1,3;1,4)\\
  \mathbf{M} &=& 2 \left( \begin{array}{ccc}
                          dx_{12} & dy_{12} & dz_{12} \\
                          dx_{13} & dy_{13} & dz_{13} \\
                          dx_{14} & dy_{14} & dz_{14} \\
                          \end{array} \right)\\
\end{eqnarray*}
(Here, the receiver with index $1$ is taken as the reference receiver,
$dx,dy,dz$ are the difference in $x,y$ and $z$ coordinates
respectively, between this receiver and the other receivers that
detect a signal)\footnote{it now becomes apparent why a random spatial
distribution of hydrophones is employed, since a regular sensor array
could lead to singular ($|\mathbf{M}|=0$) or near singular matrices,
for example if the hits were confined to a plane aligned with either
of the $x$,$y$ or $z$ axes; such a matrix would fail the inversion
process}. Therefore: 
\begin{eqnarray*}
  \vec v & = & \mathbf{M}^{-1} (\matrixR + t_0 \matrixT)\\
         & = & \mathbf{M}^{-1} \matrixR + t_0 \mathbf{M}^{-1} \matrixT\\
\end{eqnarray*}
One defines:
\begin{eqnarray}
  \matrixP & = & \mathbf{M}^{-1} \matrixR\\
  \matrixQ & = & \mathbf{M}^{-1} \matrixT
  \label{eq:linearrecon:defpq}
\end{eqnarray}
then:
\begin{equation}
  \vec v = \matrixP + t_0 \matrixQ
  \label{eq:linearrecon:vpq}
\end{equation}
The minimum number of receivers required for vertex reconstruction to
be possible is four. If there are $n$ receivers the system becomes over
constrained and matrix $\mathbf{M}$ becomes a $3 \times (n-1)$
matrix. Because $\mathbf{M}$ is no longer symmetric it must be
inverted using singular value decomposition (SVD). 
For an $(m \times n)$ matrix $\mathbf{A}$ with $m \geq n$, the
singular value decomposition is given by:
\begin{equation*}
\mathbf{A} = \mathbf{U\,L\,V}^T
\end{equation*}
where $\mathbf{U}$ and $\mathbf{V}$ are orthogonal $m \times m$ and $n
\times n$ matrices containing the column and row eigenvectors
respectively. $\mathbf{L}$ is an $m \times n$ matrix
containing the eigenvalues in increasing order. The inverse of
$\mathbf{A}$ is therefore given by:
\begin{equation*}
\mathbf{A}^{-1} = \mathbf{V\,L}^{-1}\,\mathbf{U}^T
\end{equation*}
Since $\matrixP$ and $\matrixQ$ are known, it only remains to find $t_0$. This
is done by solving the propagation time equation for any receiver
amongst the four:
\begin{equation}
  |\vec v - \vec r_i|^2 = c^2(t_i - t_0)^2 = |\matrixP + t_0 \matrixQ - 
  \vec r_i|^2
\end{equation}
which is simply a quadratic in $t_0$:
\begin{equation}
  c^2 t_i^2 + c^2 t_0^2 - 2c^2 t_i t_0 = \mathbf{P}^2 + t_0^2 \mathbf{Q}^2
  + r_i^2 + 2t_0 \matrixP \cdot \matrixQ - 2 \matrixP \cdot \vec r_i
  - 2t_0 \matrixQ \cdot \vec r_i
  \label{eq:linearrecon:propagationtime}
\end{equation}
rearranging for $t_0$ gives:
\begin{equation}
  (c^2 - \mathbf{Q}^2)t_0^2 + (2 \matrixQ \cdot \vec r_i - 2 \matrixP \cdot
  \matrixQ - 2c^2t_i)t_0 - (\mathbf{P}^2 + r_i^2 - 2 \matrixP \cdot \vec r_i
  - c^2t_i^2) = 0
  \label{eq:linearrecon:ts}
\end{equation}
Both roots to Equation \ref{eq:linearrecon:propagationtime} are
potentially viable. The correct root is sought by minimising the
following chi-squared:
\begin{equation}
  \chi^2 = \sum_{i=0}^{N_{\textrm{receivers}}}(t_{i} - t_{r})^2
\label{eq:linearrecon:chisqr}
\end{equation}
where as before $t_i$ is the arrival time of the signal at the
receiver located at $\vec r_i$, and the reconstructed time $t_r$ is
given, in terms of the signal emission time $t_{\pm s}$ and vertex
location $\vec v_{\pm}$, from the positive and negative solutions, by:
\begin{equation*}
  t_r = t_{\pm s} + \frac{(\vec r_i - \vec v_{\pm})}{c}
\end{equation*}
which yields the correct root $99.9\%$ of the time. Interestingly,
simply invoking causality is not sufficient since in many cases both
roots are allowed; the exact number of causal and non-causal roots is
dependent on the geometry of the detector and the can. The vertex
resolution of a one thousand hydrophone array, of one cubic kilometre
instrumented volume in the case of a constant sound velocity, is
plotted in Figure \ref{fig:unsmearedxres}.
\footnotetext{for historical reasons the standard deviation in \ROOT
  is actually called ``RMS'', an explanation is given at
  http://root.cern.ch/root/html/TH1.html$\#$TH1:GetRMS Accessed
  25/11/2006}
Now, given the interaction vertex, and a set of receiver signals
confined to a plane perpendicular to the direction of the neutrino,
all one has to do to resolve the cosmic source of the neutrino is
point back through the vertex with a vector normal to the pancake
plane. The neutrino pointing vector has been reconstructed in two
ways: firstly by taking the normal to the plane defined by the
vertex and the two most energetic receivers in the pancake, as
illustrated in Figure \ref{fig:pointing}; secondly by taking the cross
product of two vectors orthogonal to each other and the cascade
direction, as plotted directly from the simulation in Figure
\ref{fig:pointing}.
The procedure, subsequently called the ``Top2'' algorithm, for
determining these vectors is as follows:
\begin{enumerate}
  \renewcommand{\labelenumii}{\roman{enumii}.}
\item
  calculate the centre of mass of the system of hit receivers - this
  lies in the plane of the pancake; then, translate the coordinate
  system such that the origin is at the centre of mass.
\item
  Define a vector $\vec v_1$ that points from the centre of mass
  to the vertex.
\item
  Rotate the coordinate system such that $\vec v_1$ is aligned
  along (say) the $x$-axis, then collapse the system into $2D$
  by discarding the $x$-coordinates.
\item
  Perform a least squares fit to the hit receivers in $yz$-plane,
  because the centre of gravity is at the origin, the intercept of
  this line will be at zero, define a second vector $\vec v_2$ in
  the direction of this line.
\item
  Rotate the system back - vectors $\vec v_1$ and $\vec v_2$ are
  orthogonal to each other, the neutrino pointing vector is simply
  $\vec v_1 \times \vec v_2$.
\end{enumerate}
\begin{figure}[htb]
  \centering
  \includegraphics[width=0.45\textwidth]{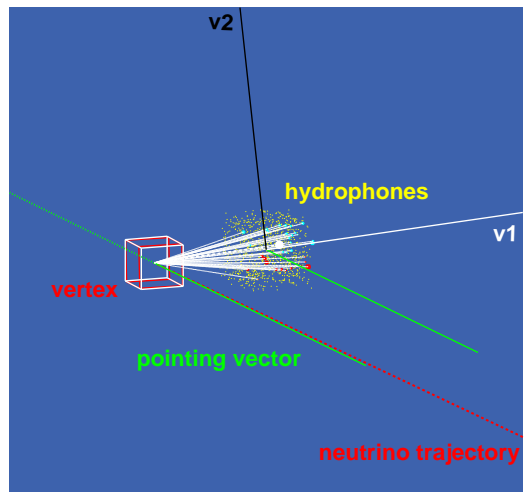}
  \caption[Pointing reconstruction]
  {
    \small
    Output from the simulation displaying vectors $\vec v_1$ and $\vec
    v_2$ and the resultant pointing vector as described in the text.
  }
  \label{fig:pointingreconstruction}
\end{figure}
	  
\begin{figure}[htb]
  \centering
  \includegraphics[width=0.75\textwidth]{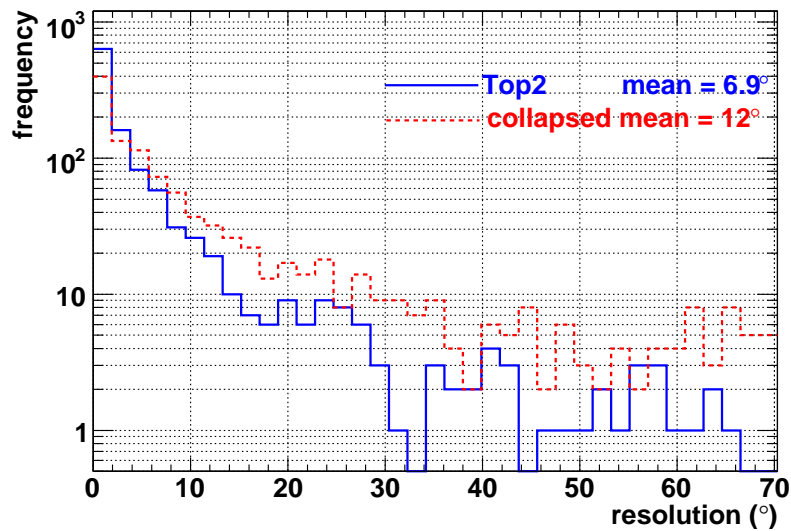}
  \caption[Pointing resolution without refraction]
      {
	\small
	The pointing resolution of both
	reconstruction techniques for an array of one thousand
	hydrophones distributed randomly in a volume of one kilometre
	cubed and a can radius of $10$\,km.
	The resolution is defined as the
	angle between the true neutrino trajectory and the
	reconstructed pointing vector.
      }
      \label{fig:pointingresolution}
\end{figure}

The pointing resolution is defined as the angle $\alpha$ between the
real neutrino trajectory and the reconstructed pointing vector. A
comparison of the resolution from both methods discussed
above is plotted in Figure \ref{fig:pointingresolution}, a mean
pointing resolution of $6.9$\,$\degrees$ is obtained from the
Top2 algorithm and $12.0$\,$\degrees$ from the collapsed mean algorithm,
in the absence of refraction, for an array of
$1000$ hydrophones distributed at random in a volume of one cubic
kilometre.
\subsection{Vertex location with curvilinear rays}
\label{subsec:circularrecon}
\begin{figure}[htb]
  \centering
  \hfill
  \begin{minipage}[!h]{0.4\textwidth}
    \includegraphics[angle=90, width=0.7\textwidth]{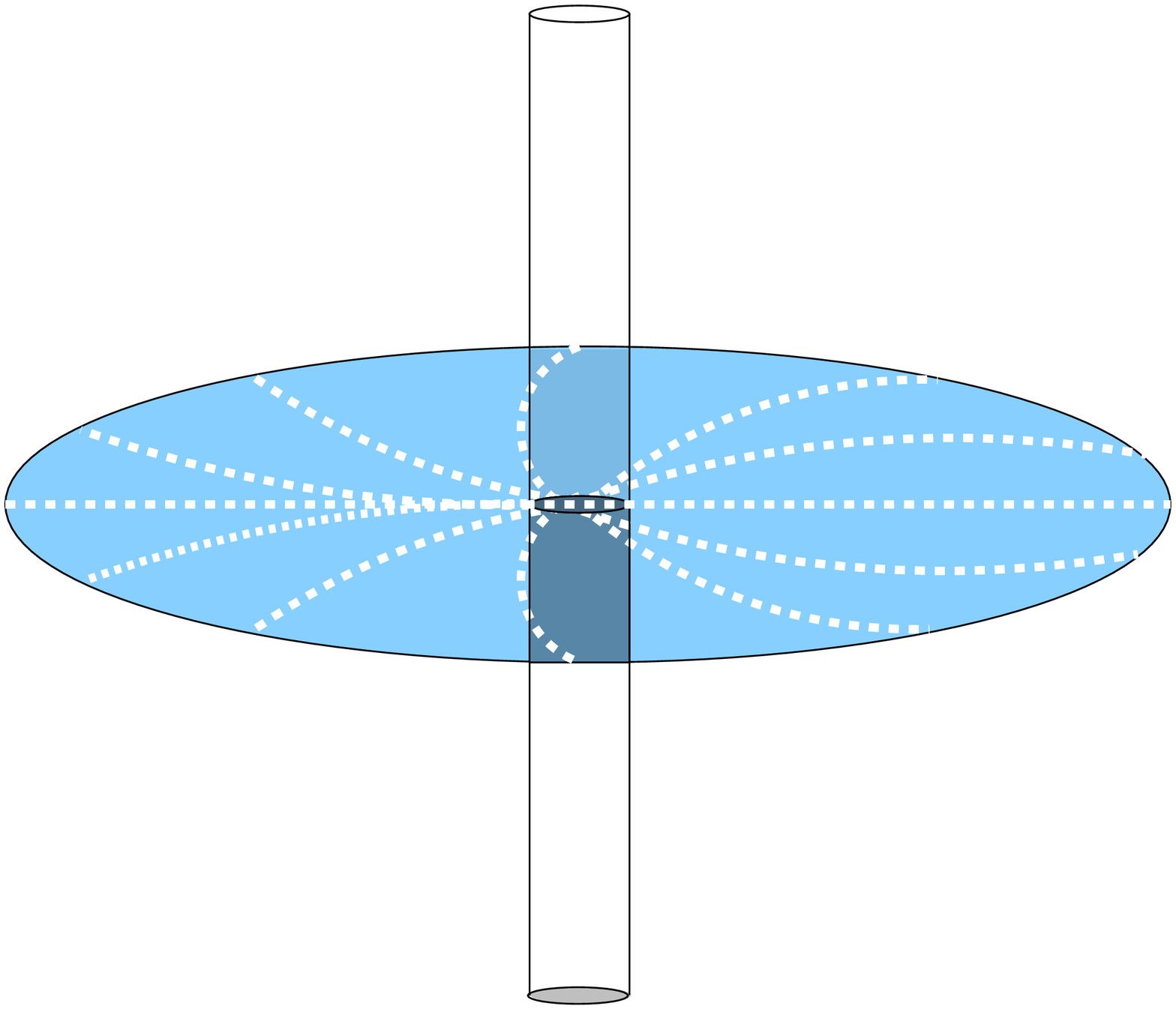}
  \end{minipage}
  \hfill
    \begin{minipage}[h!]{0.4\textwidth}
    \includegraphics[width=0.8\textwidth]{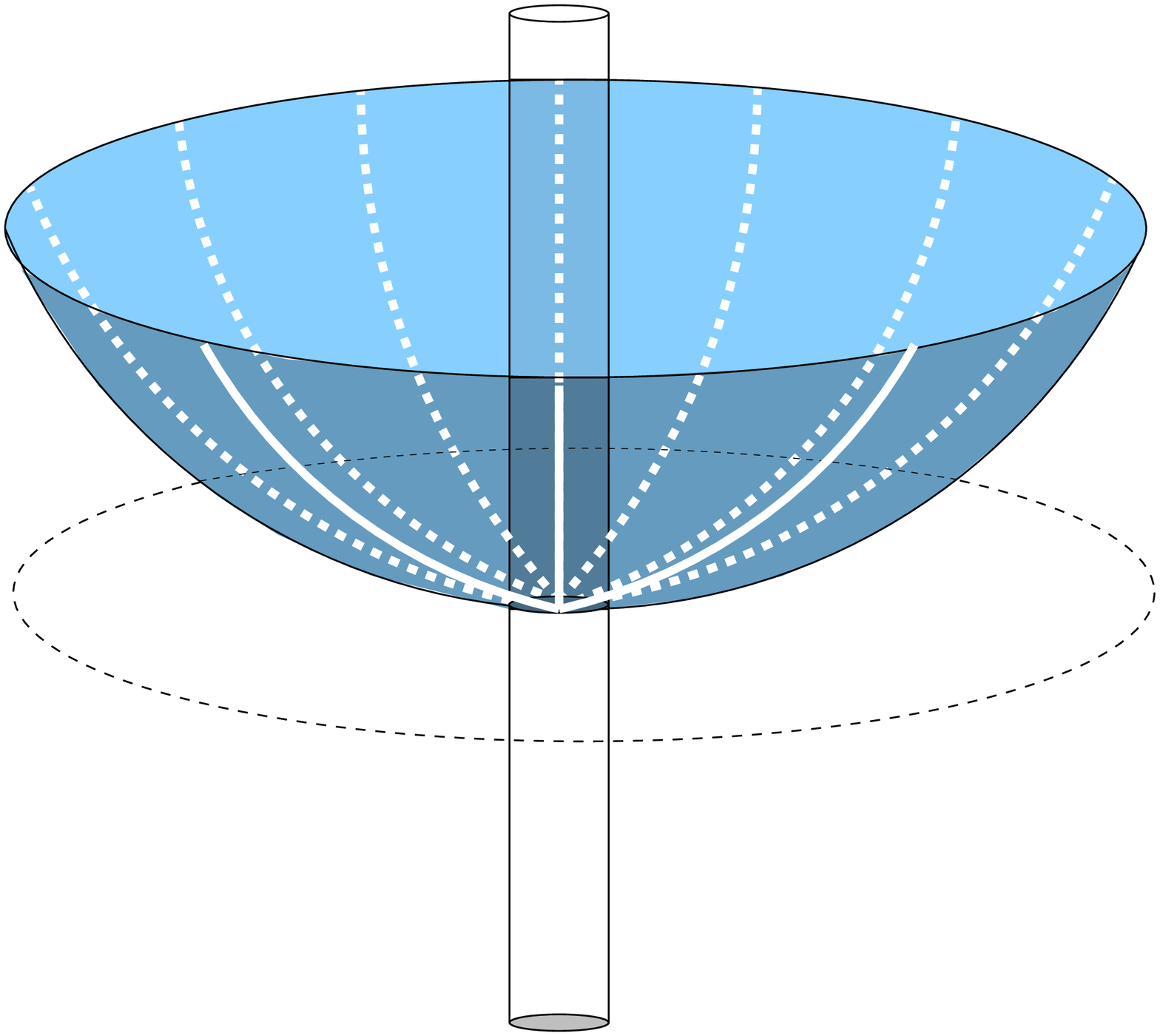}
  \end{minipage}
  \hfill    
  \caption[Schematic of refracted thermoacoustic energy deposition]
      {
	\small
	Schematic of the energy deposition. Refraction occurs only in
	the vertical plane; there are two extreme cases: that when the
	shower is horizontal ({\it Left}) and that when the shower is
	vertically inclined ({\it Right}).
	In the former case, there is no deformation of the
	pancake such that the radiation field is bent outside the
	plane orthogonal to the shower axis.
      }
      \label{fig:warped}
\end{figure}

If the velocity of sound is a function of depth, $c = c(z)$, then
Equation \ref{eq:linearrecon} is no longer valid. Sound rays follow
curved trajectories and the pancake becomes warped,
this is illustrated in Figure \ref{fig:warped}.
Unlike for constant sound velocity, there is no analytical solution to
the event vertex $\vec v$ and the signal emission time $t_0$.
One now relies on iterative minimisation to interpolate towards event
vertices in the presence of refraction. This concept has been employed
elsewhere by the RICE \cite{rice:kravchenko} and SAUND
\cite{saund:vandenbroucke} collaborations to enable
reconstruction of vertices in the data from their respective
experiments. For a given set of hits on a sensor array one can
minimise the following metric:
\begin{equation}
  \mathbf{M} =  \sum_{i=0}^{N_{\textrm{receivers}}}(t_{i} - t_{v})^2
\label{eq:curvrecon:metric}
\end{equation}
where $t_i$ is the arrival time of the signal recorded by the receiver
located at $\vec r_i$ and $t_v$ is the time it takes for a signal from
a test location $\vec v$ to reach the same receiver. In references
\cite{rice:kravchenko} and \cite{saund:vandenbroucke} minimisation
begins from a location determined by analysis of precomputed look-up
tables containing the sensor hit times for regularly spaced locations in
the fiducial volume surrounding the detector. In this approach, the
number of test locations, and hence computations, scales with the size
of the fiducial volume under consideration. When volumes of a hundred
or so cubic kilometres are being dealt with the process becomes
inefficient, taking several minutes per event (depending on can size
and lattice density). Instead, an estimate of the vertex location is
sought from which the minimisation procedure can begin.

The equations for point source location in a homogeneous medium, from
signal arrival times, can be used to approximate the
location of a point source in an inhomogeneous medium.
Starting with Equation \ref{eq:linearrecon:matrixequation}, where, in
the limiting case of four receivers:
\begin{eqnarray*}
  \matrixR   &=& {r_i^2 - r_j^2 - c^2(t_i^2 - t_j^2)} 
                 \quad (i,j = 1,2;1,3;1,4)\\
  \matrixT   &=& {2c^2(t_i - t_j)} \quad \qquad \qquad (i,j = 1,2;1,3;1,4)\\
  \mathbf{M} &=& 2 \left( \begin{array}{ccc}
                          dx_{12} & dy_{12} & dz_{12} \\
                          dx_{13} & dy_{13} & dz_{13} \\
                          dx_{14} & dy_{14} & dz_{14} \\
                          \end{array} \right)\\
\end{eqnarray*}
Again, the receiver with index $1$ is taken as the reference receiver,
$dx,dy,dz$ are the difference in $x,y$ and $z$ coordinates
respectively, between this receiver and the other receivers that
detect a signal. Additionally $c=c(z)$, where $z$ is the mean depth of the
four receivers and $\mathbf{M}$ is always a $3 \times 3$
square matrix. Solving for the vertex $\vec v$ gives:
\begin{equation*}
\vec v = \mathbf{M}^{-1}\matrixR + t_0\mathbf{M}^{-1}\matrixT
\end{equation*}
which is once more solved via the propagation time equation:
\begin{equation*}
  |\vec v - \vec r_i|^2 = c^2(t_i - t_0)^2 = |\mathbf{M}^{-1}\matrixR
  + t_0\mathbf{M}^{-1}\matrixT - \vec r_i|^2
\end{equation*}
The mean vertex $\overline{\vec v}$ from $N_q$ quadruplets of receiver
signals is thus calculated as:
\begin{equation}
\overline{\vec v} = \frac{1}{N_q} \sum_{i=0}^{i=(N_q)} \vec v_i
\end{equation}
There are:
\begin{equation}
  _NC^4 = \frac{N!}{(N-4)!4!}
\end{equation}
unique receiver quadruplets from $N$ receiver signals. This number
rapidly diverges as $N$ increases; for example, there are
$230300$ possible four-receiver combinations of $50$
receivers. Furthermore calculation of the unique permutations is
in itself computationally intensive. As such a ``reduced mean'' with
fewer elements is desired. Three ``reduced mean vertices'' have been
considered. The first is the mean vertex from each receiver and its
three spatial-nearest-neighbours; the second is the mean vertex from
each receiver and its three temporal-nearest-neighbours; and the third
is the mean vertex from each receiver and its three
pressural-nearest-neighbours\footnote{i.e. if the hydrophones are sorted
  into a list of descending pressure, the $n$th phone has pressural
  nearest neighbours $n+1$, $n+2$ and $n+3$}.
In each case there are only $N-3$ contributions to the mean, so for $50$
receivers there are only $46$ permutations compared to the $230300$
above. All three methods rely on Equation
\ref{eq:linearrecon:chisqr} to solve for the vertex, utilising the
sound velocity at the mean depth of a given receiver
quadruplet. Now, neither root of the quadratic is in fact a unique
solution to the propagation time equation (Equation
\ref{eq:linearrecon:propagationtime}) since we have substituted a
linearly variable sound velocity in place of a constant sound velocity.
As such there is an inherent instability in the estimation method, however,
an estimate of the mean is found nonetheless. Calculation of estimated
vertices takes less than one second per event on a standard $2$\,GHz
processor with $512$\,MB of RAM.

\begin{figure}[htb]
  \centering
  \includegraphics[width=0.75\textwidth]{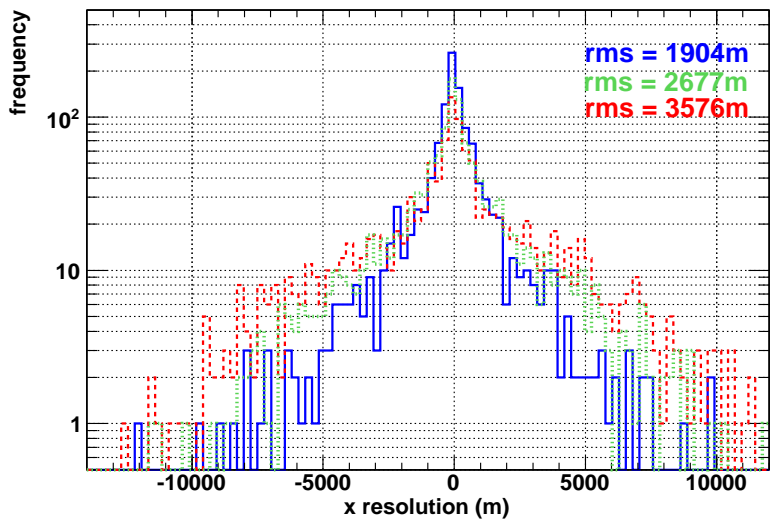}
  \caption[Reduced-mean $x$-coordinate resolution]
	  {
	    \small
	    The resolution of the reconstructed vertex $x$-coordinate.
	    The \blue{solid} curve represents the temporal mean with
	    $\sigma = 1904$\,m; the \green{dotted} curve is the
	    pressural mean with $\sigma = 2677$\,m; and, the
	    \red{dashed} curve is the spatial mean with $\sigma =
	    3576$\,m.	  
	    No smearing on the hydrophone locations was included and
	    an array of $200$ hydrophones distributed at random in a
	    volume of one kilometre cubed was used. $10,000$ neutrinos
	    were generated in a can of radius $10$\,km.
	  }
	  \label{fig:temporalvsspatialvspressural}
\end{figure}

\begin{figure}[htb]
  \centering
  \includegraphics[width=0.75\textwidth]{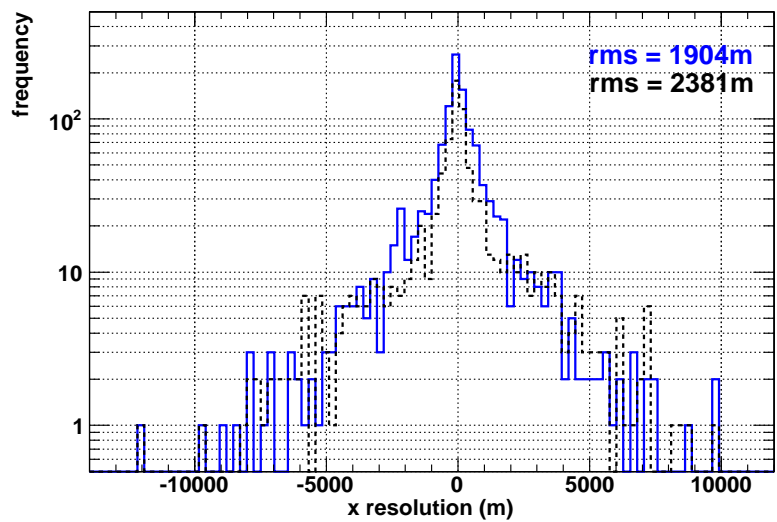}
  \caption[Reduced v.s. unreduced mean $x$-coordinate resolution]
	  {
	    \small
	    The resolution of the reconstructed vertex $x$-coordinate.
	    The \blue{solid} curve represents the temporal mean with
	    $\sigma = 1904$\,m and the dashed curve is the unreduced
	    mean with $\sigma = 2381$\,m.
	  }
	  \label{fig:temporalvsunreduced}
\end{figure}

The comparative vertex $x$-coordinate resolution from the three
reduced mean vertices is plotted in Figure
\ref{fig:temporalvsspatialvspressural}, indicating that the
temporal-nearest-neighbours gives the best estimate of the vertex. The
$x$ resolution of the temporal mean with respect to the unreduced mean
is subsequently plotted in Figure \ref{fig:temporalvsunreduced},
confirming that the vertex estimate from this method is not only a
computationally less intensive than taking the mean vertex from all
receiver combinations but more accurate. The standard deviation on the
shower time, $\sigma_{t}$, for each of the reduced mean vertices is
displayed in Table \ref{tab:rmstimeerror}; here it can be seen why
the reduced temporal mean gives the best performance, showing the
smallest intrinsic error on the shower time as returned by Equations
\ref{eq:linearrecon:propagationtime} and \ref{eq:linearrecon:chisqr}.

\begin{figure}[htb]
  \centering
  \includegraphics[width=0.75\textwidth]{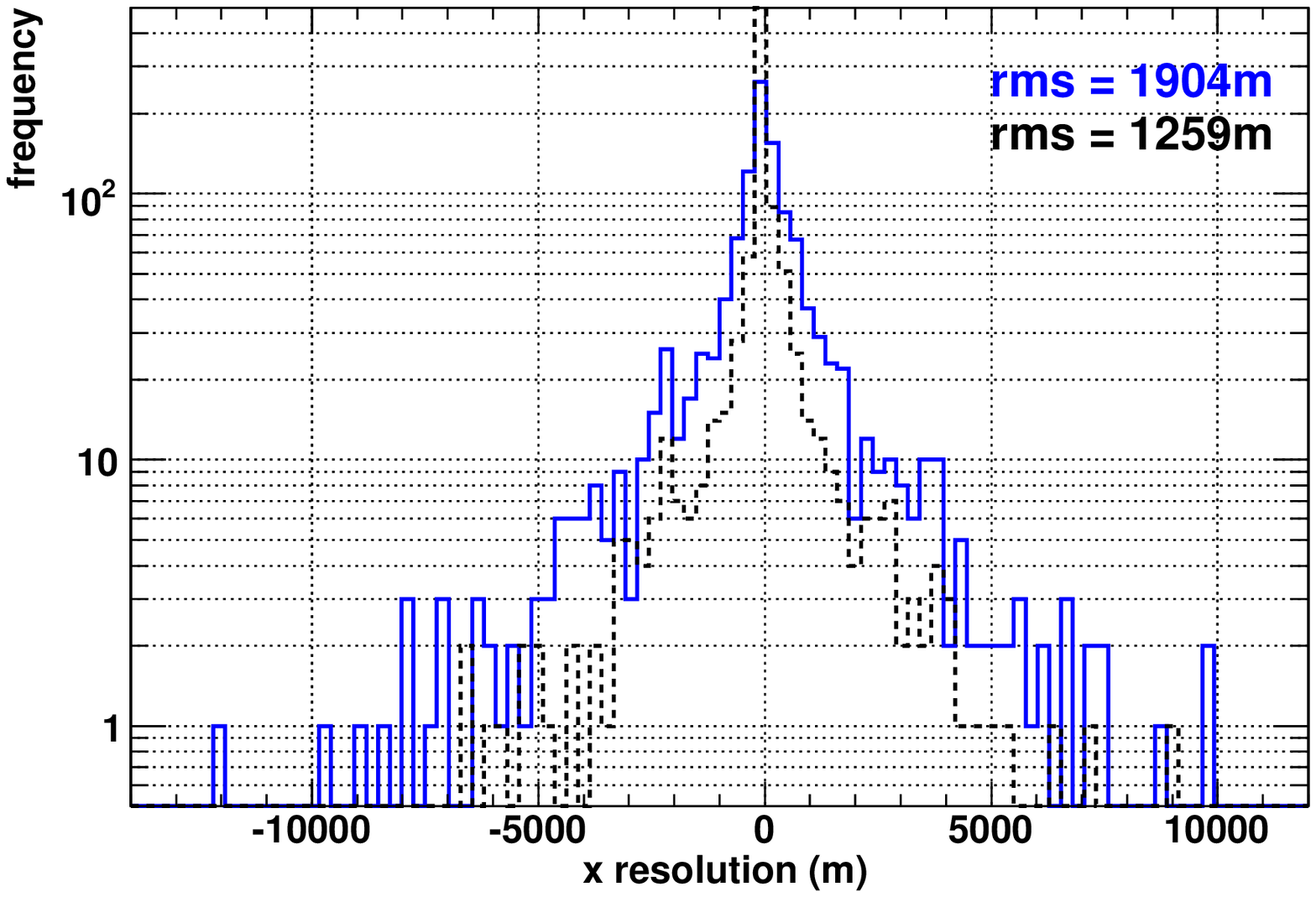}
  \caption[Minimised $x$-coordinate resolution]
	  {
	    \small
	    The resolution of the reconstructed vertex $x$-coordinate.
	    The \blue{solid} curve represents the temporal mean with
	    $\sigma = 1904$\,m and the dashed curve is the minimised
	    $x$ resolution with $\sigma = 1259$\,m.
	  }
	  \label{fig:minuitresolution}
\end{figure}

Minimisation of Equation \ref{eq:curvrecon:metric} begins from the
reduced temporal mean vertex and is performed by MINUIT \cite{minuit},
the {\it de facto} High Energy Physics minimisation package.
A comparison of the vertex resolution of the unreduced mean, the reduced
temporal mean and the vertex returned by MINUIT is plotted in Figure
\ref{fig:minuitresolution}.
\begin{table}[htb]
\centering
\begin{tabular}{ r | l }
  {\bf Reduced Mean} & {\bf $\sigma_{t}$\,(s)} \\
  \hline
  temporal      &     2.68                \\
  spatial       &     2.98                \\
  pressural     &     3.04                \\
\end{tabular}
\caption[Shower time errors]{
  \small Standard deviations of the shower times returned by Equations
  \ref{eq:linearrecon:propagationtime} and \ref{eq:linearrecon:chisqr}
  for each of the reduced mean methods in a can of radius $10$\,km.
  \label{tab:rmstimeerror}}
\end{table}

Vertex minimisation for the same $200$ hydrophones as for each of
the estimated vertices is evaluated in Table
\ref{tab:minimisation}. Eighty-seven percent of temporal minimised
vertices are found to within $1000$\,m of the true vertex compared to
$68\%$ of estimated vertices; furthermore, $30\%$ of minimised
vertices are reconstructed to within $0.1$\,m of the true vertex
compared to $1\%$ or less of estimated vertices. Naturally different
array densities will offer different reconstruction performance and
this will be discussed further in Section
\ref{sec:optimaldensity}. Furthermore increasing the radius of the
fiducial volume beyond $10$\,km will make the effects of refraction
more pronounced.
\subsubsection{A note on MINUIT performance}
\label{subsubsec:minuit}
The performance of MINUIT is not only sensitive to the starting
parameters but the topology of the $\chi^2$ hypersurface and has not
been optimised in this study. The minimisation procedure begins with
some estimate of the true vertex as returned by the
temporal-nearest-neighbour vertexing algorithm described
previously. Additionally one must specify an estimate of the error on
the associated parameters, namely the $x$, $y$ and $z$ coordinates of
the estimated vertex. Furthermore one must decide whether to constrain
the values that $x$, $y$ and $z$ can take or whether to leave them
unconstrained. In this study the vertex coordinates were constrained
to be within the fiducial volume, i.e. the can. This was decided to be
so because an unconstrained fit can lead to test locations well beyond
the fiducial volume, resulting in ray traces over thousands of
kilometres. This is problematic because the number of iterations
possible in the ray trace calculation is limited by the amount of
physical memory in the computer on which the simulation is
running. One can arbitrarily reduce the number of iterations made but
this reduces the number of straight line segments used to
approximate the circular ray path and thus causes it to appear
more and more like a straight line therefore reducing the accuracy of
the operation. An expression for the error on the estimated vertex on
an event by event basis should improve the minimisation procedure;
however, for the purpose of this study the error in $x$ and $y$ was set
at one can radius and the error in $z$ was set at
one can height for every event. Additionally the
vertex reconstruction problem in the presence of refraction is
non-linear and as such MINUIT is not guaranteed to successfully
converge on the best solution. 
\begin{table}[htb]
\centering
\footnotesize{
  \begin{tabular}{ l  r | c | c | c | c | c | c }
    & &    \multicolumn{2}{   c   }{\bf Temporal } &
           \multicolumn{2}{ | c | }{\bf Spatial  } &
           \multicolumn{2}{ | c   }{\bf Pressural} \\
    \hline
    & Res. $<\pm$ 
    & estimated & minimised & estimated & minimised & estimated & minimised \\
    \hline
    &$1000$\,m &$68\%$  &$87\%$  &$ 43\%$  &$81\%$  &$64\%$  &$84\%$\\
    & $100$\,m &$19\%$  &$64\%$  &$ 10\%$  &$59\%$  &$27\%$  &$63\%$\\
    &  $10$\,m &$ 3\%$  &$49\%$  &$ 1.8\%$ &$48\%$  &$ 7\%$  &$48\%$\\
    &   $1$\,m &$0.4\%$ &$39\%$  &$ 0.3\%$ &$34\%$  &$ 2\%$  &$39\%$\\
    & $0.1$\,m &$ 0\%$  &$30\%$  &$0.08\%$ &$26\%$  &$ 1\%$  &$29\%$\\
    &$0.01$\,m &$ 0\%$  &$14\%$  &$  0\%$  &$13\%$  &$0.7\%$ &$13\%$\\
  \end{tabular}}
\caption[Minimised vertex performance]
	{\small 
	  Minimised vertex performance. The resolution is
	  defined as $x_{recon} - x_{real}$ where $x_{recon}$ and
	  $x_{real}$ are the reconstructed and real vertex
	  $x$-coordinates
	  respectively. Minimisation beginning from the
	  temporal mean vertex can be seen to provide the best
	  performance. Here the can radius is $10$\,km.
	  \label{tab:minimisation}
	}
\end{table}
\section{The Effective Volume}
\label{sec:effectivevolume}
\begin{figure}[htb]
  \centering
  \includegraphics[width=0.75\textwidth]{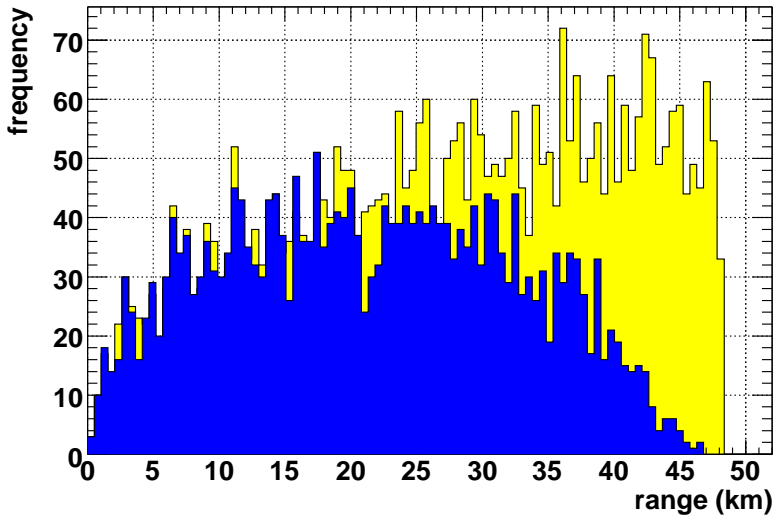}
  \caption[Range of detectable events]
	  {
	    \small
	    Range of events accessible due to the limitations imposed
	    by refraction. The lighter histogram corresponds to the
	    range of events detected if refraction is neglected and
	    shows a continuous acceptance up to $48$\,km. The darker
	    histogram is the range of detected events in consideration
	    of refraction. Beyond $24$\,km the acceptance reduces
	    until no events are detected at a range just below
	    $48$\,km.
	  }
	  \label{fig:rangeofevents}
\end{figure}

The effective volume $V_{eff}$ of a simulated array is defined as:
\begin{equation}
  V_{eff} = \frac{N_{det}}{N_{gen}} \times
  V_{gen}
  \label{eq:effective:veff}
\end{equation}
where $N_{det}$ and $N_{gen}$ are the number of events that
are detected and the number of events that are generated respectively,
and $V_{gen}$ is the volume in which events are generated. The
behaviour of Equation \ref{eq:effective:veff} is such that by
increasing the volume $V_{gen}$ in which event vertices are generated,
$V_{eff}$ will increase proportionally, however this is offset by a
reduction in the number of detections, particularly at lower energy
and in essence therefore raises the threshold.
$V_{gen}$ should certainly extend beyond the
instrumented volume and in practice one usually fixes the
size of $V_{gen}$ with the maximum range for source detection,
remembering that beyond $24$\,km parts of the sea are geometrically
inaccessible to sound rays subtending the array (see Figure
\ref{fig:sourcelimit} in Section \ref{sec:refraction}). Running the
simulation with no pressure threshold and a can radius
greater than $48$\,km determines the extent to which events can be
detected as constrained by geometry only. The range of those events
that produce a reconstructable vertex is plotted in Figure
\ref{fig:rangeofevents}. This range determines the radius of the volume
$V_{gen}$ when calculating the effective volume, $V_{eff}$.

\begin{figure}[htb]
  \centering
  \includegraphics[width=0.75\textwidth]{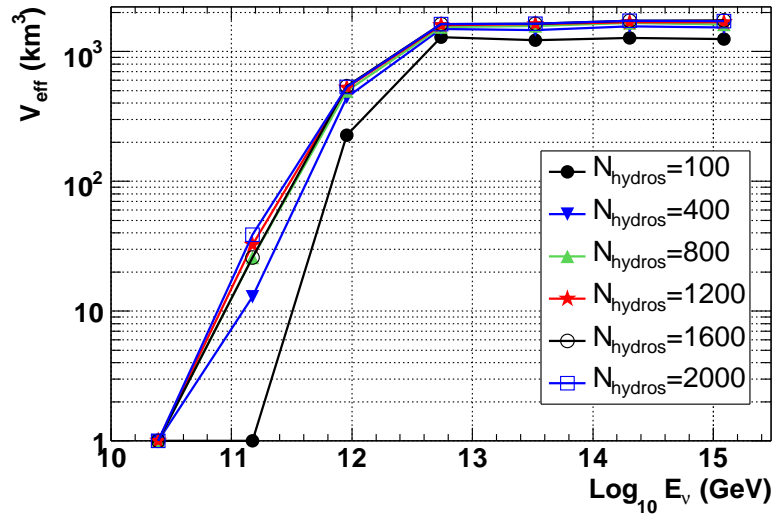}
  \caption[Effective volume as a function of hydrophone density]
	  {
	    \small
	    Effective volume as a function of hydrophone density for
	    cubic kilometre arrays.
	  }
	  \label{fig:effectivevolumes}
\end{figure}

Given an expression for the effective volume, with a $48$\,km can
radius it is possible to test the effect of different array densities.
An optimum number of hydrophones is sought, that is, the minimum
number of sensors required above which there is no significant
improvement in the effective volume. This relationship is plotted in
Figure \ref{fig:effectivevolumes}. Above a density of $400$
hydrophones per cubic kilometre the gain in effective volume is small,
especially above $10^{12}$\,GeV. At lower energies sensor density is
of more importance; at least $800$ hydrophones per cubic kilometre
are required for an optimal $v_{eff}$. The matter of hydrophone
density still requires further discussion with respect to the
resolving power of a given array.
\section{Optimal Hydrophone Density for Reconstruction in the Presence of
	  Refraction}
\label{sec:optimaldensity}
\begin{figure}[ht]
  \centering
  \hfill
  \begin{minipage}[h!]{0.47\textwidth}
    {\includegraphics[width=\textwidth]{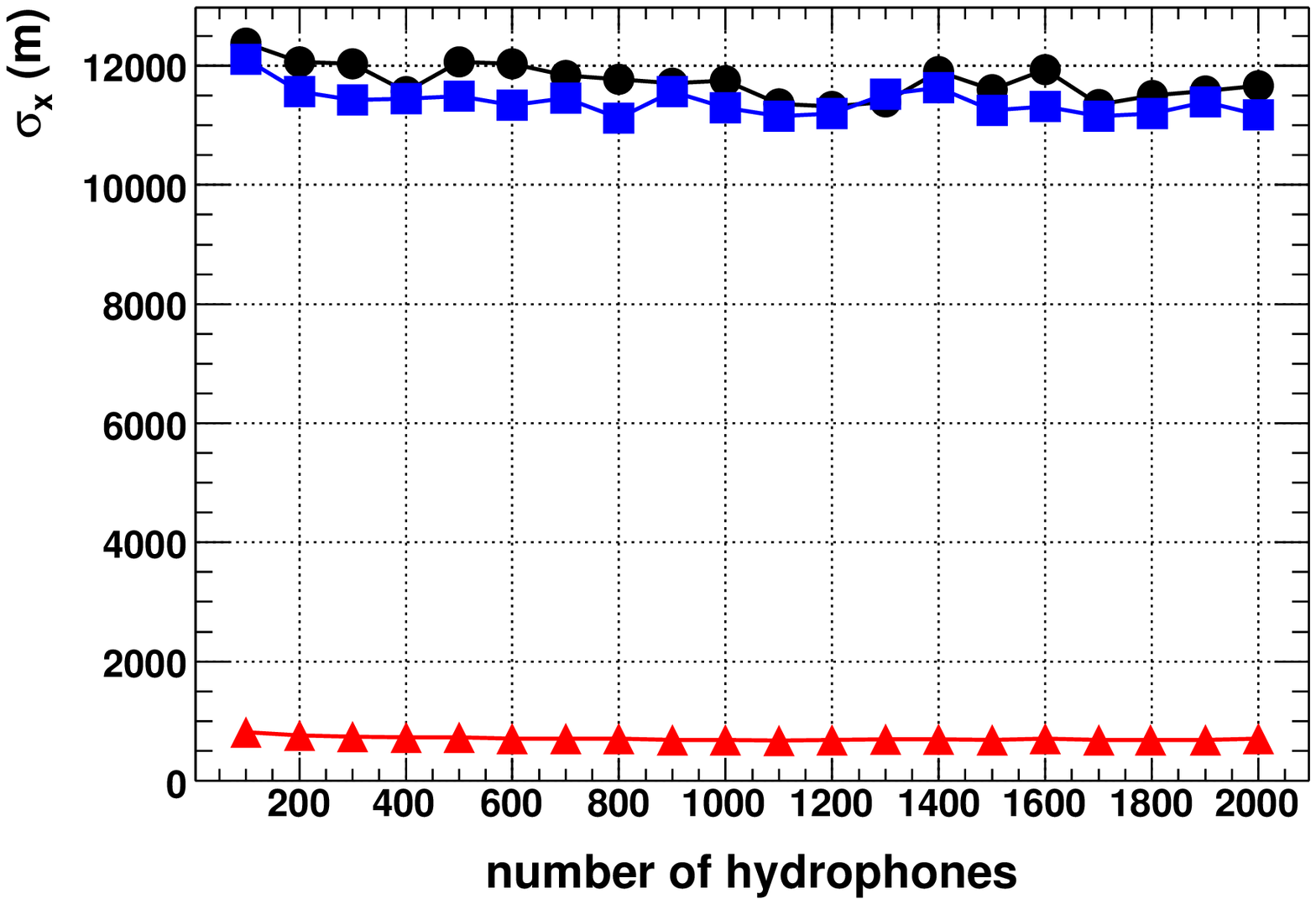}}
  \end{minipage}
  \hfill
  \begin{minipage}[h!]{0.47\textwidth}
    {\includegraphics[width=\textwidth]{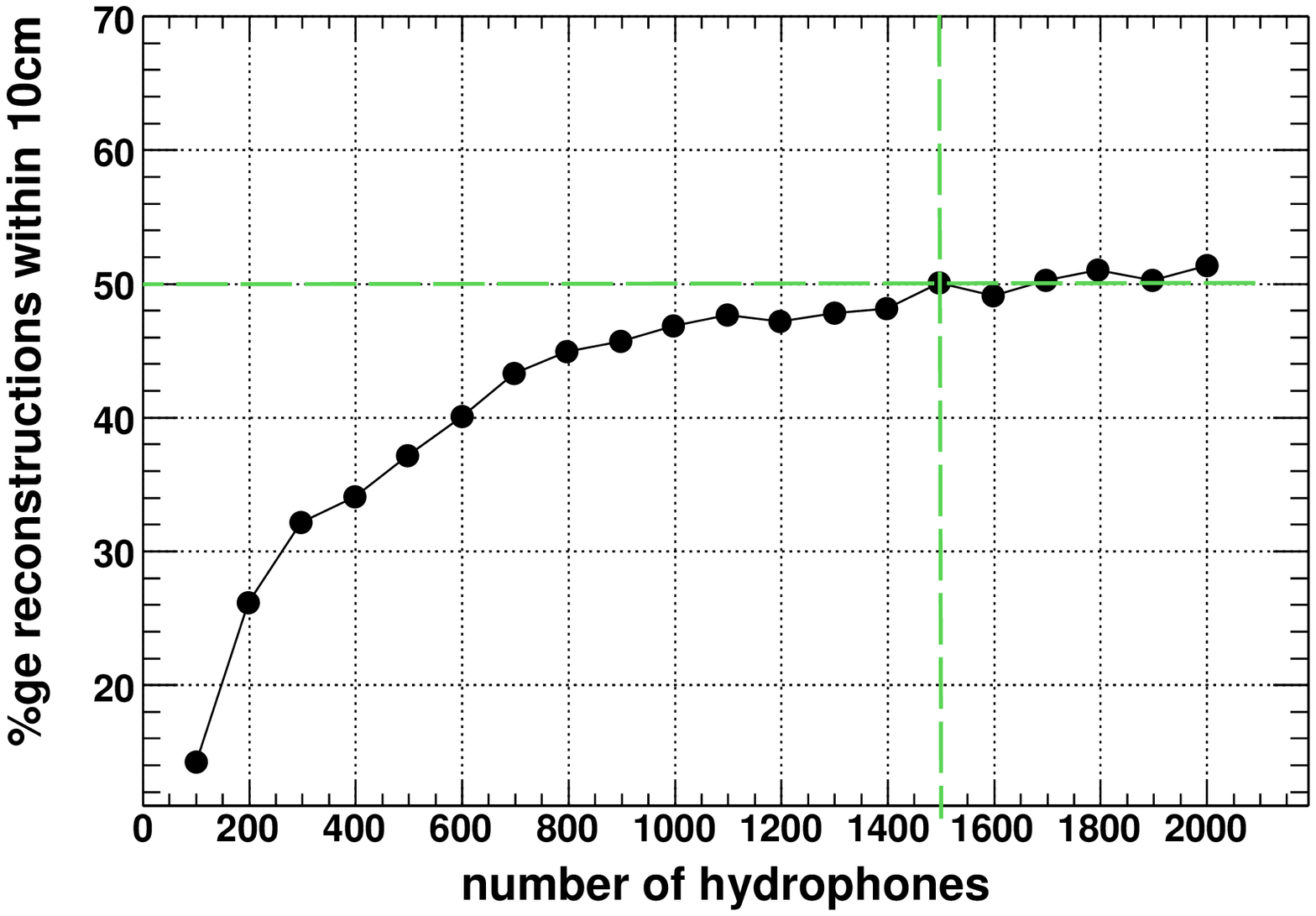}}
  \end{minipage}
  \hfill
  \caption[Vertexing performance v.s. hydrophone density]
      {
	\small
	Resolution ({\it Left}) and accuracy ({\it Right}) of vertexing as
	a function of the number of hydrophones. Resolution in $x$
	($\bullet$ marker) and $y$ (\blue{\scriptsize$\blacksquare$}
	marker) plateaus
	beyond a hydrophone density of a $300$ per cubic kilometre
	whereas the number of reconstructions up to a given accuracy
	continues to improve up to a hydrophone density of $1500$ per
	cubic kilometre. Because the can is much larger radially
	($48$\,km) than it is in height ($4$\,km) the $z$-coordinate
	(\red{$\blacktriangle$} marker) resolution is much better than
	in $x$ and $y$.}
      \hfill
      \label{fig:vertexing}
\end{figure}

\begin{figure}[ht]
  \centering
  \hfill
  \begin{minipage}[h!]{0.47\textwidth}
    {\includegraphics[width=\textwidth]{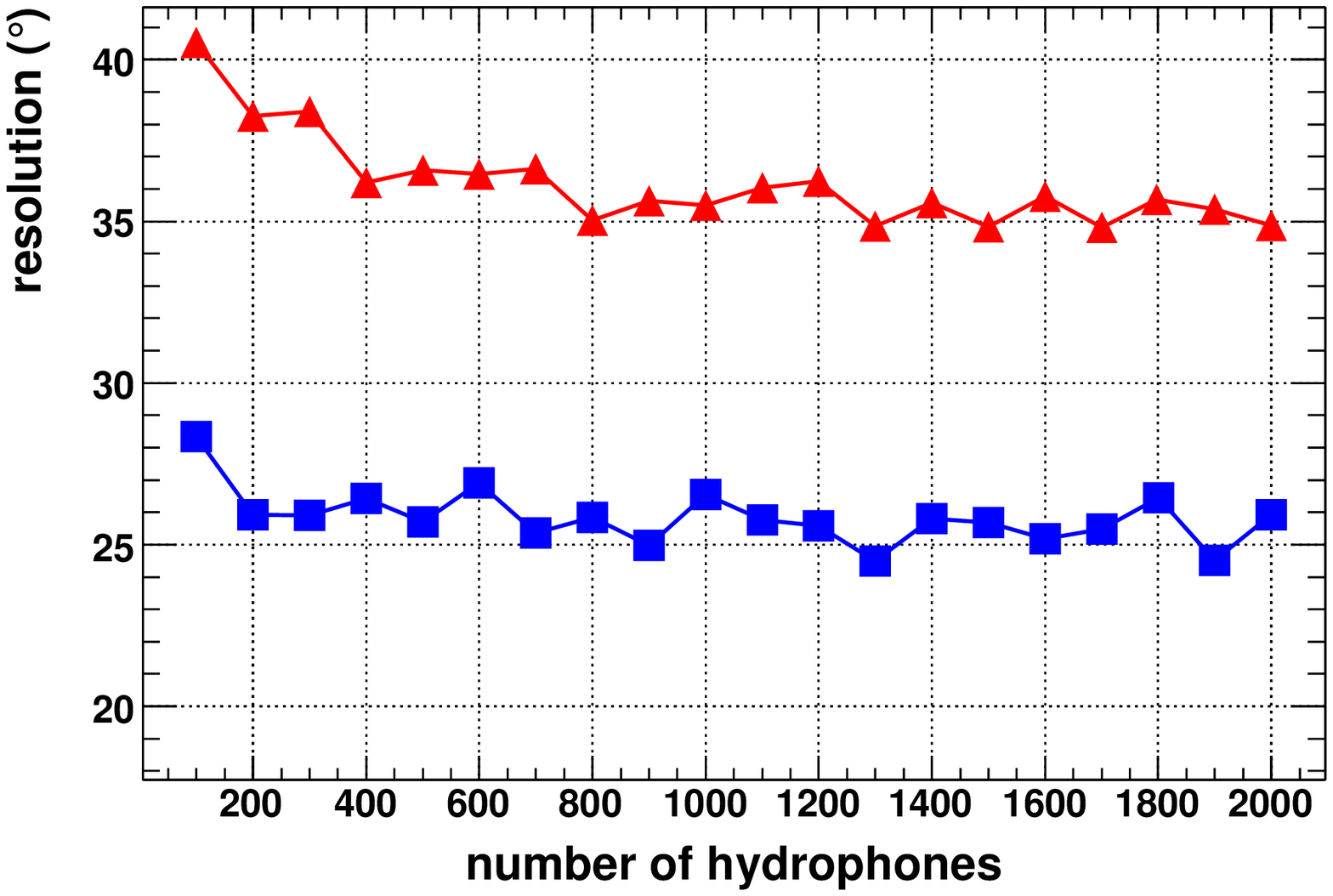}}
  \end{minipage}
  \hfill
  \begin{minipage}[h!]{0.47\textwidth}
    {\includegraphics[width=\textwidth]{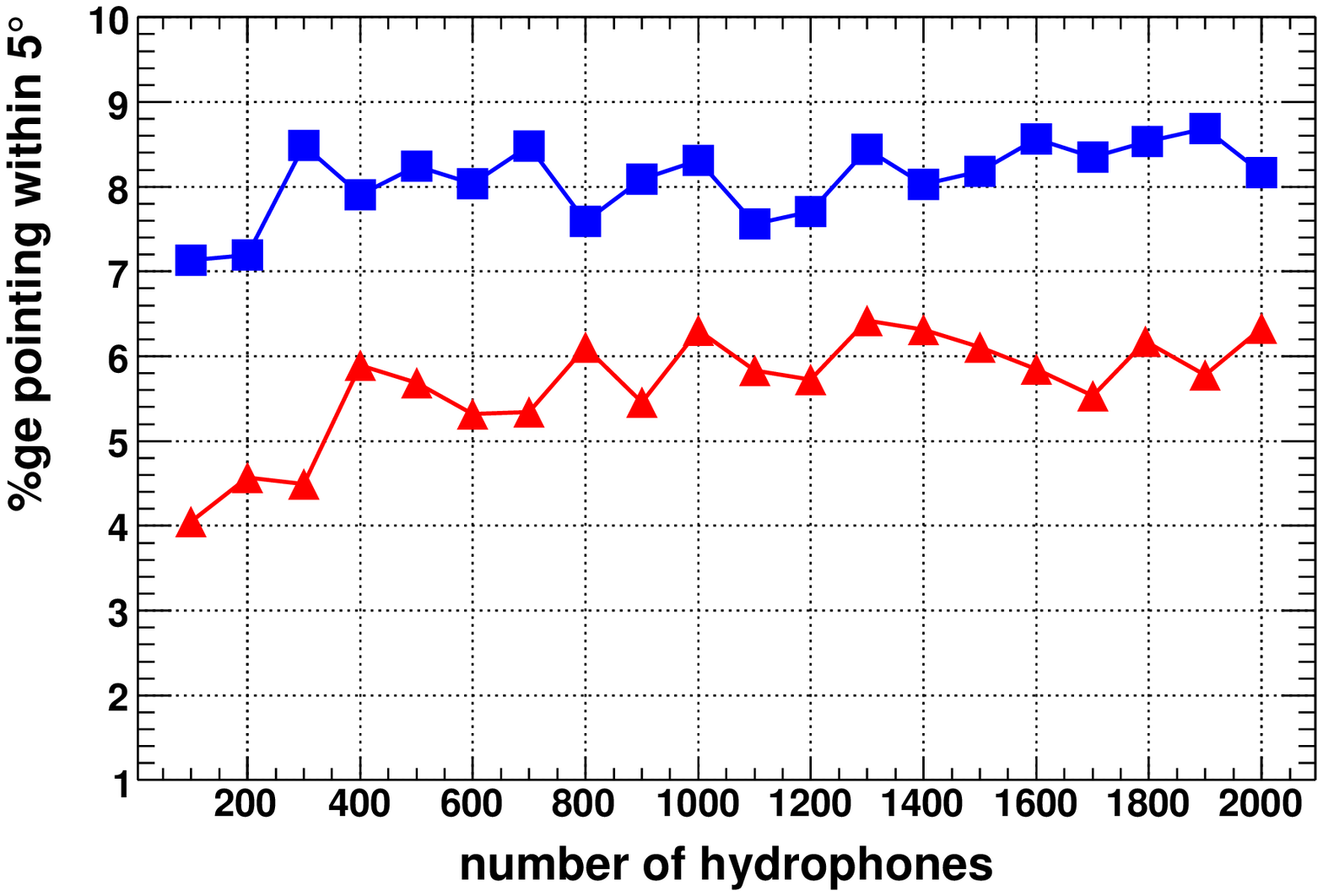}}
  \end{minipage}
  \hfill
  \caption[Pointing reconstruction v.s. hydrophone density]
	  {
	    \small
	    Top2 (\blue{\scriptsize$\blacksquare$} marker) and Collapsed
	    (\red{$\blacktriangle$} marker) pointing resolution 
	    ({\it Left}) and accuracy ({\it Right}). In each case
	    there is little improvement (within fluctuations) from
	    beyond $600$ hydrophones per cubic kilometre.	    
	  }
	  \label{fig:pointingreconstructions}
\end{figure}

The sensitivity of a hypothetical detector relies on its ability to
reconstruct the vertex of a given event. However if such a detector is
to be used as a ``telescope'' for the purpose of Neutrino Astronomy,
then it must also provide information on the trajectory and energy of
a neutrino. One may therefore pose the following question:
{``what is the optimal hydrophone density required to provide the best
  vertex and pointing resolution, via the reconstruction algorithms
  introduced thus far?''}
Typically one thousand hydrophones have been considered in a volume of
one kilometre cubed. From Section \ref{sec:effectivevolume} a minimum
number of sensors has been found for which the effective volume of an
array is maximal. An optimum sensor density for vertexing and pointing
accuracy is now desired.  The standard deviation on the vertex $x$, $y$ and
$z$ coordinates as a function of hydrophone density and the number of
reconstructions within $10$\,cm of the actual vertex location are
plotted in Figure \ref{fig:vertexing}. 
The pointing reconstruction as a function of hydrophone density is
evaluated in Figure \ref{fig:pointingreconstructions}.
\section{The Effect of Uncertainty in Hydrophone \mbox{Locations}}
\label{sec:smearing}
The ANTARES \cite{antares:becherini} neutrino telescope utilises a
High Frequency Long Base Line (HFLB) acoustic
positioning system, operating at $40-60$\,kHz over a base of $300
\times 300$\,m, combined with local tilt-meter-compass sensors to
obtain the location of its sensor elements in the depths of the
Mediterranean Sea. The system provides spatial positioning to an
accuracy of $0.1$\,m. One may therefore assume that any future
undersea detector array will have at least comparable performance.
The effect of smearing the hydrophone locations by a Gaussian
distribution with width $\sigma_h = 0.1$\,m on the standard deviation
of the vertex $x$-coordinate ($\sigma_x$) and the Top2 pointing vector
($\sigma_{\mathrm{Top2}}$) is summarised in Table \ref{tab:smearingtab}.
\begin{table}[htb]
  \centering
  \begin{tabular}{ l | c | c | c | c }
    & \multicolumn{2}{   c | }{\bf Vertexing } 
    & \multicolumn{2}{ | c   }{\bf Pointing  } \\
    \hline
      $\sigma_h$ (m)
    & $\sigma_x$ (m)                       
    & $\sigma_x<0.1$\,m 
    & $\sigma_{\mathrm{Top2}}$ ($\degrees$) 
    & $\sigma_{\mathrm{Top2}}<5$\,$\degrees$ \\
    \hline
     $0.0$ & $11230$ & $47.6\%$ &$28.0$ &$8.89\%$\\
     $0.1$ & $11800$ & $47.5\%$ &$28.6$ &$8.60\%$\\
  \end{tabular}
  \caption[The effect of $0.1$\,m uncertainty on hydrophone locations]{
    \small The effect of a $0.1$\,m hydrophone location uncertainty on
    reconstruction performance of a $1000$ hydrophone, $1$\,km$^3$
    array in a can of radius $48$\,km.
    \label{tab:smearingtab}}
\end{table}
\section{Reconstruction of the Neutrino Energy}
\label{sec:energyrecon}
As previously stated, an estimate of the energy of an
interacting neutrino is required if one is to use a hydrophone array
for neutrino astronomy. To be able to identify a source in the sky is
one thing, but to understand the processes occurring therein some kind
of calorimetry is necessary. A simple method was tested for
determining the energy resolution. The following $\chi^2$ was
evaluated:
\begin{equation} \label{eq:energychisquared}
\chi^2 = \sum_{h=1}^{N_{\mathrm{hydros}}} P_h - P_e
\end{equation}
where $N_{\mathrm{hydros}}$ is the number of hydrophones in the array,
$P_h$ is the pressure recorded on each hit hydrophone and $P_e$ is the
pressure expected at each phone. The pressure expected at each phone
is a  function of the reconstructed vertex location, the angle out of
the plane of the pancake and the incoming neutrino energy. $\chi^2$ is
calculated for a range of energies and the energy at which it is a
minimum should then correspond to the energy of the incoming neutrino.
\begin{figure}[ht]
  \centering
    \includegraphics[width=0.75\textwidth]{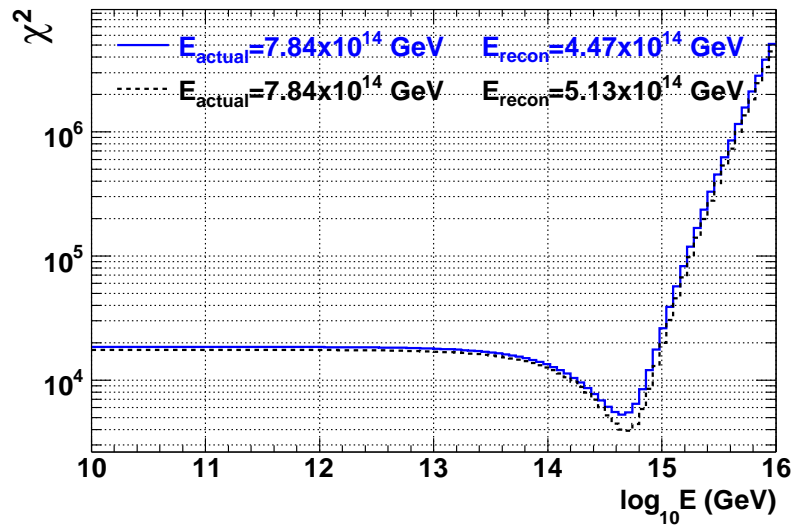}
    \caption[Energy reconstruction: $\chi^2$ minimisation]
	{
	  \small
	  Plot of $\chi^2$ as described by equation \ref{eq:energychisquared}
	  for one event with successful vertex reconstruction with
	  (\blue{solid} curve) and without (dashed curve) refraction.
	  In each case the energy is reconstructed below the true
	  value.	  
	}
	\label{fig:energyreconstructionchi}
\end{figure}

\begin{figure}[htb]
  \centering
  \hfill
  \begin{minipage}[h!]{0.47\textwidth}
    {\includegraphics[width=\textwidth]{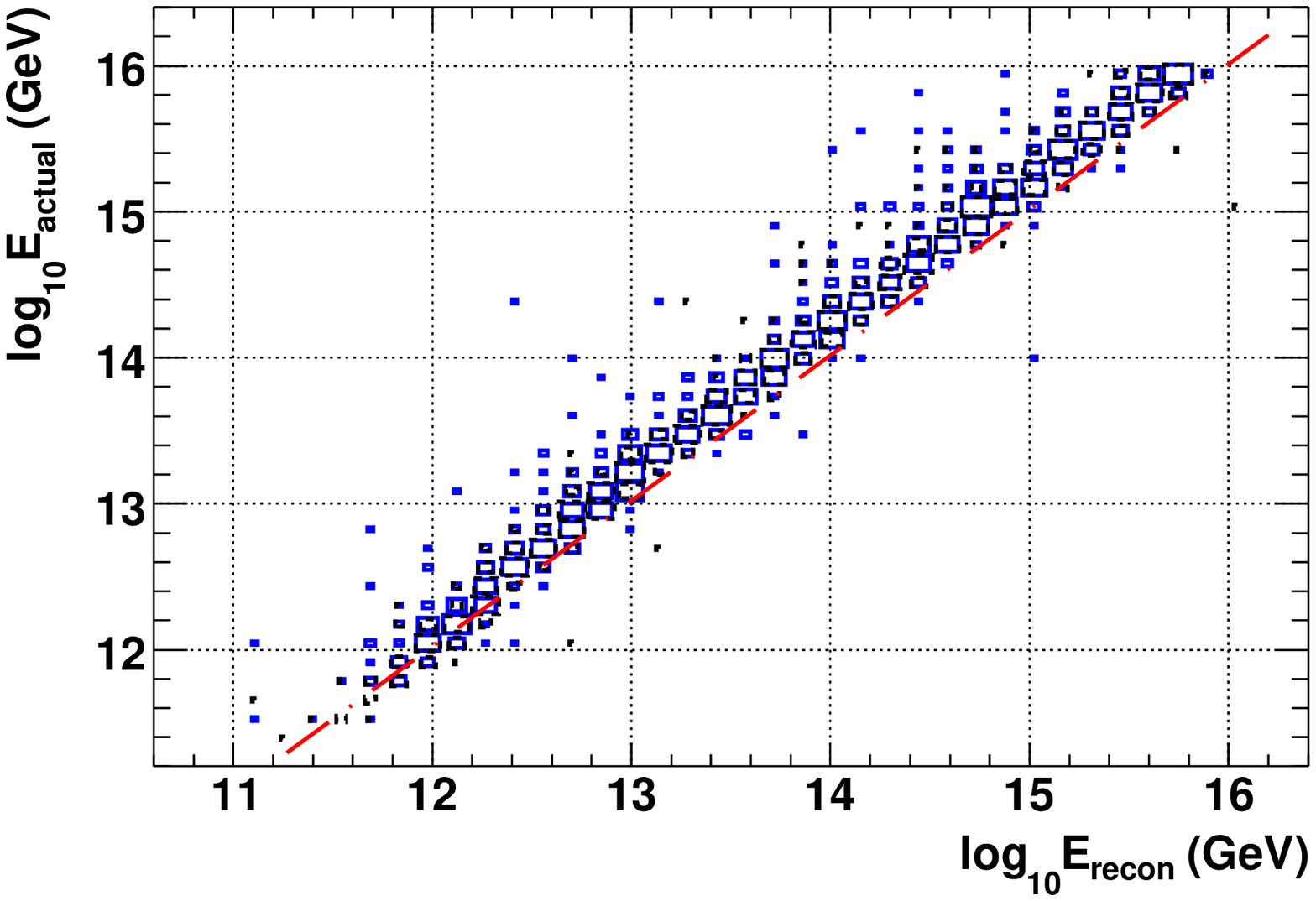}}
  \end{minipage}
  \hfill
  \begin{minipage}[h!]{0.47\textwidth}
   {\includegraphics[width=\textwidth]{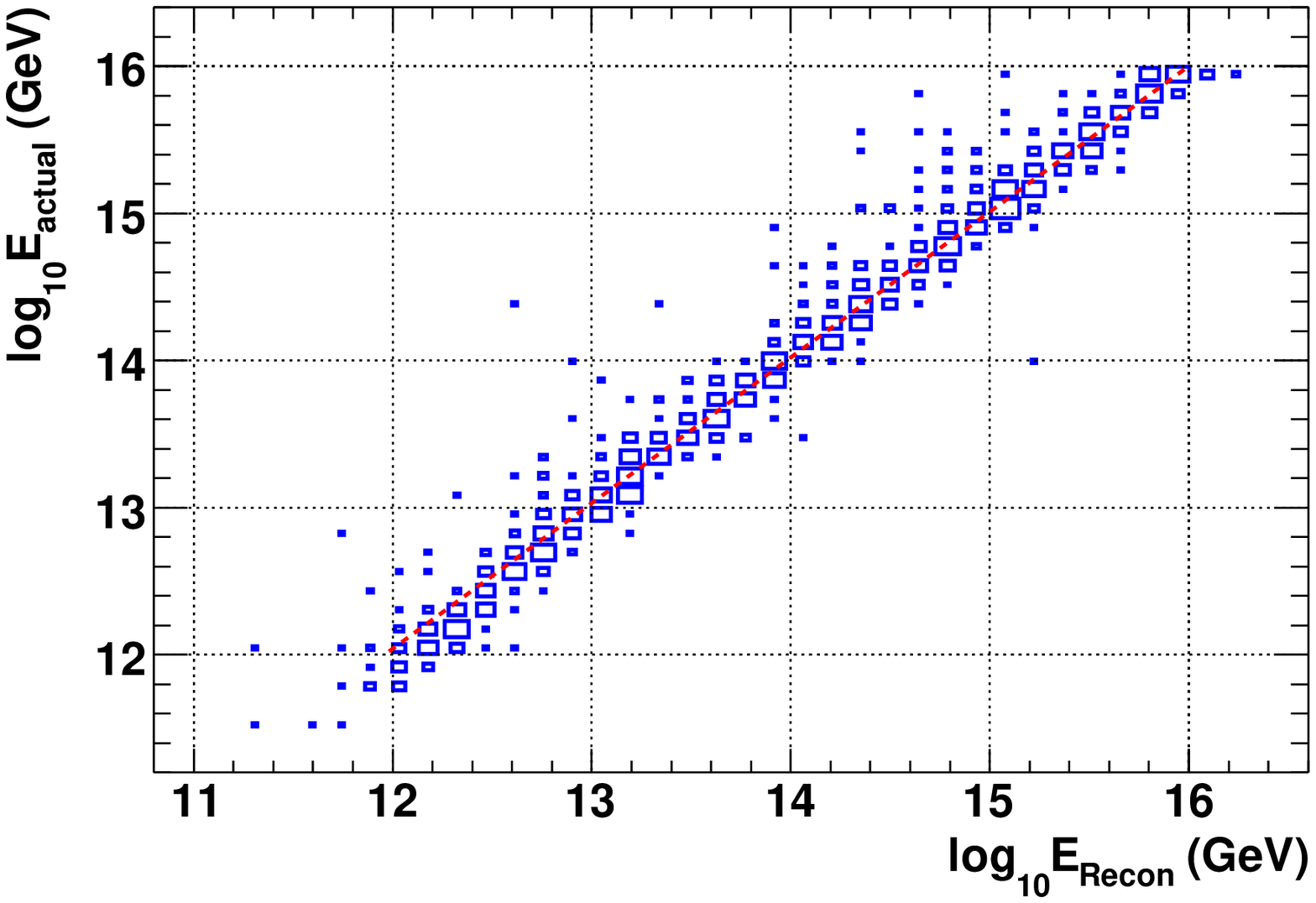}}
  \end{minipage}
  \hfill
  \caption[Reconstructed energy v.s. actual energy]
    	  {
	    \small
	    ({\it Left}) The distribution of $E_{\mathrm{recon}}$
	    v.s.$E_{\mathrm{actual}}$,it is offset from the
    	    \red{dashed} line $E_{\mathrm{recon}} = E_{\mathrm{actual}}$
    	    due to systematic underestimation of the neutrino energy.
	    ({\it Right}) The shifted distribution of $E_{\mathrm{recon}}$
	    v.s.$E_{\mathrm{actual}}$ compensating for the 
	    underestimation of the neutrino energy. Both plots include
    	    the effects of refraction.
	  }
	  \label{fig:energyreconstruction}
\end{figure}

\begin{figure}[ht]
  \centering
  \hfill
  \begin{minipage}[h!]{0.47\textwidth}
    {\includegraphics[width=\textwidth]{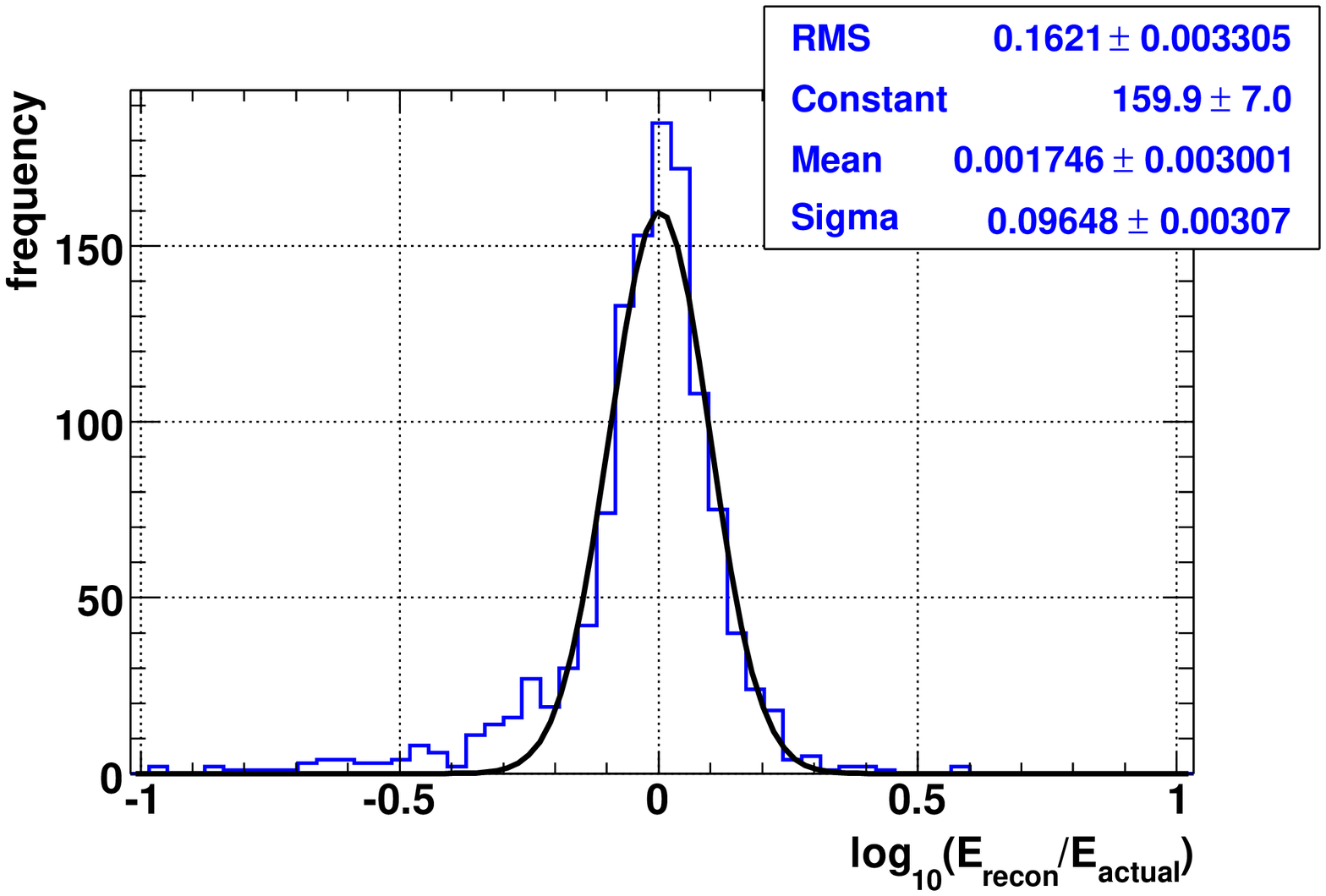}}
\end{minipage}
  \hfill
  \begin{minipage}[h!]{0.47\textwidth}
    {\includegraphics[width=\textwidth]{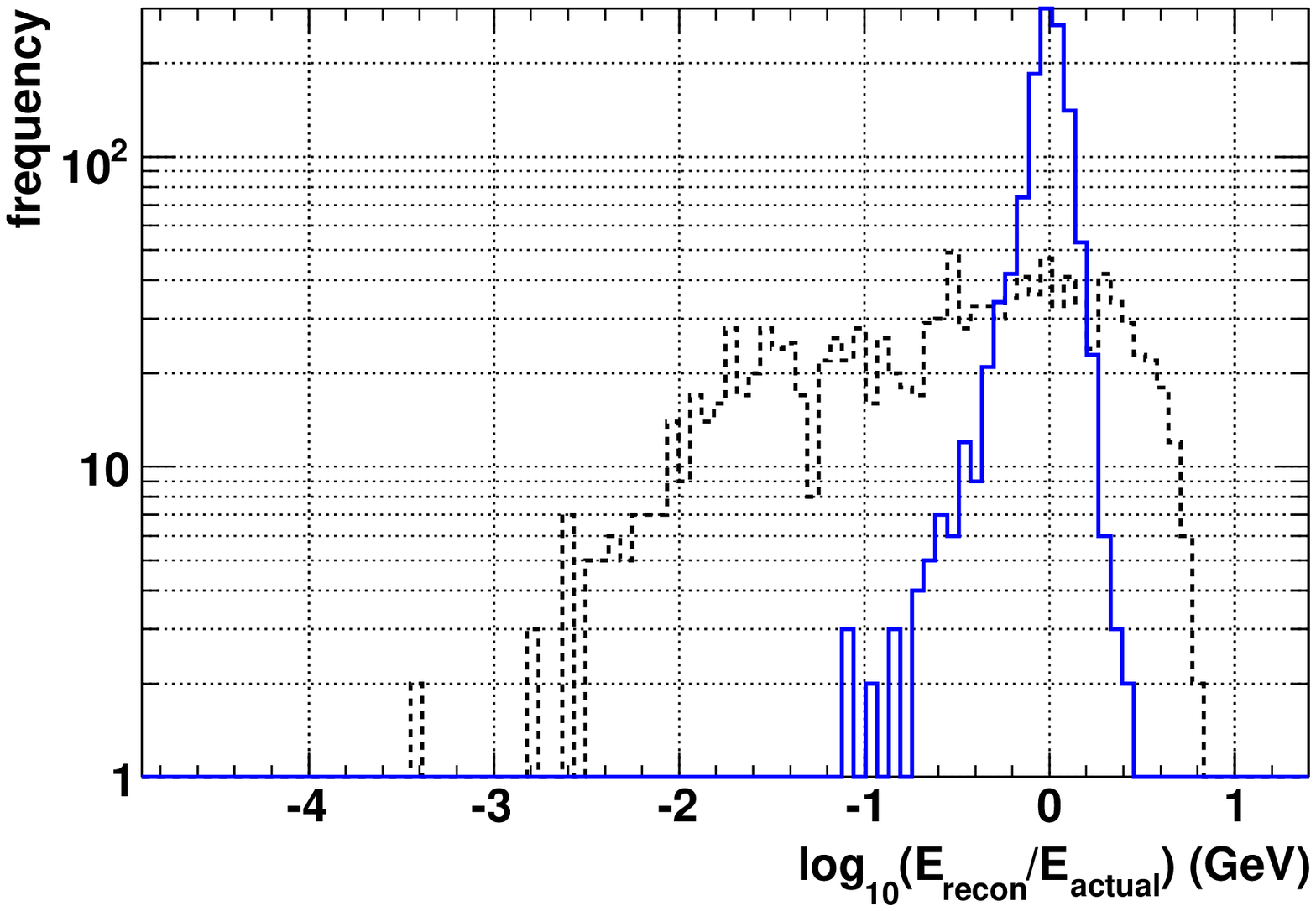}}
  \end{minipage}
  \caption[Energy resolution]
	  {
	    \small
	    ({\it Left}) Potential energy resolution with
	    $\sigma_{\log_{10}E} \sim 0.1$, assuming a constant
	    Bjorken-$y$ of $0.25$.
	    ({\it Right}) Energy resolution with $\sigma_{\log_{10}E}
	    \sim 0.8$, smeared by a non-constant Bjorken-$y$ (dashed) on top
	    of the unsmeared energy resolution (\blue{solid}). Both plots
	    include compensation for systematic underestimation of the
	    neutrino energy and include the effects of refraction.
	  }
	  \label{fig:energyresolution}
	  \hfill
\end{figure}

As can be seen from the plot in Figure \ref{fig:energyreconstructionchi}
the $\chi^2$ minimisation has yielded a result that is below the
actual neutrino energy; however, this is easily explained. Here the
expected energies were calculated merely as function of the distance
of the resolved vertex without considering the attenuation
due to being outside the plane of the pancake. So, the expected
pulses on those phones outside the pancake plane is  over-estimated,
the true neutrino energy would have to be slightly larger in reality
to produce these signals. We can see this effect more easily in Figure
\ref{fig:energyreconstruction} which displays
$log_{10}E_{\mathrm{recon}}$ on the horizontal axis and
$log_{10}E_{\mathrm{actual}}$ on the vertical axis, for over a
thousand events.
Superimposed on the plot is the line $y=x$ to aid the observation of
the systematic shift of the $E_{\mathrm{recon}}$
 towards lower energy. The mean
reconstructed energy is $(61 \pm 17)\%$ of the actual energy in the
presence of refraction and $(63 \pm 12)\%$ without. Naturally one can
assume that adding the effects of angular attenuation to the $\chi^2$
calculation will improve the reconstruction; however this has not been
undertaken in this work. Instead, to give an estimate of the potential
energy reconstruction performance $E_{\mathrm{recon}}$ has been
multiplied by a scale factor in order to systematically shift values
to higher energy, this can be seen plotted in Figure
\ref{fig:energyreconstruction}. The resulting energy resolution is
plotted in Figure \ref{fig:energyresolution}. 
%
%
%
Up until this point it has been assumed that a constant $25\%$ of
the energy of the incident neutrino is imparted upon the resulting
hadronic cascade which in turn generates the thermoacoustic pressure
wave detected a given hydrophone array. As discussed in Section
\ref{sec:nuevent} the proportion of the neutrino energy liberated by
the cascade can in fact fluctuate from $0-100\%$. Including the effects
of a non-constant Bjorken-$y$ by sampling from the distribution
plotted in Figure \ref{fig:anis:b} value reduces the effectiveness of
the energy reconstruction (by a factor orders of magnitude greater
than the underestimation error); the resulting energy resolution is
plotted in Figure \ref{fig:energyresolution}.
\section{The Sensitivity Calculation and Predicted Results}
\label{sec:sensitivitycal}
\begin{figure}[htb]
  \centering
  \includegraphics[width=0.75\textwidth]{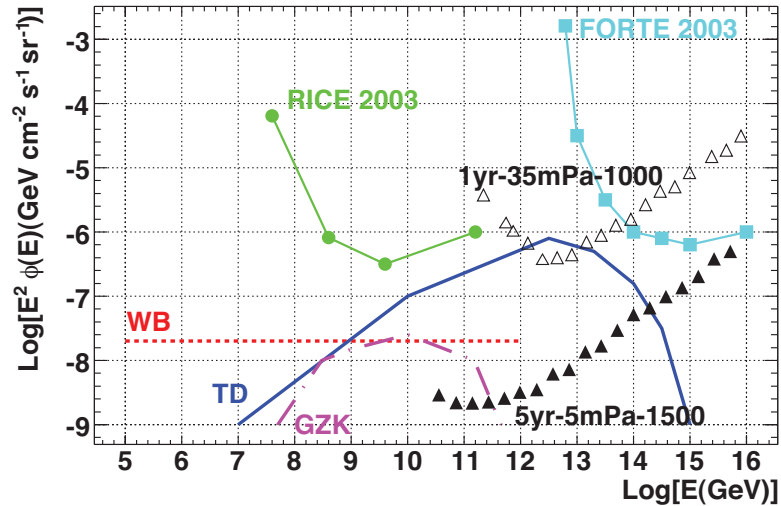}
  \caption[Sensitivity of simulated arrays]
	  {
	    \small
	    Predicted limits on the neutrino flux from this study
	    ($\triangle$ and $\blacktriangle$),
	    RICE (\green{$\bullet$}) and FORTE
	    (\textcolor{cyan}{\scriptsize$\blacksquare$}). Also shown
	    are those fluxes predicted by the Waxman-Bahcall (\red{dotted}),
	    Topological Defect (\blue{solid}) and GZK
	    (\textcolor{magenta}{dash-dotted}) models from Figure 
	    \ref{fig:saundresults}. 	    
	  }
	  \label{fig:newgzksensitivity}
\end{figure}

The simulations described thus far determine the ability of a given
hypothetical array of hydrophones to detect UHE neutrinos across a
range of energies. For a given energy, it is therefore possible to place an
upper limit on the neutrino flux, assuming that no events are detected
within a certain time window. The following calculation is based on
the formalism laid out in reference \cite{forte:lehtinen}. The
principal equation is:
\begin{equation}
\Phi(E) = \frac{S_{\mathrm{up}}}{E \times \lambda (E)}
\label{eq:differentialFlux}
\end{equation}
\begin{itemize}
\item $\Phi(E)$ is the limit on the differential flux
  $[dN_{\nu}/dEdAd\Omega dt]$ in units of \\
  GeV$^{-1}$cm$^{-2}$\,sr$^{-1}$\,s$^{-1}$.
\item $S_{\mathrm{up}}$ is the upper limit on the number of detected
  events that one places. Assuming that no events are seen, then 95\%
  confidence level upper limits correspond to setting 
  $S_{\mathrm{up}} = 3.0$\,
  (or $2.3$ for 90\% confidence level) in the absence of background
  \cite{journalofphysicsg}.
\item $\lambda (E)$ is the number of events that 
we would detect if a unit differential flux were incident on our detector at a
unique energy $E$, for time $T$. It can therefore be written as:
\begin{equation}
  \lambda (E) = \sigma_{\nu N} (E) \times \rho \times N_\mathrm{A}
  \times V \times 2\pi \times f_{MC}(E) \times T
  \label{lambda}
\end{equation}
where:
\begin{itemize}
\item $\sigma_{\nu N} (E)$ is the neutrino-nucleon cross section at energy $E$.
This is an extrapolated result (see Appendix \ref{appendix:nucrosssection}).
\item $\rho$ is the density of water.
\item $N_\mathrm{A}$ is Avogadro's number.
\item $V$ is the fiducial volume (i.e. the can) where 
$ V = \pi \times (R_{\mathrm{can}})^2 \times H_{\mathrm{can}}$. 
\item $T$ is the exposure time in seconds
\item $f_{MC}(E)$ is the fraction of Monte Carlo events generated at energy
$E$ that are detected (and reconstructed). 

%
%
\end{itemize}
\end{itemize}
The sensitivity of an array of $1000$ hydrophones, distributed at
random in a volume of one cubic kilometre, is plotted in Figure
\ref{fig:newgzksensitivity} (labelled \textsf{1yr-35mPa-1000}),
the pressure threshold is $35$\,mPa and the period of operation one year. 
Also shown is the sensitivity of $1500$\,km$^3$ array, with $100$
hydrophones per cubic kilometre operating at a pressure threshold of
$5$\,mPa for $5$ years (labelled \textsf{5yr-5mPa-1500}).
This is an indicator of the size and scale
of an array required if one is to be sensitive to the cosmogenic
neutrino flux (labelled GZK) resulting from the attenuation of UHECRs
on the $2.7$\,K CMB photons. The sensitivity curves for this study and
the other experiments included on the plot represent a limit on the
neutrino flux based on no neutrinos being detected within the period
of operation. Also shown are the neutrino fluxes predicted by some
theoretical models.
\section{Summary}
\label{sec:largescalesummary}
The details of a simulation dedicated to calculating the ability of
hypothetical large scale arrays of hydrophones to detect a flux of UHE
neutrinos, has been reported. Through the use of ray tracing
algorithms, the effects of refraction have been incorporated on an
event by event basis on cubic kilometre scales, building on previous
work (\cite{acorne:proposal},\cite{timo:thesis}) which neglected or
approximated refractive effects.
The treatment of vertex reconstruction, which in a homogeneous medium
can be solved analytically with almost perfect accuracy (see Figure
\ref{fig:unsmearedxres}), presents a non-linear problem, without an
analytically determinable solution. Numerical computation of refracted
vertices has subsequently been discussed, extending the work beyond
that which incorporates pre-computed look-up tables
(\cite{rice:kravchenko},\cite{saund:vandenbroucke}) which prove to be
computationally expensive when looking to simulate arrays of thousands
of cubic kilometres. 
Additionally the dependencies of effective volume,
vertex resolution and pointing accuracy on hydrophone density have
been investigated. A full study of array optimisation is required to
maximise performance for all three effects in unison.
Array performances have been translated into limits on the neutrino flux as
a means of comparing the sensitivity of different instrumented
volumes. In order to observe and successfully reconstruct a so-called
GZK neutrino, it has been shown that hydrophone arrays encompassing
volumes of several thousand cubic kilometres and operating at pressure
thresholds of around $5$\,mPa are required.

  \cleardoublepage

  \chapter{First Data from Rona Underwater Acoustic Range}
\label{chp:experiment}
%
%
\section{Introduction}
\label{expintro}
The ACoRNE collaboration has access to the QinetiQ Underwater Acoustic
Range at Rona in Northwest Scotland. This facility is operated and
maintained by QinetiQ on behalf of the Ministry of Defence
(MoD) and is used for the measurement of radiated acoustic signatures
from Royal Navy ships and submarines in broad equivalence to the
United States  AUTEC array in the Bahamas as used by the SAUND
collaboration. The array comprises eight
wide band hydrophones distributed along Inner Sound - a $12$\,km long
strait separating the Inner Hebridean islands of Rassay and Rona from
the Scottish mainland - with an average depth of $\sim 230$\,m. Six of
the hydrophones are located at mid-water on cables that  are anchored
at the lower end to the seabed and pulled taut by a buoy tethered at
the upper end, one is anchored to the floor and the eighth is $30$\,m
from the bottom. The array extends over about one and a half
kilometres and thus has a sound transit time of approximately one
second. Since December $2005$ data acquisition has been underway. The
first run constituted $15$ days of continuous readout at a sampling
frequency of $140$\,kHz and amounted to $2.8$\,TB of raw information.
\section{Data Acquisition}
\label{sec:DAQ}
\subsection{Recording the hydrophone signals}
\label{subsec:recording}
The eight Rona hydrophones are commercially available wide band
sensors with a flat frequency response up to resonance at
$50$\,kHz. The ACoRNE data acquisition (DAQ) philosophy is to send
all data to shore unfiltered and archive it for processing and
analysis offline. Every ten minutes each of the Rona hydrophones
produces a waveform audio format
(WAV)\footnote{http://www.microsoft.com/ Accessed 03/03/2007,
  http://www.ibm.com/ Accessed 03/03/2007} file;
the eight files are placed in a time-stamped directory and written
immediately to an $8$\,TB RAID disk array. In September $2006$ the DAQ
was upgraded to incorporate an LT03 tape drive with mechanical loader
and barcode reader. The WAV files are compressed using the Free
Lossless Audio Compression
(FLAC)\footnote{http://flac.sourceforge.net/ Accessed 03/03/2007}
codec and written onto $400$\,GB LT03 tapes nightly and then removed
from the RAID array. Compression via FLAC reduces the file size to
approximately $45\%$ of its orginal size, with no loss of
information. Because of the sensitive nature of the Rona facility
remote access to these data is impossible; as such tapes are
physically mailed to ACoRNE members by the operating staff. No
filtering whatsoever has been introduced thus far, so prior to
analysis the raw data remains intact.

The motivation for such a byte-intensive system is to enable the use of
linear-phase digital filters. Unless the phase response of a recording
system is linear, different Fourier components of the signal
can lag over different timescales and the signal becomes distorted. As
such a zero distortion, linear phase DAQ is desired. Analogue filters
that meet these criteria are readily available; however, digital signal
processing offers a greater degree of versatility and is thus employed
instead. Digital linear phase filters require a time-reversed signal
input in order to maintain zero distortion. This cannot be achieved in
real time which is why unfiltered data are retained. Furthermore
there is no limit on the number of times the unfiltered data can be
reprocessed following any further developments in the DAQ system. If a
single set of analogue filters were implemented before the data were
written it would be a one shot mechanism and the raw data would be
irreversibly changed. Moreover, it has already been pointed out that
several matched filters will be required for optimised detection of
neutrinos at all angles and distances (see Section
\ref{sec:noisefilterandthresholds}). Offline analysis of unfiltered data
thus allows several passes through multiple matched filters, giving
the greatest degree of flexibility for matched filter development.
\subsection{Reducing the data}
\label{subsec:reduction}
The current DAQ chain is composed of four triggers that are designed to
maximise retention of neutrino-like signals. The emphasis of this
study is to categorise as many types of potential background as
possible such that they can be readily identified and subtracted from
future data, in the hope that what remains is the signal observed from
UHE neutrino induced particle cascades in the sea. In order of
increasing priority, candidate events were triggered in the
initial offline data reduction by: pressure (T1=$P$); the derivative of
pressure with respect to time (T2=$dP/dt$); the second derivative of
pressure with respect to time (T3=$d^2P/dt^2$); and a Matched Filter
(T4=$\mathcal{M}$). The raw data were analysed in $10$\,s intervals. For
each interval the time stamp for the five most energetic triggers in
$P$, $dP/dt$, $d^2P/dt^2$ and $\mathcal{M}$ triggers are taken. $1000$
samples ($7.14$\,ms) either side of the event are retained for each
time stamp and an overlap is said to  occur if two triggers are within
$500$ samples. If two triggers of the  same type occur within the
$500$ sample coincidence window, then the  most energetic is taken
and the other discarded. If triggers of  different types occur within
the $500$ sample window then the trigger  with the highest priority as
described above is retained.
\subsection{Rona calibration tones}
\label{subsec:tones}
Roughly once a day a test of the data transmission between each of the
Rona hydrophones and the shore station is performed. A $1$\,kHz
sinusoid is injected electrically behind the cable-hydrophone
connection and the signal is recorded at the shore station, to verify
the integrity of data transmission along the cable.
The calibration tones are injected into the system sequentially,
that is one hydrophone at a time, and are thus easily rejected by a demand
for coincidence between events. No $1$\,kHz signals have been
observed in this analysis as would be expected since a demand for
coincidence rejects the sequential signals.
\section{A Novel Analysis of the Reduced Data}
\label{sec:analysis}
\subsection{Event discrimination}
\label{subsec:discrimination}
A total of $12,453,500$ signals were retained from the first
$15$ days of Rona data by the data reduction code. An attempt has been
made to identify and classify different types of event within this
data set. In concordance with the large scale detector simulation
discussed in previous chapters, only those events above a threshold
pressure of $35$\,mPa and with a four-fold or greater hit multiplicity
were considered. 
In order to discriminate against random noise, Fourier analysis of the
events was applied such that only those events that showed the same
peak harmonic in each coincident hydrophone were retained.

Fourier analysis facilitates identification of the frequency
composition of a signal. The FFTW
algorithm\footnote{http://www.fftw.org/ Accessed 13/03/2007}
was employed to calculate a
discrete Fourier transform (DFT) of the sampled data. An effective DFT
requires a window function ${w}(n)$ to select a finite number of
samples of the raw data $x(n)$. The sampled data can be thus expressed
by the following equation:
\begin{equation}
  {v}(n) = {w}(n) x(n)
  \label{eq:sampled}
\end{equation}
where ${w}(n)$ is the aforementioned window function. There
are many forms of window function available, a couple of popular
examples being the Hanning and Blackmann Windows as defined by
Equations \ref{eq:hanning} and \ref{eq:blackmann} respectively.
\begin{eqnarray}
  {w}(n) &=& \frac{1}{2} \big[ 1 1- \cos \frac{2 \pi n}{N-1} \big]
  \label{eq:hanning}\\
  {w}(n) &=& 0.42 - 0.5\cos\frac{2 \pi n}{N-1} + 0.08\cos\frac{4
    \pi n}{N-1}
  \label{eq:blackmann}
\end{eqnarray}
However, such window functions are best suited to analysis of
continuous periodic signals and were not implemented. The type
of signal one expects from a neutrino is much more impulsive. A
so-called Rectangular Window is better suited to transient signals,
this is expressed in Equation \ref{eq:rectangular} below:
\begin{eqnarray}
  {w}(n) &=& 1\qquad \mathrm{for} \quad 0 \leq n \leq N-1 \nonumber\\
  {w}(n) &=& 0\qquad \mathrm{elsewhere}.
  \label{eq:rectangular}
\end{eqnarray}
In general a DFT requires that the number of samples is an integer
power of two. In this analysis, an event was viewed through a window of
$N=2^7=128$ samples, corresponding to width of $128/140000=914.28$\,\textmu
s; this is the sample period. Remember, a typical bipolar signal from
a neutrino with a dominant frequency of $10$\,kHz produces a pulse
$\sim 100$\,\textmu s wide and thus fits comfortably within this
window. A smaller window could have been chosen, to better isolate
events, but at a cost of reducing the number of harmonics identifiable
in the DFT. There are $N/2=64$ fundamental harmonics of the
sampling period in a DFT of $N=128$\,samples of data; the
$n$th harmonic is at $n \times 140000/128 = n \times
1093.75$\,Hz. A window of $N=128$ samples was chosen because it
gave a good compromise between signal isolation and harmonic
identification.

\begin{figure}[htbp]
  \centering
  \includegraphics[width=0.8\textwidth]{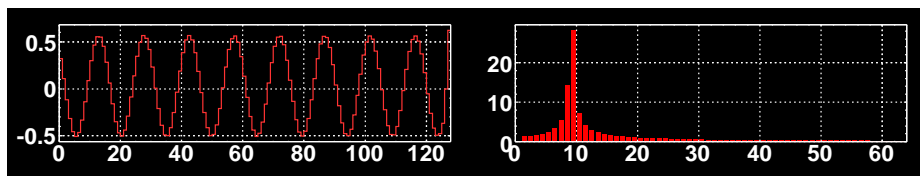}
  \caption[$10$\,kHz oscillating signal at Rona]
	  {
	    \small
	    Example of a $10$\,kHz oscillating signal.
	    The left hand display shows the
	    pressure signal in mPa and the right hand plot shows the
	    magnitude of the DFT. In the pressure signal the $x$-axis
	    corresponds to sample number, there are $128$ samples
	    in the event window producing $64$
	    harmonics in the DFT. A clear peak in the tenth harmonic
	    of the DFT is indicative of a single frequency sinusoidal
	    composition at $\sim 10$\,kHz.
	  }
	  \label{fig:oscillatingtone}
\end{figure}

Identification of events showing the same dominant frequency was
achieved by requiring that all coincident events showed a maximum in
the same bin (i.e harmonic) of the DFT histogram. This is illustrated
clearly by the identification of sinusoidally oscillating tones
present in the data, as plotted in Figure
\ref{fig:oscillatingtone}.
Discrimination of sinusoidal signals is aided by
the fact that they show distinct, isolated peaks in the DFT.
If a sinusoidally oscillating signal has a period that is an integer
number of sampling periods, a peak occupying a single bin in the DFT
would be observed. However, in the case plotted in Figure
\ref{fig:oscillatingtone}, the signal period is
$100$\,\textmu s and the sampling period is $128/140000 =
914.29$\,\textmu s. This mismatch is the source of the tails either
side of the peak in the tenth harmonic in the DFT; the effect is known
as leakage.
\subsection{Categorisation of events}
\label{subsec:categorisation}
\begin{figure}[htb]
  \centering
  \hfill
  \begin{minipage}[h!]{0.47\textwidth}
    {\includegraphics[width=\textwidth]{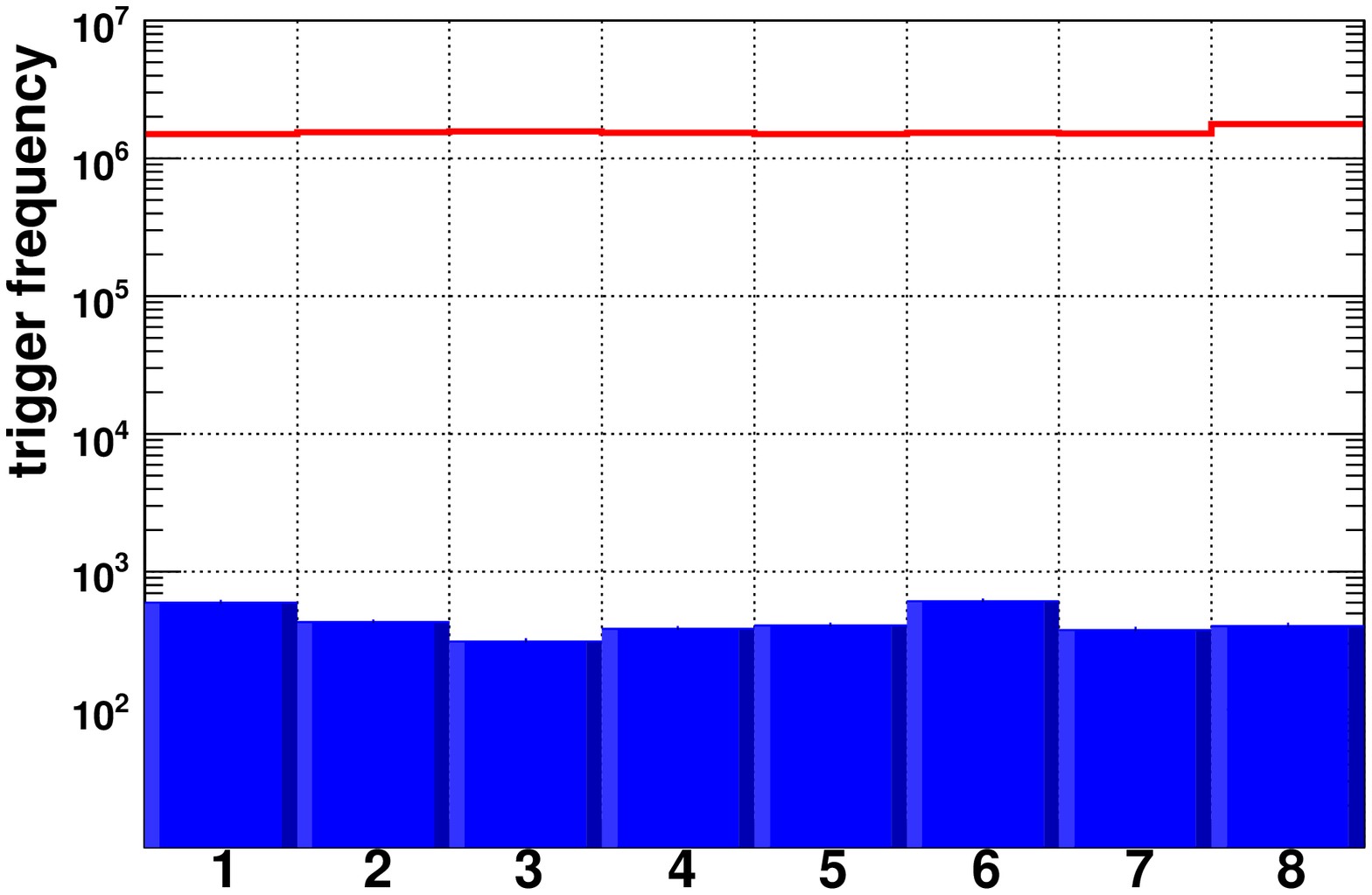}}
  \end{minipage}
  \hfill
  \begin{minipage}[h!]{0.47\textwidth}
    {\includegraphics[width=\textwidth]{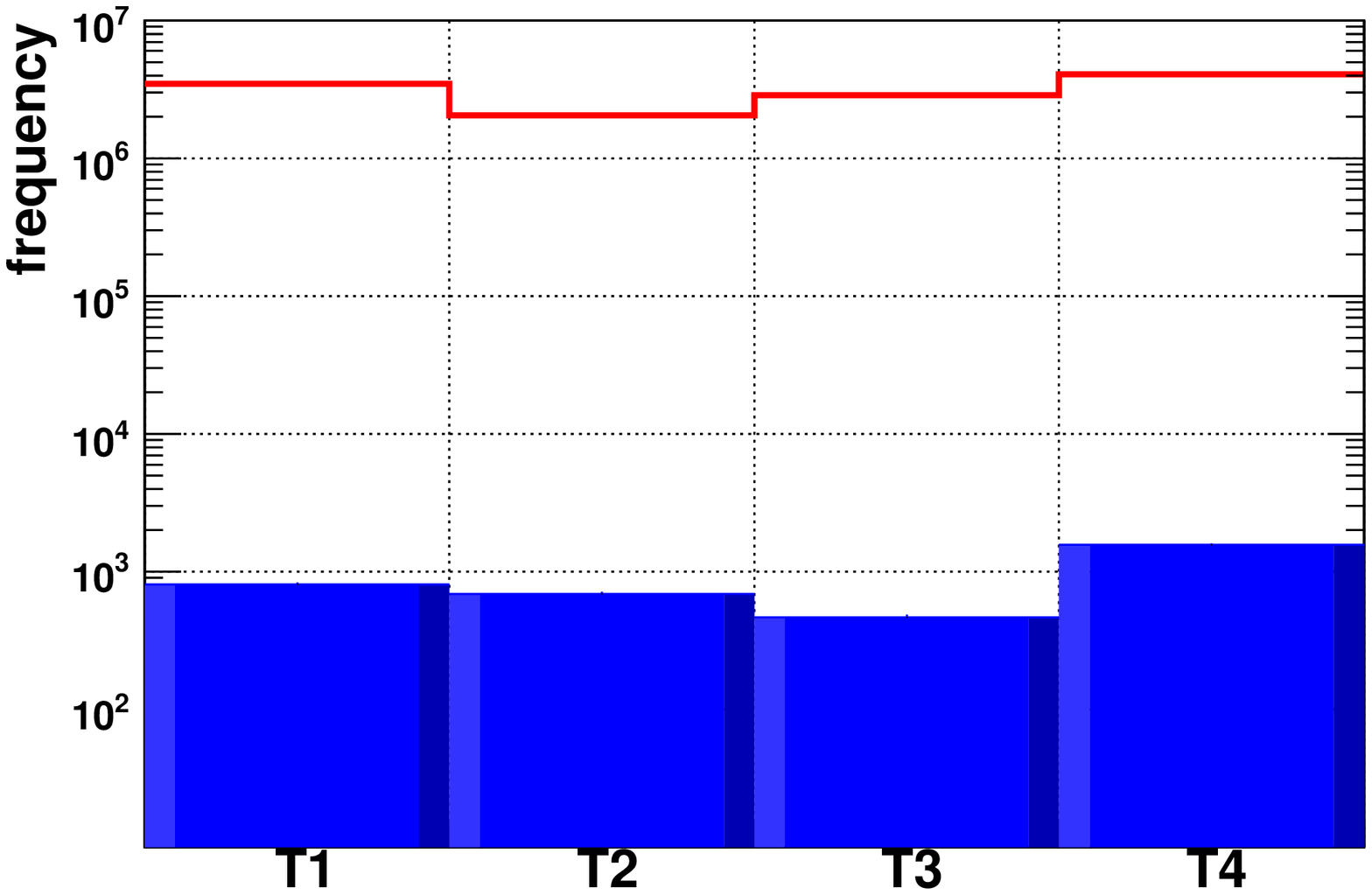}}
  \end{minipage}
  \hfill
  \caption[Rona triggers]
	  {
	    \small
	    The number of triggers before (\red{unfilled}) and after
	    (\blue{filled}) cuts on each hydrophone at Rona, as
	    retained by the data reduction code ({\it Left}).
	    The number of instances if each trigger ({\it Right}).
	  }
	  \label{fig:triggerinstances}
\end{figure}

Of the $12,453,500$ reduced events, $3526$ corresponded to multiplicities
of four or greater above the $35$\,mPa threshold with an equivalent
frequency composition as identified in the DFT.
The number of triggers per hydrophone and the trigger frequencies are
plotted in Figure \ref{fig:triggerinstances}.
\subsubsection{Oscillating signals}
\label{subsubsec:longlived}
\begin{figure}[htb]
  \centering
  \includegraphics[width=0.8\textwidth]{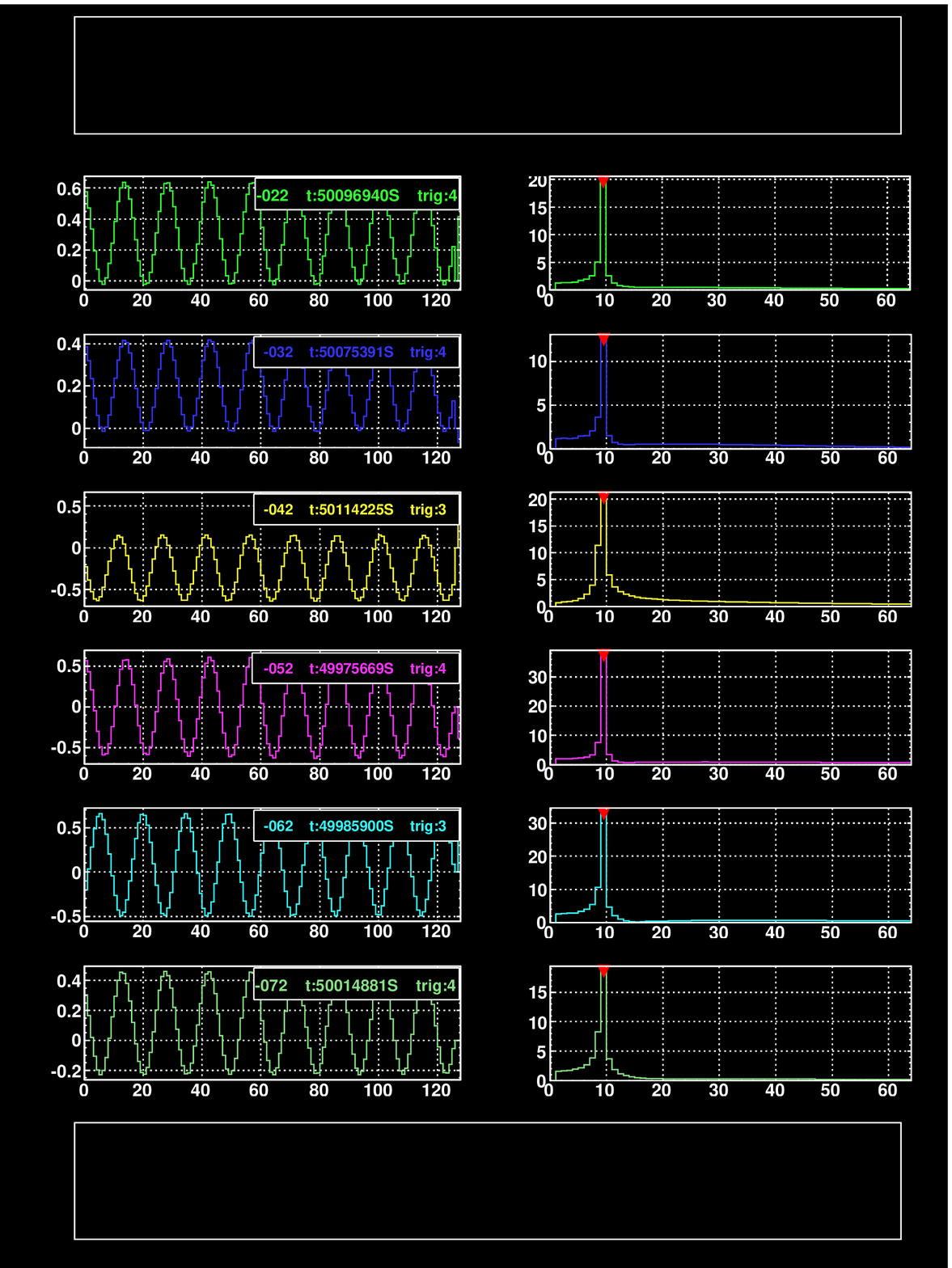}
  \caption[$10$\,kHz oscillating signal on $6$ channels]
	  {
	    \small
	    A $10$\,kHz event coincident on six hydrophones showing a
	    clear sine wave component. Traces are Pressure (Pa)
	    ({\it Left}) and DFT magnitude ({\it Right}).
	    Sample number runs along the $x$-axis.
	  }
	  \label{fig:sinusoidal}
\end{figure}

\begin{figure}[htb]
  \centering
  \includegraphics[width=0.8\textwidth]{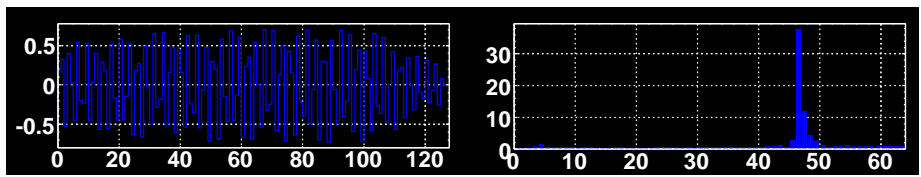}
  \caption[A high frequency signal]
	  {
	    \small
	    Example of a high frequency signal at Rona. The peak of
	    the DFT is in the $46$th harmonic, corresponding to a
	    frequency of $\sim 50$\,kHz.
	  }
	  \label{fig:highfrequencyevent}
\end{figure}

One hundred and twenty-nine of the $3526$ events that passed the
selection criteria were signals of a sinusoidally oscillating
nature. These events, whilst originally being classified as calibration
tones, did not occur at occasions when a Rona calibration tone was
scheduled to be injected into the system, as recorded in the recorder
logbook entries. Nor were the frequencies consistent with those
injected by the Rona operators. An example of such a signal is plotted
in Figure \ref{fig:sinusoidal}.
The source of these signals is currently unknown but it is suggested
that they represent an anthropogenic background as opposed to
bio-acoustical or environmental noise. All of the $10$\,kHz
oscillating signals identified in this analysis occurred on
14/12/2005 between 11:24 and 14:14. Two higher frequency
oscillating signals were recorded with a five-fold coincidence at
10:24 on 14/12/2005 at 10:24, one high frequency signal on 16/12 at
04:44 and one further high frequency oscillating signal on 24/12 at
07:33.  An example of one of the high frequency
hydrophone traces is plotted in Figure \ref{fig:highfrequencyevent},
the remainder of the oscillating signals being at $10$\,kHz.
%
%
\subsubsection{Transient events}
\label{subsubsec:transient}
\begin{landscape}
\begin{figure}[htb]
  \centering
  \hfill
  \begin{minipage}[h!]{0.7\textwidth}
    \includegraphics[width=\textwidth]{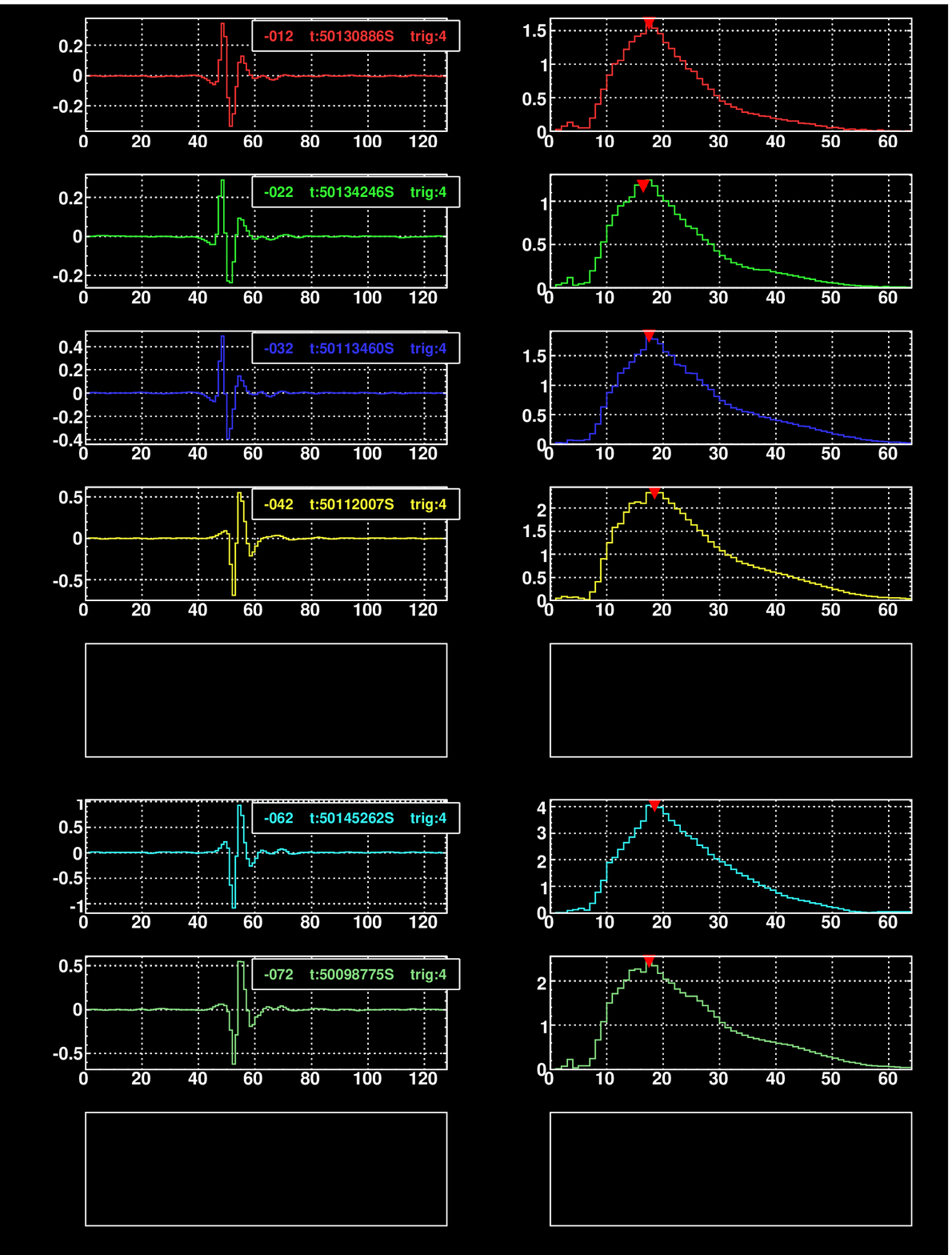}
    \caption[A {\it smooth} event at Rona]
	    {
	      \small
	      An event with a {\it smooth} DFT\\
	      {$\overline{N}_{\mathrm{peaks}} = 1$}.
	    }
	  \label{fig:smoothevent}
  \end{minipage}
  \hfill
  \begin{minipage}[h!]{0.7\textwidth}
    \includegraphics[width=\textwidth]{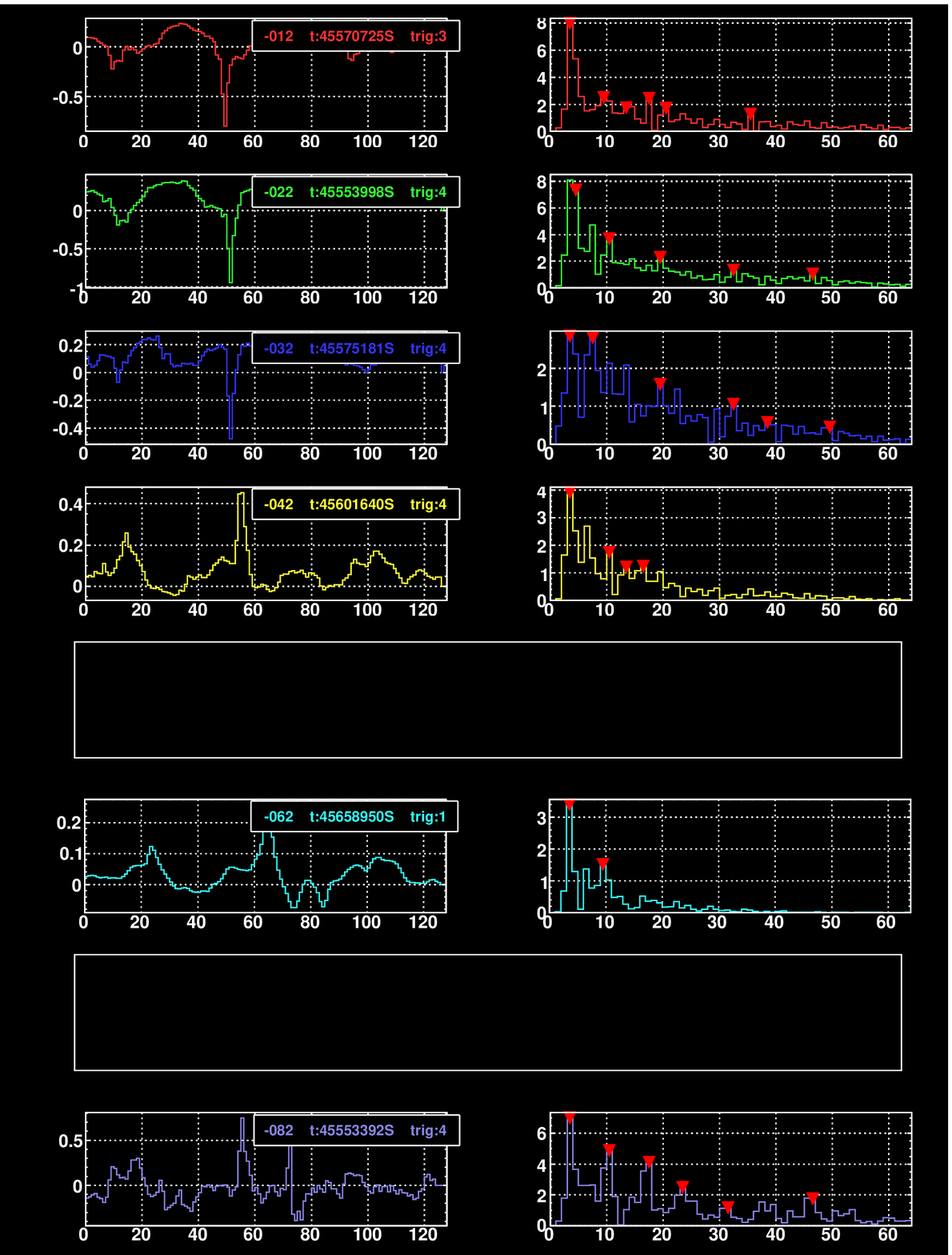}
    \caption[A {\it peaky} event at Rona]
	    {
	      \small
	      An event with a {\it peaky} DFT\\
	      {$\overline{N}_{\mathrm{peaks}} = 4.83$}.
	    }
	    \label{fig:peakyevent}
  \end{minipage}
  \hfill
\end{figure}
\end{landscape}

The remaining $3397$ events were further subcategorised in terms of their
frequency composition. Firstly the DFT of an event was analysed and
classified as one of three types: {\it smooth}, {\it mixed}, or {\it
  peaky} as according to the criteria in Table \ref{tab:peakiness}.
%
\begin{table}[htb]
  \centering
  \begin{tabular}{ l | c }
    {\bf Peakiness} & {\bf Mean Number of Peaks in DFT}            \\
    \hline
    {\it smooth}    &    $\overline{N}_{\mathrm{peaks}}<2.5$       \\
    {\it mixed}     &    $\overline{N}_{\mathrm{peaks}}=2.5$       \\
    {\it peaky}     &    $\overline{N}_{\mathrm{peaks}}>2.5$       \\    
  \end{tabular}
  \caption[{\it Peakiness} of Rona events]
	  {
	    \small
	    Events were categorised in terms of their {\it Peakiness},
	    a variable determined by the number of peaks in the DFT.
	    \label{tab:peakiness}
	  }
\end{table}
The threshold for {\it peakiness} of $2.5$ is somewhat arbitrary but
was chosen because it gave a good separation between events that
showed a continuous DFT spectrum and those that showed a relative
separation between peaks in the DFT. An example of two such events is
plotted in Figures \ref{fig:smoothevent} and \ref{fig:peakyevent}.
Observing the DFTs of the reduced Rona events, it is clear to see that
there are two extremes in the acoustic data. The first is a
periodic, oscillating signal such as the $10$\,kHz sinusoid in Figure
\ref{fig:sinusoidal}. This type of signal produces a distinct
single peak in the DFT.
The second is a type of signal evident in Figure
\ref{fig:smoothevent}. This appears to be an impulsive
event, as indicated by the continuous DFT\footnote{the DFT of an
  impulse is a constant: $DFT \{ { \delta (n) } \} = 1$}. The pulse
shape of impulsive events is dominated by the impulse response of the
hydrophone system whereas continuous oscillating signals show little
contribution from the detector elements. Studying the shape of the
DFT, i.e. their {\it peakiness}, thus gives an insight on the nature and
origin of these events.

Analysis of the remaining $3397$ events that did not categorise as
sinusoidally oscillating signals yielded $2329$\,{\it smooth},
$44$\,{\it mixed} and $969$\,{\it peaky} signals. These events were
further sub-categorised in terms of the magnitude of their frequency
components according to the criteria in Table
\ref{tab:frequencyrange}.
\begin{table}[htb]
  \centering
  \begin{tabular}{ l c | r }
    \multicolumn{2}{c|}{\bf Frequency Range (kHz)}&{\bf  Mean DFT Harmonic}\\
    \hline
    low    & ($\phantom{0}0<F< 22$) & $ 0 < \overline{\mathrm{DFT}} <20$\\
    medium & ($22          <F< 44$) & $20 < \overline{\mathrm{DFT}} <40$\\
    high   & ($44          <F< 70$) & $40 < \overline{\mathrm{DFT}} <64$\\
  \end{tabular}
  \caption[Frequency composition of Rona events]
	  {
	    \small
	    Rona events were sub-categorised according to the mean
	    value of the DFT histogram.
	    \label{tab:frequencyrange}
	  }
\end{table}
No events satisfied the criteria for classification as high
frequency. Only $55$ events classified as medium frequency events thus
indicating a clear dominance of events below $\sim 22$\,kHz.
\subsection{Time dependence of events}
\label{subsec:timedependence}
\begin{figure}[h!]
  \centering
  \hfill
  \subfigure[2D temporal distribution of Rona events]
	    {
	      \includegraphics[width=0.8\textwidth]{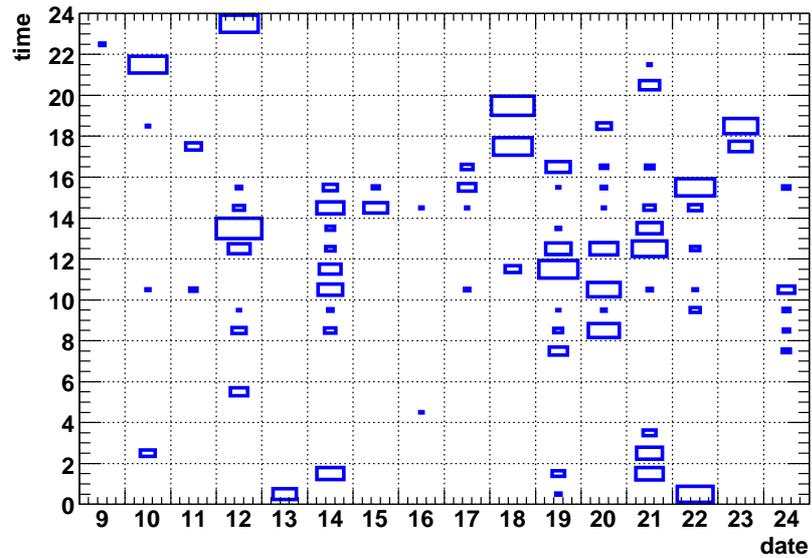}
	      \label{fig:timedate}
	    }
  \hfill
  \subfigure[$3D$ temporal distribution of Rona events]
	    {
	      \includegraphics[width=0.8\textwidth]{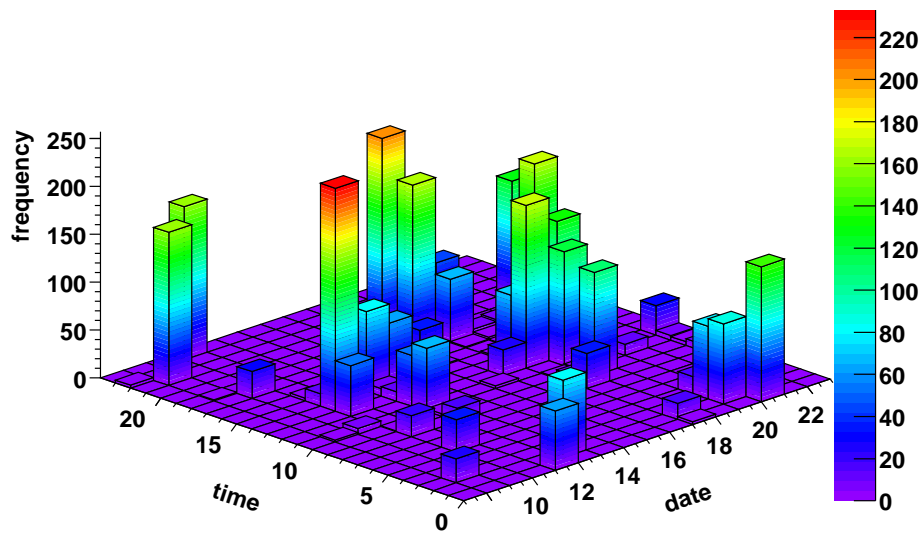}
	      \label{fig:timedatefreq}
	    }
  \hfill
  \caption[Temporal distribution of Rona events]
      {
	\small
	$2D$ and $3D$ plots of the temporal distribution of Rona
	events. Activity appears to be constrained to bursts of
	varying length and intensity. A large number of events can
	occur in a short space of time, producing isolated peaks such
	as the one at 21:00 on the 10th; or, events can be produced
	during a burst sustained over several hours such as the period
	between 08.00 and 16.00 on the 14th.      
      }
      \label{fig:temporal}
\end{figure}

Investigation into the temporal distribution of Rona events passing
pressure and coincidence cuts suggests that activity is
intermittent. There can be periods of sustained activity, lasting for
several hours, showing events in every consecutive ten minute WAV
file; and, there can be periods where a high number of events occur
within a very short space of time, occupying only a single ten minute
WAV file. The temporal variation in activity is best illustrated in a
plot of event frequency as a function of time. Figure \ref{fig:temporal}
shows this.
\subsection{Recurring event types}
\label{subsec:recurring}
Certain types of event, with distinct characteristics, are seen in the
data on a recurring basis. An attempt has been made to categorise such
events. Six distinct signal shapes have been identified. What follows
is a brief description of each event type, ending with a table
summarising the dates on which each of the events occur in the Rona
data. For each type of event a single hydrophone signal inclusive of
DFT is plotted, pressure in Pa runs along the $y$-axes and sample
number along the $x$-axes. 
The full event displays showing coincidences can be
found in Appendix \ref{appendix:recurring}.
\subsubsection{Short impulsive}
\label{subsubsec:shortimpulsive}
\begin{figure}[h!]
  \centering
  \includegraphics[width=0.9\textwidth]{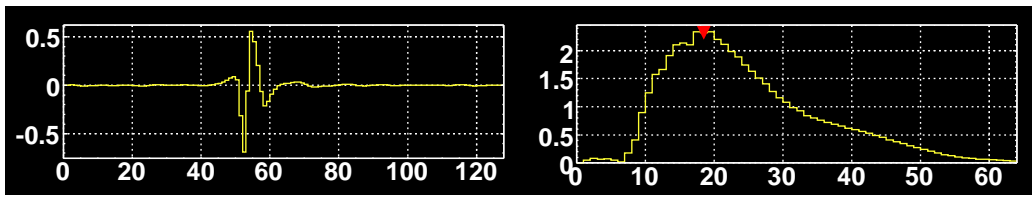}
  \caption[A short-lived impulsive event]
	  {\small A short-lived impulsive event.}
  \label{fig:shortimpulsive}
\end{figure}
As already introduced in Section \ref{subsubsec:transient}, there is
a category of event that appears to display an impulsive component. It
is categorised by a short duration multi-polar waveform and a
continuous DFT. It is however indeterminable from the DFT alone
whether the signal is actually the response of a hydrophone to an
impulse or if it really is a bipolar signal with a $\sim 20$\,kHz
peak frequency as one might expect from an UHE hadronic cascade.
\subsubsection{Low frequency oscillator}
\label{subsubsec:lowfreqoscillator}
\begin{figure}[h!]
  \centering
  \includegraphics[width=0.9\textwidth]{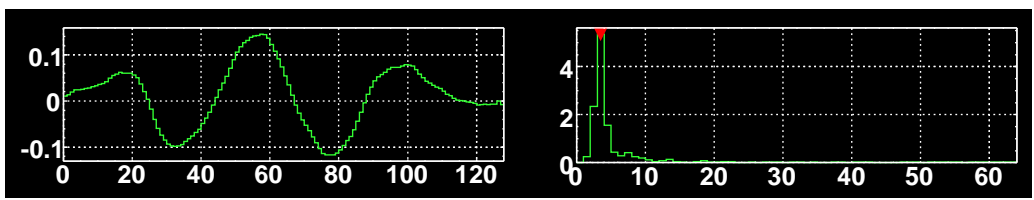}
  \caption[A low frequency oscillating event]
	  {\small A low frequency oscillating event.}
  \label{fig:lowfreqoscillator}
\end{figure}
An event categorised by a low frequency ($\leq 4$\,kHz) oscillating
waveform appears consistently throughout the first fifteen days of
Rona data. The pressure wave of these signals can often appear as in
Figure \ref{fig:lowfreqoscillator} or it can be somewhat
distorted. Nevertheless this kind of signal is present in almost every
day's worth of data (see Table \ref{tab:recurringevents}) and thus
constitutes a significant background.
\subsubsection{10kHz sinusoidal}
\label{subsubsec:10khzsinusoidal}
\begin{figure}[h!]
  \centering
  \includegraphics[width=0.9\textwidth]{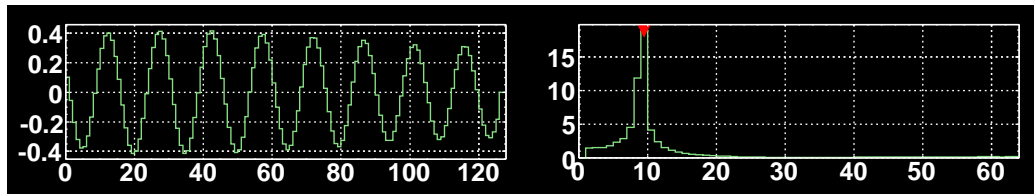}
  \caption[A $10$\,kHz sinusoidally oscillating event]
	  {\small A $10$\,kHz sinusoidally oscillating event.}
  \label{fig:10khzsinusoidal}
\end{figure}
Again, this signal has previously been introduced in Section
\ref{subsec:discrimination}. At first it was thought that these
$10$\,kHz oscillating signals were in fact calibration charges
injected into the Rona DAQ to verify the integrity of transmission
along the eight hydrophone to shore cables. However, this was later
ruled out because they do not occur at times when the calibration
tones were injected nor are they at the right frequency. Nonetheless,
it is suggested that these signals are anthropogenic in origin.
\subsubsection{High frequency oscillator}
\label{subsubsec:highfreqoscillator}
\begin{figure}[h!]
  \centering
  \includegraphics[width=0.9\textwidth]{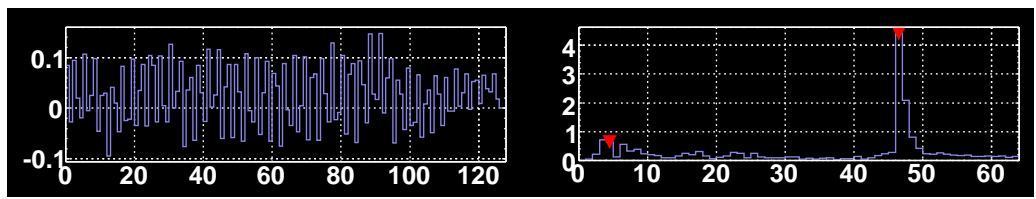}
  \caption[A high frequency event]
	  {\small A high frequency event.}
  \label{fig:hifreq}
\end{figure}
Higher frequency oscillating signals are also present in the Rona data
but with a much lower abundance than the low frequency oscillators and
the $10$\,kHz sinusoids. Typically they occur at frequencies around
$50$\,kHz.
\subsubsection{Ringer}
\label{subsubsec:ringer}
\begin{figure}[h!]
  \centering
  \includegraphics[width=0.9\textwidth]{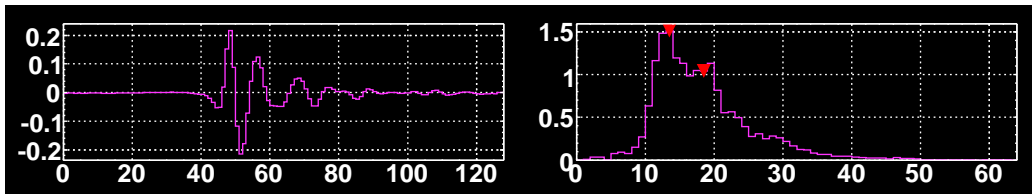}
  \caption[A `ringing' event]
	  {\small A `ringing' event.}
  \label{fig:ringer}
\end{figure}
Some events in the Rona data occur periodically that demonstrate a
decaying and oscillating pressure signal. There are two possible
reasons for the appearance of such a signal: firstly, there is a
decaying oscillating pressure wave in the water; secondly, due to a
non-linearity in the phase response of the hydrophone system an
asymmetric time delay of different Fourier components has caused the
signal to ring. It is impossible to know for certain which of the two
explanations is correct for any given case so these events have been
arbitrarily labelled as ``ringers''.
\subsubsection{Bipolar}
\label{subsubsec:bipolar}
\begin{figure}[h!]
  \centering
  \includegraphics[width=0.9\textwidth]{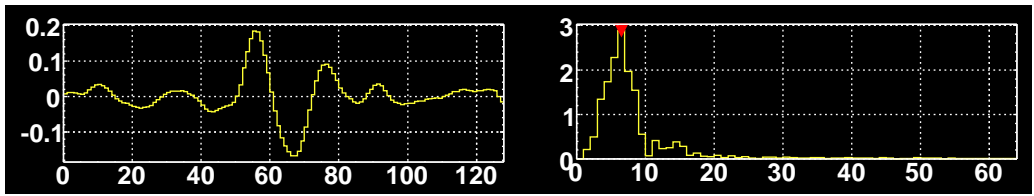}
  \caption[A `bipolar' event]
	  {\small A `bipolar' event.}
  \label{fig:idealbipolar}
\end{figure}
Finally, this analysis has identified a class of event that is
approximately bipolar in shape with a frequency close to that which we
would expect from a neutrino induced particle cascade at some
far-field distance from the Rona array. The signal is certainly not
flat either side of the bipolar part of the pulse, and may indeed
contain an oscillating component but one can imagine that if a
smaller window had been chosen a clearer bipolar signal would be
observed.

Clearly there are noise sources around the Rona array that can
portray themselves as originating from UHE hadronic cascades. Further,
more detailed work, beyond the scope of this somewhat qualitative
analysis, is required to fully understand such backgrounds.
\subsubsection{Occurrences of identified signal types}
\label{subsubsec:occurences}
The identified signal types as described above recurred on several
days throughout the data analysis. Table \ref{tab:recurringevents}
summarises this information.
\begin{table}[htb]
  \centering
  \begin{tabular}{ l | l }
    {\bf Event Type}         & {\bf Recurrence}                     \\
    \hline
    Short impulsive          & 11th, 12th, 14th, 20th               \\
    Low frequency oscillator & 9th-12th, 14th-17th, 19th-21st, 23rd \\
    $10$\,kHz sinusoidal     & 14th                                 \\
    High frequency           & 14th, 16th, 24th                     \\
    Ringer                   & 12th, 21st, 24th                     \\
    Bipolar                  & 17th, 19th-22nd, 24th                \\
  \end{tabular}
  \caption[Recurrence of events at Rona]
	  {\small Recurrence of events at Rona.
	  \label{tab:recurringevents}}
\end{table}
\section{Summary}
The DAQ as was in operation between 09/12 and
24/12/2005
at the Rona hydrophone array in Scotland has been
described. Additionally a  novel analysis of the raw $12,453,500$ events
triggered by the ACoRNE data reduction has been
reported. $3526$ passed offline cuts demanding that: there is a
coincidence between four or more hydrophones; the peak pressure is
above $35$\,mPa; and, the maximum peak in the DFT of each coincident
signal is at the same frequency. The requirement for four-fold or
greater coincidence above a threshold of $35$\,mPa puts the analysis
in concordance with the sensitivity simulation described in previous
chapters. Table \ref{tab:ncoincidences} summarises the scale of
coincidence amongst the analysed data. A total of $1059$ events showed
a five-fold coincidence or greater. In Section
\ref{sec:noisefilterandthresholds} it was stated that for a mean noise
level of $35.9$\,dB only one false trigger in ten years, due to noise,
with a five-fold coincidence should be observed. By integrating the
energy of the pulses triggering the ACoRNE DAQ in this $15$ days worth
of data, a noise level of between $90 \to 100$\,dB is observed, which
demonstrates a higher level of noise activity than the location considered
in the ACoRNE proposal. Furthermore it is suggested that there is an
anthropogenic component of the Rona events that raises the noise level
above that predicted by sea state alone. The noise at Rona provides an
abundance of ``signals'', the observation of which enables development
of the data acquisition system. The more developed DAQ can then, in
theory, be implemented elsewhere at a warmer, deeper and ultimately
quieter site more suited to the acoustic detection of UHE neutrinos.
\begin{table}[h!]
 \centering
 \begin{tabular}{ l | c | c | c | c | c }
    {\bf Level}     & 4-fold & 5-fold & 6-fold & 7-fold & 8-fold \\
    \hline
    {\bf Frequency} & 2468   & 845    & 182    & 32     & 0      \\
  \end{tabular}
 \caption[Level of coincidence at Rona]
	  {\small Coincidences at Rona. \label{tab:ncoincidences}}
\end{table}

  \cleardoublepage

  \chapter{Discussion and Outlook}
\label{chp:discussion}
\section{The Motivation for Neutrino Astronomy}
\label{sec:nuastonomy}
The motivation for Neutrino Astronomy is broad and compelling. As was
discussed in the introduction the eventual first detection of
neutrinos lagged their initial postulation by some twenty-three
years. They are an essential component of the Standard Model of
physics and are the second-most abundant known particle in the
Universe yet their properties are still shrouded in mystery. One thing
is for sure: the UHE $p p$ and $p \gamma$ interactions that follow the
production of the highest energy cosmic rays will carry with them an
associated flux of UHE neutrinos (Section \ref{subsec:bottomup}).
Due to their infinitesimal mass and
lack of electric charge they can offer a unique perspective of the
Cosmos. What is more, the scope for so-called ``new physics'' offered
by Neutrino Astronomy is abundant, whether it is from the decay of
exotics such as topological defects or relic big bang
particles, or the annihilation of dark matter, neutrinos can offer a
window into the frontier territories of quantum scale gravity, extra
dimensions, super-symmetry and dark matter (Section
\ref{subsec:topdown}).
\section{The Current Scale of Development}
\label{sec:development}
The Neutrino Astronomy effort is a global endeavour extending across
the surface of the Earth from pole-to-pole and above-and-below from
high in the atmosphere to the depths of the sea (Sections
\ref{sec:opticalcerenkov}, \ref{sec:radiocerenkov} and
\ref{sec:acousticdetectors}).  It is because of the pioneering efforts
of those such as the ill-fated DUMAND project and the long established
lake Baikal experiment that today's neutrino telescopes can look to
the future and envisage a technology on the cubic kilometre scale and
beyond. Despite such grand aspirations the field as a whole remains in
its infancy when compared, say, to the status of radio astronomy and
to a lesser extent UHECR and TeV gamma astronomy. One hopes
that in the years to come the fruits borne by this labour can match
that which is emerging from experiments such as AUGER and HESS in the
present day.

The acoustic detection of UHE neutrinos is undoubtedly the least well
developed technique. However it is arguably the most elegant in its
simplicity. The direct detection of the thermoacoustic emission
resulting from the interaction of UHE neutrinos at the Earth, with an
off the shelf technology such as commercially available wide band
hydrophones, is unrivalled in its potential to offer huge effective
volumes without the necessity of complicated detector design. 
\section{Simulation of Acoustic Arrays}
\label{sec:acousticarrays}
\subsection{Neutrino interactions}
The ACoRNE simulation work has continued to be developed following
the formation of the collaboration. From calculations of the initial
interaction of a neutrino at the Earth to the reconstruction of the
thermoacoustic emission by cubic kilometre arrays of hydrophones a
vast number of processes require consideration. Understanding the
neutrino-nucleon cross-section at EeV energies depends on some
combination of parameterised empirical data, extrapolated to higher
energies, and theoretically motivated approximations. Man-made
particle accelerators will likely never rival those of nature in their
ability to accelerate particles to the highest energies. Presently it
is from CR air shower experiments that we gain the most insight into
particle interaction cross-sections at the TeV scale and
beyond. Eventually neutrino telescopes, it is hoped, will provide the
same quality of information as experiments such as HI-RES, KASCADE,
AUGER and those alike, allowing the existing models that are relied upon
in programs such as ANIS, PYTHIA and CORSIKA to be tuned to a finer
degree of accuracy. For now we accept that the level of uncertainty in
our particle interaction simulations is at worst within the level of
individual event fluctuations (Sections \ref{subsec:thresholds} and
Appendix \ref{appendix:geantvscorsika}).
\subsection{Formation and propagation of the signal}
The formation of the bipolar acoustic signal characteristic of an UHE
neutrino induced particle cascade relies exclusively on knowledge of
the thermoacoustic energy deposited by the hadronic cascade. Efforts
both experimental and theoretical have been made to best understand
this mechanism. In the simulation work contained in this thesis,
parameterisations of the integrated thermal energy resulting from the
\geant and CORSIKA programs at various energies have been utilised to
these ends (Section \ref{sec:formation}). Furthermore, the far-field
approximation has been invoked to parameterise the peak pulse
pressure, which has been shown to vary in direct proportion to the
energy of the cascade.

The propagation of acoustic signals in the sea is as rich and diverse
a field as Neutrino Astronomy. To attempt to incorporate something
with such a magnitude of complexity into simple parameterisations of
absorption, attenuation and propagation (Sections
\ref{sec:attenuation} and \ref{sec:refraction}) is perhaps something
of an injustice. Nevertheless progress will continue to be made upon the
first generation codes that incorporated a constant velocity of sound
towards more complicated ray-traced methods as described in this work,
complex absorption models and beyond. Knowledge of the sound
velocity profile surrounding a detector is critical to our
understanding of its performance. Indeed we are fortunate that the
deep sea presents something of an idealised situation in that the SVP
is essentially linear, yet it remains to be seen exactly how accurate
our understanding of it need be. What is more the boundary conditions
at the sea surface and sea bed that have been neglected in this work
present in themselves a broad subject for study. If the bipolar acoustic
signal can undergo surface or sea bed reflections without
experiencing too much distortion then the effective volume for a given
array may extend well beyond the cutoff of $48$\,km observed in this work.
\subsection{Noise and filtering}
The effects of noise and filtering again encompass a vast amount of
research that is parameterised into the simple models utilised in the
large scale detector simulation described in this thesis (Section
\ref{sec:noisefilterandthresholds}). What is being learnt from
experimental projects such as SAUND in the Bahamas, the
acoustic branches of the ANTARES and NEMO collaborations at their
respective sites, and from the ACoRNE project at Rona (Chapter
\ref{chp:experiment}) is that the backgrounds presented to the acoustic
neutrino astronomer are numerous and multifaceted, presenting in
themselves events that emulate the type of signal one may expect from
an UHE neutrino induced particle cascade. Understanding of thresholds
continues to develop as data are gathered from arrays such as those
operated by the aforementioned groups and so does the continuing
development of matched filters.
\subsection{The large scale detector simulation}
The reconstruction algorithms developed as part of this work represent
the first iteration in what will undoubtedly mature into a vastly more
complex field of study. Sensitivity predictions for acoustic arrays at
present represent the work of a handful of people working broadly in
isolation. In addition to the development of the computerised models,
described above, that effectively sit as a front-end to the large
scale simulation there is scope for further refinement of the
techniques, already in use in this study, with respect to vertexing,
pointing and calorimetry (Sections \ref{sec:eventreconstruction},
\ref{sec:optimaldensity} and \ref{sec:energyrecon}). Better
development of the minimisation procedure has already been alluded to
(Section \ref{subsubsec:minuit}), additionally optimising the
reconstruction of the acoustic pancake remains an interesting and open
matter for study. One can envisage much improved performance by fully
exploiting the unique pancake geometry.

The nature of this work, as it was conceived in the beginning, was to
consider typically one thousand hydrophones in a volume of one cubic
kilometre. We have seen that there are different factors for which
array density can be optimised such as the effective volume and the
reconstruction performance (Sections \ref{sec:effectivevolume} and
\ref{sec:optimaldensity}). Naturally if one were to begin a proposal
to design and build a dedicated acoustic array for the purpose of UHE
neutrino astronomy extra considerations such as cost and deployability
come into play; the randomly distributed arrays investigated here may
practically be very difficult to manufacture. Clearly there is a great
deal of research potential for proper optimisation of the
geometry of an acoustic array.

With respect to the sensitivity predictions made here (Section
\ref{sec:sensitivitycal}), there is broad 
agreement with the work published in reference \cite{timo:thesis},
(which recursively agrees with predictions from the IceCube
collaboration)\footnote{see the presentations and proceedings of
  the ARENA 2005/06 workshops \cite{arena:2005},\cite{arena:2006}}.
In loose terms, as current predictions stand, one
requires a pressure threshold around $5-10$\,mPa operating across an
array of over a thousand cubic kilometres with at least one
hundred hydrophones per cubic kilometre if the proposed flux of GZK
neutrinos is to be observed. Given further study and development of
array optimisation and reconstruction algorithms, this may change.
\section{Predictions for Future Work}
\label{sec:predictions}
Within the field of acoustic neutrino astronomy there is now some idea
of the scales (thousands of cubic kilometres) to be considered if one
wants to guarantee observation of those neutrinos resulting from the
attenuation of UHECRs by the cosmic microwave background
radiation. Undoubtedly that level of commitment is years beyond
today's level of activity. Currently, acoustic projects exist by
utilising existing infrastructures, be it through military hardware
such as the AUTEC and Rona arrays used by SAUND and ACoRNE
respectively, or by piggybacking the deployment of existing sea or ice
based detectors such as ANTARES, NEMO and IceCube. This latter
strategy paves the way for hybrid detection. Because water doesn't
facilitate radio detection and there is no overlap in energy for
simultaneous optical and acoustic detection of a given neutrino (see
Figure \ref{fig:energyranges}), ice is surely the best medium for
hybrid detection. This is further enhanced by its higher density and
longer light absorption lengths than for water, coupled with the
potential for shear wave acoustics thanks to the lattice-like
structure of ice. Within the IceCube collaboration work is underway to
extend to hybrid optical/radio/acoustic detection\footnote{again, see
  for example the   proceedings of the ARENA 2005/06 workshops
  \cite{arena:2005},\cite{arena:2006}}. Any detector offering
simultaneous registration of the same event via two independent means
has intrinsically less systematic uncertainty. One as yet unanswered
question that remains for the acoustic neutrino astronomer is that of
positively identifying a signal as being that of a neutrino and
completely ruling out any other source. Naturally a coincident radio
based detection would remove a great deal of uncertainty.

As the ANTARES project nears full deployment the boreal cubic
kilometre scale neutrino telescopes begin their design
phases. Acoustic instrumentation is already a part of the ANTARES
design scheme and is almost certainly going to feature in the next
generation telescopes in the Mediterranean. If this existing
infrastructure can be utilised in such a way as to provide readout for
acoustic detection then the first cubic kilometre, sea-based acoustic
neutrino telescopes may not be far away thus completing complementary
neutrino telescopy in both the northern and southern hemispheres, a
requisite for full sky coverage. The neutrino-light in which the Earth
bathes will soon illuminate the world's largest active particle
detectors and enlighten physicists to the mysteries foretold by
nature's most enigmatic cosmic messenger.











  \cleardoublepage

  \renewcommand{\chaptermark}[1]{ \markboth{\appendixname\ \thechapter.\ #1}{}}
  \renewcommand{\sectionmark}[1]{ \markright{\thesection.\ #1}}
  
  \def\thefigure{{\Alph{chapter}}.{\arabic{figure}}}
  \def\thetable {{\Alph{chapter}}.{\arabic{table}}}

  \appendix

%
\chapter{Empirical Data}
\section{Neutrino-Nucleon Cross Section}
\label{appendix:nucrosssection}
\begin{figure}[!h]
  \begin{center}
    \includegraphics[width=0.7\textwidth]{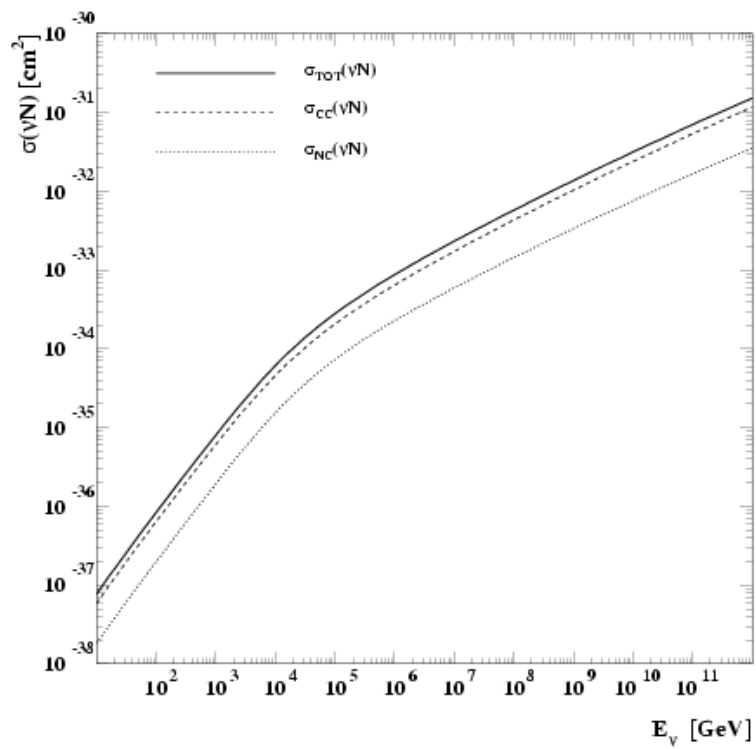}
    \caption[Neutrino-Nucleon cross section]
	    {
	      \small	    
	      The total neutrino-nucleon cross section and its charged 
	      current and neutral current contributions. Taken from 
	      \cite{uhenuphysics:kwiecinski}. The cross section for neutrino
	      interactions at UHE was extrapolated from this plot and
	      translated into the following code:
	    }
	    \label{nunucleonxsection}
  \end{center}
\end{figure}
\begin{verbatim}
  // log10(sigma/cm^2) = -32.60 + (0.362*(log10(E/GeV)-7))
  // INPUT  : energy in eV
  // OUTPUT : cross section in cm^2

  double egev    = energy/1E09;
  double log10cs = -32.60 + (0.362*(log10(egev)-7.0));
  return pow(10,log10cs);
\end{verbatim}
\section{Knudsen Noise Curves}
\label{appendix:knudsen}
\begin{figure}[htb]
  \centering
  \includegraphics[width=\textwidth]{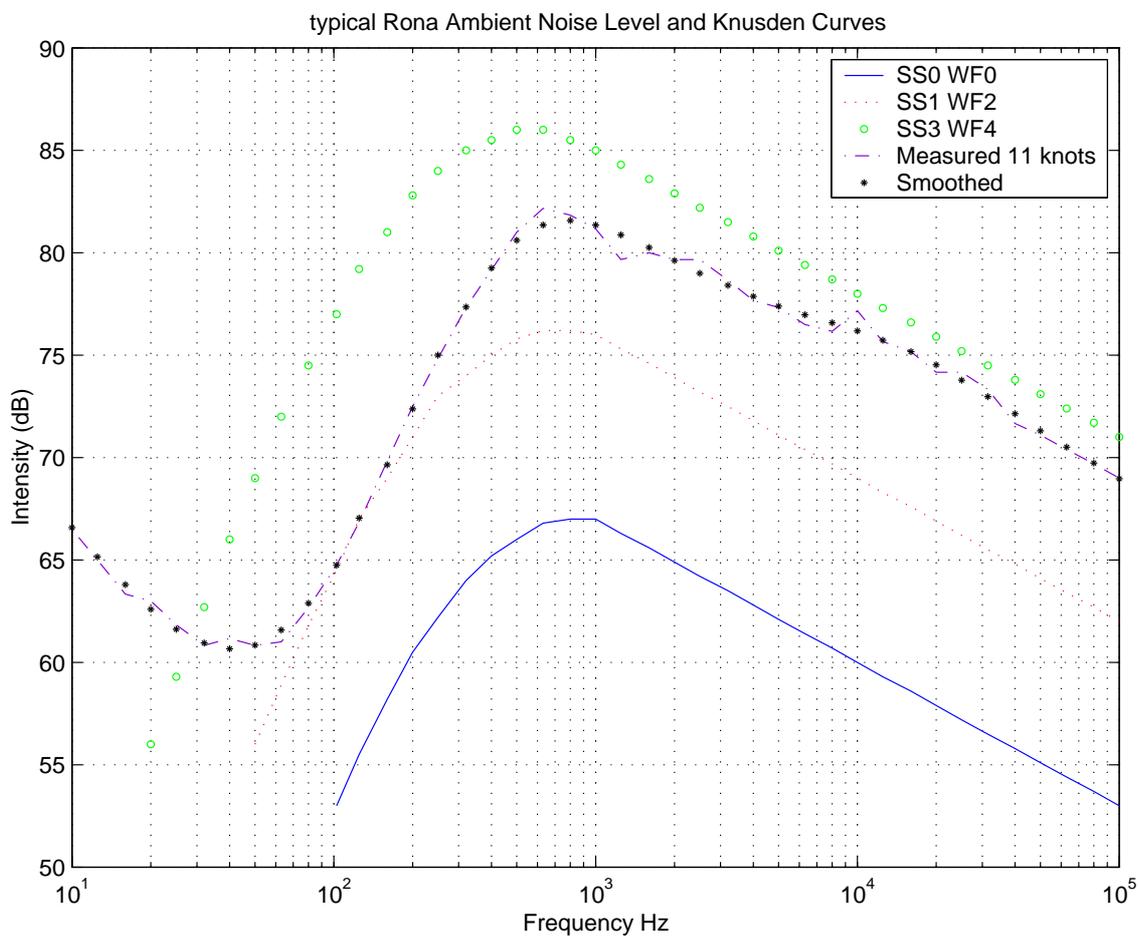}
  \caption[Knudsen Noise Curves]
    {
      \small
      Typical Knudsen noise curves (SS0, SS1, SS3) at the RONA site
      off the west Scottish coast used by the ACoRNE collaboration. These
      curves parameterise the ambient noise levels as a function of Sea
      States as defined by the World Meteorological Organisation. Also
      included is the measured noise level at Rona (Refer to legend).
    }
    \label{fig:knudsencurves}
\end{figure}
%
%
\chapter{Comparison of Hadronic Showers from \geant and CORSIKA}
\label{appendix:geantvscorsika}
The longitudinal and radial profiles of thermal energy deposition
from hadronic cascades produced by two programs: \geant and CORSIKA are
compared. The CORSIKA program, designed for the production of UHECR
induced air showers, in the upper atmosphere of Earth, has been
modified to produce hadronic cascades resulting from UHE neutrinos in
a seawater medium \cite{terry:corsika}.

In each case the average of $100$ showers is plotted to smooth out
fluctuations on an event by event basis. The agreement between the
shower shapes from \geant and CORSIKA is within the $20\% \to 30\%$
fluctuations seen between showers and the $\sim 30\%$ uncertaintity
between choice of theoretical models in each program.

The modified version of CORSIKA has been developed so thermal energy
densities for hadronic cascades initiated by primaries beyond the
$100$\,TeV limit inherent to \geant can be produced. Hence the
extrapolated pressure to energy parameterisation need not be relied
upon. A study of shower shapes and peak pulse pressure resulting from
CORSIKA showers has however validated the assumption that pressure
scales linearly with shower energy.
\begin{figure}[htb]
\centering
\includegraphics[width=0.8\textwidth]{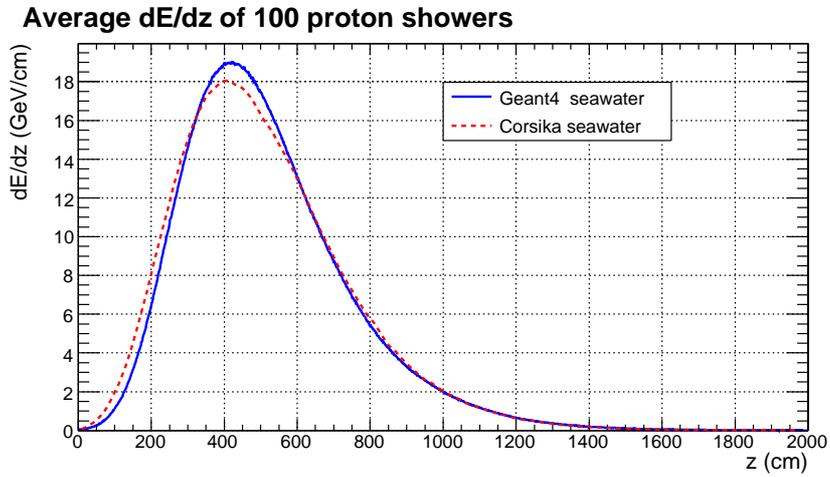}
\caption[\geant and CORSIKA longitudinal energy deposition at
$10^4$\,eV]{\small \geant and CORSIKA longitudinal energy deposition at
$10^4$\,eV.}
\end{figure}
\begin{figure}[htb]
\centering
\includegraphics[width=0.8\textwidth]{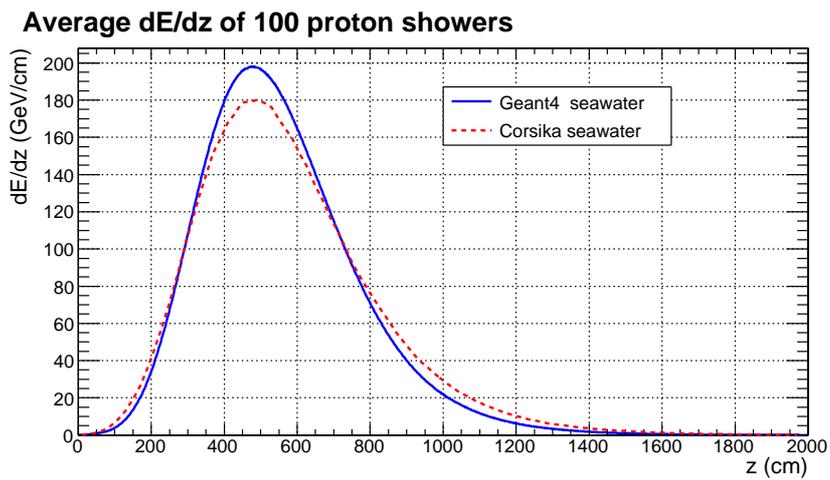}
\caption[\geant and CORSIKA longitudinal energy deposition at
$10^5$\,eV]{\small \geant and CORSIKA longitudinal energy deposition at
$10^5$\,eV.}
\end{figure}
\begin{figure}[htb]
\centering
\includegraphics[width=0.8\textwidth]{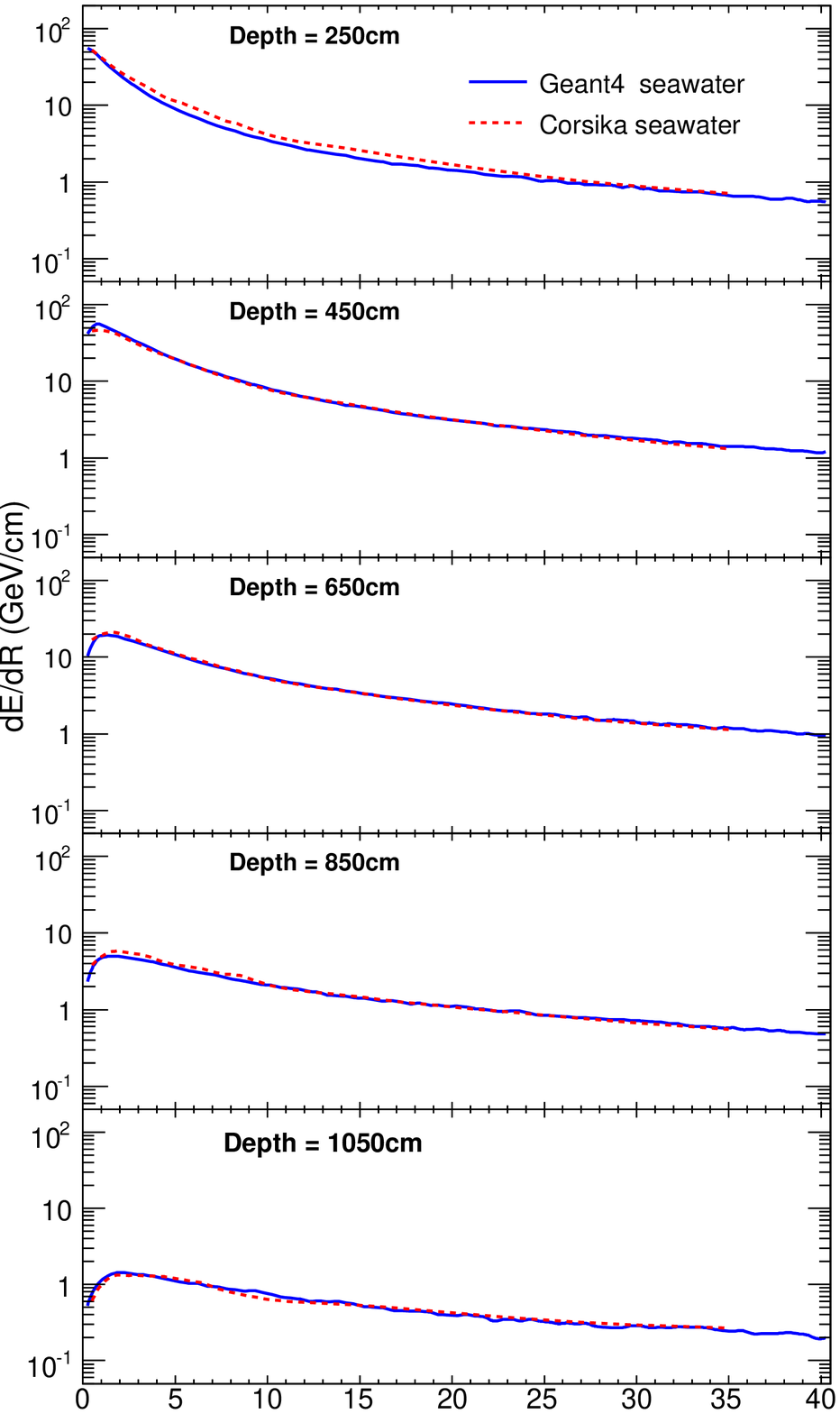}
\caption[\geant and CORSIKA radial energy deposition at
$10^4$\,eV]{\small \geant and CORSIKA radial energy deposition at
$10^4$\,eV for various depths along the shower axis.}
\end{figure}
\begin{figure}[htb]
\centering
\includegraphics[width=0.8\textwidth]{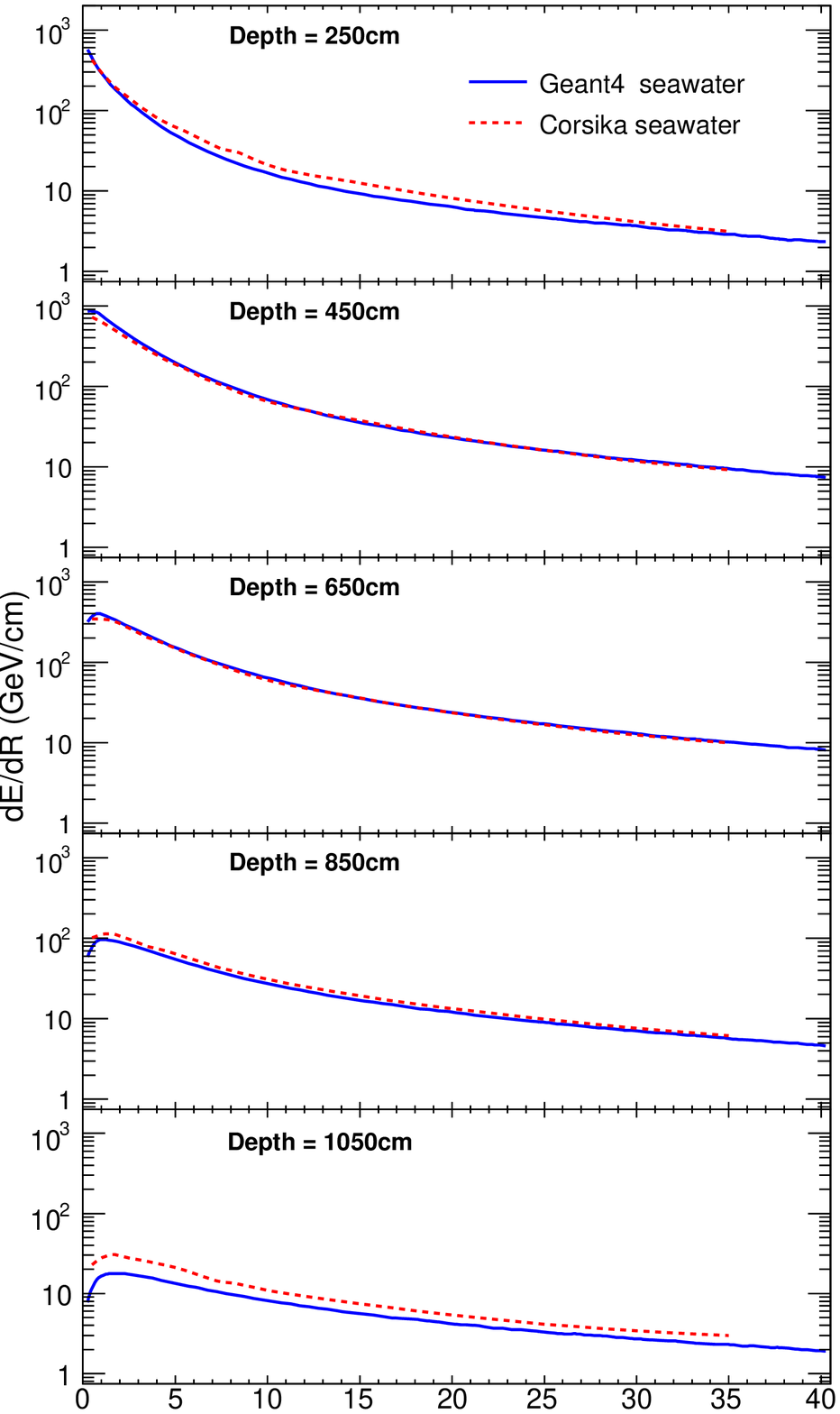}
\caption[\geant and CORSIKA radial energy deposition at
$10^5$\,eV]{\small \geant and CORSIKA radial energy deposition at
$10^5$\,eV for various depths along the shower axis.}
\end{figure}
%
%
\chapter{Formalism of Ray Tracing Algorithm}
\label{appendix:raytracing}
We start with the differential equations for ray paths in
inhomogeneous (meaning the velocity of sound is not constant) media as
given in \cite{acoustic:boyles} and proceed through the derivation
therein:
\begin{eqnarray}
\frac{d}{ds}\Big(n\frac{dx}{ds}\Big)&=&\frac{\partial n}{\partial x}
\label{eq:ray:raypathx} \\\nonumber\\
\frac{d}{ds}\Big(n\frac{dy}{ds}\Big)&=&\frac{\partial n}{\partial y}
\label{eq:ray:raypathy} \\\nonumber\\
\frac{d}{ds}\Big(n\frac{dz}{ds}\Big)&=&\frac{\partial n}{\partial z}
\label{eq:ray:raypathz}
\end{eqnarray}
where index of refraction $n = c_n / c$ and $c_n$ is the velocity when
the angle $\theta$ of the ray with respect to the horizontal is $0$.
If the velocity $c$ is a function of $z$ only then
$n=n(z)$ and Equations \ref{eq:ray:raypathx} through
\ref{eq:ray:raypathz} are reduced to:
\begin{eqnarray}
                      n\frac{dx}{ds} &=& \textrm{constant} \nonumber \\
                      n\frac{dy}{ds} &=& \textrm{constant} \nonumber \\
\frac{d}{ds}\Big(n\frac{dz}{ds}\Big) &=& \frac{dn}{dz}
\label{eq:ray:reduced}
\end{eqnarray}
We can arbitrarily collapse into $2$ dimensions by observing that the
first two equations imply the ray path lies in a plane orthogonal to the
$xy$-plane.
\begin{figure}[h]
  \centering
  \includegraphics[width=0.5\textwidth]{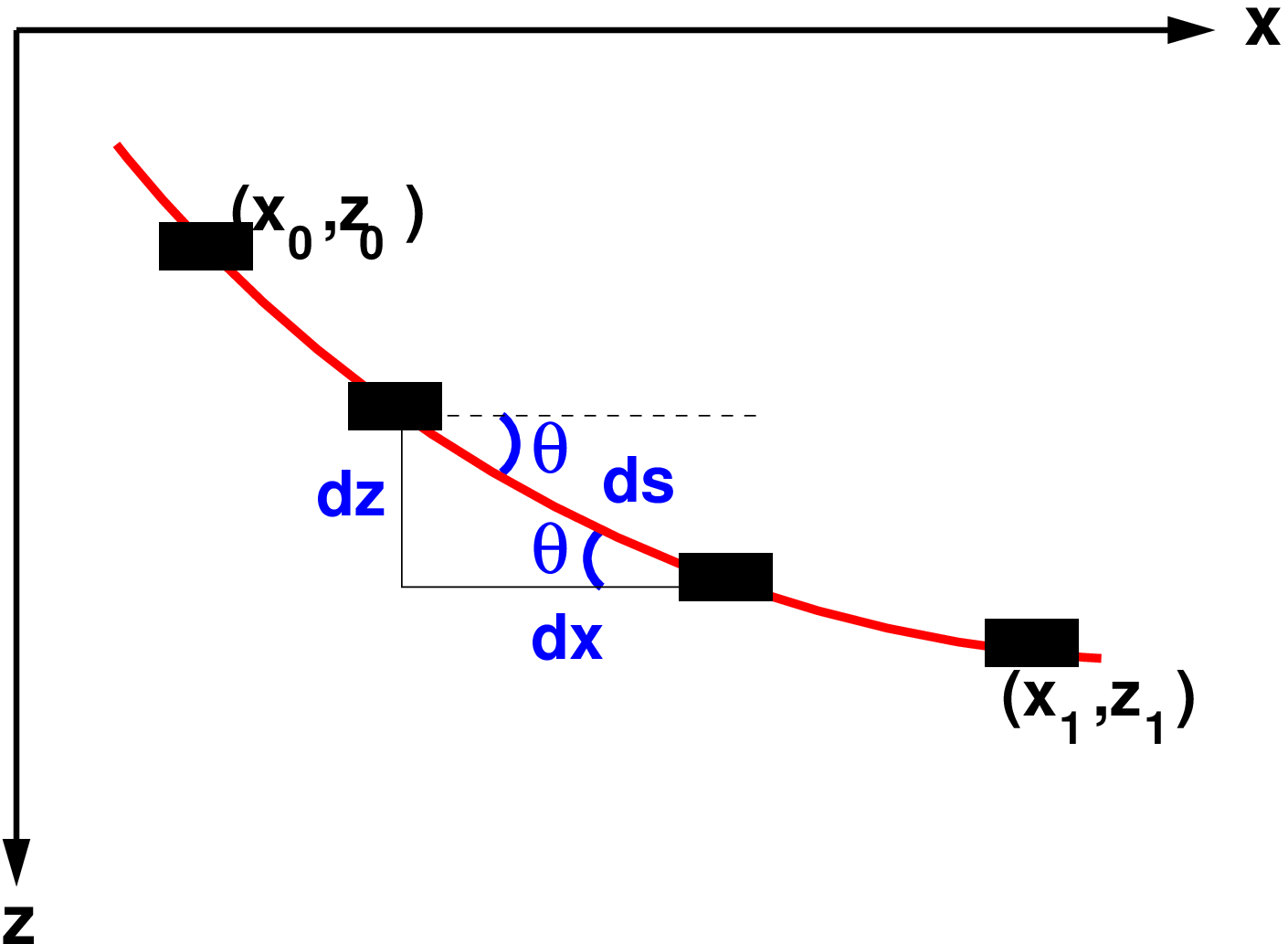}
  \caption[Refraction Geometry]
	  {
	    \small
	    Geometry of a single curved ray in a medium with varying
	    sound velocity.
	  }
	  \label{fig:refractiongeomapp}
\end{figure}
From Figure \ref{fig:refractiongeomapp} we see:
\begin{eqnarray}
\frac{dx}{ds}  &=& \cos \theta \nonumber\\\nonumber\\
\frac{dz}{ds}  &=& \sin \theta
\label{eq:ray:cosines}
\end{eqnarray}
Taking the first equation of Equation \ref{eq:ray:reduced} and
substituting the first equation from Equation \ref{eq:ray:cosines}:
\begin{equation}
\frac{\cos \theta}{c} = \frac{1}{c_n} = K
\label{eq:ray:snell}
\end{equation}
which is Snell's Law. $K$ is called the ray parameter which, for the
geometry in Figure \ref{fig:refractiongeomapp} can be conveniently
written as:
\begin{equation}
K = \frac{\cos \theta_0}{c_0} = \frac{\cos \theta_1}{c_1}
\label{eq:ray:convenient}
\end{equation}
where $c_0,\theta_0$ and  $c_1,\theta_1$ are the sound velocity and
angle at $(x_0,z_0)$ and $(x_1,z_1)$ respectively.
Substituting the second equation from Equation \ref{eq:ray:cosines}
into the second equation in Equation \ref{eq:ray:reduced} we have:
\begin{eqnarray*}
\frac{dn}{dz} 
&=&\frac{d}{ds}(n\sin\theta)\\\\
&=&n\frac{d\sin\theta}{ds}+\sin\theta\frac{dn}{ds}\\\\
&=&n\cos\theta\frac{d\theta}{ds}+\sin\theta\frac{dn}{ds}\\\\
&=&n\cos\theta\frac{d\theta}{ds}+\sin\theta\frac{dn}{dz}\frac{dz}{ds}\\\\
&=& n\cos\theta\frac{d\theta}{ds}+\sin^2\theta\frac{dn}{dz}
\label{eq:ray:dndz}
\end{eqnarray*}
which can be solved for $d\theta/ds$,
\begin{eqnarray}
\frac{d\theta}{ds}&=&\frac{\cos\theta}{n}\frac{dn}{dz}\nonumber\\\nonumber\\
                  &=&-\frac{\cos\theta}{c}\frac{dc}{dz}\nonumber\\\nonumber\\
		  &=&-K\frac{dc}{dz}
\label{eq:ray:dthetads}
\end{eqnarray}
This yields the important result that the curvature of the ray
$d\theta/ds$ is directly proportional to the velocity gradient
$dc/dz$. From the geometry in Figure \ref{fig:refractiongeomapp} we can
see that
\begin{equation}
\frac{dx}{dz} = \cot \theta
\label{eq:ray:cottheta}
\end{equation}
such that horizontal range of the ray is given by:
\begin{eqnarray}
dx        &=& \cot \theta dz \nonumber\\\nonumber\\
x_1 - x_0 &=& \int_{z_0}^{z_1} \cot \theta dz
\label{eq:ray:xrange}
\end{eqnarray}
The travel time along the ray is given by:
\begin{eqnarray}
       dt &=& \frac{ds}{c} \nonumber\\\nonumber\\
t_1 - t_0 &=& \int_{z_0}^{z_1} \frac{ds}{c} \nonumber\\\nonumber\\
t_1 - t_0 &=& \int_{z_0}^{z_1} \frac{dz}{c \sin \theta}
\label{eq:ray:time}
\end{eqnarray}
Similarly, from the geometry in Figure \ref{fig:refractiongeomapp}, the
ray path length is given by:
\begin{eqnarray}
       ds &=& \frac{dz}{\sin \theta} \nonumber\\\nonumber\\
s_1 - s_0 &=& \int_{z_0}^{z_1} \frac{dz}{\sin \theta}
\label{eq:ray:raylength}
\end{eqnarray}
%



%
In the case of a linear sound velocity gradient:
\begin{eqnarray}
            c &=& c_m + gz \nonumber\\\nonumber\\
\frac{dc}{dz} &=& c' = g
\label{eq:ray:lineargradient}
\end{eqnarray}
where the second term is the derivative with respect to $z$ of the
first. We can now substitute the differential form of Equation
\ref{eq:ray:lineargradient} into Equation \ref{eq:ray:dthetads}:
\begin{equation}
\frac{d\theta}{ds} = -Kg
\label{eq:ray:curvature}
\end{equation}
$d\theta/ds$ is the curvature of the ray, and for a linear sound
velocity gradient is constant. Hence the ray trajectory follows the
arc of a circle. The radius of curvature $R$ is given by:
\begin{equation}
R_c = \frac{1}{|d\theta/ds|} = \frac{1}{|-Kg|}
\label{eq:ray:radius}
\end{equation}
Equations \ref{eq:ray:xrange}, \ref{eq:ray:time} and
\ref{eq:ray:raylength} for horizontal range, travel time and path
length respectively can be expressed in terms of $\theta$. Starting
with Equation \ref{eq:ray:xrange} we have:
\begin{eqnarray*}
x_1-x_0&=& \int_{z_0}^{z_1}\cot \theta dz               \\\\
       &=& \int_{z_0}^{z_1}\frac{\cos\theta}{\sin\theta}\\\\
       &=& \int_{\theta_0}^{\theta_1}\frac{\cos\theta d\theta}{Kc'}
\end{eqnarray*}
where we have used $dz/\sin\theta=-d\theta/Kc'$ obtained by
differentiating Snell's Law:
\begin{eqnarray*}
                    \cos\theta &=& Kc                  \\\\
        \frac{d\cos\theta}{dz} &=& Kc'                 \\\\
\frac{\sin\theta\,d\theta}{dz} &=& -Kc'                \\\\
         \frac{dz}{\sin\theta} &=& -\frac{d\theta}{Kc'}
\end{eqnarray*}
The integral now becomes:
\begin{eqnarray*}
x_1 - x_0
&=&-\int_{\theta_0}^{\theta_1}\frac{\cos\theta\,d\theta}{Kg}\\\\
&=&-\frac{1}{Kg}[\sin\theta]_{\theta_0}^{\theta_1}          \\\\
&=& \frac{1}{Kg}(\sin\theta_0 - \sin\theta_1)               \\\\
&=&\frac{c_0}{g}\Big(\frac{\sin\theta_0-\sin\theta_1}{\cos\theta_0}\Big)
\end{eqnarray*}
Similarly, starting with Equation \ref{eq:ray:time}:
\begin{eqnarray*}
t_1 - t_0
&=&\int_{z_0}^{z_1}\frac{dz}{c\sin\theta}                \\\\
&=&-\int_{\theta_0}^{\theta_1}\frac{d\theta}{g\cos\theta}\\\\
&=&-\frac{1}{g}\ln\Bigg\{\frac{\sec\theta_1+\tan\theta_1}{\sec\theta_0+\tan\theta_0}\Bigg\}\\\\
&=&-\frac{1}{g}\ln\Bigg\{\frac{\tan\Big(\frac{\theta_1+\pi/2}{2}\Big)}{\tan\Big(\frac{\theta_0+\pi/2}{2}\Big)}\Bigg\}\\\\
&=&-\frac{1}{g}\ln\Bigg\{\frac{\cos\theta_0(1+\sin\theta_1)}{\cos\theta_1(1+\sin\theta_0)}\Bigg\}
\end{eqnarray*}
Finally from Equation \ref{eq:ray:raylength}:
\begin{eqnarray*}
s_1 - s_0
&=&\int_{z_0}^{z_1}\frac{dz}{\sin\theta}\\\\
&=&-\int_{\theta_0}^{\theta_1}\frac{d\theta}{Kg}\\\\
&=&\frac{1}{Kg}{\theta_0-\theta_1}\\\\
&=&\frac{c_0}{g\cos\theta_0}(\theta_0-\theta_1)
\end{eqnarray*}
One can formulate an equation for the change in depth of the ray
directly from Equation \ref{eq:ray:lineargradient} by using Snell's
Law as given in Equation \ref{eq:ray:convenient}. From:
\begin{equation*}
z = \frac{c - c_m}{g}
\end{equation*}
One can write:
\begin{eqnarray*}
z_1 - z_0
&=&\Bigg(\frac{c_1-c_m}{g}\Bigg)-\Bigg(\frac{c_0-c_m}{g}\Bigg)\\\\
&=&\frac{c_1-c_0}{g}\\\\
&=&\frac{1}{g}\Bigg(c_0\frac{\cos\theta_1}{\cos\theta_0}-c_0\Bigg)\\\\
&=&\frac{c_0}{g\cos\theta_0}(\cos\theta_1-\cos\theta_0)
\end{eqnarray*}
In summary, when the sound velocity gradient is constant, the
horizontal range, travel time, ray path length and change in depth are
given by:
\begin{eqnarray}
x_1 - x_0
&=&\frac{c_0}{g}\Bigg(\frac{\sin\theta_0-\sin\theta_1}{\cos\theta_0}\Bigg)\\\nonumber\\
\label{eq:ray:xconst}
t_1 - t_0
&=&-\frac{1}{g}\ln\Bigg\{\frac{\cos\theta_0(1+\sin\theta_1)}{\cos\theta_1(1+\sin\theta_0)}\Bigg\}\\\nonumber\\
\label{eq:ray:timeconst}
s_1 - s_0
&=&\frac{c_0}{g\cos\theta_0}(\theta_0-\theta_1)\\\nonumber\\
\label{eq:ray:raylengthconst}
z_1 - z_0
&=&\frac{c_0}{g\cos\theta_0}(\cos\theta_1-\cos\theta_0)
\label{eq:ray:depthchangeconst}
\end{eqnarray}
Equations \ref{eq:ray:xconst}, \ref{eq:ray:timeconst},
\ref{eq:ray:raylengthconst} and \ref{eq:ray:depthchangeconst} form the
basis for the ray tracing algorithm.
%
%
\chapter{Neutrino Source Data}
\label{appendix:nusourcedata}
The neutrino spectrum $E_{\nu}$, source distribution
$(\rho,\theta,\phi)_{\nu}$ and interaction coordinates
$(r,z,\phi)_{\mathrm{int}}$ as generated in the large scale detector
simulation (Chapter \ref{chp:largescale}) is summarised by the
following plots (with energy in GeV, angles in radians and distances
in metres):
\begin{figure}[htb]
\centering
\begin{minipage}[!h]{0.4\textwidth}
\includegraphics[width=\textwidth]{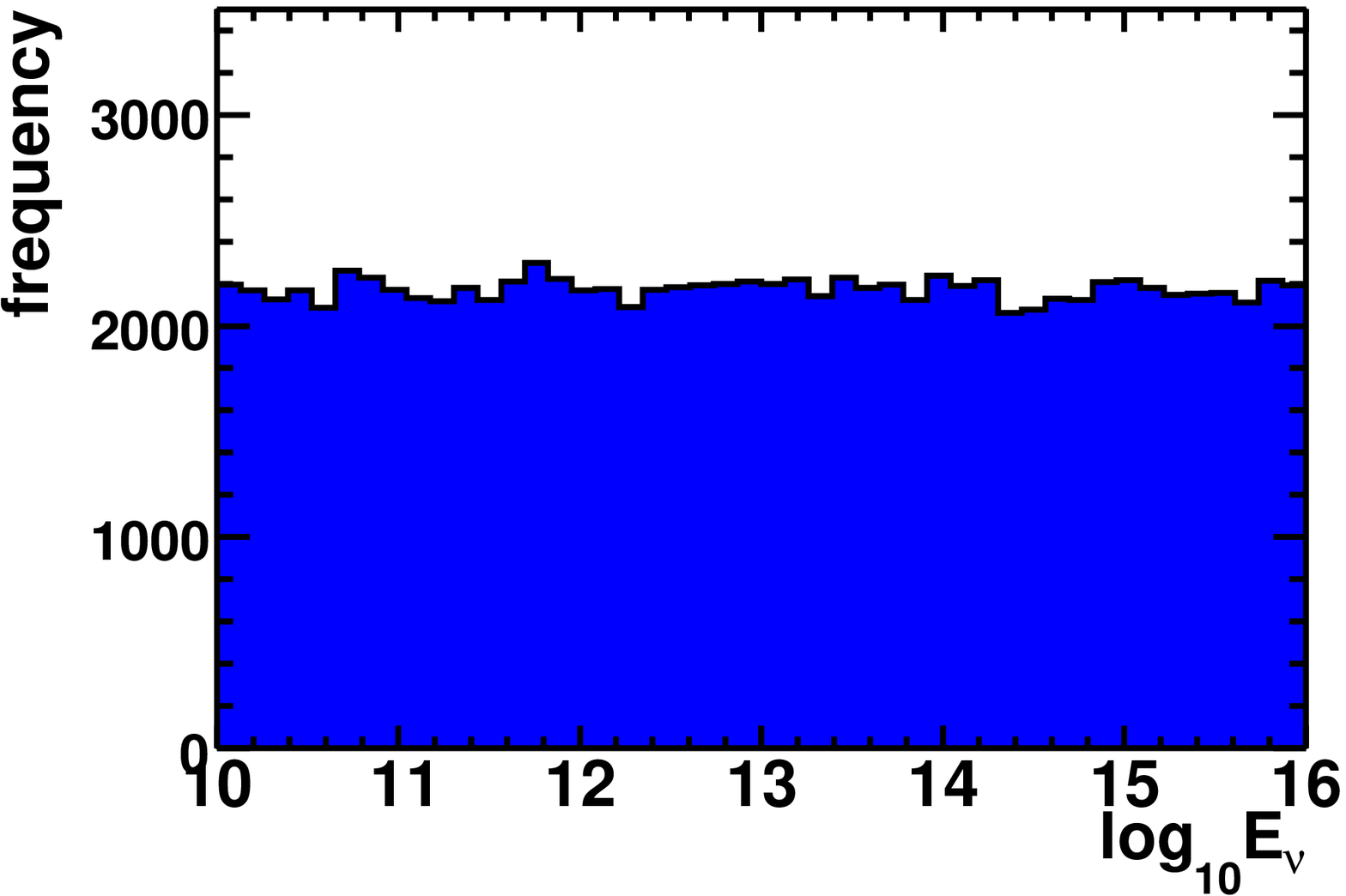}
\end{minipage}
\hfill
\begin{minipage}[!h]{0.4\textwidth}
\includegraphics[width=\textwidth]{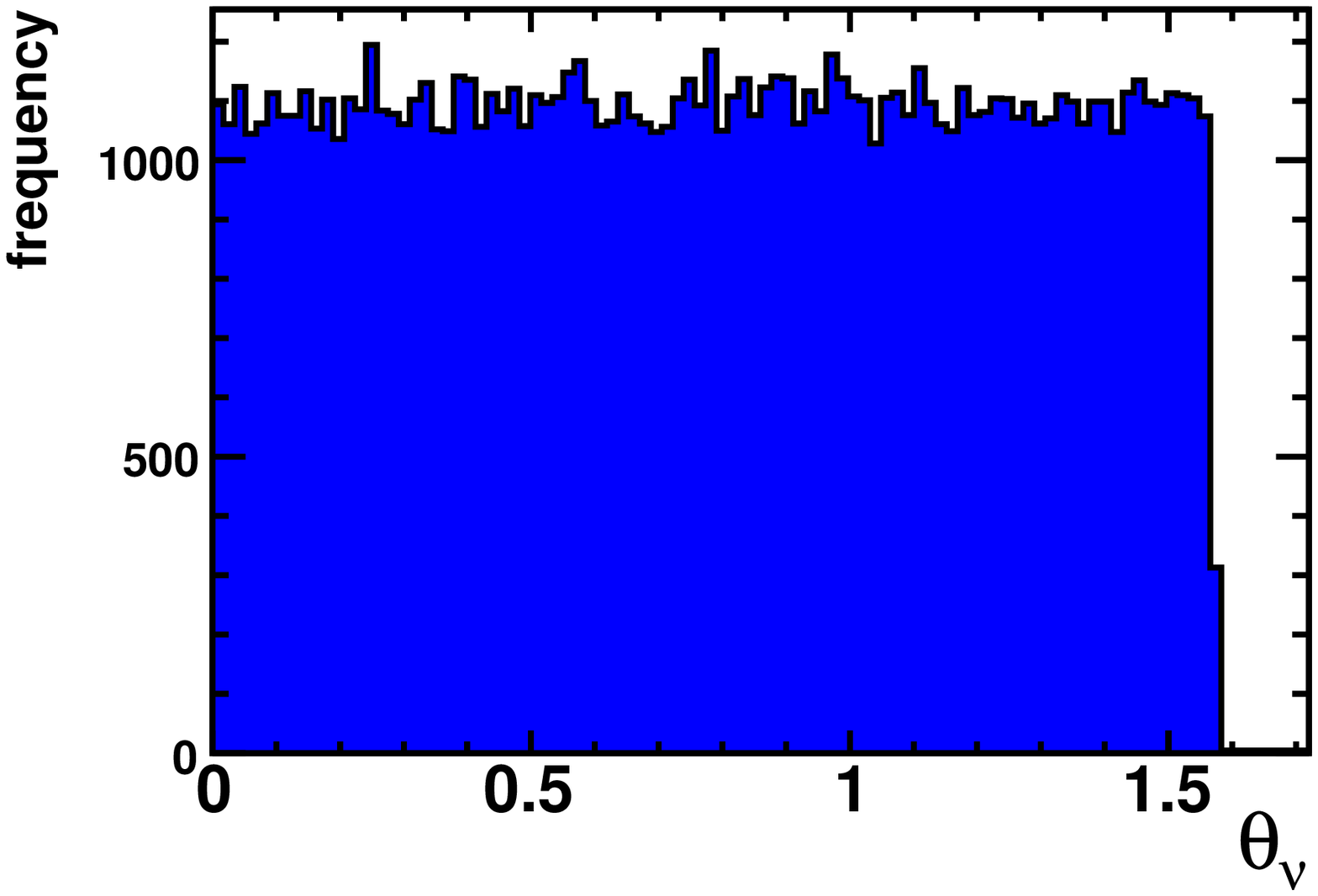}
\end{minipage}
\hfill
\begin{minipage}[!h]{0.4\textwidth}
\includegraphics[width=\textwidth]{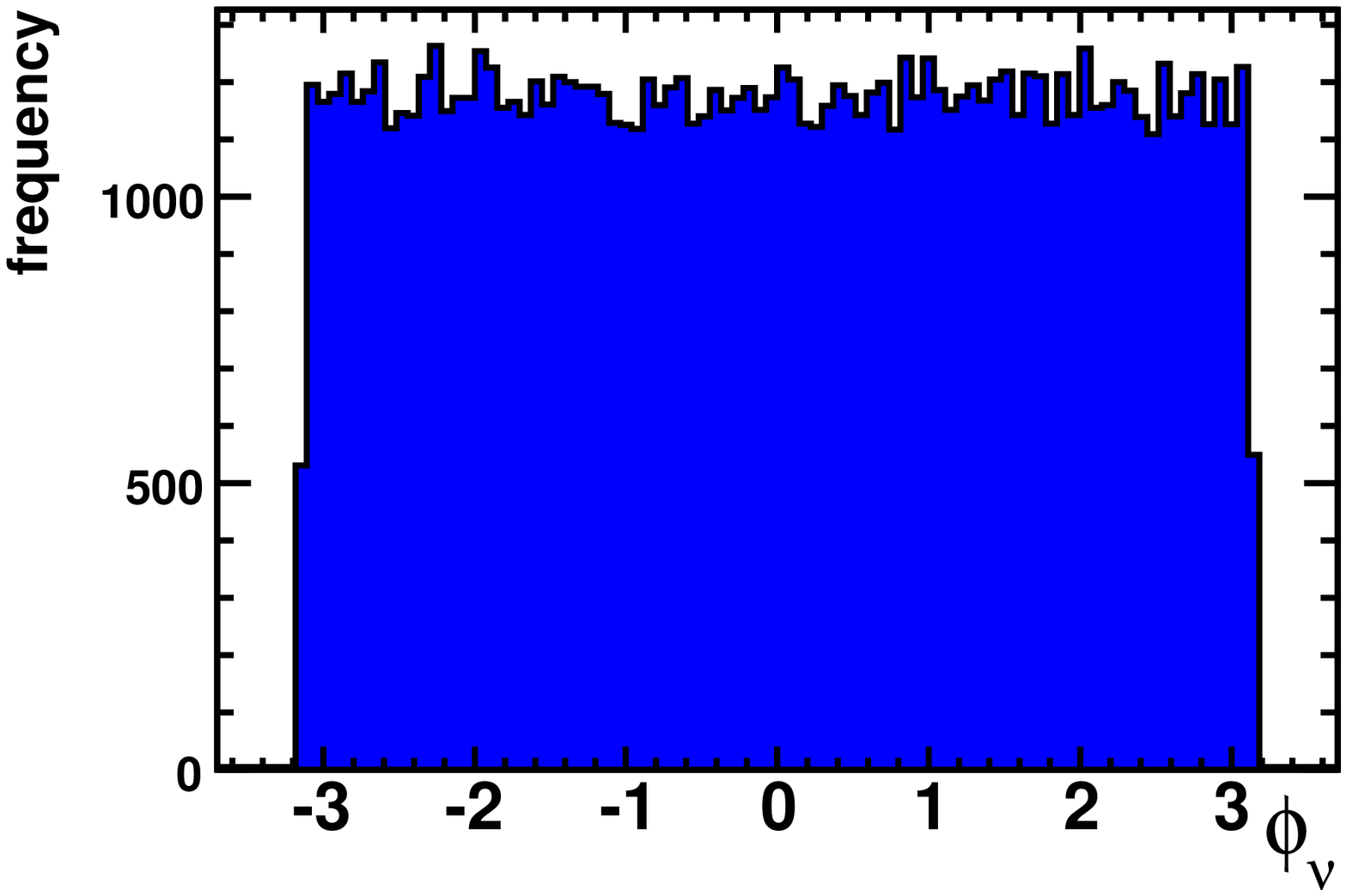}
\end{minipage}
\hfill
\begin{minipage}[!h]{0.4\textwidth}
\includegraphics[width=\textwidth]{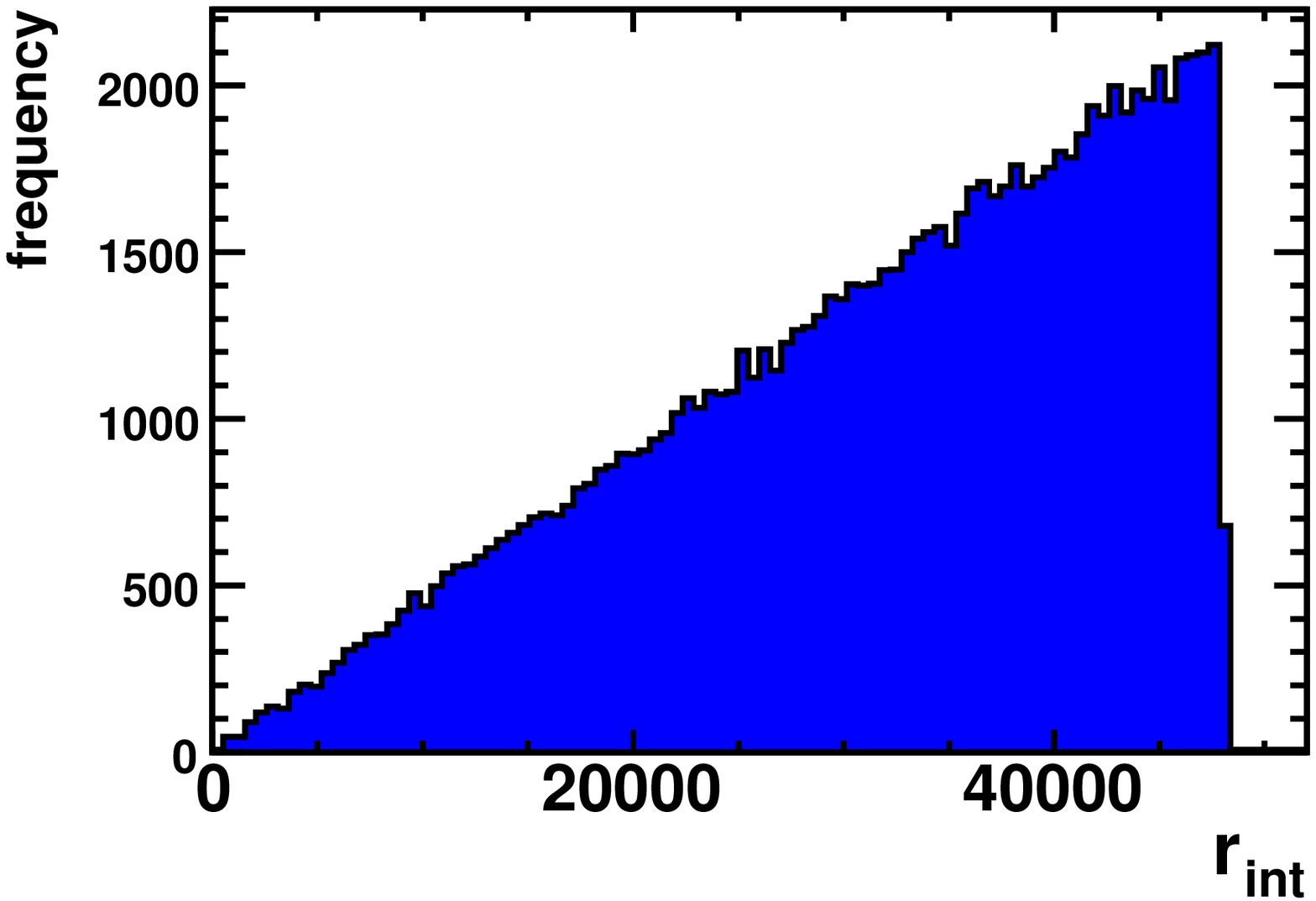}
\end{minipage}
\hfill
\begin{minipage}[!h]{0.4\textwidth}
\includegraphics[width=\textwidth]{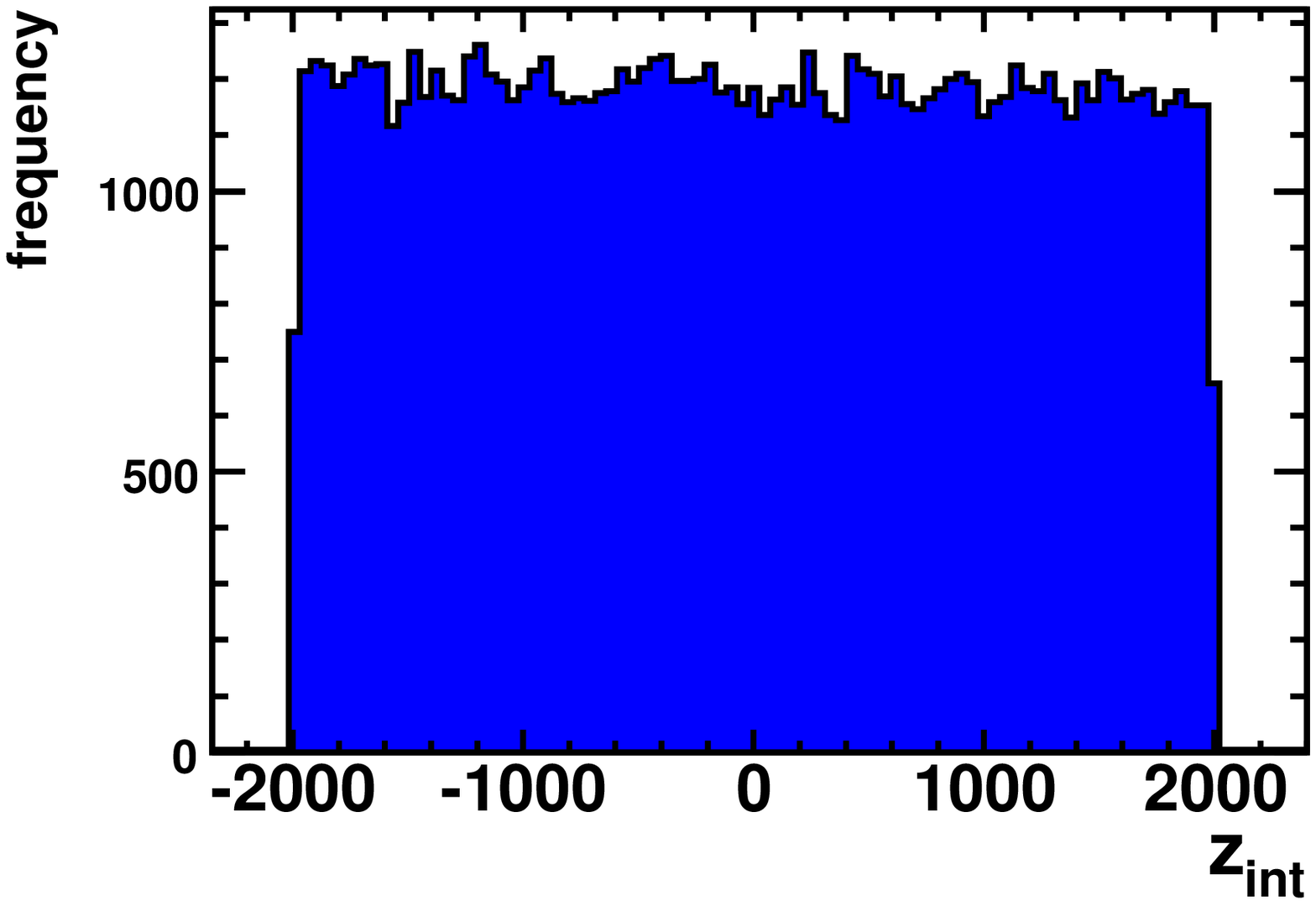}
\end{minipage}
\hfill
\begin{minipage}[!h]{0.4\textwidth}
\includegraphics[width=\textwidth]{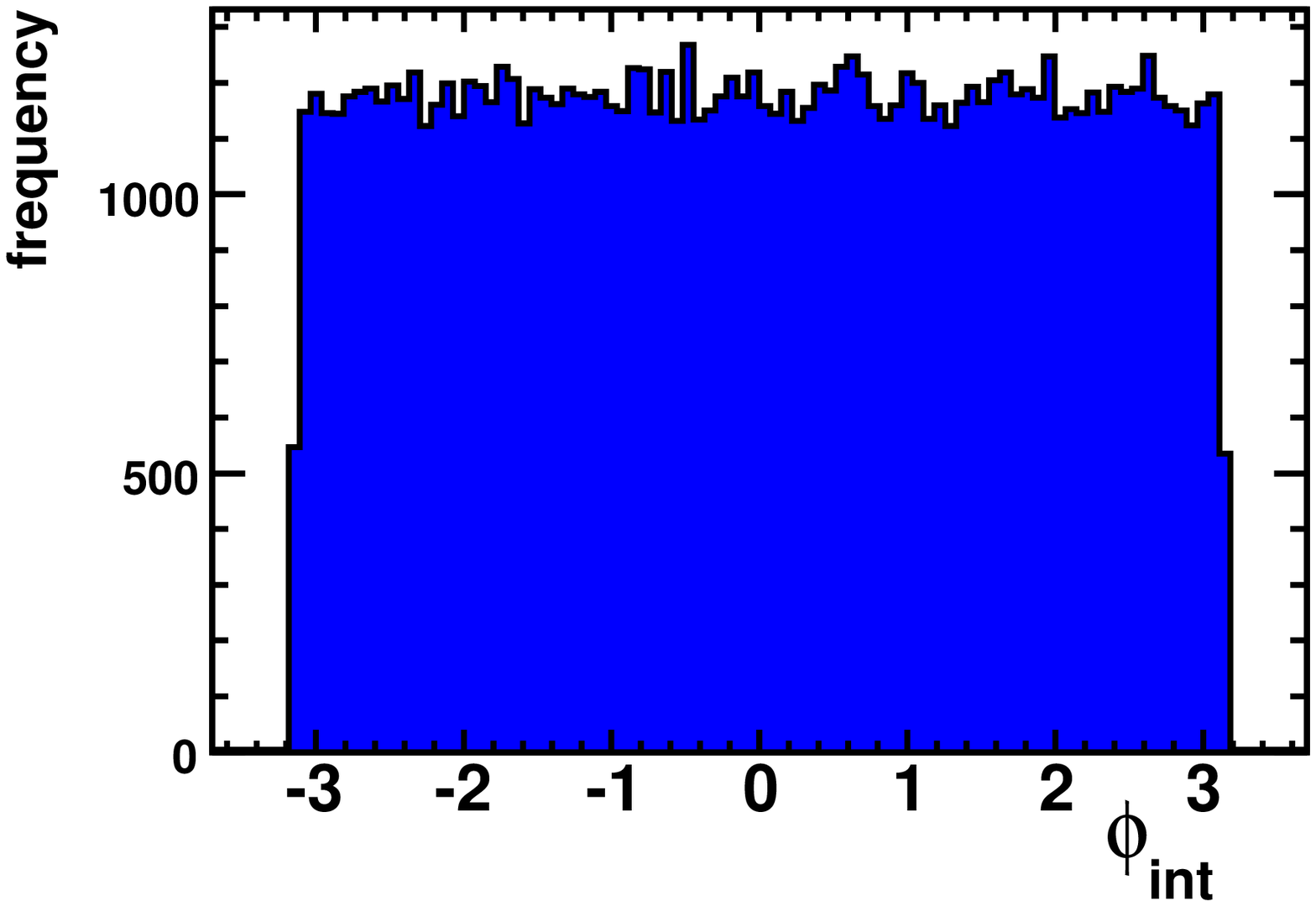}
\end{minipage}
\hfill
\end{figure}
%
%
\chapter{Recurring Event Types at Rona}
\label{appendix:recurring}
The coincident signals for each event introduced in Section
\ref{subsec:recurring} are as follows:
\begin{figure}[htbp]
  \centering
  \includegraphics[width=\textwidth]{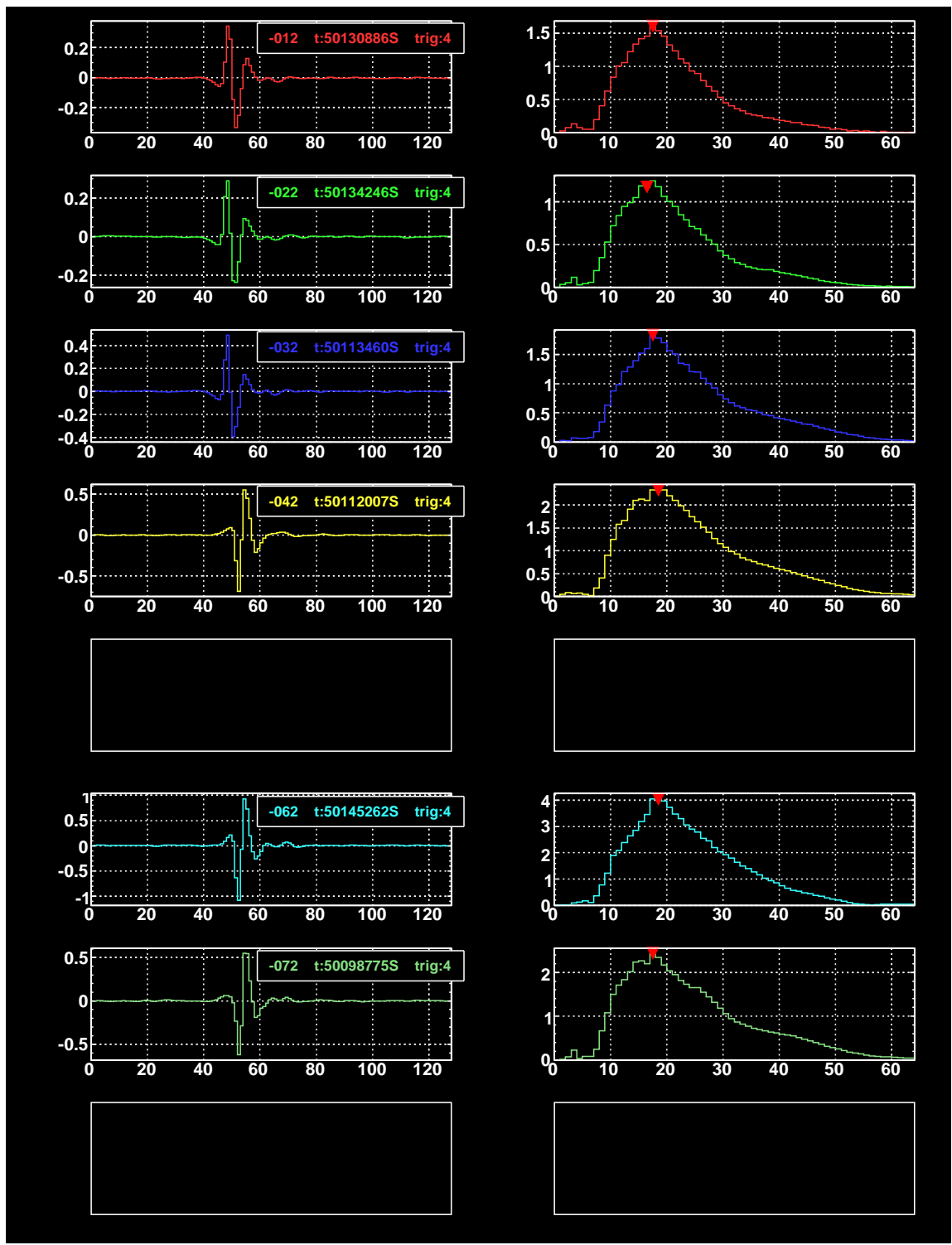}
  \caption[A short-lived impulsive event]
	  {\small A short-lived impulsive event.}
  \label{fig:shortimpulsive}
\end{figure}
\begin{figure}[htbp]
  \centering
  \includegraphics[width=\textwidth]{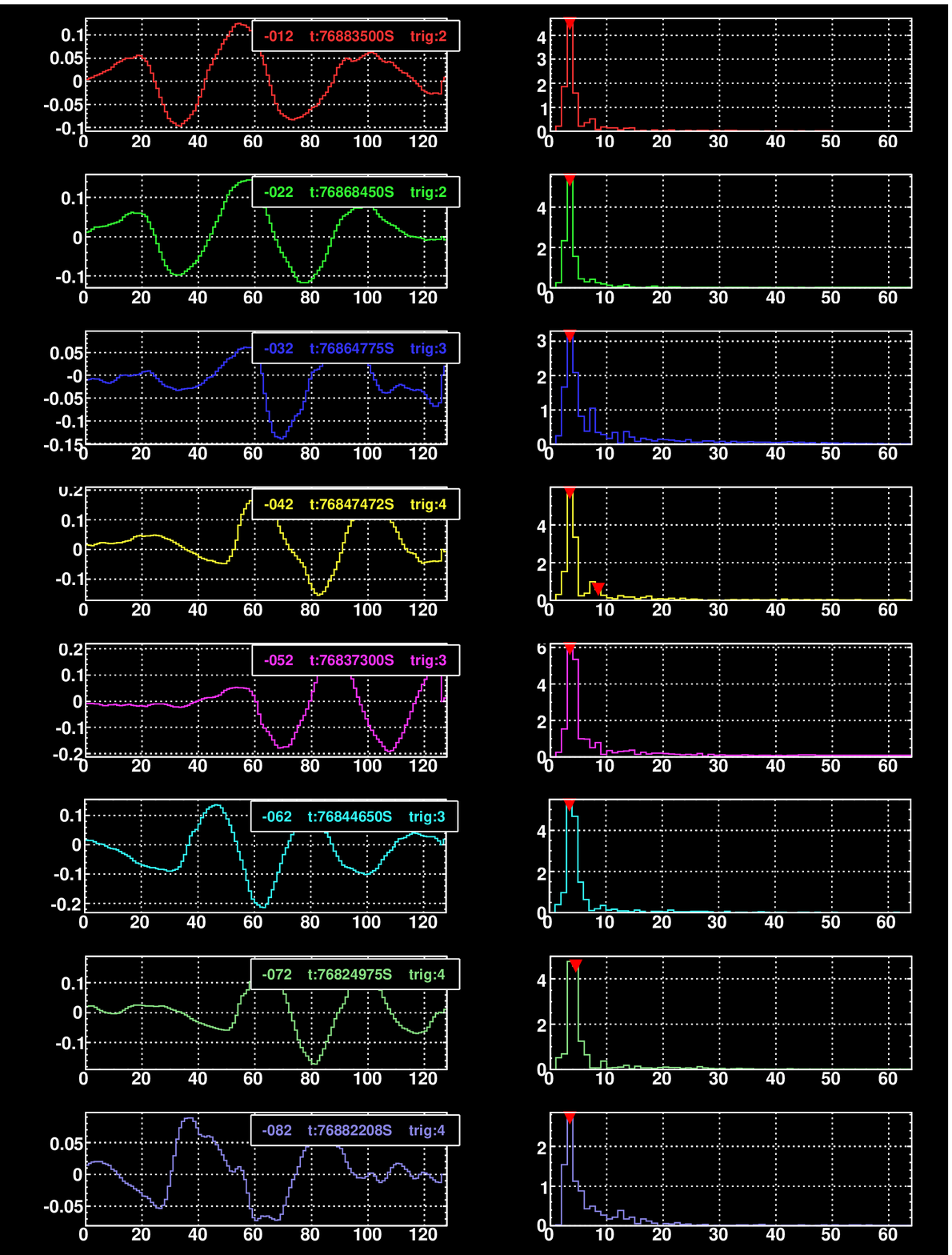}
  \caption[A low frequency oscillating event]
	  {\small A low frequency oscillating event.}
  \label{fig:lowfreqoscillator}
\end{figure}
\begin{figure}[htbp]
  \centering
  \includegraphics[width=\textwidth]{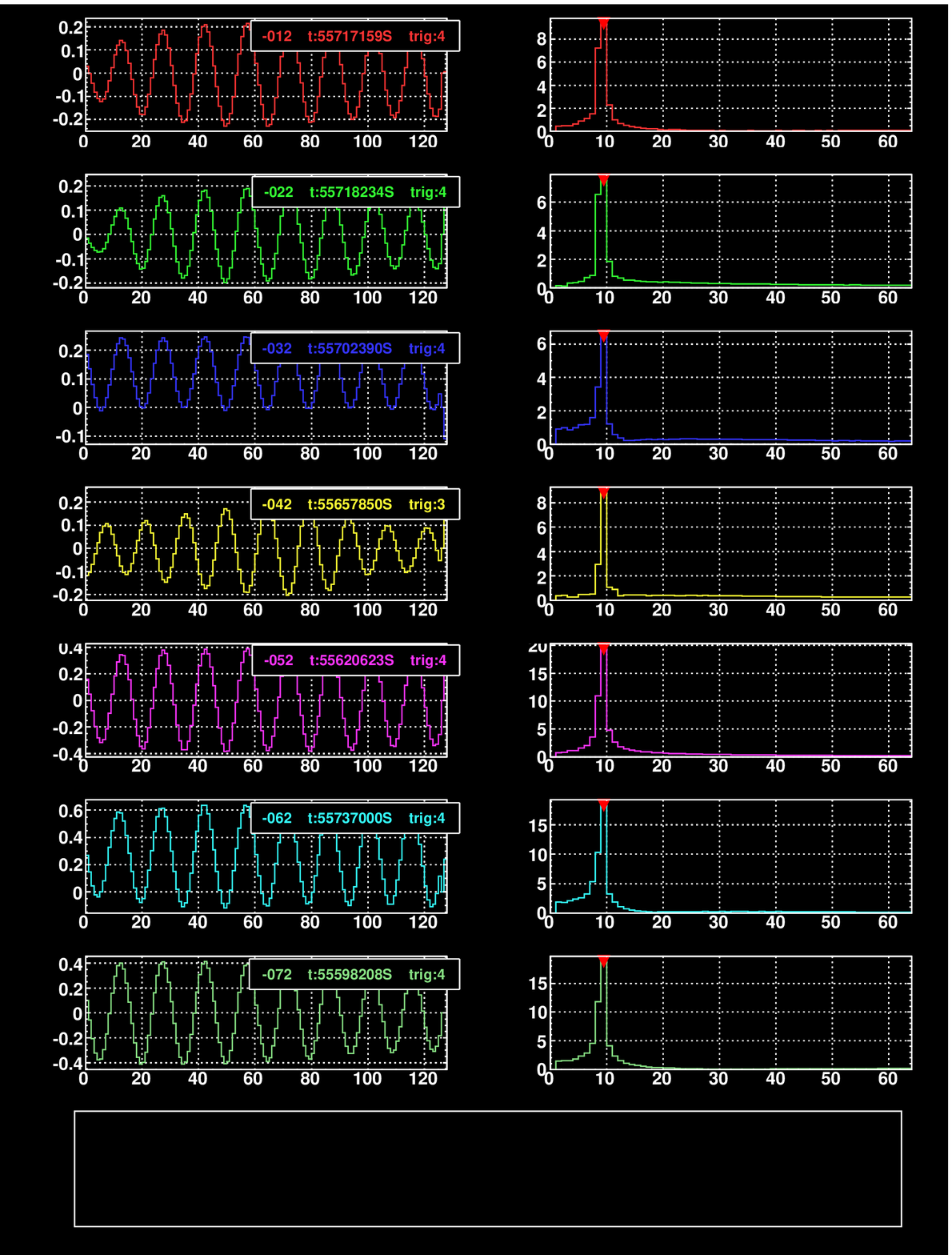}
  \caption[A $10$\,kHz sinusoidally oscillating event]
	  {\small A $10$\,kHz sinusoidally oscillating event.}
  \label{fig:10khzsinusoidal}
\end{figure}
\begin{figure}[htbp]
  \centering
  \includegraphics[width=\textwidth]{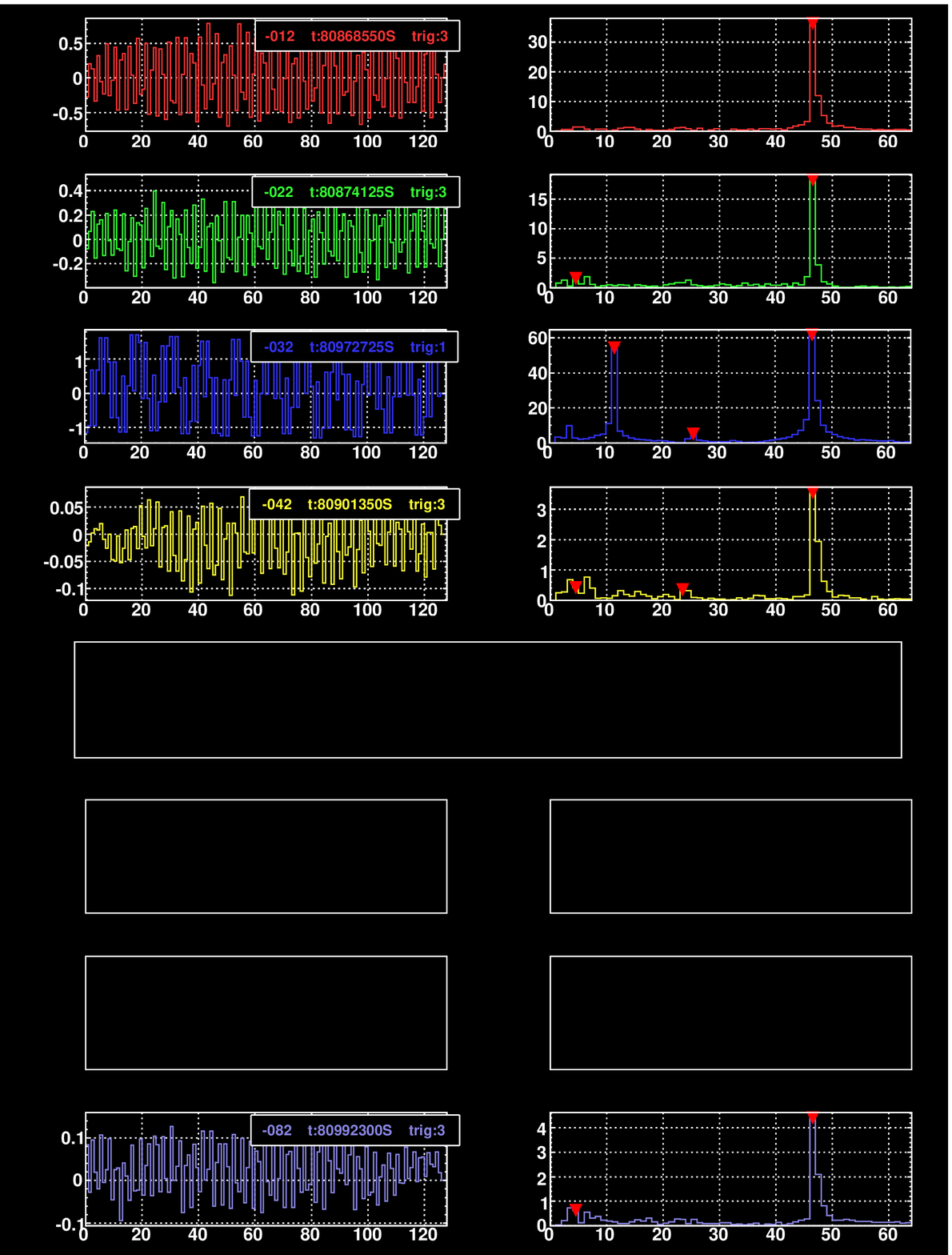}
  \caption[A high frequency event]
	  {\small A high frequency event.}
  \label{fig:hifreq}
\end{figure}
\begin{figure}[htbp]
  \centering
  \includegraphics[width=\textwidth]{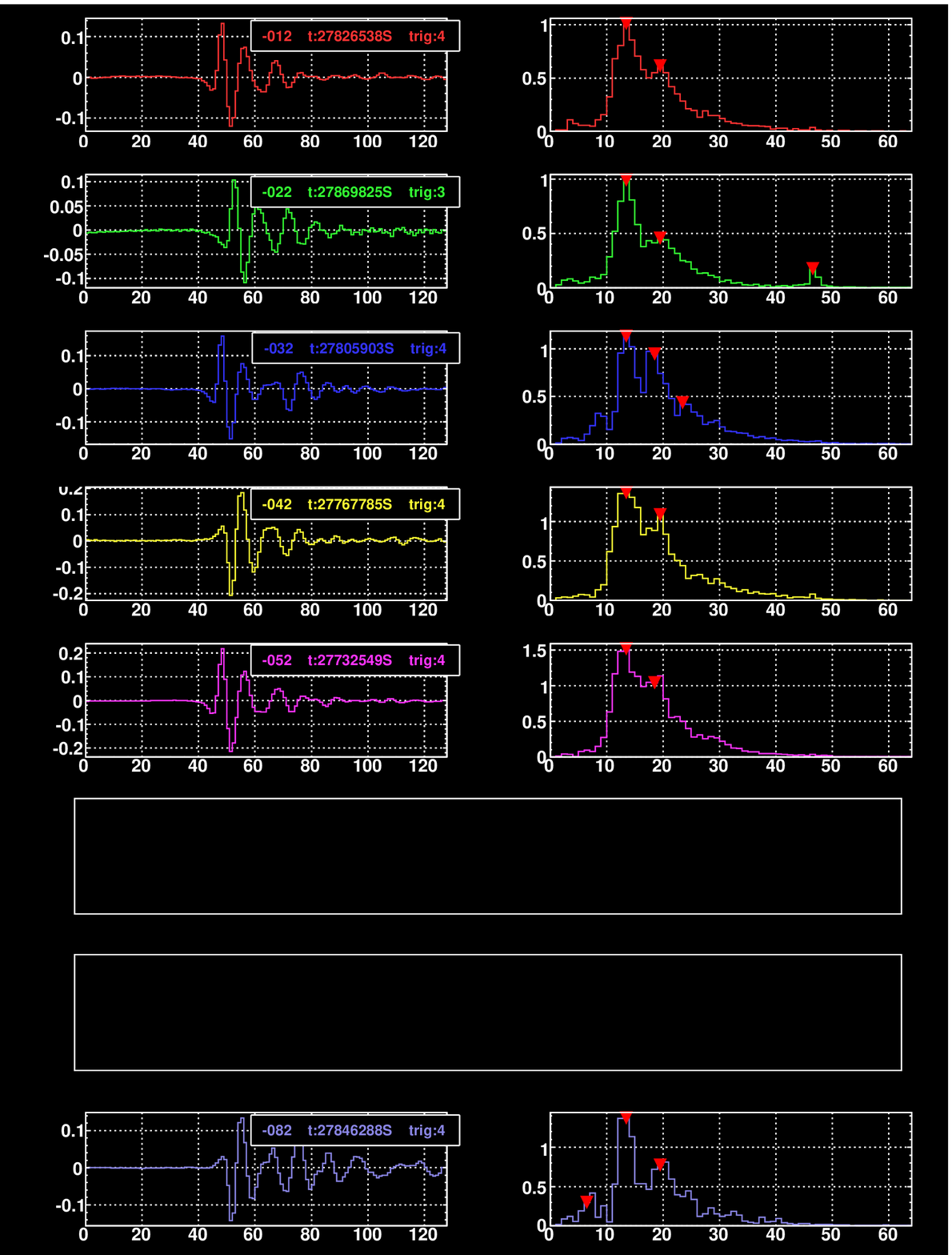}
  \caption[A `ringing' event]
	  {\small A `ringing' event.}
  \label{fig:ringer}
\end{figure}
\begin{figure}[htbp]
  \centering
  \includegraphics[width=\textwidth]{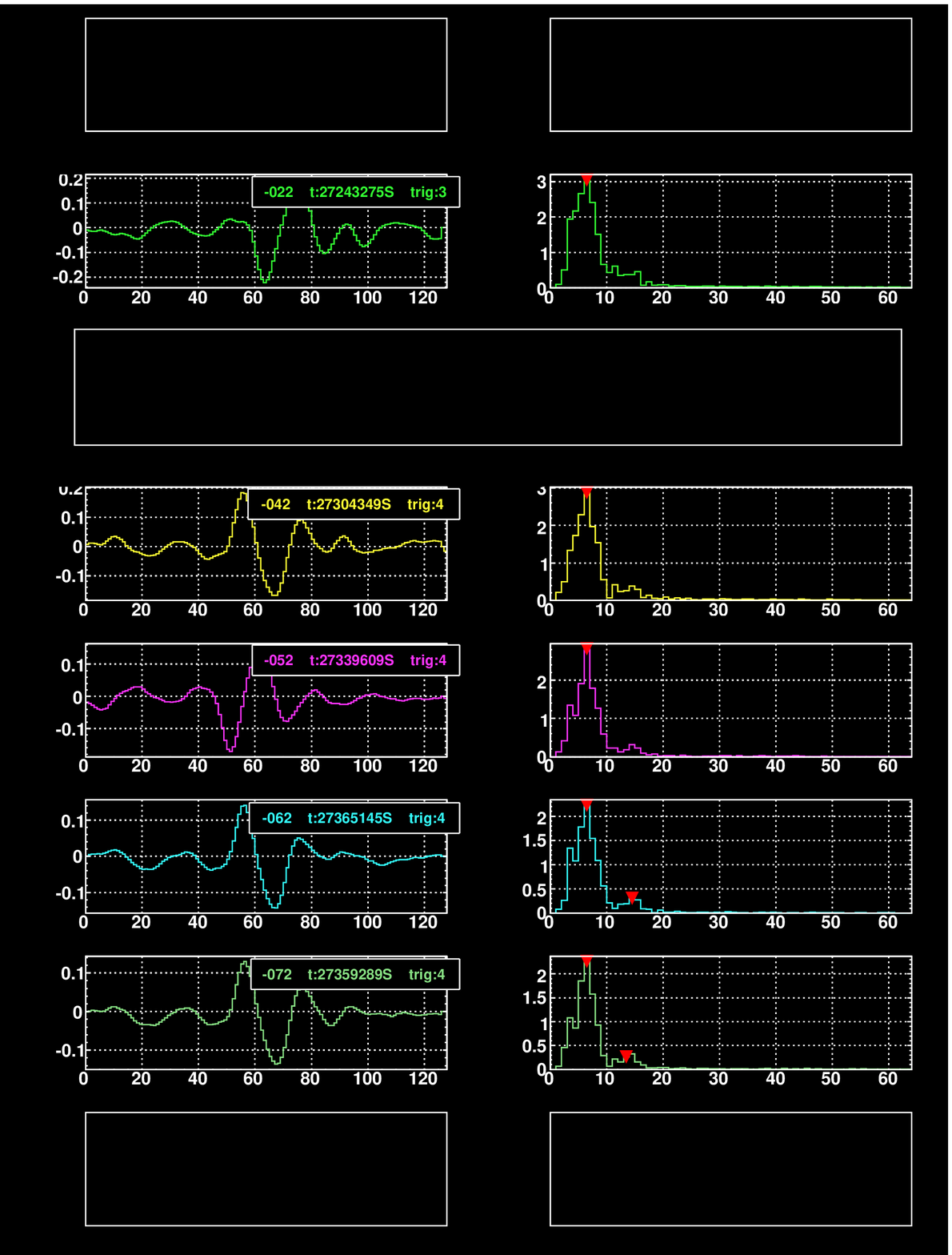}
  \caption[A `bipolar' event]
	  {\small A `bipolar' event.}
  \label{fig:idealbipolar}
\end{figure}
%

  \cleardoublepage


  \bibliography{refs}
  \bibliographystyle{h-physrev3}

\end{document}